\documentclass[twoside]{article}
\usepackage{amssymb}
\usepackage{amsmath}
\usepackage[mathscr]{eucal}
\usepackage{amsfonts}
\usepackage{latexsym}
\usepackage{amsxtra}
\usepackage{amsbsy}
\usepackage{amsthm}
\usepackage{amscd}
\usepackage{amsopn}
\usepackage{amstext}
\newcounter{z0}
\newcounter{z1}
\newcounter{z2}
\newcounter{z3}
\newtheorem{ay}{Lemma}[section]

\newtheorem{by}{Proposition}[section]

\newtheorem{cy}{Definition}[section]

\newtheorem{dy}{Theorem}[section]

\newtheorem{aaaa}{Definition}[subsection]
\newtheorem{bbbb}{Proposition}[subsection]
\newtheorem{cccc}{Lemma}[subsection]
\newtheorem{dddd}{Theorem}[subsection]
\newtheorem{ffff}{Corollary}[subsection]
\newtheorem{aaaaa}{Definition}[section]
\newtheorem{bbbbb}{Proposition}[section]
\newtheorem{ccccc}{Lemma}[section]

\theoremstyle{definition}

\theoremstyle{definition}
\newtheorem{eeee}{Remark}[subsection]

\theoremstyle{definition}
\newtheorem{eeeee}{Remark}[section]

\theoremstyle{definition}
\newtheorem{ey}{Remark}[section]

\newcommand{\me}{\mathrm{e}}
\newcommand{\mi}{\mathrm{i}}
\newcommand{\md}{\mathrm{d}}
\renewcommand{\Im}{\mathrm{Im}}
\renewcommand{\Re}{\mathrm{Re}}
\newcommand{\id}{\boldsymbol{\mi \md}}
\begin{document}
\baselineskip=12pt
\frenchspacing
\title{Long-Time Asymptotics of Solutions to the Cauchy Problem for the 
Defocusing Non-Linear Schr\"{o}dinger Equation with Finite-Density Initial 
Data.~II.~Dark Solitons on Continua}
\author{A.~H.~Vartanian\thanks{\texttt{E-mail: vartaniana@winthrop.edu}} \\
Department of Mathematics \\
Winthrop University \\
Rock Hill, South Carolina 29733 \\
U.~S.~A}
\date{21 October 2002}
\maketitle
\begin{abstract}
\noindent
For Lax-pair isospectral deformations whose associated spectrum, for given 
initial data, consists of the disjoint union of a finitely denumerable 
discrete spectrum (solitons) and a continuous spectrum (continuum), the 
matrix Riemann-Hilbert problem approach is used to derive the leading-order 
asymptotics as $\vert t \vert \! \to \! \infty$ $(x/t \! \sim \! \mathcal{
O}(1))$ of solutions $(u \! = \! u(x,t))$ to the Cauchy problem for the 
defocusing non-linear Schr\"{o}dinger equation (D${}_{f}$NLSE), $\mi 
\partial_{t}u \! + \! \partial_{x}^{2}u\! - \! 2(\vert u \vert^{2} \! - \! 
1)u \! = \! 0$, with finite-density initial data $u(x,0) \! =_{x \to \pm 
\infty} \! \exp (\tfrac{\mi (1 \mp 1) \theta}{2})(1 \! + \! o(1))$, $\theta 
\! \in \! [0,2\pi)$. The D${}_{f}$NLSE dark soliton position shifts in the 
presence of the continuum are also obtained.

\vspace{1.35cm}

37K15: (Secondary) 30E20, 30E25, 81U40

\vspace{0.50cm}
{\bf Abbreviated Title.} Asymptotics of the D${}_{f}$NLSE Dark Solitons on 
Continua

\vspace{0.50cm}

a\-n\-n-H\-i\-l\-b\-e\-r\-t problems, singular integral equations
\end{abstract}
\clearpage
\section{Introduction}
In direct detection systems making use of polarisation-preserving single-mode 
(PPSM) optical fibres, return-to-zero bright soliton (strictly speaking, 
soliton-like) pulses, which propagate in the anomalous group velocity 
dispersion (GVD) regime (wavelengths $> \! 1.3 \mu \mathrm{m}$ in standard 
telecommunications fibres), have been shown to be effective toward the 
partial resolution of the deleterious problem of performance degradation 
caused by, for example, dispersive pulse spreading \cite{a1}. For coherent 
communications systems, non-return-to-zero dark soliton pulses, which 
propagate in the normal GVD regime (wavelengths $< \! 1.3 \mu \mathrm{m}$) 
and consist of a rapid dip in the intensity of a broad pulse of a continuous 
wave background, offer an analogous benefit \cite{a2,a3,a4}.

A model for dark soliton pulse propagation in PPSM optical fibres in the 
picosecond time scale, which describes the slowly varying amplitude of the 
complex field envelope, $u \! = \! u(x,t)$, in normalised and dimensionless 
form, is the Cauchy problem for the defocusing non-linear Schr\"{o}dinger 
equation (D${}_{f}$NLSE) with finite-density, or non-vanishing, initial data 
\cite{a1,a2,a3,a4},
\begin{equation}
\begin{split}
\mi \partial_{t}u \! + \! \partial_{x}^{2}u \! - \! 2(\vert u \vert^{2} \! - 
\! 1)u \! = \! 0, \qquad (x,t) \! \in \! \mathbb{R} \times \mathbb{R}, \\
u(x,0) \! := \! u_{o}(x) \! \underset{x \to \pm \infty}{=} \! \exp (\tfrac{
\mi (1 \mp 1) \theta}{2})(1 \! + \! o(1)),
\end{split}
\end{equation}
where $u_{o}(x) \! \in \! \mathbf{C}^{\infty}(\mathbb{R})$, $\theta \! \in \! 
[0,2 \pi)$ (see Eq.~(3)), and $o(1)$ is to be understood in the sense that, 
$\forall \, (k,l) \! \in \! \mathbb{Z}_{\geqslant 0} \! \times \! \mathbb{
Z}_{\geqslant 0}$, $\vert x \vert^{k}(\tfrac{\md}{\md x})^{l}(u_{o}(x) \! - 
\! \exp (\tfrac{\mi (1 \mp 1) \theta}{2})) \! =_{x \to \pm \infty} \! 0$. It 
is shown in \cite{a5} that, for initial data satisfying $\vert x \vert^{k}
(\tfrac{\md}{\md x})^{l}(u_{o}(x) \! - \! \exp (\tfrac{\mi (1 \mp 1) \theta}
{2})) \! =_{x \to \pm \infty} \! 0$, $(k,l) \! \in \! \mathbb{Z}_{\geqslant 
0} \! \times \! \mathbb{Z}_{\geqslant 0}$, the closure of the set of soliton, 
or reflectionless, potentials of the D${}_{f}$NLSE in the topology of uniform 
convergence of functions on compact sets of $\mathbb{R}$ remains an invariant 
set of the model $\forall \, \, t \! \in \! \mathbb{R}$ and not just for $t 
\! = \! 0$ (see, also, \cite{a6}).

When (temporal) dark solitons are launched sufficiently close together in 
optical fibres, they interact not only through soliton-soliton interactions, 
but also through soliton-radiation-tail interactions. Such interactions 
manifest as a jitter in the arrival times of dark solitons, potentially 
resulting in their shift outside of some predetermined timing window and 
giving rise to errors in the detected information \cite{a4}. Physically, the 
optical pulse adjusts its width as it propagates along the optical fibre to 
evolve into a (multi-) dark soliton pulse/mode, and a part, however small, 
of the pulse energy is shed in the form of an asymptotically decaying 
dispersive wavetrain, manifesting as a low-level broadband background 
radiation (a continuum of linear-like radiative waves/modes). Modulo an 
$\mathcal{O}(1)$ position shift due to cummulative interactions with 
other dark solitons and the (dispersive) continuum, the dark soliton 
pulse/mode maintains its robust/stable properties. {}From the physical and 
theoretical point of view, therefore, it is important to understand how the 
dark solitons and continuum interact, and to be able to derive an explicit 
functional form for this process, namely, to study the asymptotics as $\vert 
t \vert \! \to \! \infty$ $(x/t \! \sim \! \mathcal{O}(1))$ of solutions to 
the Cauchy problem for the D${}_{f}$NLSE with finite-density initial data 
having a (not the only one possible) decomposition of the form $u_{o}(x) 
\! := \! u_{\mathrm{sol}}(x) \! + \! u_{\mathrm{rad}}(x)$, where $u_{o}(x)$ 
satisfies the conditions stated heretofore, $u_{\mathrm{sol}}(x)$ 
``generates'' the multi- or $N$-dark soliton component of the solution, and 
$u_{\mathrm{rad}}(x)$ is the ``small'' non-dark-soliton part giving rise 
to the dispersive component of the solution. In this paper, the leading- 
$(\mathcal{O}(1))$ and next-to-leading-order $(\mathcal{O}(\vert t \vert^{
-1/2}))$ terms of the asymptotic expansion as $\vert t \vert \! \to \! 
\infty$ $(x/t \! \sim \! \mathcal{O}(1))$ of the solution to the Cauchy 
problem for the D${}_{f}$NLSE with finite-density initial data are derived: 
they represent, respectively, the $N$-dark soliton component, and the 
dispersive continuum and non-trivial interaction/overlap of the $N$-dark 
solitons with the continuum.

Within the framework of the inverse scattering method (ISM) \cite{a7,a8,a9} 
(see, also, \cite{a10}), it is well known that the D${}_{f}$NLSE is a 
completely integrable non-linear evolution equation (NLEE) having a 
representation as an infinite-dimensional Hamiltonian system \cite{a11,a12}. 
Even though the analysis of completely integrable NLEEs with rapidly 
decaying, e.g., Schwartz class, initial data on $\mathbb{R}$ have received 
the vast majority of the attention within the ISM framework, there have been 
a handful of works devoted exclusively to the direct and inverse scattering 
analysis of completely integrable NLEEs belonging to the ZS-AKNS class with 
non-vanishing (as $\vert x \vert \! \to \! \infty)$ values of the initial 
data \cite{a13,a14,a15} (see, also, \cite{a16,a17}). Other, very interesting 
classes of finite-density-type initial data for completely integrable NLEEs 
have also been considered 
\cite{a18,a19,a20,a21,a22,a23,a24,a25,a26,a27,a28,a29,a30,a31,a32}.

Within the ISM framework, the asymptotic analysis of the solution to the 
Cauchy problem for the D${}_{f}$NLSE with finite-density initial data is 
divided into two steps: (1) the analysis of the \emph{solitonless} (pure 
radiative, or continuous) component of the solution; and (2) the inclusion 
of the $N$-dark soliton component via the application of a ``dressing'' 
procedure to the solitonless background \cite{a33,a34,a35,a36,a37}. The 
complete details of the asymptotic analysis that constitutes stage~(1) of 
the two-step asymptotic paradigm above, which is quite technical and whose 
results are essential in order to obtain those of the present paper, can be 
found in \cite{a38}: this paper addresses stage~(2) of the above programme 
via the matrix Riemann-Hilbert problem (RHP) approach \cite{a7,a12,a39,a40,
a41,a42,a43,a44,a45,a46,a47}. It is important to note that, to the best of 
the author's knowledge as at the time of the presents, the first to obtain 
asymptotics of solutions to the Cauchy problem for the D${}_{f}$NLSE with 
finite-density initial data in the solitonless sector were Its and Ustinov 
\cite{a48,a49}.

This paper is organized as follows. In Section~2, the necessary facts {}from 
the direct and inverse scattering analysis for the D${}_{f}$NLSE with 
finite-density initial data are given, the (matrix) RHP analysed 
asymptotically as $\vert t \vert \! \to \! \infty$ $(x/t \! \sim \! 
\mathcal{O}(1))$ is stated, and the results of this paper are summarised in 
Theorems~2.2.1--2.2.4 (and Corollaries~2.2.1 and~2.2.2). In Section~3, an 
augmented RHP, which is equivalent to the original one stated in Section~2, 
is formulated, and it is shown that, as $t \! \to \! +\infty$, modulo 
exponentially small terms, the solution of the augmented RHP converges to 
the solution of an explicitly solvable, model RHP. In Section~4, the model 
RHP is solved asymptotically as $t \! \to \! +\infty$, {}from which the 
asymptotics of $u(x,t)$ and $\int_{\pm \infty}^{x}(\vert u(x^{\prime},t) 
\vert^{2} \! - \! 1) \, \md x^{\prime}$ are derived, and, in Appendix~A, 
the---analogous---asymptotic analysis is succinctly reworked for the case 
when $t \! \to \! -\infty$. In Appendices~B and~C, respectively, formulae 
which are necessary in order to obtain the remaining asymptotic results of 
this paper are presented, and a panoramic view of the matrix RH theory in 
the $L^{2}$-Sobolev space is given \cite{a44,a45,a46,a50}.
\section{The Riemann-Hilbert Problem and Summary of Results}
In this section, a synopsis of the direct/inverse spectral analysis for 
Eq.~(1) is given, the matrix RHP studied asymptotically as $\vert t \vert 
\! \to \! \infty$ $(x/t \! \sim \! \mathcal{O}(1))$ is stated, and the 
results of this paper are summarised in Theorems~2.2.1--2.2.4. Before doing 
so, however, it will be convenient to introduce the notation used throughout 
this work.
\begin{center}
\underline{\textsc{Notational Conventions}}
\end{center}
\begin{enumerate}
\item[(1)] $\mathrm{I} \! = \! 
\left(
\begin{smallmatrix}
1 & 0 \\
0 & 1
\end{smallmatrix}
\right)$ is the $2 \times 2$ identity matrix, $\mathbf{0} \! = \! 
\left(
\begin{smallmatrix}
0 & 0 \\
0 & 0
\end{smallmatrix}
\right)$, $\sigma_{1} \! = \! 
\left(
\begin{smallmatrix}
0 & 1 \\
1 & 0
\end{smallmatrix}
\right)$, $\sigma_{2} \! = \! 
\left(
\begin{smallmatrix}
0 & -\mi \\
\mi & 0
\end{smallmatrix}
\right)$, and $\sigma_{3} \! = \! 
\left(
\begin{smallmatrix}
1 & 0 \\
0 & -1
\end{smallmatrix}
\right)$ are the Pauli matrices, $\sigma_{+} \! = \! 
\left(
\begin{smallmatrix}
0 & 1 \\
0 & 0
\end{smallmatrix}
\right)$ and $\sigma_{-} \! = \! 
\left(
\begin{smallmatrix}
0 & 0 \\
1 & 0
\end{smallmatrix}
\right)$ are, respectively, the raising and lowering matrices, $\mathrm{sgn}
(x) \! := \! 0$ if $x \! = \! 0$ and $x \vert x \vert^{-1}$ if $x \! \not= 
\! 0$, and $\mathbb{R}_{\pm} \! := \! \{\mathstrut x; \, \pm x \! > \! 0\}$;
\item[(2)] for a scalar $\omega$ and a $2 \! \times \! 2$ matrix $\Upsilon$, 
$\omega^{\mathrm{ad}(\sigma_{3})} \Upsilon \! := \! \omega^{\sigma_{3}} 
\Upsilon \omega^{-\sigma_{3}}$;
\item[(3)] for each segment of an oriented contour $\mathcal{D}$, according 
to the given orientation, the ``+'' side is to the left and the ``-'' side 
is to the right as one traverses the contour in the direction of orientation, 
that is, for a matrix $\mathcal{A}_{ij}(\cdot)$, $i,j \! \in \! \{1,2\}$, 
$(\mathcal{A}_{ij}(\cdot))_{\pm}$ denote the non-tangential limits 
$(\mathcal{A}_{ij}(z))_{\pm} \! := \! \lim_{\genfrac{}{}{0pt}{2}{z^{\prime} 
\, \to \, z}{z^{\prime} \, \in \, \pm \, \mathrm{side} \, \mathrm{of} \, 
\mathcal{D}}} \mathcal{A}_{ij}(z^{\prime})$;
\item[(4)] for a matrix $\mathcal{A}_{ij}(\cdot)$, $i,j \! \in \! \{1,2\}$, 
to have boundary values in the $\mathcal{L}^{2}$ sense on an oriented contour 
$\mathcal{D}$, it is meant that $\lim_{\genfrac{}{}{0pt}{2}{z^{\prime} \, 
\to \, z}{z^{\prime} \, \in \, \pm \, \mathrm{side} \, \mathrm{of} \, 
\mathcal{D}}} \int_{\mathcal{D}} \vert \mathcal{A}(z^{\prime}) \! - \! 
(\mathcal{A}(z))_{\pm} \vert^{2} \, \vert \md z \vert \! = \! 0$, where 
$\vert \mathcal{A}(\cdot) \vert$ denotes the Hilbert-Schmidt norm, $\vert 
\mathcal{A}(\cdot) \vert \! := \! (\sum_{i,j=1}^{2} \overline{\mathcal{A}_{
ij}(\cdot)} \, \mathcal{A}_{ij}(\cdot))^{1/2}$, with $\overline{(\bullet)}$ 
denoting complex conjugation of $(\bullet)$;
\item[(5)] for $1 \! \leqslant \! p \! < \! \infty$ and $\mathcal{D}$ some 
point set,
\begin{equation*}
\mathcal{L}^{p}_{\mathrm{M}_{2}(\mathbb{C})}(\mathcal{D}) \! := \! \{
\mathstrut f \colon \mathcal{D} \! \to \! \mathrm{M}_{2}(\mathbb{C}); \, 
\vert \vert f(\cdot) \vert \vert_{\mathcal{L}^{p}_{\mathrm{M}_{2}(\mathbb{
C})}(\mathcal{D})} \! := \! (\smallint\nolimits_{\mathcal{D}} \vert f(z) 
\vert^{p} \, \vert \md z \vert)^{1/p} \! < \! \infty\},
\end{equation*}
and, for $p \! = \! \infty$,
\begin{equation*}
\mathcal{L}^{\infty}_{\mathrm{M}_{2}(\mathbb{C})}(\mathcal{D}) \! := \! \{
\mathstrut g \colon \mathcal{D} \! \to \! \mathrm{M}_{2}(\mathbb{C}); \, 
\vert \vert g(\cdot) \vert \vert_{\mathcal{L}^{\infty}_{\mathrm{M}_{2}
(\mathbb{C})}(\mathcal{D})} \! := \! \max_{i,j \in \{1,2\}} \sup_{z \in 
\mathcal{D}} \vert g_{ij}(z) \vert \! < \! \infty\};
\end{equation*}
\item[(6)] for $D$ an unbounded domain of $\mathbb{R}$, $\mathcal{S}_{\mathbb{
C}}(D)$ (respectively, $\mathcal{S}_{\mathrm{M}_{2}(\mathbb{C})}(D))$ denotes 
the Schwartz space on $D$, namely, the space of all infinitely continuously 
differentiable (smooth) $\mathbb{C}$-valued (respectively, $\mathrm{M}_{2}
(\mathbb{C})$-valued) functions which together with all their derivatives 
tend to zero faster than any positive power of $\vert \bullet \vert^{-1}$ as 
$\vert \bullet \vert \! \to \! \infty$, that is, $\mathcal{S}_{\mathbb{C}}
(D) \! := \! \mathbf{C}^{\infty}(D) \cap \{ \mathstrut f \colon D \! \to \! 
\mathbb{C}; \, \vert \vert f(\cdot) \vert \vert_{k,l} \! := \! \sup_{x \in 
\mathbb{R}} \vert x^{k}(\tfrac{\md}{\md x})^{l}f(x) \vert \! < \! \infty, \, 
(k,l) \! \in \! \mathbb{Z}_{\geqslant 0} \times \mathbb{Z}_{\geqslant 0}\}$ 
and $\mathcal{S}_{\mathrm{M}_{2}(\mathbb{C})}(D) \! := \! \{\mathstrut F 
\colon D \! \to \! \mathrm{M}_{2}(\mathbb{C}); \, F_{ij}(\cdot) \! \in \! 
\mathbf{C}^{\infty}(D), \, i,j \! \in \! \{1,2\}\} \cap \{\mathstrut G \colon 
D \! \to \! \mathrm{M}_{2}(\mathbb{C}); \, \vert \vert G_{ij}(\cdot) \vert 
\vert_{k,l} \! := \! \max_{i,j \in \{1,2\}} \sup_{x \in \mathbb{R}} \vert 
x^{k}(\tfrac{\md}{\md x})^{l}G_{ij}(x) \vert \! < \! \infty, \, (k,l) \! 
\in \! \mathbb{Z}_{\geqslant 0} \! \times \! \mathbb{Z}_{\geqslant 0}\}$, 
and $\mathbf{C}^{\infty}_{0}(\ast) \! := \! \cap_{k=0}^{\infty} \mathbf{
C}^{k}_{0}(\ast)$;
\item[(7)] for $D$ an unbounded domain of $\mathbb{R}$, $\mathcal{S}_{\mathbb{
C}}^{1}(D) \! := \! \mathcal{S}_{\mathbb{C}}(D) \cap \{\mathstrut h(z); \, 
\vert \vert h(\cdot) \vert \vert_{\mathcal{L}^{\infty}(D)} \! := \! \sup_{z 
\in D} \vert h(z) \vert \! < \! 1\}$;
\item[(8)] $\vert \vert \mathscr{F}(\cdot) \vert \vert_{\cap_{p \in J} 
\mathcal{L}^{p}_{\mathrm{M}_{2}(\mathbb{C})}(\ast)} \! := \! \sum_{p \in J} 
\vert \vert \mathscr{F}(\cdot) \vert \vert_{\mathcal{L}^{p}_{\mathrm{M}_{2}
(\mathbb{C})}(\ast)}$, with $\mathrm{card}(J) \! < \! \infty$;
\item[(9)] for $(\mu,\widetilde{\nu}) \! \in \! \mathbb{R} \times \mathbb{
R}$, the function $(\bullet \! - \! \mu)^{\mi \widetilde{\nu}} \colon \mathbb{
C} \setminus (-\infty,\mu) \! \to \! \mathbb{C} \colon \bullet \! \mapsto \! 
\me^{\mi \widetilde{\nu} \ln (\bullet -\mu)}$, with the branch cut taken 
along $(-\infty,\mu)$ and the principal branch of the logarithm chosen, 
$\ln (\bullet \! - \! \mu) \! := \! \ln \! \vert \! \bullet -\mu \vert \! 
+ \! \mi \arg (\bullet \! - \! \mu)$, $\arg (\bullet \! - \! \mu) \! \in \! 
(-\pi,\pi)$;
\item[(10)] a contour, $\mathcal{D}$, say, which is the finite union of 
piecewise-smooth, simple, closed curves, is said to be \emph{orientable} if 
its complement, $\mathbb{C} \setminus \mathcal{D}$, can always be divided 
into two, possibly disconnected, disjoint open sets $\mho^{+}$ and $\mho^{-}$, 
either of which has finitely many components, such that $\mathcal{D}$ admits 
an orientation so that it can either be viewed as a positively oriented 
boundary $\mathcal{D}^{+}$ for $\mho^{+}$ or as a negatively oriented 
boundary $\mathcal{D}^{-}$ for $\mho^{-}$ \cite{a45}, i.e., the (possibly 
disconnected) components of $\mathbb{C} \setminus \mathcal{D}$ can be 
coloured by $+$ or $-$ in such a way that the $+$ regions do not share 
boundary with the $-$ regions, except, possibly, at finitely many points 
\cite{a46};
\item[(11)] for $\boldsymbol{\gamma}$ a nullhomologous path in a region 
$\mathscr{D} \subset \mathbb{C}$, $\mathrm{int}(\boldsymbol{\gamma}) \! := \! 
\{\mathstrut \zeta \! \in \! \mathscr{D} \setminus \boldsymbol{\gamma}; \, 
\mathrm{ind}_{\boldsymbol{\gamma}}(\zeta) \! := \! \tfrac{1}{2 \pi \mi} 
\int_{\boldsymbol{\gamma}} \tfrac{\md \zeta^{\prime}}{\zeta^{\prime}-\zeta} 
\! \not=\! 0\}$.
\end{enumerate}
\subsection{The RHP for the D${}_{f}$NLSE}
In this subsection, the main results {}from the direct/inverse scattering 
analysis of the Cauchy problem for the D${}_{f}$NLSE are succinctly 
recapitulated: since the proofs of these results are given in \cite{a38}, 
only final results are stated.
\begin{bbbb}
The necessary and sufficient condition for the compatibility of the following 
linear system (Lax-pair), for arbitrary $\zeta \! \in \! \mathbb{C}$,
\begin{equation}
\partial_{x} \Psi (x,t;\zeta)=\mathcal{U}(x,t;\zeta) \Psi (x,t;\zeta), 
\qquad \quad \partial_{t} \Psi (x,t;\zeta)=\mathcal{V}(x,t;\zeta) \Psi 
(x,t;\zeta),
\end{equation}
where
\begin{align*}
\mathcal{U}(x,t;\zeta) &= \! -\mi \lambda (\zeta) \sigma_{3} \! + \! 
\begin{pmatrix}
0 & u \\
\overline{u} & 0
\end{pmatrix}, \\
\mathcal{V}(x,t;\zeta) &= \! -2 \mi (\lambda(\zeta))^{2} \sigma_{3} \! 
+ \! 2 \lambda (\zeta) \! 
\begin{pmatrix}
0 & u \\
\overline{u} & 0
\end{pmatrix} \! - \! \mi \! 
\begin{pmatrix}
u \overline{u}-1 & \partial_{x}u \\
\partial_{x} \overline{u} & u \overline{u}-1
\end{pmatrix} \! \sigma_{3},
\end{align*}
and $\lambda (\zeta) \! = \! \tfrac{1}{2}(\zeta \! + \! \zeta^{-1})$, with 
$\partial_{\ast} \zeta \! = \! 0$, $\ast \! \in \! \{x,t\}$, is that $u \! 
= \! u(x,t)$ satisfies the {\rm D${}_{f}$NLSE}.
\end{bbbb}
One proves Proposition~2.1.1 via the isospectral deformation condition 
$(\partial_{\ast} \zeta \! = \! 0$, $\ast \! \in \! \{x,t\})$, and invoking 
the Frobenius compatibility condition, $\partial_{t} \partial_{x} \Psi (x,
t;\zeta) \! = \! \partial_{x} \partial_{t} \Psi (x,t;\zeta) \! \Rightarrow 
\! \partial_{t} \mathcal{U}(x,t;\zeta) \! - \! \partial_{x} \mathcal{V}(x,t;
\zeta) \linebreak[4]
+ \! [\mathcal{U}(x,t;\zeta),\mathcal{V}(x,t;\zeta)] \! = \! \mathbf{0}$, 
$\zeta \! \in \! \mathbb{C}$, where $[\mathscr{A},\mathscr{B}] \! := \! 
\mathscr{A} \mathscr{B} \! - \! \mathscr{B} \mathscr{A}$ is the matrix 
commutator (note that $\mathrm{tr}(\mathcal{U}(x,t;\zeta)) \! = \! 
\mathrm{tr}(\mathcal{V}(x,t;\zeta)) \! = \! 0)$.
\begin{eeee}
Note that, if $u(x,t)$ is a solution of the D${}_{f}$NLSE with $\Psi (x,
t;\zeta)$ the corresponding solution of system~(2), $\Psi (x,t;\zeta) 
\mathcal{Q}(\zeta)$, with $\mathcal{Q}(\zeta) \! \in \! \mathrm{M}_{2}
(\mathbb{C})$, also solves system~(2).
\end{eeee}
The ISM analysis for the D${}_{f}$NLSE is based on the direct scattering 
problem for the (self-adjoint) operator (cf.~Proposition~2.1.1) 
$\mathcal{O}^{\mathcal{D}} \! := \! \mi \sigma_{3} \partial_{x} \! - \! 
\left(
\begin{smallmatrix}
\frac{1}{2}(\zeta + \zeta^{-1}) & \mi u_{o}(x) \\
\overline{\mi u_{o}(x)} & \frac{1}{2}(\zeta + \zeta^{-1})
\end{smallmatrix}
\right)$, where $u(x,0) \! := \! u_{o}(x)$ satisfies $u_{o}(x) \! =_{x \to 
\pm \infty} \! u_{o}(\pm \infty)(1 \! + \! o(1))$, with $u_{o}(\pm \infty) \! 
:= \! \exp (\tfrac{\mi (1 \mp 1) \theta}{2})$, $\theta \! \in \! [0,2 \pi)$ 
(see Eq.~(3)), $u_{o}(x) \! \in \! \mathbf{C}^{\infty}(\mathbb{R})$, and $u_{
o}(x) \! - \! u_{o}(\pm \infty) \! \in \! \mathcal{S}_{\mathbb{C}}(\mathbb{
R}_{\pm})$.
\begin{aaaa}
Let $u(x,t)$ be the solution of the {\rm D${}_{f}$NLSE} with $u(x,0) \! := \! 
u_{o}(x) \! =_{x \to \pm \infty} \! u_{o}(\pm \infty)(1 \! + \! o(1))$, where 
$u_{o}(\pm \infty) \! := \! \exp (\tfrac{\mi (1 \mp 1) \theta}{2})$, $\theta 
\! \in \! [0,2 \pi)$ (see Eq.~{\rm (3))}, $u_{o}(x) \! \in \! \mathbf{C}^{
\infty}(\mathbb{R})$, and $u_{o}(x) \! - \! u_{o}(\pm \infty) \! \in \! 
\mathcal{S}_{\mathbb{C}}(\mathbb{R}_{\pm})$. Define $\Psi^{\pm}(x,0;\zeta)$ 
as the (Jost) solutions of the first equation of system~{\rm (2)}, $\mathcal{
O}^{\mathcal{D}} \Psi^{\pm}(x,0;\zeta) \! = \! \mathbf{0}$, with the 
following asymptotics:
\begin{equation*}
\Psi^{\pm}(x,0;\zeta) \! \underset{x \, \to \, \pm \infty}{=} \! \left( 
\me^{\frac{\mi (1 \mp 1) \theta}{4} \sigma_{3}} \! 
\begin{pmatrix}
1 & -\mi \zeta^{-1} \\
\mi \zeta^{-1} & 1
\end{pmatrix} \! + \! o(1) \right) \! \me^{-\mi k(\zeta)x \sigma_{3}},
\end{equation*}
where $k(\zeta) \! = \! \tfrac{1}{2}(\zeta \! - \! \zeta^{-1})$.
\end{aaaa}
\begin{ffff}
Let $u(x,t)$ be the solution of the Cauchy problem for the 
{\rm D${}_{f}$NLSE} and $\Psi (x,t;\zeta)$ the corresponding solution of 
system~{\rm (2)} with the asymptotics stated in Definition~{\rm 2.1.1}. 
Then $\Psi (x,t;\zeta)$ satisfies the symmetry reductions $\sigma_{1} 
\overline{\Psi (x,t;\overline{\zeta})} \, \sigma_{1} \! = \! \Psi (x,t;
\zeta)$ and $\Psi (x,t;\zeta^{-1}) \! = \! \zeta \Psi (x,t;\zeta) 
\sigma_{2}$.
\end{ffff}
\begin{bbbb}
Set $\Psi^{\pm}(x,0;\zeta) \! := \! 
\left(
\begin{smallmatrix}
\Psi^{\pm}_{11}(\zeta) & \Psi^{\pm}_{12}(\zeta) \\
\Psi^{\pm}_{21}(\zeta) & \Psi^{\pm}_{22}(\zeta)
\end{smallmatrix}
\right)$. Then 
$\left(
\begin{smallmatrix}
\Psi_{12}^{+}(\zeta) \\
\Psi^{+}_{22}(\zeta)
\end{smallmatrix}
\right)$ and 
$\left(
\begin{smallmatrix}
\Psi_{11}^{-}(\zeta) \\
\Psi^{-}_{21}(\zeta)
\end{smallmatrix}
\right)$ have analytic continuation to $\mathbb{C}_{+}$ 
(respectively, $\left( 
\begin{smallmatrix}
\Psi_{11}^{+}(\zeta) \\
\Psi^{+}_{21}(\zeta)
\end{smallmatrix}
\right)$ and~$\left(
\begin{smallmatrix}
\Psi_{12}^{-}(\zeta) \\
\Psi^{-}_{22}(\zeta)
\end{smallmatrix}
\right)$ have analytic continuation to $\mathbb{C}_{-})$, the monodromy 
(scattering) matrix, $\mathrm{T}(\zeta)$, is defined by $\Psi^{-}(x,0;\zeta) 
\! := \! \Psi^{+}(x,0;\zeta) \mathrm{T}(\zeta)$, $\Im (\zeta) \! = \! 0$, 
where $\mathrm{T}(\zeta) \! = \! \left(
\begin{smallmatrix}
a(\zeta) & \overline{b(\overline{\zeta})} \\
b(\zeta) & \overline{a(\overline{\zeta})}
\end{smallmatrix}
\right)$, with $a(\zeta) \! = \! (1 \! - \! \zeta^{-2})^{-1}(\Psi_{22}^{+}
(\zeta) \Psi_{11}^{-}(\zeta) \! - \! \Psi_{12}^{+}(\zeta) \Psi_{21}^{-}
(\zeta))$, $b(\zeta) \! = \! (1 \! - \! \zeta^{-2})^{-1}(\overline{\Psi_{
22}^{+}(\overline{\zeta})} \, \Psi_{21}^{-}(\zeta) \! - \! \overline{\Psi
_{12}^{+}(\overline{\zeta})} \, \Psi_{11}^{-}(\zeta))$, $\vert a(\zeta) 
\vert^{2} \! - \! \vert b(\zeta) \vert^{2} \! = \! 1$, $a(\zeta^{-1}) \! = 
\! \overline{a(\overline{\zeta})}$, $b(\zeta^{-1}) \! = \! -\overline{b
(\overline{\zeta})}$, and $\det (\Psi^{\pm}(x,0;\zeta)) \vert_{\zeta =\pm 
1} \! = \! 0$.
\end{bbbb}
\begin{ffff}
Let the reflection coefficient associated with the direct scattering problem 
for the operator $\mathcal{O}^{\mathcal{D}}$ be defined by $r(\zeta) \! := 
\! b(\zeta)/a(\zeta)$. Then $r(\zeta^{-1}) \! = \! -\overline{r(\overline{
\zeta})}$.
\end{ffff}
\begin{eeee}
Note that, even though $a(\zeta)$ (respectively, $\overline{a(\overline{
\zeta})})$ has an analytic continuation off $\Im (\zeta) \! = \! 0$ to 
$\mathbb{C}_{+}$ (respectively, $\mathbb{C}_{-})$ and is continuous on 
$\overline{\mathbb{C}}_{+}$ (respectively, $\overline{\mathbb{C}}_{-})$, 
$b(\zeta)$ does not, in general, have an analytic continuation to 
$\mathbb{C} \setminus \mathbb{R}$. Furthermore, for the finite-density 
initial data considered here, it is shown in \cite{a14} that, using 
Volterra-type integral representations for the elements of $\Psi^{\pm}
(x,0;\zeta)$ and a successive approximations argument, $r(\zeta) \! 
\in \! \mathcal{S}_{\mathbb{C}}(\mathbb{R})$ (see, also, Part~1 of 
\cite{a12}).
\end{eeee}
\begin{cccc}
Let $u(x,t)$ be the solution of the Cauchy problem for the 
{\rm D${}_{f}$NLSE} and $\Psi^{\pm}(x,0;\zeta)$ the corresponding (Jost) 
solutions of $\mathcal{O}^{\mathcal{D}} \Psi^{\pm}(x,0;\zeta) \! = \! 
\mathbf{0}$ given in Definition~{\rm 2.1.1}. Then $\Psi^{\pm}(x,0;\zeta)$ 
have the following asymptotics:
\begin{align}
\Psi^{-}(x,0;\zeta) \! &\underset{\zeta \, \to \, \infty}{=} \! \me^{\frac{
\mi \theta}{2} \sigma_{3}} \! \left( \mathrm{I} \! + \! \tfrac{1}{\zeta} 
\left(
\begin{smallmatrix}
\mi \int_{-\infty}^{x}(\vert u_{o}(x^{\prime}) \vert^{2}-1) \, \md x^{\prime} 
& -\mi u_{o}(x) \me^{-\mi \theta} \\
\mi \overline{u_{o}(x)} \, \me^{\mi \theta} & -\mi \int_{-\infty}^{x}
(\vert u_{o}(x^{\prime}) \vert^{2}-1) \, \md x^{\prime}
\end{smallmatrix}
\right) \! + \! \mathcal{O}(\zeta^{-2}) \right) \! \me^{-\mi k(\zeta)x 
\sigma_{3}}, \nonumber \\
\Psi^{+}(x,0;\zeta) \! &\underset{\zeta \, \to \, \infty}{=} \! \left( 
\mathrm{I} \! + \! \tfrac{1}{\zeta} 
\left(
\begin{smallmatrix}
\mi \int_{+\infty}^{x}(\vert u_{o}(x^{\prime}) \vert^{2}-1) \, \md x^{\prime} 
& -\mi u_{o}(x) \\
\mi \overline{u_{o}(x)} & -\mi \int_{+\infty}^{x}(\vert u_{o}(x^{\prime}) 
\vert^{2}-1) \, \md x^{\prime}
\end{smallmatrix}
\right) \! + \! \mathcal{O}(\zeta^{-2}) \right) \! \me^{-\mi k(\zeta)x 
\sigma_{3}}, \nonumber \\
\Psi^{-}(x,0;\zeta) &\underset{\zeta \, \to \, 0}{=} \! \left(\zeta^{-1} 
\sigma_{2} \me^{-\frac{\mi \theta}{2} \sigma_{3}} \! + \! \mathcal{O}(1) 
\right) \! \me^{-\mi k(\zeta)x \sigma_{3}}, \qquad \Psi^{+}(x,0;\zeta) 
\! \underset{\zeta \, \to \, 0}{=} \! \left(\zeta^{-1} \sigma_{2} \! + 
\! \mathcal{O}(1) \right) \! \me^{-\mi k(\zeta)x \sigma_{3}}. \nonumber
\end{align}
\end{cccc}
\begin{ffff}
The following asymptotics are valid:
\begin{gather*}
a(\zeta) \! \underset{\zeta \, \to \, \infty}{=} \! \me^{\frac{\mi \theta}{
2}} \left(1 \! + \! \left(\mi \int\nolimits_{-\infty}^{+\infty}(\vert u_{o}
(x^{\prime}) \vert^{2} \! - \! 1) \, \md x^{\prime} \right) \zeta^{-1} \! 
+ \! \mathcal{O}(\zeta^{-2}) \right), \qquad \quad a(\zeta) \! \underset{
\zeta \, \to \, 0}{=} \! \me^{-\frac{\mi \theta}{2}}(1 \! + \! \mathcal{O}
(\zeta)), \\
r(\zeta) \! \underset{\zeta \, \to \, \infty}{=} \! \mathcal{O}(\zeta^{-1}), 
\qquad \quad \quad r(\zeta) \! \underset{\zeta \, \to \, 0}{=} \! \mathcal{O}
(\zeta);
\end{gather*}
in particular, $r(0) \! = \! 0$.
\end{ffff}
In \cite{a38} it is shown that, for $u(x,0) \! := \! u_{o}(x)$ satisfying 
$u_{o}(x) \! =_{x \to \pm \infty} \! u_{o}(\pm \infty)(1 \! + \! o(1))$, with 
$u_{o}(\pm \infty) \! := \! \exp (\tfrac{\mi (1 \mp 1) \theta}{2})$, $\theta 
\! \in \! [0,2 \pi)$ (see Eq.~(3)), $u_{o}(x) \! \in \! \mathbf{C}^{\infty}
(\mathbb{R})$, and $u_{o}(x) \! - \! u_{o}(\pm \infty) \! \in \! \mathcal{
S}_{\mathbb{C}}(\mathbb{R}_{\pm})$, $\sigma_{\mathcal{O}^{\mathcal{D}}} \! := 
\! \mathrm{spec}(\mathcal{O}^{\mathcal{D}}) \! = \! \sigma_{d} \cup \sigma_{
c}$ $(\sigma_{d} \cap \sigma_{c} \! = \! \emptyset)$, where $\sigma_{d}$ is 
the finitely denumerable ``discrete'' spectrum given by $\sigma_{d} \! = 
\! \Delta_{a} \cup \overline{\Delta_{a}}$, where $\Delta_{a} \! := \! \{
\mathstrut \varsigma_{n}; \, a(\zeta) \vert_{\zeta =\varsigma_{n}} \! = \! 
0, \, \varsigma_{n} \! = \! \me^{\mi \phi_{n}}, \, \phi_{n} \! \in \! (0,
\pi), \, n \! \in \! \{1,2,\ldots,N\}\}$, with
\begin{gather}
a(\zeta) \! = \! \me^{\frac{\mi \theta}{2}} \prod_{n=1}^{N} \dfrac{(\zeta \! 
- \! \varsigma_{n})}{(\zeta \! - \! \overline{\varsigma_{n}})} \exp \! \left(
-\int\nolimits_{-\infty}^{+\infty} \dfrac{\ln (1 \! - \! \vert r(\mu) \vert^{
2})}{(\mu \! - \! \zeta)} \, \dfrac{\md \mu}{2 \pi \mi} \right), \quad \zeta 
\! \in \! \mathbb{C}_{+}, \nonumber \\
0 \leqslant \theta \! = \! -2 \sum_{n=1}^{N} \phi_{n} \! - \! \int\nolimits_{
-\infty}^{+\infty} \dfrac{\ln (1 \! - \! \vert r(\mu) \vert^{2})}{\mu} \, 
\dfrac{\md \mu}{2 \pi} < 2 \pi,
\end{gather}
and $\Delta_{a} \cap \overline{\Delta_{a}} \! = \! \emptyset$ $(\mathrm{card}
(\sigma_{d}) \! = \! 2N)$, and $\sigma_{c}$ is the ``continuous'' spectrum 
given by $\sigma_{c} \! = \! \{\mathstrut \zeta; \, \Im (\zeta) \! = \! 0\}$, 
with orientation {}from $-\infty$ to $+\infty$ $(\mathrm{card}(\sigma_{c}) 
\! = \! \infty)$. Furthermore, it is shown in \cite{a38} that, for $r(\zeta) 
\! \in \! \mathcal{S}_{\mathbb{C}}^{1}(\mathbb{R})$ and $\vert r(\pm 1) 
\vert \! \not= \! 1$,
\begin{equation*}
a(s \! + \! \mi \varepsilon) \! \underset{\varepsilon \downarrow 0}{=} \! 
\dfrac{(-s)^{N} \exp \! \left(\mi \! \left(\frac{\theta}{2} \! + \! \sum_{
n=1}^{N} \phi_{n} \! + \! \mathrm{P.V.} \int_{\mathbb{R} \setminus \{s\}} 
\frac{\ln (1-\vert r(\mu) \vert^{2})}{(\mu-s)} \, \frac{\md \mu}{2 \pi} 
\right) \! \right)}{(1 \! - \! \vert r(s) \vert^{2})^{\kappa_{\mathrm{sgn}
(s)}}}(1 \! + \! o(1)), \quad s \! \in \! \{\pm 1\},
\end{equation*}
where $\mathrm{P.V.} \int$ denotes the principal value integral, with 
$\kappa_{\pm}$ real, possibly zero, constants, and (trace identity)
\begin{equation}
\int_{-\infty}^{+\infty}(\vert u(x^{\prime},t) \vert^{2} \! - \! 1) \, \md 
x^{\prime} \! = \! -2 \sum_{n=1}^{N} \sin (\phi_{n}) \! - \! \int_{-\infty}^{
+\infty} \ln (1 \! - \! \vert r(\mu) \vert^{2}) \, \tfrac{\md \mu}{2 \pi}.
\end{equation}

The ``inverse part'' of the ISM analysis is invoked by re-introducing the 
$t$-dependence, namely, studying the $\partial_{t} \Psi (x,t;\zeta) \! = 
\! \mathcal{V}(x,t;\zeta) \Psi (x,t;\zeta)$ component of system~(2). The 
scattering map $(\mathscr{S})$ $u_{o}(x) \! \mapsto \! r(\zeta) \! = \! 
\mathscr{R}(u_{o}(\cdot))$, which is a bijection for $u_{o}(x)$ satisfying 
the finite-density initial conditions and $r(\zeta) \! \in \! \mathcal{S}_{
\mathbb{C}}^{1}(\mathbb{R})$, linearises the D${}_{f}$NLSE flow in the sense 
that, since $a(\zeta,t) \! = \! a(\zeta)$ is the ``generator'' of the 
integrals of motion and $b(\zeta,t) \! = \! b(\zeta) \exp (4 \mi k(\zeta) 
\lambda (\zeta)t)$ \cite{a12}, $r(\zeta,t) \! := \! b(\zeta,t)/a(\zeta,t)$ 
evolves in the scattering data phase space according to the rule $r(\zeta,
t) \! = \! r(\zeta) \exp (4 \mi k(\zeta) \lambda (\zeta)t)$. Set \cite{a38}
\begin{equation*}
\widetilde{\Phi}(x,t;\zeta) \! := \! 
\begin{cases}
\left(
\begin{smallmatrix}
\frac{\Psi^{-}_{11}(x,t;\zeta)}{a(\zeta)} & \, \, \Psi^{+}_{12}(x,t;\zeta) \\
\frac{\Psi^{-}_{21}(x,t;\zeta)}{a(\zeta)} & \, \, \Psi^{+}_{22}(x,t;\zeta)
\end{smallmatrix}
\right), &\text{$\zeta \! \in \! \mathbb{C}_{+}$,} \\
\left(
\begin{smallmatrix}
\Psi^{+}_{11}(x,t;\zeta) & \, \, \frac{\Psi^{-}_{12}(x,t;\zeta)}{\overline{
a(\overline{\zeta})}} \\
\Psi^{+}_{21}(x,t;\zeta) & \, \, \frac{\Psi^{-}_{22}(x,t;\zeta)}{\overline{
a(\overline{\zeta})}}
\end{smallmatrix}
\right), &\text{$\zeta \! \in \! \mathbb{C}_{-}$,}
\end{cases}
\end{equation*}
with $\Psi^{\pm}(x,t;\zeta)$ the solutions of system~(2): $\widetilde{\Phi}
(x,t;\zeta)$ has the asymptotics \cite{a38}
\begin{gather*}
\widetilde{\Phi}(x,t;\zeta) \! \underset{\zeta \to \infty}{=} \! 
\left( 
\mathrm{I} \! + \! \tfrac{1}{\zeta} \! 
\left(
\begin{smallmatrix}
\mi \int_{+\infty}^{x}(\vert u(x^{\prime},t) \vert^{2}-1) \, \md x^{\prime} 
& -\mi u(x,t) \\
\mi \overline{u(x,t)} & -\mi \int_{+\infty}^{x}(\vert u(x^{\prime},t) \vert^{
2}-1) \, \md x^{\prime}
\end{smallmatrix}
\right) \! + \! \mathcal{O}(\zeta^{-2}) \right) \! \me^{-\mi k(\zeta)(x+2 
\lambda (\zeta)t) \sigma_{3}}, \\
\widetilde{\Phi}(x,t;\zeta) \! \underset{\zeta \to 0}{=} \! \left(\zeta^{-1} 
\sigma_{2} \! + \! \mathcal{O}(1) \right) \! \me^{-\mi k(\zeta)(x+2 \lambda 
(\zeta)t) \sigma_{3}}.
\end{gather*}
\begin{cccc}[{\rm \cite{a38}}]
Let $u(x,t)$ be the solution of the Cauchy problem for the 
{\rm D${}_{f}$NLSE} with finite-density initial data $u(x,0) \! := \! u_{o}
(x) \! =_{x \to \pm \infty} \! u_{o}(\pm \infty)(1 \! + \! o(1))$, where $u_{
o}(\pm \infty) \! := \! \exp (\tfrac{\mi (1 \mp 1) \theta}{2})$, $0 \! 
\leqslant \! \theta \! = \! -2 \sum_{n=1}^{N} \sin (\phi_{n}) \! - \! \int_{
-\infty}^{+\infty} \tfrac{\ln (1-\vert r(\mu) \vert^{2})}{\mu} \, \tfrac{\md 
\mu}{2 \pi} \! < \! 2 \pi$, $u_{o}(x) \! \in \! \mathbf{C}^{\infty}(\mathbb{
R})$, and $u_{o}(x) \! - \! u_{o}(\pm \infty) \! \in \! \mathcal{S}_{\mathbb{
C}}(\mathbb{R}_{\pm})$. Set
\begin{equation*}
m(x,t;\zeta) \! := \! \widetilde{\Phi}(x,t;\zeta) \exp (\mi k(\zeta)(x \! 
+ \! 2 \lambda (\zeta) t) \sigma_{3}).
\end{equation*}
Then: {\rm (1)} the bounded discrete set $\sigma_{d}$ is finite; {\rm (2)} 
the poles of $m(x,t;\zeta)$ are simple; {\rm (3)} the first (respectively, 
second) column of $m(x,t;\zeta)$ has poles in $\mathbb{C}_{+}$ (respectively, 
$\mathbb{C}_{-})$ at $\{\varsigma_{n}\}_{n=1}^{N}$ (respectively, 
$\{\overline{\varsigma_{n}}\}_{n=1}^{N});$ and {\rm (4)} $m(x,t;\zeta) 
\colon \mathbb{C} \setminus (\sigma_{d} \cup \sigma_{c}) \! \to \! 
\mathrm{M}_{2}(\mathbb{C})$ solves the following {\rm RHP:}
\begin{enumerate}
\item[(i)] $m(x,t;\zeta)$ is piecewise (sectionally) meromorphic $\forall \, 
\zeta \! \in \! \mathbb{C} \setminus \sigma_{c};$
\item[(ii)] $m_{\pm}(x,t;\zeta) \! := \! \lim_{\genfrac{}{}{0pt}{2}{\zeta^{
\prime} \, \to \, \zeta}{\pm \Im (\zeta^{\prime})>0}}m(x,t;\zeta^{\prime})$ 
satisfy the jump condition
\begin{equation*}
m_{+}(x,t;\zeta) \! = \! m_{-}(x,t;\zeta) \mathcal{G}(x,t;\zeta), \quad \zeta 
\! \in \! \mathbb{R},
\end{equation*}
where $\mathcal{G}(x,t;\zeta) \! = \! \exp (-\mi k(\zeta)(x \! + \! 2 \lambda 
(\zeta)t) \mathrm{ad}(\sigma_{3})) \! 
\left(
\begin{smallmatrix}
1+r(\zeta)r(\zeta^{-1}) & \, \, r(\zeta^{-1}) \\
r(\zeta) & \, \, 1
\end{smallmatrix}
\right)$, and $r(\zeta)$, the reflection coefficient associated with the 
direct scattering problem for the operator $\mathcal{O}^{\mathcal{D}}$, 
satisfies $r(\zeta) \! =_{\zeta \to 0} \! \mathcal{O}(\zeta)$, $r(\zeta) \! 
=_{\zeta \to \infty} \! \mathcal{O}(\zeta^{-1})$, $r(\zeta^{-1}) \! = \! 
-\overline{r(\overline{\zeta})}$, and $r(\zeta) \! \in \! \mathcal{S}_{
\mathbb{C}}^{1}(\mathbb{R});$
\item[(iii)] for the simple poles of $m(x,t;\zeta)$ at $\{\varsigma_{n}\}_{n
=1}^{N}$ and $\{\overline{\varsigma_{n}}\}_{n=1}^{N}$, there exist nilpotent 
matrices, with degree of nilpotency 2, such that $m(x,t;\zeta)$ satisfies the 
polar conditions
\begin{align*}
\mathrm{Res}(m(x,t;\zeta);\varsigma_{n}) &= \! \lim_{\zeta \, \to \, 
\varsigma_{n}}m(x,t;\zeta)g_{n}(x,t) \sigma_{-}, & n \! &\in \! \{1,2,
\ldots,N\}, \\
\mathrm{Res}(m(x,t;\zeta);\overline{\varsigma_{n}}) &= \! \sigma_{1} 
\overline{\mathrm{Res}(m(x,t;\zeta);\varsigma_{n})} \, \sigma_{1}, & n 
\! &\in \! \{1,2,\ldots,N\},
\end{align*}
where $g_{n}(x,t) \! = \! g_{n} \exp \! \left(2 \mi k(\varsigma_{n})(x \! + 
\! 2 \lambda (\varsigma_{n})t) \right)$, with
\begin{equation*}
g_{n} \! := \! \vert \gamma_{n} \vert \me^{\mi \theta_{\gamma_{n}}}(\varsigma_{
n} \! - \! \overline{\varsigma_{n}}) \exp \! \left(-\dfrac{\mi \theta}{2} \! 
+ \! \int_{-\infty}^{+\infty} \! \dfrac{\ln (1 \! - \! \vert r(\mu) \vert^{
2})}{(\mu \! - \! \varsigma_{n})} \, \dfrac{\md \mu}{2 \pi \mi} \right) \! 
\prod_{\genfrac{}{}{0pt}{2}{k=1}{k \not= n}}^{N} \! \left(\dfrac{\varsigma_{
n} \! - \! \overline{\varsigma_{k}}}{\varsigma_{n} \! - \! \varsigma_{k}} 
\right), \qquad \theta_{\gamma_{n}} \! = \! \pm \tfrac{\pi}{2};
\end{equation*}
\item[(iv)] $\det (m(x,t;\zeta)) \vert_{\zeta = \pm 1} \! = \! 0;$
\item[(v)] $m(x,t;\zeta) \! =_{\zeta \to 0} \! \zeta^{-1} \sigma_{2} \! + 
\! \mathcal{O}(1);$
\item[(vi)] $m(x,t;\zeta) \! =_{\genfrac{}{}{0pt}{2}{\zeta \to \infty}{\zeta 
\in \mathbb{C} \setminus (\sigma_{d} \cup \sigma_{c})}} \! \mathrm{I} \! + 
\! \mathcal{O}(\zeta^{-1});$
\item[(vii)] $m(x,t;\zeta)$ possesses the symmetry reductions $m(x,t;\zeta) 
\! = \! \sigma_{1} \overline{m(x,t;\overline{\zeta})} \, \sigma_{1}$ and 
$m(x,t;\zeta^{-1}) \! = \! \zeta m(x,\linebreak[4]
t;\zeta) \sigma_{2}$.
\end{enumerate}
For $r(\zeta) \! \in \! \mathcal{S}_{\mathbb{C}}^{1}(\mathbb{R})$: (i) the 
{\rm RHP} for $m(x,t;\zeta)$ formulated above is uniquely asymptotically 
solvable; and (ii) $\widetilde{\Phi}(x,t;\zeta) \! = \! m(x,t;\zeta) \exp 
(-\mi k(\zeta)(x \! + \! 2 \lambda (\zeta)t) \sigma_{3})$ solves 
system~{\rm (2)} with
\begin{equation}
u(x,t) \! := \! \mi \lim_{\genfrac{}{}{0pt}{2}{\zeta \to \infty}{\zeta \in 
\mathbb{C} \, \setminus (\sigma_{d} \cup \sigma_{c})}}(\zeta (m(x,t;\zeta) 
\! - \! \mathrm{I}))_{12}
\end{equation}
the solution of the Cauchy problem for the {\rm D${}_{f}$NLSE}, and
\begin{equation}
\int\nolimits_{+\infty}^{x}(\vert u(x^{\prime},t) \vert^{2} \! - \! 1) \, \md 
x^{\prime} \! := \! -\mi \lim_{\genfrac{}{}{0pt}{2}{\zeta \to \infty}{\zeta 
\in \mathbb{C} \, \setminus (\sigma_{d} \cup \sigma_{c})}}(\zeta (m(x,t;
\zeta) \! - \! \mathrm{I}))_{11}.
\end{equation}
\end{cccc}
\begin{eeee}
In this paper, for $r(\zeta) \! \in \! \mathcal{S}_{\mathbb{C}}^{1}(\mathbb{
R})$, the solvability of the RHP for $m(x,t;\zeta)$ formulated in Lemma~2.1.2 
is proved, via explicit construction, for all sufficiently large $\vert t 
\vert$ $(x/t \! \sim \! \mathcal{O}(1))$: the solvability of the RHP in the 
solitonless sector, $\sigma_{d} \! \equiv \! \emptyset$, for $r(\zeta) \! 
\in \! \mathcal{S}_{\mathbb{C}}^{1}(\mathbb{R})$, as $\vert t \vert \! \to 
\! \infty$ and $\vert x \vert \! \to \! \infty$ such that $z_{o} \! := \! 
x/t \! \sim \! \mathcal{O}(1)$ and $\in \! \mathbb{R} \setminus \{-2,0,2\}$, 
was proved in \cite{a38}.
\end{eeee}
\subsection{Summary of Results}
In this subsection, the results of this work are summarised in 
Theorems~2.2.1--2.2.4: before doing so, however, the following preamble 
is necessary. Recall {}from Subsection~2.1 that $\varsigma_{n} \! := \! \me^{
\mi \phi_{n}}$, $\phi_{n} \! \in \! (0,\pi)$, $n \! \in \! \{1,2,\ldots,N\}$. 
Set $\varsigma_{n} \! := \! \xi_{n} \! + \! \mi \eta_{n}$, where $\xi_{n} \! 
= \! \Re (\varsigma_{n}) \! = \! \cos (\phi_{n}) \! \in \! (-1,1)$, and 
$\eta_{n} \! = \! \Im (\varsigma_{n}) \! = \! \sin (\phi_{n}) \! \in \! (0,
1)$. Throughout this paper, it is assumed that: (1) $\xi_{i} \! \not= \! \xi_{
j}$ $\forall \, \, i \! \not= \! j \! \in \! \{1,2,\ldots,N\}$; and (2) the 
following ordering (enumeration) for the elements of the discrete spectrum 
(solitons), $\sigma_{d}$, is taken, $\xi_{1} \! > \! \xi_{2} \! > \! \cdots 
\! > \! \xi_{N}$.
\begin{eeee}
Throughout this paper, the ``symbols'' $c^{\mathcal{S}}(\diamondsuit)$, 
$\underline{c}(\flat,\natural,\sharp)$, $\underline{c}(z_{1},z_{2},z_{3},z_{
4})$, $\underline{c}(\bullet)$, and $\underline{c}$, appearing in the various 
error estimates, are to be understood as follows: (1) for $\pm \diamondsuit \! 
> \! 0$, $c^{\mathcal{S}}(\diamondsuit) \! \in \! \mathcal{S}_{\mathbb{C}}
(\mathbb{R}_{\pm})$; (2) for $\pm \flat \! > \! 0$, $\underline{c}(\flat,
\natural,\sharp) \! \in \! \mathcal{L}^{\infty}_{\mathbb{C}}(\mathbb{R}_{\pm} 
\! \times \! \mathbb{C}^{\ast} \! \times \! \overline{\mathbb{C}^{\ast}})$, 
where $\mathbb{C}^{\ast} \! := \! \mathbb{C} \setminus \{0\}$; (3) for 
$(z_{1},z_{2}) \! \in \! \mathbb{R}_{\pm} \times \mathbb{R}_{\pm}$, 
$\underline{c}(z_{1},z_{2},z_{3},z_{4}) \! \in \! \mathcal{L}^{\infty}_{
\mathbb{C}}(\mathbb{R}^{2}_{\pm} \! \times \! \mathbb{C}^{\ast} \! \times \! 
\overline{\mathbb{C}^{\ast}})$; (4) for $\pm \bullet \! > \! 0$, $\underline{
c}(\bullet) \! \in \! \mathcal{L}^{\infty}_{\mathbb{C}}(\mathrm{D}_{\pm})$, 
where $\mathrm{D}_{+} \! := \! (0,2)$ and $\mathrm{D}_{-} \! := \! (-2,0)$; 
and (5) $\underline{c} \! \in \! \mathbb{C}^{\ast}$. Even though the symbols 
$c^{\mathcal{S}}(\diamondsuit)$, $\underline{c}(\flat,\natural,\sharp)$, 
$\underline{c}(z_{1},z_{2},z_{3},z_{4})$, $\underline{c}(\bullet)$, and 
$\underline{c}$ are not, in general, equal, and should properly be denoted 
as $c_{1}(\cdot)$, $c_{2}(\cdot)$, etc., the simplified notations $c^{
\mathcal{S}}(\diamondsuit)$, $\underline{c}(\flat,\natural,\sharp)$, 
$\underline{c}(z_{1},z_{2},z_{3},z_{4})$, $\underline{c}(\bullet)$, and 
$\underline{c}$ are retained throughout in order to eschew a flood of 
superfluous notation as well as to maintain consistency with the main theme 
of this work, namely, to derive explicitly the leading-order asymptotics and 
the classes to which the errors belong without regard to their precise 
$z_{o}$-dependence.
\end{eeee}
\begin{eeee}
In Theorems~2.2.1---2.2.4 below, one should keep, everywhere, the upper 
(respectively, l\-o\-w\-e\-r) signs as $t \! \to \! +\infty$ (respectively, 
$t \! \to \! -\infty)$.
\end{eeee}
\begin{dddd}
For $r(\zeta) \! \in \! \mathcal{S}_{\mathbb{C}}^{1}(\mathbb{R})$, let 
$m(x,t;\zeta)$ be the solution of the Riemann-Hilbert problem formulated 
in Lemma~{\rm 2.1.2}. Let $u(x,t)$, the solution of the Cauchy problem for 
the {\rm D${}_{f}$NLSE} with finite-density initial data $u(x,0) \! := \! u_{
o}(x) \! =_{x \to \pm \infty} \! u_{o}(\pm \infty)(1 \! + \! o(1))$, where 
$u_{o}(\pm \infty) \! := \! \exp (\tfrac{\mi (1 \mp 1) \theta}{2})$, $0 \! 
\leqslant \! \theta \! = \! -2 \sum_{n=1}^{N} \sin (\phi_{n}) \! - \! \int_{
-\infty}^{+\infty} \tfrac{\ln (1-\vert r(\mu) \vert^{2})}{\mu} \, \tfrac{\md 
\mu}{2 \pi} \! < \! 2 \pi$, $u_{o}(x) \! \in \! \mathbf{C}^{\infty}(\mathbb{
R})$, and $u_{o}(x) \! - \! u_{o}(\pm \infty) \! \in \! \mathcal{S}_{\mathbb{
C}}(\mathbb{R}_{\pm})$, be defined by Eq.~{\rm (5)}. Then, for $\theta_{
\gamma_{m}} \! = \! \varepsilon_{b} \pi/2$, $\varepsilon_{b} \! \in \! \{\pm 
1\}$, $m \! \in \! \{1,2,\ldots,N\}$, as $t \! \to \! \pm \infty$ and $x \! 
\to \! \mp \infty$ such that $z_{o} \! := \! x/t \! < \! -2$ and $(x,t) \! 
\in \! \{\mathstrut (x,t); \, x \! + \! 2t \cos (\phi_{m}) \! = \! 
\mathcal{O}(1), \, \phi_{m} \! \in \! (0,\pi)\}$,
\begin{align}
u(x,t) \! &= \! \me^{-\mi (\theta^{\pm}(1)+s^{\pm})} \! \left(\widetilde{
u}_{\mathcal{S}}(x,t) \! + \! \dfrac{\sqrt{\nu (\lambda_{1})}}{\sqrt{\vert t 
\vert (\lambda_{1} \! - \! \lambda_{2})} \, (z_{o}^{2} \! + \! 32)^{1/4}} \! 
\left(\widetilde{u}_{\mathcal{C}}(x,t) \! + \! \widetilde{u}_{\mathcal{SC}}
(x,t) \right) \! \right. \nonumber \\
 &+ \! \left. \mathcal{O} \! \left( \! \left( \dfrac{c^{\mathcal{S}}
(\lambda_{1}) \underline{c}(\lambda_{2},\lambda_{3},\lambda_{4})}{\sqrt{
\lambda_{1}(z_{o}^{2} \! + \! 32)}} \! + \! \dfrac{c^{\mathcal{S}}
(\lambda_{2}) \underline{c}(\lambda_{1},\lambda_{3},\lambda_{4})}{\sqrt{
\lambda_{2}(z_{o}^{2} \! + \! 32)}} \right) \! \dfrac{\ln \vert t \vert}
{(\lambda_{1} \! - \! \lambda_{2}) t} \right) \! \right),
\end{align}
where
\begin{align}
\theta^{+}(j) \! =& \! \left(\int_{-\infty}^{0} \! + \! \int_{\lambda_{
2}}^{\lambda_{1}} \right) \! \dfrac{\ln (1 \! - \! \vert r(\mu) \vert^{
2})}{\mu^{j}} \, \dfrac{\md \mu}{2 \pi}, \quad j \! \in \! \{0,1\}, \\
\theta^{-}(l) \! =& \! \left(\int_{0}^{\lambda_{2}} \! + \! \int_{
\lambda_{1}}^{+\infty} \right) \dfrac{\ln (1 \! - \! \vert r(\mu) \vert^{
2})}{\mu^{l}} \, \dfrac{\md \mu}{2 \pi}, \quad l \! \in \! \{0,1\},
\end{align}
\begin{equation}
\begin{split}
\lambda_{1} \! = \! -\tfrac{1}{2}(a_{1} \! - &(a_{1}^{2} \! - \! 4)^{1/2}), 
\qquad \, \lambda_{2} \! = \! \lambda_{1}^{-1}, \qquad \, \lambda_{3} \! = 
\! -\tfrac{1}{2}(a_{2} \! - \! \mi (4 \! - \! a_{2}^{2})^{1/2}), \qquad \, 
\lambda_{4} \! = \! \overline{\lambda_{3}}, \\
a_{1} \! &= \! \tfrac{1}{4}(z_{o} \! - \! (z_{o}^{2} \! + \! 32)^{1/2}), \, 
\qquad a_{2} \! = \! \tfrac{1}{4}(z_{o} \! + \! (z_{o}^{2} \! + \! 32)^{
1/2}),
\end{split}
\end{equation}
$0 \! < \! \lambda_{2} \! < \! \lambda_{1}$, $\vert \lambda_{3} \vert^{2} \! 
= \! 1$, $a_{1}a_{2} \! = \! -2$,
\begin{gather}
s^{+} \! = \! 2\sum_{k=m+1}^{N} \phi_{k}, \qquad \, \, s^{-} \! = \! 2
\sum_{k=1}^{m-1} \phi_{k}, \qquad \, \, \nu (z) \! = \! -\tfrac{1}{2 \pi} 
\ln (1 \! - \! \vert r(z) \vert^{2}), \\
\widetilde{u}_{\mathcal{S}}(x,t) \! = \! \dfrac{1 \! + \! \varepsilon_{b} 
\widetilde{\varepsilon}_{\mathscr{P}} \me^{-2 \mi \phi_{m}+\Omega^{\pm}
(x,t)}}{(1 \! + \! \varepsilon_{b} \widetilde{\varepsilon}_{\mathscr{P}} 
\me^{\Omega^{\pm}(x,t)})}, \\
\widetilde{\varepsilon}_{\mathscr{P}} \! = \! \mathrm{sgn} \! \left( \! 
\left( \prod_{k=1}^{m-1} \dfrac{\sin (\frac{1}{2}(\phi_{m} \! + \! \phi_{
k}))}{\sin (\frac{1}{2}(\phi_{m} \! - \! \phi_{k}))} \right) \! \! \left( 
\prod_{k=m+1}^{N} \dfrac{\sin (\frac{1}{2}(\phi_{m} \! + \! \phi_{k}))}{\sin 
(\frac{1}{2}(\phi_{m} \! - \! \phi_{k}))} \right)^{-1} \right) \! = \! (-1)^{
N-m}, \\
\Omega^{\pm}(x,t) \! = \! -2 \sin (\phi_{m})(x \! + \! 2t \cos (\phi_{m}) \! 
- \! \widetilde{x}^{\pm}_{m}),
\end{gather}
\begin{align}
\widetilde{x}^{\pm}_{m} \! =& \, \dfrac{\ln (\vert \gamma_{m} \vert)}{2 \sin 
(\phi_{m})} \! \pm \! \sum_{k=1}^{N} \dfrac{\mathrm{sgn}(m \! - \! k)}{2 \sin 
(\phi_{m})} \ln \! \left( \left\vert \dfrac{\sin (\frac{1}{2}(\phi_{m} \! + 
\! \phi_{k}))}{\sin (\frac{1}{2}(\phi_{m} \! - \! \phi_{k}))} \right\vert 
\right) \nonumber \\
 \pm& \, \dfrac{1}{2} \! \left(\int\nolimits_{0}^{\lambda_{2}} \! + \! 
\int\nolimits_{\lambda_{1}}^{+\infty} \! - \! \int\nolimits_{-\infty}^{0} \! 
- \! \int\nolimits_{\lambda_{2}}^{\lambda_{1}} \right) \! \dfrac{\ln (1 \! - 
\! \vert r(\mu) \vert^{2})}{(\mu^{2} \! - \! 2 \mu \cos (\phi_{m}) \! + \! 
1)} \, \dfrac{\md \mu}{2 \pi},
\end{align}
\begin{gather}
\widetilde{u}_{\mathcal{C}}(x,t) \! = \! \mi \me^{\mi s^{\pm}} \! \left( 
\lambda_{1} \me^{\mp \mi \left(\Theta^{\pm}(z_{o},t) \pm (2 \mp 1) \frac{\pi}
{4} \right)} \! + \! \lambda_{2} \me^{\pm \mi \left(\Theta^{\pm}(z_{o},t) 
\pm (2 \mp 1) \frac{\pi}{4} \right)} \right),
\end{gather}
\begin{align}
\Theta^{\pm}(z_{o},t) \! =& \! \pm \arg r(\lambda_{1}) \! \pm \! 4 \sum_{k 
\in J^{\pm}} \arg (\lambda_{1} \! - \! \me^{\mi \phi_{k}}) \! - \! \arg 
\Gamma (\mi \nu (\lambda_{1})) \! \pm \! t(\lambda_{1} \! - \! \lambda_{2})
(z_{o} \! + \! \lambda_{1} \! + \! \lambda_{2}) \nonumber \\
 +& \, \nu (\lambda_{1}) \ln \vert t \vert \! + \! 3 \nu (\lambda_{1}) \ln 
(\lambda_{1} \! - \! \lambda_{2}) \! + \! \tfrac{1}{2} \nu (\lambda_{1}) \ln 
\! \left(z_{o}^{2} \! + \! 32 \right) \! \mp \! \Xi^{\pm}(\lambda_{1}) \! 
\pm \! \tfrac{1}{2} \Xi^{\pm}(0),
\end{align}
\begin{align}
\Xi^{+}(z) \! =& \, \dfrac{1}{\pi} \! \left(\int\nolimits_{-\infty}^{0} \! + 
\! \int\nolimits_{\lambda_{2}}^{\lambda_{1}} \right) \! \ln \! \vert \mu \! 
- \! z \vert \md \ln (1 \! - \! \vert r(\mu) \vert^{2}), \\
\Xi^{-}(z) \! =& \, \dfrac{1}{\pi} \! \left(\int\nolimits_{0}^{\lambda_{2}} 
\! + \! \int_{\lambda_{1}}^{+\infty} \right) \! \ln \! \vert \mu \! - \! 
z \vert \md \ln (1 \! - \! \vert r(\mu) \vert^{2}),
\end{align}
$\sum_{k \in J^{+}} \! := \! \sum_{k=m+1}^{N}$, $\sum_{k \in J^{-}} \! := \! 
\sum_{k=1}^{m-1}$, $\Gamma (\cdot)$ is the gamma function {\rm \cite{a51}}, 
and
\begin{equation}
\widetilde{u}_{\mathcal{SC}}(x,t) \! = \! \sum_{k=1}^{7} \widetilde{u}_{
\mathcal{SC}}^{(k)}(x,t),
\end{equation}
with
\begin{align*}
\widetilde{u}_{\mathcal{SC}}^{(1)}(x,t) &= -2 \mi \varepsilon_{b} 
\widetilde{\varepsilon}_{\mathscr{P}} \csc (\phi_{m}) \sin (s^{\pm}) \cos 
(\Theta^{\pm}(z_{o},t) \! \pm \! (2 \! \mp \! 1) \tfrac{\pi}{4}) \sinh 
(\Omega^{\pm}(x,t)), \\
\widetilde{u}_{\mathcal{SC}}^{(2)}(x,t) &= 2 \mi \varepsilon_{b} 
\widetilde{\varepsilon}_{\mathscr{P}} \! \left(\cos (\phi_{m}) \me^{\mi 
s^{\pm}} \! + \! 2 \sin (\phi_{m}) \sin (s^{\pm}) \right) \! \cos 
(\Theta^{\pm}(z_{o},t) \! \pm \! (2 \! \mp \! 1) \tfrac{\pi}{4}) \me^{
\Omega^{\pm}(x,t)}, \\
\widetilde{u}_{\mathcal{SC}}^{(3)}(x,t) &= \dfrac{4 \mi \varepsilon_{b} 
\widetilde{\varepsilon}_{\mathscr{P}} \lambda_{1}^{2} \sin (\phi_{m}) \sin 
(s^{\pm}) \me^{\Omega^{\pm}(x,t)}}{(\lambda_{1}^{2} \! - \! 2 \lambda_{1} 
\cos (\phi_{m}) \! + \! 1)^{2}} \! \left(((\lambda_{1} \! + \! \lambda_{2}) 
\cos (\phi_{m}) \! - \! 2) \cos (\Theta^{\pm}(z_{o},t) \right. \\
&\left. \pm \, (2 \! \mp \! 1) \tfrac{\pi}{4}) \! \pm \! (\lambda_{1} \! - 
\! \lambda_{2}) \sin (\phi_{m}) \sin (\Theta^{\pm}(z_{o},t) \! \pm \! (2 \! 
\mp \! 1) \tfrac{\pi}{4}) \right),
\end{align*}
\begin{align*}
\widetilde{u}_{\mathcal{SC}}^{(4)}(x,t) &= \dfrac{2 \mi \varepsilon_{
b} \widetilde{\varepsilon}_{\mathscr{P}} \lambda_{1} \cos (\phi_{m}) \me^{
\Omega^{\pm}(x,t)}}{(\lambda_{1}^{2} \! - \! 2 \lambda_{1} \cos (\phi_{m}) 
\! + \! 1)} \! \left(2 \cos (s^{\pm} \! - \! \phi_{m}) \cos (\Theta^{\pm}
(z_{o},t) \! \pm \! (2 \! \mp \! 1) \tfrac{\pi}{4}) \! - \! (\lambda_{1} \! 
+ \! \lambda_{2}) \right. \nonumber \\
&\times \left. \cos (s^{\pm}) \cos (\Theta^{\pm}(z_{o},t) \! \pm \! (2 \! 
\mp \! 1) \tfrac{\pi}{4}) \! \mp \! (\lambda_{1} \! - \! \lambda_{2}) \sin 
(s^{\pm}) \sin (\Theta^{\pm}(z_{o},t) \! \pm \! (2 \! \mp \! 1) \tfrac{
\pi}{4}) \right), \\
\widetilde{u}_{\mathcal{SC}}^{(5)}(x,t) &= -\dfrac{4 \varepsilon_{b} 
\widetilde{\varepsilon}_{\mathscr{P}} \sin (\phi_{m}) \me^{\Omega^{\pm}(x,
t)}}{(1 \! - \! \me^{2 \Omega^{\pm}(x,t)})} \cos (\Theta^{\pm}(z_{o},t) \! 
\pm \! (2 \! \mp \! 1) \tfrac{\pi}{4}) \! \left(\me^{-\mi s^{\pm}} \! + \! 
\cos (s^{\pm} \! - \! \phi_{m}) \me^{-\mi \phi_{m}+2 \Omega^{\pm}(x,t)} 
\right), \\
\widetilde{u}_{\mathcal{SC}}^{(6)}(x,t) &= \dfrac{4 \widetilde{
\varepsilon}_{\mathscr{P}} \lambda_{1} \sin (\phi_{m}) \me^{\Omega^{\pm}
(x,t)}}{(1 \! - \! \me^{2 \Omega^{\pm}(x,t)})(\lambda_{1}^{2} \! - \! 2 
\lambda_{1} \cos (\phi_{m}) \! + \! 1)} \! \left(\! \left(\me^{\Omega^{\pm}
(x,t)}(1 \! + \! \varepsilon_{b} \widetilde{\varepsilon}_{\mathscr{P}} \cos 
(\phi_{m}) \me^{-\mi \phi_{m}+\Omega^{\pm}(x,t)}) \right) \right. \\
&\left. \times \left(-2 \widetilde{\varepsilon}_{\mathscr{P}} \cos (s^{\pm} 
\! - \! \phi_{m}) \cos (\Theta^{\pm}(z_{o},t) \! \pm \! (2 \! \mp \! 1) 
\tfrac{\pi}{4}) \! + \! \widetilde{\varepsilon}_{\mathscr{P}}(\lambda_{1} \! 
+ \! \lambda_{2}) \cos (s^{\pm}) \cos (\Theta^{\pm}(z_{o},t) \right. \right. 
\\
&\left. \left. \pm \, (2 \! \mp \! 1) \tfrac{\pi}{4}) \! \pm \! \widetilde{
\varepsilon}_{\mathscr{P}}(\lambda_{1} \! - \! \lambda_{2}) \sin (s^{\pm}) 
\sin (\Theta^{\pm}(z_{o},t) \! \pm \! (2 \! \mp \! 1) \tfrac{\pi}{4}) \right) 
\! + \! \left(1 \! - \! \varepsilon_{b} \widetilde{\varepsilon}_{\mathscr{P}} 
\me^{\Omega^{\pm}(x,t)} \right) \! \left(2 \mi \varepsilon_{b} \right. 
\right. \\
&\left. \left. \times \sin (s^{\pm} \! - \! \phi_{m}) \cos (\Theta^{\pm}(z_{
o},t) \! \pm \! (2 \! \mp \! 1) \tfrac{\pi}{4}) \! - \! \mi \varepsilon_{b}
(\lambda_{1} \! + \! \lambda_{2}) \sin (s^{\pm}) \cos (\Theta^{\pm}(z_{o},t) 
\! \pm \! (2 \! \mp \! 1) \tfrac{\pi}{4}) \right. \right. \\
&\left. \left. \pm \, \mi \varepsilon_{b}(\lambda_{1} \! - \! \lambda_{2}) 
\cos (s^{\pm}) \sin (\Theta^{\pm}(z_{o},t) \! \pm \! (2 \! \mp \! 1) \tfrac{
\pi}{4}) \right) \right), \\
\widetilde{u}_{\mathcal{SC}}^{(7)}(x,t) &= -\dfrac{8 \lambda_{1}^{2} \sin^{2}
(\phi_{m}) \me^{-\mi \phi_{m}+2 \Omega^{\pm}(x,t)}}{(1 \! - \! \me^{2 
\Omega^{\pm}(x,t)})(\lambda_{1}^{2} \! - \! 2 \lambda_{1} \cos (\phi_{m}) 
\! + \! 1)^{2}} \! \left(((\lambda_{1} \! + \! \lambda_{2}) \cos (\phi_{m}) 
\! - \! 2) \cos (\Theta^{\pm}(z_{o},t) \! \pm \! (2 \! \mp \! 1) \tfrac{
\pi}{4}) \right. \\
&\left. \pm \, (\lambda_{1} \! - \! \lambda_{2}) \sin (\phi_{m}) \sin 
(\Theta^{\pm}(z_{o},t) \! \pm \! (2 \! \mp \! 1) \tfrac{\pi}{4}) \right) \! 
\! \left( \! (1 \! + \! \varepsilon_{b} \widetilde{\varepsilon}_{\mathscr{
P}} \me^{\Omega^{\pm}(x,t)}) \sin (s^{\pm}) \! + \! \mi \! \left(\tfrac{1-
\varepsilon_{b} \widetilde{\varepsilon}_{\mathscr{P}} \me^{\Omega^{\pm}(x,
t)}}{1+\varepsilon_{b} \widetilde{\varepsilon}_{\mathscr{P}} \me^{\Omega^{
\pm}(x,t)}} \right) \! \right. \\
&\left. \times \cos (s^{\pm}) \right).
\end{align*}

For the conditions stated in the formulation of the Theorem, as $t \! \to \! 
\pm \infty$ and $x \! \to \! \pm \infty$ such that $z_{o} \! > \! 2$ and 
$(x,t) \! \in \! \{\mathstrut (x,t); \, x \! + \! 2t \cos (\phi_{m}) \! = \! 
\mathcal{O}(1), \, \phi_{m} \! \in \! (-\pi,0)\}$,
\begin{align}
u(x,t) \! &= \! -\me^{-\mi (\psi^{\pm}(1)+s^{\pm})} \! \left(\widehat{u}_{
\mathcal{S}}(x,t) \! + \! \dfrac{\sqrt{\nu (\aleph_{4})}}{\sqrt{\vert t 
\vert (\aleph_{3} \! - \! \aleph_{4})} \, (z_{o}^{2} \! + \! 32)^{1/4}} 
\! \left(\widehat{u}_{\mathcal{C}}(x,t) \! + \! \widehat{u}_{\mathcal{SC}}
(x,t) \right) \! \right. \nonumber \\
 &+ \! \left. \mathcal{O} \! \left( \! \left( \dfrac{c^{\mathcal{S}}(\aleph_{
3}) \underline{c}(\aleph_{4},\aleph_{1},\aleph_{2})}{\sqrt{\vert \aleph_{3} 
\vert (z_{o}^{2} \! + \! 32)}} \! + \! \dfrac{c^{\mathcal{S}}(\aleph_{4}) 
\underline{c}(\aleph_{3},\aleph_{1},\aleph_{2})}{\sqrt{\vert \aleph_{4} \vert 
(z_{o}^{2} \! + \! 32)}} \right) \! \dfrac{\ln \vert t \vert}{(\aleph_{3} 
\! - \! \aleph_{4}) t} \right) \! \right),
\end{align}
where
\begin{align}
\psi^{+}(j) \! =& \! \left(\int_{-\infty}^{\aleph_{4}} \! + \! \int_{\aleph_{
3}}^{0} \right) \! \dfrac{\ln (1 \! - \! \vert r(\mu) \vert^{2})}{\mu^{j}} \, 
\dfrac{\md \mu}{2 \pi}, \quad j \! \in \! \{0,1\}, \\
\psi^{-}(l) \! =& \! \left(\int_{\aleph_{4}}^{\aleph_{3}} \! + \! \int_{0}^{
+\infty} \right) \! \dfrac{\ln (1 \! - \! \vert r(\mu) \vert^{2})}{\mu^{l}} 
\, \dfrac{\md \mu}{2 \pi}, \quad l \! \in \! \{0,1\},
\end{align}
\begin{equation}
\aleph_{1} \! = \! -\tfrac{1}{2}(a_{1} \! - \! \mi (4 \! - \! a_{1}^{2})^{
1/2}), \qquad \, \aleph_{2} \! = \! \overline{\aleph_{1}}, \qquad \, 
\aleph_{3} \! = \! -\tfrac{1}{2}(a_{2} \! - \! (a_{2}^{2} \! - \! 4)^{1/2}), 
\qquad \, \aleph_{4} \! = \! \aleph_{3}^{-1},
\end{equation}
$\aleph_{4} \! < \! \aleph_{3} \! < \! 0$, $\vert \aleph_{1} \vert^{2} \! = 
\! 1$,
\begin{gather}
\widehat{u}_{\mathcal{S}}(x,t) \! = \! \dfrac{1 \! + \! \varepsilon_{b} 
\widehat{\varepsilon}_{\mathscr{P}} \me^{-2 \mi \phi_{m}+\mho^{\pm}(x,
t)}}{(1 \! + \! \varepsilon_{b} \widehat{\varepsilon}_{\mathscr{P}} \me^{
\mho^{\pm}(x,t)})}, \\
\widehat{\varepsilon}_{\mathscr{P}} \! = \! \mathrm{sgn} \! \left( \! \left(
\prod_{k=1}^{m-1} \dfrac{(-\sin (\frac{1}{2}(\phi_{m} \! + \! \phi_{k})))}{
\sin (\frac{1}{2}(\phi_{m} \! - \! \phi_{k}))} \right) \! \! \left(\prod_{k=
m+1}^{N} \dfrac{(-\sin (\frac{1}{2}(\phi_{m} \! + \! \phi_{k})))}{\sin (\frac{
1}{2}(\phi_{m} \! - \! \phi_{k}))} \right)^{-1} \right) \! = \! (-1)^{m-1}, \\
\mho^{\pm}(x,t) \! = \! -2 \sin (\phi_{m})(x \! + \! 2t \cos (\phi_{m}) \! - 
\! \widehat{x}^{\pm}_{m}),
\end{gather}
\begin{align}
\widehat{x}^{\pm}_{m} \! =& \, \dfrac{\ln (\vert \gamma_{m} \vert)}{2 \sin 
(\phi_{m})} \! \pm \! \sum_{k=1}^{N} \dfrac{\mathrm{sgn}(m \! - \! k)}{2 \sin 
(\phi_{m})} \ln \! \left( \left\vert \dfrac{\sin (\frac{1}{2}(\phi_{m} \! + 
\! \phi_{k}))}{\sin (\frac{1}{2}(\phi_{m} \! - \! \phi_{k}))} \right\vert 
\right) \nonumber \\
 \pm& \, \dfrac{1}{2} \! \left(\int\nolimits_{-\infty}^{\aleph_{4}} \! + \! 
\int\nolimits_{\aleph_{3}}^{0} \! - \! \int\nolimits_{\aleph_{4}}^{\aleph_{
3}} \! - \! \int\nolimits_{0}^{+\infty} \right) \! \dfrac{\ln (1 \! - \! 
\vert r(\mu) \vert^{2})}{(\mu^{2} \! + \! 2 \mu \cos (\phi_{m}) \! + \! 1)} 
\, \dfrac{\md \mu}{2 \pi},
\end{align}
\begin{gather}
\widehat{u}_{\mathcal{C}}(x,t) \! = \! \mi \me^{\mi s^{\pm}} \! \left(\aleph_{
3} \me^{\pm \mi \left(\Phi^{\pm}(z_{o},t) \pm (2 \mp 1) \frac{\pi}{4} 
\right)} \! + \! \aleph_{4} \me^{\mp \mi \left(\Phi^{\pm}(z_{o},t) \pm (2 
\mp 1) \frac{\pi}{4} \right)} \right),
\end{gather}
\begin{align}
\Phi^{\pm}(z_{o},t) \! =& \! \pm \arg r(\aleph_{4}) \! \pm \! 4 \sum_{k \in 
J^{\pm}} \arg (\aleph_{4} \! + \! \me^{\mi \phi_{k}}) \! - \! \arg \Gamma 
(\mi \nu (\aleph_{4})) \! \pm \! t(\aleph_{4} \! - \! \aleph_{3})(z_{o} \! + 
\! \aleph_{3} \! + \! \aleph_{4}) \nonumber \\
 +& \, \nu (\aleph_{4}) \ln \vert t \vert \! + \! 3 \nu (\aleph_{4}) \ln 
(\aleph_{3} \! - \! \aleph_{4}) \! + \! \tfrac{1}{2} \nu (\aleph_{4}) \ln \! 
\left(z_{o}^{2} \! + \! 32 \right) \! \mp \! \Lambda^{\pm}(\aleph_{4}) \! 
\pm \! \tfrac{1}{2} \Lambda^{\pm}(0),
\end{align}
\begin{align}
\Lambda^{+}(z) \! =& \, \dfrac{1}{\pi} \! \left(\int\nolimits_{-\infty}^{
\aleph_{4}} \! + \! \int\nolimits_{\aleph_{3}}^{0} \right) \! \ln \! \vert 
\mu \! - \! z \vert \md \ln (1 \! - \! \vert r(\mu) \vert^{2}), \\
\Lambda^{-}(z) \! =& \, \dfrac{1}{\pi} \! \left(\int\nolimits_{\aleph_{4}}^{
\aleph_{3}} \!+ \! \int_{0}^{+\infty} \right) \! \ln \! \vert \mu \! - \! z 
\vert \md \ln (1 \! - \! \vert r(\mu) \vert^{2}),
\end{align}
and
\begin{equation}
\widehat{u}_{\mathcal{SC}}(x,t) \! = \! \sum_{k=1}^{7} \widehat{u}_{
\mathcal{SC}}^{(k)}(x,t),
\end{equation}
with
\begin{align*}
\widehat{u}_{\mathcal{SC}}^{(1)}(x,t) &= 2 \mi \varepsilon_{b} \widehat{
\varepsilon}_{\mathscr{P}} \csc (\phi_{m}) \sin (s^{\pm}) \cos (\Phi^{\pm}
(z_{o},t) \! \pm \! (2 \! \mp \! 1) \tfrac{\pi}{4}) \sinh (\mho^{\pm}(x,t)), 
\\
\widehat{u}_{\mathcal{SC}}^{(2)}(x,t) &= -2 \mi \varepsilon_{b} 
\widehat{\varepsilon}_{\mathscr{P}} \! \left(\cos (\phi_{m}) \me^{\mi 
s^{\pm}} \! + \! 2 \sin (\phi_{m}) \sin (s^{\pm}) \right) \! \cos (\Phi^{\pm}
(z_{o},t) \! \pm \! (2 \! \mp \! 1) \tfrac{\pi}{4}) \me^{\mho^{\pm}(x,t)}, \\
\widehat{u}_{\mathcal{SC}}^{(3)}(x,t) &= \dfrac{4 \mi \varepsilon_{b} 
\widehat{\varepsilon}_{\mathscr{P}} \aleph_{4}^{2} \sin (\phi_{m}) \sin 
(s^{\pm}) \me^{\mho^{\pm}(x,t)}}{(\aleph_{4}^{2} \! + \! 2 \aleph_{4} \cos 
(\phi_{m}) \! + \! 1)^{2}} \! \left(((\aleph_{4} \! + \! \aleph_{3}) \cos 
(\phi_{m}) \! + \! 2) \cos (\Phi^{\pm}(z_{o},t) \right. \\
&\left. \pm \, (2 \! \mp \! 1) \tfrac{\pi}{4}) \! \pm \! (\aleph_{4} \! - \! 
\aleph_{3}) \sin (\phi_{m}) \sin (\Phi^{\pm}(z_{o},t) \! \pm \! (2 \! \mp 
\! 1) \tfrac{\pi}{4}) \right), \\
\widehat{u}_{\mathcal{SC}}^{(4)}(x,t) &= \dfrac{2 \mi \varepsilon_{b} 
\widehat{\varepsilon}_{\mathscr{P}} \aleph_{4} \cos (\phi_{m}) \me^{\mho^{
\pm}(x,t)}}{(\aleph_{4}^{2} \! + \! 2 \aleph_{4} \cos (\phi_{m}) \! + \! 1)} 
\! \left(2 \cos (s^{\pm} \! - \! \phi_{m}) \cos (\Phi^{\pm}(z_{o},t) \! \pm 
\! (2 \! \mp \! 1) \tfrac{\pi}{4}) \! + \! (\aleph_{4} \! + \! \aleph_{3}) 
\cos (s^{\pm}) \right. \nonumber \\
&\times \left. \cos (\Phi^{\pm}(z_{o},t) \! \pm \! (2 \! \mp \! 1) \tfrac{
\pi}{4}) \! \pm \! (\aleph_{4} \! - \! \aleph_{3}) \sin (s^{\pm}) \sin 
(\Phi^{\pm}(z_{o},t) \! \pm \! (2 \! \mp \! 1) \tfrac{\pi}{4}) \right), \\
\widehat{u}_{\mathcal{SC}}^{(5)}(x,t) &= \dfrac{4 \varepsilon_{b} 
\widehat{\varepsilon}_{\mathscr{P}} \sin (\phi_{m}) \me^{\mho^{\pm}(x,t)}}{
(1 \! - \! \me^{2 \mho^{\pm}(x,t)})} \cos (\Phi^{\pm}(z_{o},t) \! \pm \! (2 
\! \mp \! 1) \tfrac{\pi}{4}) \! \left(\me^{-\mi s^{\pm}} \! + \! \cos (s^{
\pm} \! - \! \phi_{m}) \me^{-\mi \phi_{m}+2 \mho^{\pm}(x,t)} \right), \\
\widehat{u}_{\mathcal{SC}}^{(6)}(x,t) &= -\dfrac{4 \widehat{
\varepsilon}_{\mathscr{P}} \aleph_{4} \sin (\phi_{m}) \me^{\mho^{\pm}(x,t)}}
{(1 \! - \! \me^{2 \mho^{\pm}(x,t)})(\aleph_{4}^{2} \! + \! 2 \aleph_{4} \cos 
(\phi_{m}) \! + \! 1)} \! \left(\! \left(\me^{\mho^{\pm}(x,t)}(1 \! + \! 
\varepsilon_{b} \widehat{\varepsilon}_{\mathscr{P}} \cos (\phi_{m}) \me^{-
\mi \phi_{m}+\mho^{\pm}(x,t)}) \! \right) \right. \\
&\left. \times \left(2 \widehat{\varepsilon}_{\mathscr{P}} \cos (s^{\pm} \! 
- \! \phi_{m}) \cos (\Phi^{\pm}(z_{o},t) \! \pm \! (2 \! \mp \! 1) \tfrac{
\pi}{4}) \! + \! \widehat{\varepsilon}_{\mathscr{P}}(\aleph_{4} \! + \! 
\aleph_{3}) \cos (s^{\pm}) \cos (\Phi^{\pm}(z_{o},t) \right. \right. \\
&\left. \left. \pm \, (2 \! \mp \! 1) \tfrac{\pi}{4}) \! \pm \! \widehat{
\varepsilon}_{\mathscr{P}}(\aleph_{4} \! - \! \aleph_{3}) \sin (s^{\pm}) 
\sin (\Phi^{\pm}(z_{o},t) \! \pm \! (2 \! \mp \! 1) \tfrac{\pi}{4}) \right) 
\! - \! \left(1 \! - \! \varepsilon_{b} \widehat{\varepsilon}_{\mathscr{P}} 
\me^{\mho^{\pm}(x,t)} \right) \! \left(2 \mi \varepsilon_{b} \right. 
\right. \\
&\left. \left. \times \sin (s^{\pm} \! - \! \phi_{m}) \cos (\Phi^{\pm}(z_{
o},t) \! \pm \! (2 \! \mp \! 1) \tfrac{\pi}{4}) \! + \! \mi \varepsilon_{b}
(\aleph_{4} \! + \! \aleph_{3}) \sin (s^{\pm}) \cos (\Phi^{\pm}(z_{o},t) \! 
\pm \! (2 \! \mp \! 1) \tfrac{\pi}{4}) \right. \right. \\
&\left. \left. \mp \, \mi \varepsilon_{b}(\aleph_{4} \! - \! \aleph_{3}) 
\cos (s^{\pm}) \sin (\Phi^{\pm}(z_{o},t) \! \pm \! (2 \! \mp \! 1) \tfrac{
\pi}{4}) \right) \right), \\
\widehat{u}_{\mathcal{SC}}^{(7)}(x,t) &= -\dfrac{8 \aleph_{4}^{2} 
\sin^{2}(\phi_{m}) \me^{-\mi \phi_{m}+2 \mho^{\pm}(x,t)}}{(1 \! - \! \me^{
2 \mho^{\pm}(x,t)})(\aleph_{4}^{2} \! + \! 2 \aleph_{4} \cos (\phi_{m}) \! 
+ \! 1)^{2}} \! \left(((\aleph_{4} \! + \! \aleph_{3}) \cos (\phi_{m}) \! + 
\! 2) \cos (\Phi^{\pm}(z_{o},t) \! \pm \! (2 \! \mp \! 1) \tfrac{\pi}{4}) 
\right. \\
&\left. \pm \, (\aleph_{4} \! - \! \aleph_{3}) \sin (\phi_{m}) \sin (\Phi^{
\pm}(z_{o},t) \! \pm \! (2 \! \mp \! 1) \tfrac{\pi}{4}) \right) \! \! \left(
(1 \! + \! \varepsilon_{b} \widehat{\varepsilon}_{\mathscr{P}} \me^{\mho^{
\pm}(x,t)}) \sin (s^{\pm}) \! + \! \mi \! \left(\tfrac{1-\varepsilon_{b} 
\widehat{\varepsilon}_{\mathscr{P}} \me^{\mho^{\pm}(x,t)}}{1+\varepsilon_{b} 
\widehat{\varepsilon}_{\mathscr{P}} \me^{\mho^{\pm}(x,t)}} \right) \right. \\
&\left. \times \cos (s^{\pm}) \right).
\end{align*}
\end{dddd}
\begin{dddd}
For $r(\zeta) \! \in \! \mathcal{S}_{\mathbb{C}}^{1}(\mathbb{R})$, let 
$m(x,t;\zeta)$ be the solution of the Riemann-Hilbert problem formulated 
in Lemma~{\rm 2.1.2}. Let $u(x,t)$, the solution of the Cauchy problem for 
the {\rm D${}_{f}$NLSE} with finite-density initial data $u(x,0) \! := \! 
u_{o}(x) \! =_{x \to \pm \infty} \! u_{o}(\pm \infty)(1 \! + \! o(1))$, 
where $u_{o}(\pm \infty) \! := \! \exp (\tfrac{\mi (1 \mp 1) \theta}{2})$, 
$0 \! \leqslant \! \theta \! = \! -2 \sum_{n=1}^{N} \sin (\phi_{n}) \! - 
\! \int_{-\infty}^{+\infty} \tfrac{\ln (1-\vert r(\mu) \vert^{2})}{\mu} \, 
\tfrac{\md \mu}{2 \pi} \! < \! 2 \pi$, $u_{o}(x) \! \in \! \mathbf{C}^{
\infty}(\mathbb{R})$, and $u_{o}(x) \! - \! u_{o}(\pm \infty) \! \in \! 
\mathcal{S}_{\mathbb{C}}(\mathbb{R}_{\pm})$, be defined by Eq.~{\rm (5)}, 
and $\int_{+\infty}^{x}(\vert u(x^{\prime},t) \vert^{2} \! - \! 1) \, \md 
x^{\prime}$ be defined by Eq.~{\rm (6)}. Let $\epsilon \! \in \! \{\pm 1\}$. 
Then, for $\theta_{\gamma_{m}} \! = \! \varepsilon_{b} \pi/2$, $\varepsilon_{
b} \! \in \! \{\pm 1\}$, $m \! \in \! \{1,2,\ldots,N\}$, as $t \! \to \! 
\pm \infty$ and $x \! \to \! \mp \infty$ such that $z_{o} \! < \! -2$ and 
$(x,t) \! \in \! \{\mathstrut (x,t); \, x \! + \! 2t \cos (\phi_{m}) \! = 
\! \mathcal{O}(1), \, \phi_{m} \! \in \! (0,\pi)\}$,
\begin{align}
\int\nolimits_{\mathrm{sgn}(\epsilon) \infty}^{x}(\vert u(x^{\prime},t) 
\vert^{2} \! - \! 1) \, \md x^{\prime} \! &= \widetilde{\mathscr{S}}^{\pm}_{
\epsilon} \! + \! \widetilde{\mathscr{H}}^{\pm}_{\epsilon} \! + \! 
\widetilde{\mathscr{E}}_{\mathcal{S}}(x,t) \! + \! \dfrac{\sqrt{\nu 
(\lambda_{1})}}{\sqrt{\vert t \vert (\lambda_{1} \! - \! \lambda_{2})} \, 
(z_{o}^{2} \! + \! 32)^{1/4}} \! \left(\widetilde{\mathscr{E}}_{\mathcal{
C}}(x,t) \! + \! \widetilde{\mathscr{E}}_{\mathcal{SC}}(x,t) \right) 
\nonumber \\
 &+\mathcal{O} \! \left( \! \left( \dfrac{c^{\mathcal{S}}(\lambda_{1}) 
\underline{c}(\lambda_{2},\lambda_{3},\lambda_{4})}{\sqrt{\lambda_{1}(z_{
o}^{2} \! + \! 32)}} \! + \! \dfrac{c^{\mathcal{S}}(\lambda_{2}) \underline{
c}(\lambda_{1},\lambda_{3},\lambda_{4})}{\sqrt{\lambda_{2}(z_{o}^{2} \! + \! 
32)}} \right) \! \dfrac{\ln \vert t \vert}{(\lambda_{1} \! - \! \lambda_{2}) 
t} \right),
\end{align}
where
\begin{align}
\widetilde{\mathscr{S}}^{+}_{\epsilon} \! &= \! 
\begin{cases}
2 \sum_{k=m+1}^{N} \sin (\phi_{k}), &\text{$\epsilon \! = \! +1$,} \\
-2 \sum_{k=1}^{m} \sin (\phi_{k}), &\text{$\epsilon \! = \! -1$,}
\end{cases} & \widetilde{\mathscr{S}}^{-}_{\epsilon} \! &= \! 
\begin{cases}
2 \sum_{k=1}^{m-1} \sin (\phi_{k}), &\text{$\epsilon \! = \! +1$,} \\
-2 \sum_{k=m}^{N} \sin (\phi_{k}), &\text{$\epsilon \! = \! -1$,}
\end{cases} \\
\widetilde{\mathscr{H}}^{+}_{\epsilon} \! &= \! 
\begin{cases}
\theta^{+}(0), &\text{$\epsilon \! = \! +1$,} \\
-\theta^{-}(0), &\text{$\epsilon \! = \! -1$,}
\end{cases} & \widetilde{\mathscr{H}}^{-}_{\epsilon} \! &= \! 
\begin{cases}
\theta^{-}(0), &\text{$\epsilon \! = \! +1$,} \\
-\theta^{+}(0), &\text{$\epsilon \! = \! -1$,}
\end{cases}
\end{align}
\begin{align}
\widetilde{\mathscr{E}}_{\mathcal{S}}(x,t) \! = \! \dfrac{2 \varepsilon_{b} 
\widetilde{\varepsilon}_{\mathscr{P}} \sin (\phi_{m}) \me^{\Omega^{\pm}(x,
t)}}{(1 \! + \! \varepsilon_{b} \widetilde{\varepsilon}_{\mathscr{P}} \me^{
\Omega^{\pm}(x,t)})},
\end{align}
\begin{align}
\widetilde{\mathscr{E}}_{\mathcal{C}}(x,t) \! = \! -2 \cos (s^{\pm}) \cos 
\! \left(\Theta^{\pm}(z_{o},t) \! \pm \! (2 \! \mp \! 1) \tfrac{\pi}{4} 
\right),
\end{align}
and
\begin{equation}
\widetilde{\mathscr{E}}_{\mathcal{SC}}(x,t) \! = \! \sum_{k=1}^{7} 
\widetilde{\mathscr{E}}_{\mathcal{SC}}^{(k)}(x,t),
\end{equation}
with
\begin{align*}
\widetilde{\mathscr{E}}_{\mathcal{SC}}^{(1)}(x,t) &= \dfrac{8 \lambda_{
1}^{2} \sin^{2}(\phi_{m}) \cos (s^{\pm}) \me^{2 \Omega^{\pm}(x,t)}}{(1 
\! + \! \varepsilon_{b} \widetilde{\varepsilon}_{\mathscr{P}} \me^{
\Omega^{\pm}(x,t)})^{2}(\lambda_{1}^{2} \! - \! 2 \lambda_{1} \cos 
(\phi_{m}) \! + \! 1)^{2}} \! \left(((\lambda_{1} \! + \! \lambda_{2}) 
\cos (\phi_{m}) \! - \! 2) \cos (\Theta^{\pm}(z_{o},t) \! \pm \! (2 \! 
\mp \! 1) \tfrac{\pi}{4}) \right. \\
&\left. \pm \, (\lambda_{1} \! - \! \lambda_{2}) \sin (\phi_{m}) \sin 
(\Theta^{\pm}(z_{o},t) \! \pm \! (2 \! \mp \! 1) \tfrac{\pi}{4}) \right), \\
\widetilde{\mathscr{E}}_{\mathcal{SC}}^{(2)}(x,t) &= \dfrac{4 
\lambda_{1} \widetilde{\varepsilon}_{\mathscr{P}} \sin (\phi_{m}) \me^{
\Omega^{\pm}(x,t)}}{(1 \! - \! \me^{2 \Omega^{\pm}(x,t)})(\lambda_{1}^{2} 
\! - \! 2 \lambda_{1} \cos (\phi_{m}) \! + \! 1)} \! \left(2(\widetilde{
\varepsilon}_{\mathscr{P}} \cos (\phi_{m}) \sin (s^{\pm} \! - \! \phi_{m}) 
\me^{\Omega^{\pm}(x,t)} \! - \! \varepsilon_{b} \sin (s^{\pm})) \right. \\
&\left. \times \cos (\Theta^{\pm}(z_{o},t) \! \pm \! (2 \! \mp \! 1) 
\tfrac{\pi}{4}) \! - \! (\lambda_{1} \! + \! \lambda_{2})(\widetilde{
\varepsilon}_{\mathscr{P}} \cos (\phi_{m}) \sin (s^{\pm}) \me^{\Omega^{\pm}
(x,t)} \! - \! \varepsilon_{b} \sin (s^{\pm} \! + \! \phi_{m})) \right. \\
&\left. \times \cos (\Theta^{\pm}(z_{o},t) \! \pm \! (2 \! \mp \! 1) 
\tfrac{\pi}{4}) \! \pm \! (\lambda_{1} \! - \! \lambda_{2})(\widetilde{
\varepsilon}_{\mathscr{P}} \! \cos (\phi_{m}) \! \cos (s^{\pm}) \me^{
\Omega^{\pm}(x,t)} \! - \! \varepsilon_{b} \! \cos (s^{\pm} \! + \! 
\phi_{m})) \right. \\
&\left. \times \sin (\Theta^{\pm}(z_{o},t) \! \pm \! (2 \! \mp \! 1) 
\tfrac{\pi}{4}) \right), \\
\widetilde{\mathscr{E}}_{\mathcal{SC}}^{(3)}(x,t) &= \dfrac{2 
\varepsilon_{b} \widetilde{\varepsilon}_{\mathscr{P}} \lambda_{1} \cos 
(\phi_{m}) \me^{\Omega^{\pm}(x,t)}}{(\lambda_{1}^{2} \! - \! 2 \lambda_{1} 
\cos (\phi_{m}) \! + \! 1)} \! \left(2 \cos (s^{\pm} \! - \! \phi_{m}) \cos 
(\Theta^{\pm}(z_{o},t) \! \pm \! (2 \! \mp \! 1) \tfrac{\pi}{4}) \! - \! 
(\lambda_{1} \! + \! \lambda_{2}) \cos (s^{\pm}) \right. \\
&\left. \times \cos (\Theta^{\pm}(z_{o},t) \! \pm \! (2 \! \mp \! 1) 
\tfrac{\pi}{4}) \! \mp \! (\lambda_{1} \! - \! \lambda_{2}) \sin (s^{\pm}) 
\sin (\Theta^{\pm}(z_{o},t) \! \pm \! (2 \! \mp \! 1) \tfrac{\pi}{4}) 
\right), \\
\widetilde{\mathscr{E}}_{\mathcal{SC}}^{(4)}(x,t) &= \dfrac{4 
\varepsilon_{b} \widetilde{\varepsilon}_{\mathscr{P}} \lambda_{1}^{2} \sin 
(\phi_{m}) \sin (s^{\pm}) \me^{\Omega^{\pm}(x,t)}}{(\lambda_{1}^{2} \! - \! 
2 \lambda_{1} \cos (\phi_{m}) \! + \! 1)^{2}} \! \left(((\lambda_{1} \! + \! 
\lambda_{2}) \cos (\phi_{m}) \! - \! 2) \cos (\Theta^{\pm}(z_{o},t) \! \pm \! 
(2 \! \mp \! 1) \tfrac{\pi}{4}) \right. \\
&\left. \pm \, (\lambda_{1} \! - \! \lambda_{2}) \sin (\phi_{m}) \sin 
(\Theta^{\pm}(z_{o},t) \! \pm \! (2 \! \mp \! 1) \tfrac{\pi}{4}) \right), \\
\widetilde{\mathscr{E}}_{\mathcal{SC}}^{(5)}(x,t) &= -2 \varepsilon_{
b} \widetilde{\varepsilon}_{\mathscr{P}} \csc (\phi_{m}) \sin (s^{\pm}) \cos 
(\Theta^{\pm}(z_{o},t) \! \pm \! (2 \! \mp \! 1) \tfrac{\pi}{4}) \sinh 
(\Omega^{\pm}(x,t)), \\
\widetilde{\mathscr{E}}_{\mathcal{SC}}^{(6)}(x,t) &= \dfrac{4 \sin 
(\phi_{m}) \sin (s^{\pm} \! - \! \phi_{m})}{(1 \! - \! \me^{2 \Omega^{\pm}
(x,t)})} \cos (\Theta^{\pm}(z_{o},t) \! \pm \! (2 \! \mp \! 1) \tfrac{\pi}
{4}) \me^{2 \Omega^{\pm}(x,t)}, \\
\widetilde{\mathscr{E}}_{\mathcal{SC}}^{(7)}(x,t) &= 2 \varepsilon_{b} 
\widetilde{\varepsilon}_{\mathscr{P}} \cos (s^{\pm} \! - \! \phi_{m}) 
\cos (\Theta^{\pm}(z_{o},t) \! \pm \! (2 \! \mp \! 1) \tfrac{\pi}{4}) 
\me^{\Omega^{\pm}(x,t)},
\end{align*}
and $\theta^{\pm}(\cdot)$, $\{\lambda_{n}\}_{n=1}^{4}$, $s^{\pm}$ and $\nu 
(\cdot)$, $\widetilde{\varepsilon}_{\mathscr{P}}$, $\Omega^{\pm}(x,t)$, 
and $\Theta^{\pm}(z_{o},t)$ given in Theorem~{\rm 2.2.1}, Eqs.~{\rm (8)--(9)}, 
{\rm (10)}, {\rm (11)}, {\rm (13)}, {\rm (14)--(15)}, and {\rm (17)--(19)}, 
respectively.

For the conditions stated in the formulation of the Theorem, as $t \! \to 
\! \pm \infty$ and $x \! \to \! \pm \infty$ such that $z_{o} \! > \! 2$ and 
$(x,t) \! \in \! \{\mathstrut (x,t); \, x \! + \! 2t \cos (\phi_{m}) \! = \! 
\mathcal{O}(1), \, \phi_{m} \! \in \! (-\pi,0)\}$,
\begin{align}
\int\nolimits_{\mathrm{sgn}(\epsilon) \infty}^{x}(\vert u(x^{\prime},t) 
\vert^{2} \! - \! 1) \, \md x^{\prime} \! &= \widehat{\mathscr{S}}^{\pm}_{
\epsilon} \! + \! \widehat{\mathscr{H}}^{\pm}_{\epsilon} \! + \! \widehat{
\mathscr{E}}_{\mathcal{S}}(x,t) \! + \! \dfrac{\sqrt{\nu (\aleph_{4})}}{
\sqrt{\vert t \vert (\aleph_{3} \! - \! \aleph_{4})} \, (z_{o}^{2} \! + \! 
32)^{1/4}} \! \left(\widehat{\mathscr{E}}_{\mathcal{C}}(x,t) \! + \! 
\widehat{\mathscr{E}}_{\mathcal{SC}}(x,t) \right) \nonumber \\
 &+\mathcal{O} \! \left( \! \left( \dfrac{c^{\mathcal{S}}(\aleph_{3}) 
\underline{c}(\aleph_{4},\aleph_{1},\aleph_{2})}{\sqrt{\vert \aleph_{3} 
\vert (z_{o}^{2} \! + \! 32)}} \! + \! \dfrac{c^{\mathcal{S}}(\aleph_{4}) 
\underline{c}(\aleph_{3},\aleph_{1},\aleph_{2})}{\sqrt{\vert \aleph_{4} 
\vert (z_{o}^{2} \! + \! 32)}} \right) \! \dfrac{\ln \vert t \vert}{
(\aleph_{3} \! - \! \aleph_{4}) t} \right),
\end{align}
where
\begin{align}
\widehat{\mathscr{S}}^{+}_{\epsilon} \! &= \! 
\begin{cases}
-2 \sum_{k=1}^{m} \sin (\phi_{k}), &\text{$\epsilon \! = \! +1$,} \\
2 \sum_{k=m+1}^{N} \sin (\phi_{k}), &\text{$\epsilon \! = \! -1$,}
\end{cases} & \widehat{\mathscr{S}}^{-}_{\epsilon} \! &= \! 
\begin{cases}
-2 \sum_{k=m}^{N} \sin (\phi_{k}), &\text{$\epsilon \! = \! +1$,} \\
2 \sum_{k=1}^{m-1} \sin (\phi_{k}), &\text{$\epsilon \! = \! -1$,}
\end{cases} \\
\widehat{\mathscr{H}}^{+}_{\epsilon} \! &= \! 
\begin{cases}
\psi^{-}(0), &\text{$\epsilon \! = \! +1$,} \\
-\psi^{+}(0), &\text{$\epsilon \! = \! -1$,}
\end{cases} & \widehat{\mathscr{H}}^{-}_{\epsilon} \! &= \! 
\begin{cases}
\psi^{+}(0), &\text{$\epsilon \! = \! +1$,} \\
-\psi^{-}(0), &\text{$\epsilon \! = \! -1$,}
\end{cases}
\end{align}
\begin{align}
\widehat{\mathscr{E}}_{\mathcal{S}}(x,t) \! = \! \dfrac{2 \varepsilon_{b} 
\widehat{\varepsilon}_{\mathscr{P}} \sin (\phi_{m}) \me^{\mho^{\pm}(x,t)}}
{(1 \! + \! \varepsilon_{b} \widehat{\varepsilon}_{\mathscr{P}} \me^{\mho^{
\pm}(x,t)})},
\end{align}
\begin{align}
\widehat{\mathscr{E}}_{\mathcal{C}}(x,t) \! = \! 2 \cos (s^{\pm}) \cos \! 
\left(\Phi^{\pm}(z_{o},t) \! \pm \! (2 \! \mp \! 1) \tfrac{\pi}{4} \right),
\end{align}
and
\begin{equation}
\widehat{\mathscr{E}}_{\mathcal{SC}}(x,t) \! = \! \sum_{k=1}^{7} \widehat{
\mathscr{E}}_{\mathcal{SC}}^{(k)}(x,t),
\end{equation}
with
\begin{align*}
\widehat{\mathscr{E}}_{\mathcal{SC}}^{(1)}(x,t) &= \dfrac{8 \aleph_{4}^{2} 
\sin^{2}(\phi_{m}) \cos (s^{\pm}) \me^{2 \mho^{\pm}(x,t)}}{(1 \! + \! 
\varepsilon_{b} \widehat{\varepsilon}_{\mathscr{P}} \me^{\mho^{\pm}(x,t)})^{
2}(\aleph_{4}^{2} \! + \! 2 \aleph_{4} \cos (\phi_{m}) \! + \! 1)^{2}} \! 
\left(((\aleph_{4} \! + \! \aleph_{3}) \cos (\phi_{m}) \! + \! 2) \cos 
(\Phi^{\pm}(z_{o},t) \! \pm \! (2 \! \mp \! 1) \tfrac{\pi}{4}) \right. \\
&\left. \pm \, (\aleph_{4} \! - \! \aleph_{3}) \sin (\phi_{m}) \sin (\Phi^{
\pm}(z_{o},t) \! \pm \! (2 \! \mp \! 1) \tfrac{\pi}{4}) \right), \\
\widehat{\mathscr{E}}_{\mathcal{SC}}^{(2)}(x,t) &= \dfrac{4 \aleph_{4} 
\widehat{\varepsilon}_{\mathscr{P}} \sin (\phi_{m}) \me^{\mho^{\pm}(x,
t)}}{(1 \! - \! \me^{2 \mho^{\pm}(x,t)})(\aleph_{4}^{2} \! + \! 2 \aleph_{4} 
\cos (\phi_{m}) \! + \! 1)} \! \left(2(\widehat{\varepsilon}_{\mathscr{P}} 
\cos (\phi_{m}) \sin (s^{\pm} \! - \! \phi_{m}) \me^{\mho^{\pm}(x,t)} \! - 
\! \varepsilon_{b} \sin (s^{\pm})) \right. \\
&\left. \times \cos (\Phi^{\pm}(z_{o},t) \! \pm \! (2 \! \mp \! 1) \tfrac{
\pi}{4}) \! + \! (\aleph_{4} \! + \! \aleph_{3})(\widehat{\varepsilon}_{
\mathscr{P}} \cos (\phi_{m}) \! \sin (s^{\pm}) \me^{\mho^{\pm}(x,t)} \! - 
\! \varepsilon_{b} \sin (s^{\pm} \! + \! \phi_{m})) \right. \\
&\left. \times \cos (\Phi^{\pm}(z_{o},t) \! \pm \! (2 \! \mp \! 1) \tfrac{
\pi}{4}) \! \mp \! (\aleph_{4} \! - \! \aleph_{3})(\widehat{\varepsilon}_{
\mathscr{P}} \! \cos (\phi_{m}) \! \cos (s^{\pm}) \me^{\mho^{\pm}(x,t)} \! 
- \! \varepsilon_{b} \! \cos (s^{\pm} \! + \! \phi_{m})) \right. \\
&\left. \times \sin (\Phi^{\pm}(z_{o},t) \! \pm \! (2 \! \mp \! 1) \tfrac{
\pi}{4}) \right), \\
\widehat{\mathscr{E}}_{\mathcal{SC}}^{(3)}(x,t) &= -\dfrac{2 
\varepsilon_{b} \widehat{\varepsilon}_{\mathscr{P}} \aleph_{4} \cos (\phi_{
m}) \me^{\mho^{\pm}(x,t)}}{(\aleph_{4}^{2} \! + \! 2 \aleph_{4} \cos (\phi_{
m}) \! + \! 1)} \! \left(2 \cos (s^{\pm} \! - \! \phi_{m}) \cos (\Phi^{\pm}
(z_{o},t) \! \pm \! (2 \! \mp \! 1) \tfrac{\pi}{4}) \! + \! (\aleph_{4} \! 
+ \! \aleph_{3}) \cos (s^{\pm}) \right. \\
&\left. \times \cos (\Phi^{\pm}(z_{o},t) \! \pm \! (2 \! \mp \! 1) \tfrac{
\pi}{4}) \! \pm \! (\aleph_{4} \! - \! \aleph_{3}) \sin (s^{\pm}) \sin 
(\Phi^{\pm}(z_{o},t) \! \pm \! (2 \! \mp \! 1) \tfrac{\pi}{4}) \right), \\
\widehat{\mathscr{E}}_{\mathcal{SC}}^{(4)}(x,t) &= -\dfrac{4 \varepsilon_{b} 
\widehat{\varepsilon}_{\mathscr{P}} \aleph_{4}^{2} \sin (\phi_{m}) \sin 
(s^{\pm}) \me^{\mho^{\pm}(x,t)}}{(\aleph_{4}^{2} \! + \! 2 \aleph_{4} \cos 
(\phi_{m}) \! + \! 1)^{2}} \! \left(((\aleph_{4} \! + \! \aleph_{3}) \cos 
(\phi_{m}) \! + \! 2) \cos (\Phi^{\pm}(z_{o},t) \! \pm \! (2 \! \mp \! 1) 
\tfrac{\pi}{4}) \right. \\
&\left. \pm \, (\aleph_{4} \! - \! \aleph_{3}) \sin (\phi_{m}) \! \sin 
(\Phi^{\pm}(z_{o},t) \! \pm \! (2 \! \mp \! 1) \tfrac{\pi}{4}) \right), \\
\widehat{\mathscr{E}}_{\mathcal{SC}}^{(5)}(x,t) &= -2 \varepsilon_{b} 
\widehat{\varepsilon}_{\mathscr{P}} \csc (\phi_{m}) \! \sin (s^{\pm}) \! 
\cos (\Phi^{\pm}(z_{o},t) \! \pm \! (2 \! \mp \! 1) \tfrac{\pi}{4}) \! 
\sinh (\mho^{\pm}(x,t)), \\
\widehat{\mathscr{E}}_{\mathcal{SC}}^{(6)}(x,t) &= -\dfrac{4 \sin (\phi_{m}) 
\sin (s^{\pm} \! - \! \phi_{m})}{(1 \! - \! \me^{2 \mho^{\pm}(x,t)})} \cos 
(\Phi^{\pm}(z_{o},t) \! \pm \! (2 \! \mp \! 1) \tfrac{\pi}{4}) \me^{2 
\mho^{\pm}(x,t)}, \\
\widehat{\mathscr{E}}_{\mathcal{SC}}^{(7)}(x,t) &= 2 \varepsilon_{b} 
\widehat{\varepsilon}_{\mathscr{P}} \cos (s^{\pm} \! - \! \phi_{m}) \cos 
(\Phi^{\pm}(z_{o},t) \! \pm \! (2 \! \mp \! 1) \tfrac{\pi}{4}) \me^{\mho^{
\pm}(x,t)},
\end{align*}
and $\psi^{\pm}(\cdot)$, $\{\aleph_{n}\}_{n=1}^{4}$, $\widehat{\varepsilon}_{
\mathscr{P}}$, $\mho^{\pm}(x,t)$, and $\Phi^{\pm}(z_{o},t)$ given in 
Theorem~{\rm 2.2.1}, Eqs.~{\rm (22)--(23)}, {\rm (24)}, {\rm (26)}, 
{\rm (27)--(28)}, and {\rm (30)--(32)}, respectively.
\end{dddd}
One important application of the asymptotic results obtained in this paper is 
related to the so-called $N$-\emph{dark soliton scattering}, namely, the 
explicit calculation of the $n$th dark soliton position shift in the presence 
of the (non-trivial) continuous spectrum. Note that, unlike bright solitons 
of the focusing NLSE (with rapidly decaying, in the sense of Schwartz, initial 
data), which undergo both position and phase shifts \cite{a7,a12,a52}, dark 
solitons of the D${}_{f}$NLSE (for the finite-density initial data considered 
here) only undergo a position shift \cite{a13}. This leads to the following 
(see, also, Corollary~2.2.2)
\begin{ffff}
Set
\begin{equation*}
\Delta \widetilde{x}_{n} \! := \! \widetilde{x}_{n}^{+} \! - \! \widetilde{
x}_{n}^{-} \qquad \mathrm{and} \qquad \Delta \widehat{x}_{n} \! := \! 
\widehat{x}_{n}^{+} \! - \! \widehat{x}_{n}^{-}, \qquad n \! \in \! \{1,2,
\ldots,N\}.
\end{equation*}
As $t \! \to \! \pm \infty$ and $x \! \to \! \mp \infty$ such that $z_{o} \! 
:= \! x/t \! < \! -2$ and $(x,t) \! \in \! \{\mathstrut (x,t); \, x \! + \! 
2t \cos (\phi_{n}) \! = \! \mathcal{O}(1), \, \phi_{n} \! \in \! (0,\pi)\}$,
\begin{equation*}
\Delta \widetilde{x}_{n} \! = \! \sum_{k=1}^{N} \dfrac{\mathrm{sgn}(n \! - 
\! k)}{\sin (\phi_{n})} \ln \! \left( \left\vert \dfrac{\sin (\frac{1}{2}
(\phi_{n} \! + \! \phi_{k}))}{\sin (\frac{1}{2}(\phi_{n} \! - \! \phi_{k}))} 
\right\vert \right) \! + \! \left(\int\nolimits_{0}^{\lambda_{2}} \! + \! 
\int\nolimits_{\lambda_{1}}^{+\infty} \! - \! \int\nolimits_{-\infty}^{0} \! 
- \! \int\nolimits_{\lambda_{2}}^{\lambda_{1}} \right) \! \dfrac{\ln (1 \! - 
\! \vert r(\mu) \vert^{2})}{(\mu^{2} \! - \! 2 \mu \cos (\phi_{n}) \! + \! 
1)} \, \dfrac{\md \mu}{2 \pi},
\end{equation*}
and, as $t \! \to \! \pm \infty$ and $x \! \to \! \pm \infty$ such that $z_{
o} \! > \! 2$ and $(x,t) \! \in \! \{\mathstrut (x,t); \, x \! + \! 2t \cos 
(\phi_{n}) \! = \! \mathcal{O}(1), \, \phi_{n} \! \in \! (-\pi,0)\}$,
\begin{equation*}
\Delta \widehat{x}_{n} \! = \! \sum_{k=1}^{N} \dfrac{\mathrm{sgn}(n \! - 
\! k)}{\sin (\phi_{n})} \ln \! \left( \left\vert \dfrac{\sin (\frac{1}{2}
(\phi_{n} \! + \! \phi_{k}))}{\sin (\frac{1}{2}(\phi_{n} \! - \! \phi_{k}))} 
\right\vert \right) \! + \! \left( \int\nolimits_{-\infty}^{\aleph_{4}} \! 
+ \! \int\nolimits_{\aleph_{3}}^{0} \! - \! \int\nolimits_{\aleph_{4}}^{
\aleph_{3}} \! - \! \int\nolimits_{0}^{+\infty} \right) \! \dfrac{\ln (1 \! 
- \! \vert r(\mu) \vert^{2})}{(\mu^{2} \! + \! 2 \mu \cos (\phi_{n}) \! + \! 
1)} \, \dfrac{\md \mu}{2 \pi}.
\end{equation*}
\end{ffff}

\emph{Proof.} Follows {}from the definition of $\Delta \widetilde{x}_{n}$ 
and $\Delta \widehat{x}_{n}$, and Theorem~2.2.1, Eqs.~(15) and~(28). 
\hfill $\square$
\begin{dddd}
For $r(\zeta) \! \in \! \mathcal{S}_{\mathbb{C}}^{1}(\mathbb{R})$, let $m(x,
t;\zeta)$ be the solution of the Riemann-Hilbert problem formulated in 
Lemma~{\rm 2.1.2}. Let $u(x,t)$, the solution of the Cauchy problem for the 
{\rm D${}_{f}$NLSE} with finite-density initial data $u(x,0) \! := \! u_{o}
(x) \! =_{x \to \pm \infty} \! u_{o}(\pm \infty)(1 \! + \! o(1))$, where 
$u_{o}(\pm \infty) \! := \! \exp (\tfrac{\mi (1 \mp 1) \theta}{2})$, $0 \! 
\leqslant \! \theta \! = \! -2 \sum_{n=1}^{N} \sin (\phi_{n}) \! - \! \int_{
-\infty}^{+\infty} \tfrac{\ln (1-\vert r(\mu) \vert^{2})}{\mu} \, \tfrac{\md 
\mu}{2 \pi} \! < \! 2 \pi$, $u_{o}(x) \! \in \! \mathbf{C}^{\infty}(\mathbb{
R})$, and $u_{o}(x) \! - \! u_{o}(\pm \infty) \! \in \! \mathcal{S}_{\mathbb{
C}}(\mathbb{R}_{\pm})$, be defined by Eq.~{\rm (5)}. Then, for $\theta_{
\gamma_{m}} \! = \! \varepsilon_{b} \pi/2$, $\varepsilon_{b} \! \in \! \{\pm 
1\}$, $m \! \in \! \{1,2,\ldots,N\}$, as $t \! \to \! \pm \infty$ and $x \! 
\to \! \mp \infty$ such that $z_{o} \! := \! x/t \! \in \! (-2,0)$ and $(x,
t) \! \in \! \{\mathstrut (x,t); \, x \! + \! 2t \cos (\phi_{m}) \! = \! 
\mathcal{O}(1), \, \phi_{m} \! \in \! (0,\pi)\}$,
\begin{align}
u(x,t) \! &= \me^{-\mi (\varkappa^{\pm}(1)+s^{\pm})} \! \left(\dfrac{(1 \! 
+ \! \varepsilon_{b} \widetilde{\varepsilon}_{\mathscr{P}} \me^{-2 \mi \phi_{
m}+\Omega^{\pm}_{\sharp}(x,t)})}{(1 \! + \! \varepsilon_{b} \widetilde{
\varepsilon}_{\mathscr{P}} \me^{\Omega^{\pm}_{\sharp}(x,t)})} \right. 
\nonumber \\
&\left. + \, \mathcal{O} \! \left(\me^{-4 \vert t \vert \min\limits_{k \not= 
m \in \{1,2,\ldots,N\}}\{\sin (\phi_{k}) \vert \cos (\phi_{k})-\cos (\phi_{m}) 
\vert\}} \right) \right),
\end{align}
where $s^{\pm}$ and $\widetilde{\varepsilon}_{\mathscr{P}}$, respectively, 
are given in Theorem~{\rm 2.2.1}, Eqs.~{\rm (11)} and~{\rm (13)},
\begin{gather}
\varkappa^{+}(j) \! = \! \int\nolimits_{-\infty}^{0} \dfrac{\ln (1 \! - \! 
\vert r(\mu) \vert^{2})}{\mu^{j}} \, \dfrac{\md \mu}{2 \pi}, \quad \quad 
\varkappa^{-}(j) \! = \! \int\nolimits_{0}^{+\infty} \dfrac{\ln (1 \! - \! 
\vert r(\mu) \vert^{2})}{\mu^{j}} \, \dfrac{\md \mu}{2 \pi}, \quad \, j \! 
\in \! \{0,1\}, \\
\Omega^{\pm}_{\sharp}(x,t) \! = \! -2 \sin (\phi_{m})(x \! + \! 2t \cos 
(\phi_{m}) \! - \! \widetilde{x}^{\pm}_{m,\sharp}),
\end{gather}
and
\begin{align}
\widetilde{x}^{\pm}_{m,\sharp} \! =& \pm \! \sum_{k=1}^{N} \dfrac{\mathrm{
sgn}(m \! - \! k)}{2 \sin (\phi_{m})} \ln \! \left( \left\vert \dfrac{\sin 
(\frac{1}{2}(\phi_{m} \! + \! \phi_{k}))}{\sin (\frac{1}{2}(\phi_{m} \! - \! 
\phi_{k}))} \right\vert \right) \! \pm \! \dfrac{1}{2} \! \left(
\int\nolimits_{0}^{+\infty} \! - \! \int\nolimits_{-\infty}^{0} \right) \! 
\dfrac{\ln (1 \! - \! \vert r(\mu) \vert^{2})}{(\mu^{2} \! - \! 2 \mu \cos 
(\phi_{m}) \! + \! 1)} \, \dfrac{\md \mu}{2 \pi} \nonumber \\
 +& \, \dfrac{\ln (\vert \gamma_{m} \vert)}{2 \sin (\phi_{m})},
\end{align}
and, as $t \! \to \! \pm \infty$ and $x \! \to \! \pm \infty$ such that 
$z_{o} \! \in \! (0,2)$ and $(x,t) \! \in \! \{\mathstrut (x,t); \, x \! + 
\! 2t \cos (\phi_{m}) \! = \! \mathcal{O}(1), \, \phi_{m} \! \in \! (-\pi,
0)\}$,
\begin{align}
u(x,t) \! &= -\me^{-\mi (\varkappa^{\pm}(1)+s^{\pm})} \! \left(\dfrac{(1 \! 
+ \! \varepsilon_{b} \widehat{\varepsilon}_{\mathscr{P}} \me^{-2 \mi \phi_{
m}+\Omega^{\pm}_{\natural}(x,t)})}{(1 \! + \! \varepsilon_{b} \widehat{
\varepsilon}_{\mathscr{P}} \me^{\Omega^{\pm}_{\natural}(x,t)})} \right. 
\nonumber \\
&\left. + \, \mathcal{O} \! \left(\me^{-4 \vert t \vert \min\limits_{k \not= 
m \in \{1,2,\ldots,N\}}\{\vert \sin (\phi_{k}) \vert \vert \cos (\phi_{k})-
\cos (\phi_{m}) \vert\}} \right) \! \right),
\end{align}
where $\widehat{\varepsilon}_{\mathscr{P}}$ is given in Theorem~{\rm 2.2.1}, 
Eq.~{\rm (26)},
\begin{equation}
\Omega^{\pm}_{\natural}(x,t) \! = \! -2 \sin (\phi_{m})(x \! + \! 2t \cos 
(\phi_{m}) \! - \! \widehat{x}^{\pm}_{m,\natural}),
\end{equation}
and
\begin{align}
\widehat{x}^{\pm}_{m,\natural} \! =& \! \pm \! \sum_{k=1}^{N} \dfrac{\mathrm{
sgn}(m \! - \! k)}{2 \sin (\phi_{m})} \ln \! \left( \left\vert \dfrac{\sin 
(\frac{1}{2}(\phi_{m} \! + \! \phi_{k}))}{\sin (\frac{1}{2}(\phi_{m} \! 
- \! \phi_{k}))} \right\vert \right) \! \pm \! \dfrac{1}{2} \! \left( 
\int\nolimits_{-\infty}^{0} \! - \! \int\nolimits_{0}^{+\infty} \right) \! 
\dfrac{\ln (1 \! - \! \vert r(\mu) \vert^{2})}{(\mu^{2} \! + \! 2 \mu \cos 
(\phi_{m}) \! + \! 1)} \, \dfrac{\md \mu}{2 \pi} \nonumber \\
 +& \, \dfrac{\ln (\vert \gamma_{m} \vert)}{2 \sin (\phi_{m})}.
\end{align}
\end{dddd}
\begin{dddd}
For $r(\zeta) \! \in \! \mathcal{S}_{\mathbb{C}}^{1}(\mathbb{R})$, let $m
(x,t;\zeta)$ be the solution of the Riemann-Hilbert problem formulated in 
Lemma~{\rm 2.1.2}. Let $u(x,t)$, the solution of the Cauchy problem for the 
{\rm D${}_{f}$NLSE} with finite-density initial data $u(x,0) \! := \! u_{o}
(x) \! =_{x \to \pm \infty} \! u_{o}(\pm \infty)(1 \! + \! o(1))$, where 
$u_{o}(\pm \infty) \! := \! \exp (\tfrac{\mi (1 \mp 1) \theta}{2})$, $0 \! 
\leqslant \! \theta \! = \! -2 \sum_{n=1}^{N} \sin (\phi_{n}) \! - \! \int_{
-\infty}^{+\infty} \tfrac{\ln (1-\vert r(\mu) \vert^{2})}{\mu} \, \tfrac{\md 
\mu}{2 \pi} \! < \! 2 \pi$, $u_{o}(x) \! \in \! \mathbf{C}^{\infty}(\mathbb{
R})$, and $u_{o}(x) \! - \! u_{o}(\pm \infty) \! \in \! \mathcal{S}_{\mathbb{
C}}(\mathbb{R}_{\pm})$, be defined by Eq.~{\rm (5)}, and $\int_{+\infty}^{x}
(\vert u(x^{\prime},t) \vert^{2} \! - \! 1) \, \md x^{\prime}$ be defined by 
Eq.~{\rm (6)}. Let $\epsilon \! \in \! \{\pm 1\}$. Then, for $\theta_{\gamma_{
m}} \! = \! \varepsilon_{b} \pi/2$, $\varepsilon_{b} \! \in \! \{\pm 1\}$, 
$m \! \in \! \{1,2,\ldots,N\}$, as $t \! \to \! \pm \infty$ and $x \! \to 
\! \mp \infty$ such that $z_{o} \! \in \! (-2,0)$ and $(x,t) \! \in \! \{
\mathstrut (x,t); \, x \! + \! 2t \cos (\phi_{m}) \! = \! \mathcal{O}(1), 
\, \phi_{m} \! \in \! (0,\pi)\}$,
\begin{align}
\int\nolimits_{\mathrm{sgn}(\epsilon) \infty}^{x}(\vert u(x^{\prime},t) 
\vert^{2} \! - \! 1) \, \md x^{\prime} \! &= \widetilde{\mathscr{S}}^{\pm}_{
\epsilon} \! + \! \mathscr{H}^{\pm}_{\sharp,\epsilon} \! + \! \dfrac{2 
\varepsilon_{b} \widetilde{\varepsilon}_{\mathscr{P}} \sin (\phi_{m}) 
\me^{\Omega^{\pm}_{\sharp}(x,t)}}{(1 \! + \! \varepsilon_{b} \widetilde{
\varepsilon}_{\mathscr{P}} \me^{\Omega^{\pm}_{\sharp}(x,t)})} \nonumber \\
 &+\mathcal{O} \! \left(\me^{-4 \vert t \vert \min\limits_{k \not= m \in \{1,
2,\ldots,N\}}\{\sin (\phi_{k}) \vert \cos (\phi_{k})-\cos (\phi_{m}) \vert\}} 
\right),
\end{align}
where $\widetilde{\varepsilon}_{\mathscr{P}}$ is given in 
Theorem~{\rm 2.2.1}, Eq.~{\rm (13)}, $\widetilde{\mathscr{S}}^{\pm}_{
\epsilon}$ are given in Theorem~{\rm 2.2.2}, Eq.~{\rm (35)}, $\Omega^{\pm}_{
\sharp}(x,t)$ are given in Theorem~{\rm 2.2.3}, Eqs.~{\rm (48)--(49)},
\begin{equation}
\mathscr{H}^{+}_{\sharp,\epsilon} \! = \! 
\begin{cases}
\varkappa^{+}(0), &\text{$\epsilon \! = \! +1$,} \\
-\varkappa^{-}(0), &\text{$\epsilon \! = \! -1$,}
\end{cases} \qquad \quad \mathscr{H}^{-}_{\sharp,\epsilon} \! = \! 
\begin{cases}
\varkappa^{-}(0), &\text{$\epsilon \! = \! +1$,} \\
-\varkappa^{+}(0), &\text{$\epsilon \! = \! -1$,}
\end{cases}
\end{equation}
and $\varkappa^{\pm}(\cdot)$ are given in Theorem~{\rm 2.2.3}, 
Eq.~{\rm (47)}, and, as $t \! \to \! \pm \infty$ and $x \! \to \! \pm \infty$ 
such that $z_{o} \! \in \! (0,2)$ and $(x,t) \! \in \! \{\mathstrut (x,t); 
\, x \! + \! 2t \cos (\phi_{m}) \! = \! \mathcal{O}(1), \, \phi_{m} \! \in 
\! (-\pi,0)\}$,
\begin{align}
\int\nolimits_{\mathrm{sgn}(\epsilon) \infty}^{x}(\vert u(x^{\prime},t) 
\vert^{2} \! - \! 1) \, \md x^{\prime} \! &=\widehat{\mathscr{S}}^{\pm}_{
\epsilon} \! + \! \mathscr{H}^{\pm}_{\natural,\epsilon} \! + \! \dfrac{2 
\varepsilon_{b} \widehat{\varepsilon}_{\mathscr{P}} \sin (\phi_{m}) \me^{
\Omega^{\pm}_{\natural}(x,t)}}{(1 \! + \! \varepsilon_{b} \widehat{
\varepsilon}_{\mathscr{P}} \me^{\Omega^{\pm}_{\natural}(x,t)})} \nonumber \\
 &+\mathcal{O} \! \left(\me^{-4 \vert t \vert \min\limits_{k \not= m \in \{1,
2,\ldots,N\}}\{\vert \sin (\phi_{k}) \vert \vert \cos (\phi_{k})-\cos (\phi_{
m}) \vert\}} \right),
\end{align}
where $\widehat{\varepsilon}_{\mathscr{P}}$ is given in Theorem~{\rm 2.2.1}, 
Eq.~{\rm (26)}, $\widehat{\mathscr{S}}^{\pm}_{\epsilon}$ are given in 
Theorem~{\rm 2.2.2}, Eq.~{\rm (41)}, $\Omega^{\pm}_{\natural}(x,t)$ are 
given in Theorem~{\rm 2.2.3}, Eqs.~{\rm (51)--(52)}, and
\begin{equation}
\mathscr{H}^{+}_{\natural,\epsilon} \! = \! 
\begin{cases}
\varkappa^{-}(0), &\text{$\epsilon \! = \! +1$,} \\
-\varkappa^{+}(0), &\text{$\epsilon \! = \! -1$,}
\end{cases} \qquad \quad \mathscr{H}^{-}_{\natural,\epsilon} \! = \! 
\begin{cases}
\varkappa^{+}(0), &\text{$\epsilon \! = \! +1$,} \\
-\varkappa^{-}(0), &\text{$\epsilon \! = \! -1$.}
\end{cases}
\end{equation}
\end{dddd}
\begin{ffff}
Set
\begin{equation*}
\Delta x_{n}^{\sharp} \! := \! \widetilde{x}_{n,\sharp}^{+} \! - \! 
\widetilde{x}_{n,\sharp}^{-} \qquad \mathrm{and} \qquad \Delta x_{n}^{
\natural} \! := \! \widehat{x}_{n,\natural}^{+} \! - \! \widehat{x}_{n,
\natural}^{-}, \qquad n \! \in \! \{1,2,\ldots,N\}.
\end{equation*}
As $t \! \to \! \pm \infty$ and $x \! \to \! \mp \infty$ such that $z_{o} \! 
:= \! x/t \! \in \! (-2,0)$ and $(x,t) \! \in \! \{\mathstrut (x,t); \, x \! 
+ \! 2t \cos (\phi_{n}) \! = \! \mathcal{O}(1), \, \phi_{n} \! \in \! (0,
\pi)\}$,
\begin{equation*}
\Delta x_{n}^{\sharp} \! = \! \sum_{k=1}^{N} \dfrac{\mathrm{sgn}(n \! - 
\! k)}{\sin (\phi_{n})} \ln \! \left( \left\vert \dfrac{\sin (\frac{1}{2}
(\phi_{n} \! + \! \phi_{k}))}{\sin (\frac{1}{2}(\phi_{n} \! - \! \phi_{
k}))} \right\vert \right) \! + \! \left(\int\nolimits_{0}^{+\infty} \! - 
\! \int\nolimits_{-\infty}^{0} \right) \! \dfrac{\ln (1 \! - \! \vert r(\mu) 
\vert^{2})}{(\mu^{2} \! - \! 2 \mu \cos (\phi_{n}) \! + \! 1)} \, \dfrac{\md 
\mu}{2 \pi},
\end{equation*}
and, as $t \! \to \! \pm \infty$ and $x \! \to \! \pm \infty$ such that $z_{
o} \in (0,2)$ and $(x,t) \in \{\mathstrut (x,t); \, x+2t \cos (\phi_{n}) = 
\mathcal{O}(1), \, \phi_{n} \! \in \! (-\pi,0)\}$,
\begin{equation*}
\Delta x_{n}^{\natural} \! = \! \sum_{k=1}^{N} \dfrac{\mathrm{sgn}(n \! - 
\! k)}{\sin (\phi_{n})} \ln \! \left( \left\vert \dfrac{\sin (\frac{1}{2}
(\phi_{n} \! + \! \phi_{k}))}{\sin (\frac{1}{2}(\phi_{n} \! - \! \phi_{
k}))} \right\vert \right) \! + \! \left(\int\nolimits_{-\infty}^{0} \! - 
\! \int\nolimits_{0}^{+\infty} \right) \! \dfrac{\ln (1 \! - \! \vert r(\mu) 
\vert^{2})}{(\mu^{2} \! + \! 2 \mu \cos (\phi_{n}) \! + \! 1)} \, \dfrac{\md 
\mu}{2 \pi}.
\end{equation*}
\end{ffff}

\emph{Proof.} Follows {}from the definition of $\Delta x_{n}^{\sharp}$ and 
$\Delta x_{n}^{\natural}$, and Theorem~2.2.3, Eqs.~(49) and~(52). \hfill 
$\square$
\begin{eeee}
In this paper, the complete details of the asymptotic analysis are presented 
for the case $t \! \to \! +\infty$ and $x \! \to \! -\infty$ such that $z_{o} 
\! := \! x/t \! < \! -2$ and $(x,t) \! \in \! \{\mathstrut (x,t); \, x \! 
+ \! 2t \cos (\phi_{n}) \! = \! \mathcal{O}(1), \, \phi_{n} \! \in \! (0,
\pi)\}$, and the final results for the analogous asymptotic analysis as 
$t \! \to \! -\infty$ and $x \! \to \! +\infty$ such that $z_{o} \! < \! 
-2$ and $(x,t) \! \in \! \{\mathstrut (x,t); \, x \! + \! 2t \cos (\phi_{
n}) \! = \! \mathcal{O}(1), \, \phi_{n} \! \in \! (0,\pi)\}$ are given in 
Appendix~A. The remaining cases are treated similarly, and one uses the 
results of Appendix~B to obtain the corresponding leading-order asymptotic 
expansions.
\end{eeee}
\section{The Model RHP}
In this section, the RHP studied asymptotically (as $t \! \to \! +\infty)$ 
in Section~4, the so-called model RHP, is derived: it is obtained {}from the 
(normalised at $\infty)$ RHP for $m(x,t;\zeta)$ formulated in Lemma~2.1.2 
via an ingenius method due to Deift \emph{et al.} \cite{a34} (see below). 
Set $\daleth_{m} \! := \! \{\mathstrut (x,t); \, x \! + \! 2t \cos 
(\phi_{m}) \! = \! \mathcal{O}(1), \, \phi_{m} \! \in \! (0,\pi)\}$, $m \! 
\in \! \{1,2,\ldots,N\}$: note that the $m$th dark soliton ``trajectory'' 
in the $(x,t)$-plane, $\mathbb{R}^{2}$, belongs to $\daleth_{m}$. {}From 
Lemma~2.1.2, \emph{(iii)}, and the dark soliton ordering adopted in 
Subsection~2.2, one notes that, as $t \! \to \! +\infty$ and $x \! \to \! 
-\infty$ such that $z_{o} \! := \! x/t \! < \! -2$ and $(x,t) \! \in \! 
\daleth_{m}$: (1) for $n \! = \! m$, $g_{n}(x,t) \! \! \upharpoonright_{
\daleth_{m}} \! = \! \mathcal{O}(1)$; (2) for $n \! < \! m$, $g_{n}(x,t) 
\! \! \upharpoonright_{\daleth_{m}} \! = \! \mathcal{O}(\exp (-4t \sin 
(\phi_{n}) \vert \cos (\phi_{n}) \! - \! \cos (\phi_{m}) \vert)) \! \to \! 
0$; and (3) for $n \! > \! m$, $g_{n}(x,t) \! \! \upharpoonright_{\daleth_{
m}} \! = \! \mathcal{O}(\exp (4t \sin (\phi_{n}) \vert \cos (\phi_{n}) \! 
- \! \cos (\phi_{m}) \vert)) \! \to \! \infty$. Thus (cf.~Remark~2.1.3), 
since the RHP for $m(x,t;\zeta)$ formulated in Lemma~2.1.2 is asymptotically 
solvable for the $(x,t)$-sector stated above, one deduces that, along the 
trajectory of the (arbitrarily fixed) $m$th dark soliton: (1) for $n \! = 
\! m$, $\mathrm{Res}(m(x,t;\zeta);
\varsigma_{n}) \! = \! 
\left(
\begin{smallmatrix}
\mathcal{O}(1) & 0 \\
\mathcal{O}(1) & 0
\end{smallmatrix}
\right)$ and $\mathrm{Res}(m(x,t;\zeta);\overline{\varsigma_{n}}) \! = \! 
\left(
\begin{smallmatrix}
0 & \mathcal{O}(1) \\
0 & \mathcal{O}(1)
\end{smallmatrix}
\right)$; (2) for $n \! < \! m$, $\mathrm{Res}(m(x,t;\zeta);\varsigma_{n}) 
\! = \! 
\left(
\begin{smallmatrix}
\mathcal{O}(\blacklozenge) & 0 \\
\mathcal{O}(\blacklozenge) & 0
\end{smallmatrix}
\right) \! \to \! \mathbf{0}$ and $\mathrm{Res}(m(x,t;\zeta);
\overline{\varsigma_{n}}) \! = \! 
\left(
\begin{smallmatrix}
0 & \mathcal{O}(\blacklozenge) \\
0 & \mathcal{O}(\blacklozenge)
\end{smallmatrix}
\right) \! \to \! \mathbf{0}$, where $\blacklozenge \! := \! \exp (-4t \sin 
(\phi_{n}) \vert \cos (\phi_{n}) \! - \! \cos (\phi_{m}) \vert)$; and (3) 
for $n \! > \! m$, $\mathrm{Res}(m(x,t;\zeta);
\varsigma_{n}) \! = \! 
\left(
\begin{smallmatrix}
\mathcal{O}(\blacklozenge^{-1}) & 0 \\
\mathcal{O}(\blacklozenge^{-1}) & 0
\end{smallmatrix}
\right) \! \to \! 
\left(
\begin{smallmatrix}
\infty & 0 \\
\infty & 0
\end{smallmatrix}
\right)$ and $\mathrm{Res}(m(x,t;\zeta);\overline{\varsigma_{n}}) \! = \! 
\left(
\begin{smallmatrix}
0 & \mathcal{O}(\blacklozenge^{-1}) \\
0 & \mathcal{O}(\blacklozenge^{-1})
\end{smallmatrix}
\right) \! \to \! 
\left(
\begin{smallmatrix}
0 & \infty \\
0 & \infty
\end{smallmatrix}
\right)$. Hence, along the trajectory of the (arbitrarily fixed) $m$th 
dark soliton, there are exponentially growing polar (residue) conditions 
for solitons $n$ with $n \! \in \! \{m \! + \! 1,m \! + \! 2,\ldots,N\}$. 
In a paper dealing with the Toda Rarefaction Problem \cite{a34}, Deift 
\emph{et al.} showed how this problem could be dealt with. Proceeding {}from 
the construction of Zhou \cite{a44,a45,a46} related to the singular RHP (see 
the synopsis below Theorem~C.1.4 in Appendix~C), one uses the method of 
Deift \emph{et al.} to ``replace'' the poles which give rise to the 
exponentially growing residue conditions by jump matrices on mutually 
disjoint, and disjoint with respect to $\sigma_{c}$, ``small'' circles 
(see \cite{a46}, Section~2, Remark~2.18, for a discussion about the radii of 
these circles) in such a way that the jump matrices on these small circles 
behave like $\mathrm{I}$ $+$ exponentially decreasing terms (as $t \! \to 
\! +\infty)$, thus constructing the \emph{augmented} contour $\sigma_{
\mathrm{augmented}} \! := \! \sigma_{c} \cup (\cup_{n=m+1}^{N} \partial 
(\mathrm{small} \, \, \mathrm{circles}))$. Thus, instead of the original 
RHP, one obtains an augmented (and normalised at $\infty)$ RHP with $2(N \! 
- \! m)$ fewer poles and $2(N \! - \! m)$ additional circles with jump 
conditions stated on them. Finally, by ``removing'' the $2(N \! - \! m)$ 
small circles {}from the augmented RHP, one arrives at an asymptotically 
solvable, equivalent, ``model'' RHP, in the sense that a solution of the 
equivalent RHP gives a solution of the augmented RHP and \emph{vice versa}; 
in particular, if there are two RHPs, $(\mathcal{X}_{1}(\lambda),\upsilon_{1}
(\lambda),\Gamma_{1})$ and $(\mathcal{X}_{2}(\lambda),\upsilon_{2}(\lambda),
\Gamma_{2})$, say, with $\Gamma_{2} \subset \Gamma_{1}$ and $\upsilon_{1}
(\lambda) \! \! \upharpoonright_{\Gamma_{1} \setminus \Gamma_{2}}=_{t \to 
+\infty} \! \mathrm{I} \! + \! o(1)$, then, modulo $o(1)$ estimates, their 
solutions, $\mathcal{X}_{1}(\lambda)$ and $\mathcal{X}_{2}(\lambda)$, 
respectively, are asymptotically equal. Actually, as will be shown below 
(see Lemma~3.5), the solution of the model RHP approximates, up to terms 
that are exponentially small (as $t \! \to \! +\infty)$, the solution of 
the augmented RHP (hence the original RHP). 

The reason for introducing the factor $\delta (\zeta)$ in Lemma~3.1 below is 
given in Section~4 of \cite{a38}.
\begin{eeeee}
For notational convenience, all explicit $x,t$ dependencies are hereafter 
suppressed, except where absolutely necessary and/or where confusion may 
arise.
\end{eeeee}
\begin{ccccc}
For $r(\zeta) \! \in \! \mathcal{S}_{\mathbb{C}}^{1}(\mathbb{R})$, let 
$m(\zeta) \colon \mathbb{C} \setminus (\sigma_{d} \cup \sigma_{c}) \! \to \! 
\mathrm{M}_{2}(\mathbb{C})$ be the solution of the {\rm RHP} formulated in 
Lemma~{\rm 2.1.2}. Set
\begin{equation*}
\widehat{m}(\zeta) \! := \! m(\zeta)(\delta (\zeta))^{-\sigma_{3}},
\end{equation*}
where $\delta (\zeta) \! = \! \exp \! \left( \! \left( \int_{-\infty}^{0} \! 
+ \! \int_{\lambda_{2}}^{\lambda_{1}} \right) \! \tfrac{\ln (1-\vert r(\mu) 
\vert^{2})}{(\mu -\zeta)} \tfrac{\md \mu}{2 \pi \mi} 
\right)$, with $\lambda_{1}$ and $\lambda_{2}$ given in Theorem~{\rm 2.2.1}, 
Eq.~{\rm (10)}, $\delta (\zeta) \overline{\delta (\overline{\zeta})} \! = \! 
1$, $\delta (\zeta) \delta (\zeta^{-1}) \! = \! \delta (0)$, and $\vert \vert 
(\delta (\cdot))^{\pm 1} \vert \vert_{\mathcal{L}^{\infty}(\mathbb{C})} \! 
\! := \! \sup_{\zeta \in \mathbb{C}} \vert (\delta (\zeta))^{\pm 1} \vert 
\! < \! \infty$. Then $\widehat{m}(\zeta) \colon \mathbb{C} \setminus 
(\sigma_{d} \cup \sigma_{c}) \! \to \! \mathrm{M}_{2}(\mathbb{C})$ solves 
the following {\rm RHP:}
\begin{enumerate}
\item[(i)] $\widehat{m}(\zeta)$ is piecewise (sectionally) meromorphic 
$\forall \, \zeta \! \in \! \mathbb{C} \setminus \sigma_{c};$
\item[(ii)] $\widehat{m}_{\pm}(\zeta) \! := \! \lim_{\genfrac{}{}{0pt}{2}
{\zeta^{\prime} \, \to \, \zeta}{\pm \Im (\zeta^{\prime})>0}} \widehat{m}
(\zeta^{\prime})$ satisfy the jump condition
\begin{equation*}
\widehat{m}_{+}(\zeta) \! = \! \widehat{m}_{-}(\zeta) \exp (-\mi k(\zeta)(x 
\! + \! 2 \lambda (\zeta)t) \mathrm{ad}(\sigma_{3})) \widehat{\mathcal{G}}
(\zeta), \quad \zeta \! \in \! \mathbb{R},
\end{equation*}
where
\begin{equation*}
\widehat{\mathcal{G}}(\zeta) \! = \! 
\begin{pmatrix}
(1 \! - \! r(\zeta) \overline{r(\overline{\zeta})}) \delta_{-}(\zeta)
(\delta_{+}(\zeta))^{-1} & \, \, \, -\overline{r(\overline{\zeta})} \, 
\delta_{-}(\zeta) \delta_{+}(\zeta) \\
r(\zeta)(\delta_{-}(\zeta) \delta_{+}(\zeta))^{-1} & \, \, \, (\delta_{-}
(\zeta))^{-1} \delta_{+}(\zeta)
\end{pmatrix};
\end{equation*}
\item[(iii)] $\widehat{m}(\zeta)$ has simple poles in $\sigma_{d} \! = \! 
\cup_{n=1}^{N}(\{\varsigma_{n}\} \cup \{\overline{\varsigma_{n}}\})$ with
\begin{align*}
\mathrm{Res}(\widehat{m}(\zeta);\varsigma_{n}) &= \! \lim_{\zeta \to 
\varsigma_{n}} \widehat{m}(\zeta)g_{n}(\delta (\varsigma_{n}))^{-2} \sigma_{
-}, & n \! &\in \! \{1,2,\ldots,N\}, \\
\mathrm{Res}(\widehat{m}(\zeta);\overline{\varsigma_{n}}) &= \! \sigma_{1} 
\overline{\mathrm{Res}(\widehat{m}(\zeta);\varsigma_{n})} \, \sigma_{1}, & 
n \! &\in \! \{1,2,\ldots,N\},
\end{align*}
where $g_{n} \! := \! \vert g_{n} \vert \me^{\mi \theta_{g_{n}}} \exp (2 \mi 
k(\varsigma_{n})(x \! + \! 2 \lambda (\varsigma_{n})t))$, with
\begin{gather*}
\vert g_{n} \vert \! = \! 2 \vert \gamma_{n} \vert \sin (\phi_{n}) \exp \! 
\left(\int\nolimits_{-\infty}^{+\infty} \! \dfrac{\sin (\phi_{n}) \ln (1 \! 
- \! \vert r(\mu) \vert^{2})}{(\mu^{2} \! - \! 2 \mu \cos (\phi_{n}) \! + \! 
1)} \, \dfrac{\md \mu}{2 \pi} \right) \! \prod\nolimits_{\genfrac{}{}{0pt}
{2}{k=1}{k \not= n}}^{N} \! \dfrac{\sin (\frac{1}{2}(\phi_{n} \! + \! 
\phi_{k}))}{\sin (\frac{1}{2}(\phi_{n} \! - \! \phi_{k}))}, \\
\theta_{g_{n}} \! = \! \theta_{\gamma_{n}} \! + \! \dfrac{\pi}{2} \! - \! 
\dfrac{\theta}{2} \! - \! \int\nolimits_{-\infty}^{+\infty} \! \dfrac{(\mu 
\! - \! \cos \phi_{n}) \ln (1 \! - \! \vert r(\mu) \vert^{2})}{(\mu^{2} \! 
- \! 2 \mu \cos (\phi_{n}) \! + \! 1)} \, \dfrac{\md \mu}{2 \pi} \! - \! 
\sum_{\genfrac{}{}{0pt}{2}{k=1}{k \not= n}}^{N} \phi_{k}, \qquad \theta_{
\gamma_{n}} \! = \! \pm \tfrac{\pi}{2};
\end{gather*}
\item[(iv)] $\det (\widehat{m}(\zeta)) \vert_{\zeta = \pm 1} \! = \! 0;$
\item[(v)] $\widehat{m}(\zeta) \! =_{\zeta \to 0} \! \zeta^{-1}(\delta (0))^{
\sigma_{3}} \sigma_{2} \! + \! \mathcal{O}(1);$
\item[(vi)] $\widehat{m}(\zeta) \! =_{\genfrac{}{}{0pt}{2}{\zeta \to \infty}
{\zeta \in \mathbb{C} \setminus (\sigma_{d} \cup \sigma_{c})}} \! \mathrm{I} 
\! + \! \mathcal{O}(\zeta^{-1});$
\item[(vii)] $\widehat{m}(\zeta) \! = \! \sigma_{1} \overline{\widehat{m}
(\overline{\zeta})} \, \sigma_{1}$ and $\widehat{m}(\zeta^{-1}) \! = \! 
\zeta \widehat{m}(\zeta)(\delta (0))^{\sigma_{3}} \sigma_{2}$.
\end{enumerate}
Let
\begin{gather}
u(x,t) \! := \! \mi \lim_{\genfrac{}{}{0pt}{2}{\zeta \to \infty}{\zeta \in 
\mathbb{C} \setminus (\sigma_{d} \cup \sigma_{c})}}(\zeta (\widehat{m}
(\zeta)(\delta (\zeta))^{\sigma_{3}} \! - \! \mathrm{I}))_{12},
\end{gather}
and
\begin{gather}
\int\nolimits_{+\infty}^{x}(\vert u(x^{\prime},t) \vert^{2} \! - \! 1) \, \md 
x^{\prime} \! := \! -\mi \lim_{\genfrac{}{}{0pt}{2}{\zeta \to \infty}{\zeta 
\in \mathbb{C} \setminus (\sigma_{d} \cup \sigma_{c})}}(\zeta (\widehat{m}
(\zeta)(\delta (\zeta))^{\sigma_{3}} \! - \! \mathrm{I}))_{11}.
\end{gather}
Then $u(x,t)$ is the solution of the Cauchy problem for the 
{\rm D${}_{f}$NLSE}.
\end{ccccc}

\emph{Proof.} The RHP for $\widehat{m}(\zeta)$ (respectively, Eqs.~(57) 
and~(58)) follows {}from the RHP for $m(\zeta)$ formulated in Lemma~2.1.2 
(respectively, Eqs.~(5) and~(6)) upon using $\widehat{m}(\zeta) \! := \! 
m(\zeta)(\delta (\zeta))^{-\sigma_{3}}$, with $\delta (\zeta)$ given in 
the Lemma. \hfill $\square$
\begin{aaaaa}
For $m \! \in \! \{1,2,\ldots,N\}$ and $\{\varsigma_{n}\}_{n=m+1}^{N} \subset 
\mathbb{C}_{+}$ (respectively, $\{\overline{\varsigma_{n}}\}_{n=m+1}^{N} 
\subset \mathbb{C}_{-})$, define the clockwise (respectively, 
counter-clockwise) oriented circles $\widehat{\mathscr{K}}_{n} \! := \! 
\{\mathstrut \zeta; \, \vert \zeta \! - \! \varsigma_{n} \vert \! = \! 
\widehat{\varepsilon}_{n}^{\mathscr{K}}\}$ (respectively, $\widehat{
\mathscr{L}}_{n} \! := \! \{\mathstrut \zeta; \, \vert \zeta \! - \! 
\overline{\varsigma_{n}} \vert \! = \! \widehat{\varepsilon}_{n}^{
\mathscr{L}}\})$, with $\widehat{\varepsilon}_{n}^{\mathscr{K}}$ 
(respectively, $\widehat{\varepsilon}_{n}^{\mathscr{L}})$ chosen sufficiently 
small such that $\widehat{\mathscr{K}}_{n} \cap \widehat{\mathscr{K}}_{n^{
\prime}} \! = \! \widehat{\mathscr{L}}_{n} \cap \widehat{\mathscr{L}}_{n^{
\prime}} \! = \! \widehat{\mathscr{K}}_{n} \cap \widehat{\mathscr{L}}_{n} 
\! = \! \widehat{\mathscr{K}}_{n} \cap \sigma_{c} \! = \! \widehat{\mathscr{
L}}_{n} \cap \sigma_{c} \! = \! \emptyset$ $\forall \, \, n \! \not= \! 
n^{\prime} \! \in \! \{m \! + \! 1,m \! + \! 2,\ldots,N\}$.
\end{aaaaa}
\begin{eeeee}
Note that the orientation for $\widehat{\mathscr{K}}_{n}$ $(\subset \! 
\mathbb{C}_{+})$ and $\widehat{\mathscr{L}}_{n}$ $(\subset \! \mathbb{
C}_{-})$ is consistent with Eq.~(C.1) (see Appendix~C).
\end{eeeee}
\begin{ccccc}
For $r(\zeta) \! \in \! \mathcal{S}_{\mathbb{C}}^{1}(\mathbb{R})$, let 
$\widehat{m}(\zeta) \colon \mathbb{C} \setminus (\sigma_{d} \cup \sigma_{c}) 
\! \to \! \mathrm{M}_{2}(\mathbb{C})$ be the solution of the {\rm RHP} 
formulated in Lemma~{\rm 3.1}. Set
\begin{equation*}
\widehat{m}^{\flat}(\zeta) \! := \! 
\begin{cases}
\widehat{m}(\zeta), &\text{$\zeta \! \in \! \mathbb{C} \setminus (\sigma_{c} 
\cup (\cup_{n=m+1}^{N}(\widehat{\mathscr{K}}_{n} \cup \mathrm{int}(\widehat{
\mathscr{K}}_{n}) \cup \widehat{\mathscr{L}}_{n} \cup \mathrm{int}(\widehat{
\mathscr{L}}_{n}))))$,} \\
\widehat{m}(\zeta) \! \left(\mathrm{I} \! - \! \tfrac{g_{n}(\delta 
(\varsigma_{n}))^{-2}}{(\zeta -\varsigma_{n})} \sigma_{-} \right), 
&\text{$\zeta \! \in \! \mathrm{int}(\widehat{\mathscr{K}}_{n}), \quad n \! 
\in \! \{m \! + \! 1,m \! + \! 2,\ldots,N\}$,} \\
\widehat{m}(\zeta) \! \left(\mathrm{I} \! + \! \tfrac{\overline{g_{n}
(\delta (\varsigma_{n}))^{-2}}}{(\zeta -\overline{\varsigma_{n}})} \sigma_{
+} \right), &\text{$\zeta \! \in \! \mathrm{int}(\widehat{\mathscr{L}}_{n}), 
\quad n \! \in \! \{m \! + \! 1,m \! + \! 2,\ldots,N\}$.}
\end{cases}
\end{equation*}
Then $\widehat{m}^{\flat}(\zeta) \colon \mathbb{C} \setminus ((\sigma_{d} 
\setminus \cup_{n=m+1}^{N}(\{\varsigma_{n}\} \cup \{\overline{\varsigma_{n}}
\})) \cup (\sigma_{c} \cup (\cup_{n=m+1}^{N}(\widehat{\mathscr{K}}_{n} \cup 
\widehat{\mathscr{L}}_{n})))) \! \to \! \mathrm{M}_{2}(\mathbb{C})$ solves 
the following {\rm RHP:}
\begin{enumerate}
\item[(i)] $\widehat{m}^{\flat}(\zeta)$ is piecewise (sectionally) meromorphic 
$\forall \, \zeta \! \in \! \mathbb{C} \setminus (\sigma_{c} \cup (\cup_{n=m
+1}^{N}(\widehat{\mathscr{K}}_{n} \cup \widehat{\mathscr{L}}_{n})));$
\item[(ii)] $\widehat{m}_{\pm}^{\flat}(\zeta) \! := \! \lim_{\genfrac{}{}
{0pt}{2}{\zeta^{\prime} \, \to \, \zeta}{\zeta^{\prime} \, \in \, \pm \, 
\mathrm{side} \, \mathrm{of} \, \sigma_{c} \cup (\cup_{n=m+1}^{N}(\widehat{
\mathscr{K}}_{n} \cup \widehat{\mathscr{L}}_{n}))}} \widehat{m}^{\flat}
(\zeta^{\prime})$ satisfy the jump condition
\begin{equation*}
\widehat{m}_{+}^{\flat}(\zeta) \! = \! \widehat{m}_{-}^{\flat}(\zeta) 
\widehat{\upsilon}^{\flat}(\zeta), \quad \zeta \! \in \! \sigma_{c} \cup 
(\cup_{n=m+1}^{N}(\widehat{\mathscr{K}}_{n} \cup \widehat{\mathscr{L}}_{n})),
\end{equation*}
where
\begin{equation*}
\widehat{\upsilon}^{\flat}(\zeta) \! = \! 
\begin{cases}
\exp (-\mi k(\zeta)(x \! + \! 2 \lambda (\zeta)t) \mathrm{ad}(\sigma_{3})) 
\widehat{\mathcal{G}}(\zeta), &\text{$\zeta \! \in \! \mathbb{R}$,} \\
\mathrm{I} \! + \! \tfrac{g_{n}(\delta (\varsigma_{n}))^{-2}}{(\zeta -
\varsigma_{n})} \sigma_{-}, &\text{$\zeta \! \in \! \widehat{\mathscr{
K}}_{n}, \quad n \! \in \! \{m \! + \! 1,m \! + \! 2,\ldots,N\}$,} \\
\mathrm{I} \! + \! \tfrac{\overline{g_{n}(\delta (\varsigma_{n}))^{-2}}}
{(\zeta -\overline{\varsigma_{n}})} \sigma_{+}, &\text{$\zeta \! \in \! 
\widehat{\mathscr{L}}_{n}, \quad n \! \in \! \{m \! + \! 1,m \! + \! 2,
\ldots,N\}$,}
\end{cases}
\end{equation*}
with $\widehat{\mathcal{G}}(\zeta)$ given in Lemma~{\rm 3.1}, (ii);
\item[(iii)] $\widehat{m}^{\flat}(\zeta)$ has simple poles in $\sigma_{d} 
\setminus \cup_{n=m+1}^{N}(\{\varsigma_{n}\} \cup \{\overline{\varsigma_{n}}
\})$ with
\begin{align*}
\mathrm{Res}(\widehat{m}^{\flat}(\zeta);\varsigma_{n}) &= \! \lim_{\zeta \to 
\varsigma_{n}} \widehat{m}^{\flat}(\zeta)g_{n}(\delta (\varsigma_{n}))^{-2} 
\sigma_{-}, & n \! &\in \! \{1,2,\ldots,m\}, \\
\mathrm{Res}(\widehat{m}^{\flat}(\zeta);\overline{\varsigma_{n}}) &= \! 
\sigma_{1} \overline{\mathrm{Res}(\widehat{m}^{\flat}(\zeta);\varsigma_{
n})} \, \sigma_{1}, & n \! &\in \! \{1,2,\ldots,m\};
\end{align*}
\item[(iv)] $\det (\widehat{m}^{\flat}(\zeta)) \vert_{\zeta = \pm 1} \! = \! 
0;$
\item[(v)] $\widehat{m}^{\flat}(\zeta) \! =_{\zeta \to 0} \! \zeta^{-1}
(\delta (0))^{\sigma_{3}} \sigma_{2} \! + \! \mathcal{O}(1);$
\item[(vi)] as $\zeta \! \to \! \infty$, $\zeta \! \in \! \mathbb{C} 
\setminus ((\sigma_{d} \setminus \cup_{n=m+1}^{N}(\{\varsigma_{n}\} \cup 
\{\overline{\varsigma_{n}}\})) \cup (\sigma_{c} \cup (\cup_{n=m+1}^{N}
(\widehat{\mathscr{K}}_{n} \cup \widehat{\mathscr{L}}_{n}))))$, $\widehat{
m}^{\flat}(\zeta) \! = \! \mathrm{I} \! + \! \mathcal{O}(\zeta^{-1});$
\item[(vii)] $\widehat{m}^{\flat}(\zeta) \! = \! \sigma_{1} \overline{
\widehat{m}^{\flat}(\overline{\zeta})} \, \sigma_{1}$ and $\widehat{m}^{
\flat}(\zeta^{-1}) \! = \! \zeta \widehat{m}^{\flat}(\zeta)(\delta (0))^{
\sigma_{3}} \sigma_{2}$.
\end{enumerate}
For $\zeta \! \in \! \mathbb{C} \setminus ((\sigma_{d} \setminus \cup_{n=m+
1}^{N}(\{\varsigma_{n}\} \cup \{\overline{\varsigma_{n}}\})) \cup (\sigma_{
c} \cup (\cup_{n=m+1}^{N}(\widehat{\mathscr{K}}_{n} \cup \widehat{\mathscr{
L}}_{n}))))$, let
\begin{gather}
u(x,t) \! := \! \mi \lim_{\zeta \to \infty}(\zeta (\widehat{m}^{\flat}(\zeta)
(\delta (\zeta))^{\sigma_{3}} \! - \! \mathrm{I}))_{12},
\end{gather}
and
\begin{gather}
\int\nolimits_{+\infty}^{x}(\vert u(x^{\prime},t) \vert^{2} \! - \! 1) \, \md 
x^{\prime} \! := \! -\mi \lim_{\zeta \to \infty}(\zeta (\widehat{m}(\zeta)
(\delta (\zeta))^{\sigma_{3}} \! - \! \mathrm{I}))_{11}.
\end{gather}
Then $u(x,t)$ is the solution of the Cauchy problem for the 
{\rm D${}_{f}$NLSE}.
\end{ccccc}

\emph{Proof.} The RHP for $\widehat{m}^{\flat}(\zeta)$ (respectively, 
Eqs.~(59) and~(60)) follows {}from the RHP for $\widehat{m}(\zeta)$ 
formulated in Lemma~3.1 (respectively, Eqs.~(57) and~(58)) upon using 
the definition of $\widehat{m}^{\flat}(\zeta)$ in terms of $\widehat{m}
(\zeta)$ given in the Lemma. \hfill $\square$
\begin{eeeee}
Even though the set (of first-order poles) $\cup_{n=m+1}^{N}(\{\varsigma_{
n}\} \cup \{\overline{\varsigma_{n}}\})$, giving rise to the exponentially 
growing residue conditions, has been removed {}from the specification of 
the RHP and replaced by jump matrices on $\cup_{n=m+1}^{N}(\widehat{
\mathscr{K}}_{n} \cup \widehat{\mathscr{L}}_{n})$, it should be noted that 
these jump matrices are also exponentially growing (as $t \! \to \! 
+\infty)$. These lower/upper diagonal, exponentially growing jump matrices 
are now replaced, via a finite sequence of transformations, by upper/lower 
diagonal jump matrices which converge to $\mathrm{I}$ as $t \! \to \! 
+\infty$.
\end{eeeee}
\begin{ccccc}
For $m \! \in \! \{1,2,\ldots,N\}$, let $\sigma^{\prime}_{d} \! := \! \sigma_{
d} \setminus \cup_{n=m+1}^{n}(\{\varsigma_{n}\} \cup \{\overline{\varsigma_{
n}}\})$, $\sigma^{\prime}_{c} \! := \! \sigma_{c} \cup (\cup_{n=m+1}^{N}
(\widehat{\mathscr{K}}_{n} \cup \widehat{\mathscr{L}}_{n}))$, where $\widehat{
\mathscr{K}}_{n}$ and $\widehat{\mathscr{L}}_{n}$ are given in 
Definition~{\rm 3.1}, and $\sigma^{\prime}_{\mathcal{O}^{\mathcal{D}}} \! := 
\! \sigma^{\prime}_{d} \cup \sigma^{\prime}_{c}$ $(\sigma_{d}^{\prime} \cap 
\sigma_{c}^{\prime} \! = \! \emptyset)$. Set
\begin{equation*}
\widehat{m}^{\sharp}(\zeta) \! := \! 
\begin{cases}
\widehat{m}^{\flat}(\zeta) \prod_{k=m+1}^{N}(d_{k}^{+}(\zeta))^{-\sigma_{3}}, 
&\text{$\zeta \! \in \! \mathbb{C} \setminus (\sigma_{c}^{\prime} \cup (\cup_{
n=m+1}^{N}(\mathrm{int}(\widehat{\mathscr{K}}_{n}) \cup \mathrm{int}(\widehat{
\mathscr{L}}_{n}))))$}, \\
\widehat{m}^{\flat}(\zeta)(J_{\widehat{\mathscr{K}}_{n}}(\zeta))^{-1} \prod_{
k=m+1}^{N}(d_{k}^{-}(\zeta))^{-\sigma_{3}}, &\text{$\zeta \! \in \! \mathrm{
int}(\widehat{\mathscr{K}}_{n}), \quad n \! \in \! \{m \! + \! 1,m \! + \! 2,
\ldots,N\}$,} \\
\widehat{m}^{\flat}(\zeta)(J_{\widehat{\mathscr{L}}_{n}}(\zeta))^{-1} \prod_{
k=m+1}^{N}(d_{k}^{-}(\zeta))^{-\sigma_{3}}, &\text{$\zeta \! \in \! \mathrm{
int}(\widehat{\mathscr{L}}_{n}), \quad n \! \in \! \{m \! + \! 1,m \! + \! 2,
\ldots,N\}$,}
\end{cases}
\end{equation*}
where
\begin{align*}
d_{n}^{+}(\zeta) \! =& \, \dfrac{\zeta \! - \! \overline{\varsigma_{n}}}{
\zeta \! - \! \varsigma_{n}}, \quad \zeta \! \in \! \mathbb{C} \setminus 
(\sigma_{c}^{\prime} \cup (\cup_{n=m+1}^{N}(\mathrm{int}(\widehat{\mathscr{
K}}_{n}) \cup \mathrm{int}(\widehat{\mathscr{L}}_{n})))), \\
d_{n}^{-}(\zeta) \! =& 
\begin{cases}
\zeta \! - \! \overline{\varsigma_{n}}, &\text{$\zeta \! \in \! \mathrm{int}
(\widehat{\mathscr{K}}_{n})$,} \\
(\zeta \! - \! \varsigma_{n})^{-1}, &\text{$\zeta \! \in \! \mathrm{int}
(\widehat{\mathscr{L}}_{n})$,}
\end{cases}
\end{align*}
$J_{\widehat{\mathscr{K}}_{n}}(\zeta)$ $(\in \! \mathrm{SL}(2,\mathbb{C}))$ 
and $J_{\widehat{\mathscr{L}}_{n}}(\zeta)$ $(\in \! \mathrm{SL}(2,\mathbb{
C}))$, respectively, are holomorphic in $\cup_{k=m+1}^{N} \mathrm{int}
(\widehat{\mathscr{K}}_{k})$ and $\cup_{l=m+1}^{N} \mathrm{int}(\widehat{
\mathscr{L}}_{l})$, with
\begin{align*}
J_{\widehat{\mathscr{K}}_{n}}(\zeta) \! =& \! 
\begin{pmatrix}
\frac{\prod_{\genfrac{}{}{0pt}{2}{k=m+1}{k \not= n}}^{N} \frac{d_{k}^{+}
(\zeta)}{d_{k}^{-}(\zeta)}-\frac{C_{n}^{\mathscr{K}}g_{n}(\delta (\varsigma_{
n}))^{-2}}{(\zeta -\overline{\varsigma_{n}})^{2}} \prod_{\genfrac{}{}{0pt}{2}
{k=m+1}{k \not= n}}^{N} \frac{(d_{k}^{+}(\zeta))^{-1}}{d_{k}^{-}(\zeta)}}{
(\zeta -\varsigma_{n})} & \, \, \frac{C_{n}^{\mathscr{K}}}{(\zeta -\overline{
\varsigma_{n}})^{2}} \prod_{\genfrac{}{}{0pt}{2}{k=m+1}{k \not= n}}^{N} \frac{
(d_{k}^{+}(\zeta))^{-1}}{d_{k}^{-}(\zeta)} \\
-g_{n}(\delta (\varsigma_{n}))^{-2} \prod_{\genfrac{}{}{0pt}{2}{k=m+1}{k \not= 
n}}^{N} \frac{d_{k}^{-}(\zeta)}{d_{k}^{+}(\zeta)} & \, \, (\zeta \! - \! 
\varsigma_{n}) \prod_{\genfrac{}{}{0pt}{2}{k=m+1}{k \not= n}}^{N} \frac{d_{
k}^{-}(\zeta)}{d_{k}^{+}(\zeta)}
\end{pmatrix}, \\
J_{\widehat{\mathscr{L}}_{n}}(\zeta) \! =& \! 
\begin{pmatrix}
(\zeta \! - \! \overline{\varsigma_{n}}) \prod_{\genfrac{}{}{0pt}{2}{k=m+1}
{k \not= n}}^{N} \frac{d_{k}^{+}(\zeta)}{d_{k}^{-}(\zeta)} & \, \, \overline{
g_{n}(\delta (\varsigma_{n}))^{-2}} \prod_{\genfrac{}{}{0pt}{2}{k=m+1}{k 
\not= n}}^{N} \frac{d_{k}^{+}(\zeta)}{d_{k}^{-}(\zeta)} \\
-\frac{C_{n}^{\mathscr{L}}}{(\zeta -\varsigma_{n})^{2}} \prod_{\genfrac{}{}
{0pt}{2}{k=m+1}{k \not= n}}^{N} \frac{d_{k}^{-}(\zeta)}{(d_{k}^{+}(\zeta))^{
-1}} & \, \, \frac{\prod_{\genfrac{}{}{0pt}{2}{k=m+1}{k \not= n}}^{N} \frac{
d_{k}^{-}(\zeta)}{d_{k}^{+}(\zeta)}-\frac{C_{n}^{\mathscr{L}} \overline{g_{n}
(\delta (\varsigma_{n}))^{-2}}}{(\zeta -\varsigma_{n})^{2}} \prod_{\genfrac{}
{}{0pt}{2}{k=m+1}{k \not= n}}^{N} \frac{d_{k}^{-}(\zeta)}{(d_{k}^{+}(\zeta))^{
-1}}}{(\zeta -\overline{\varsigma_{n}})}
\end{pmatrix},
\end{align*}
and
\begin{equation*}
C_{n}^{\mathscr{K}} \! = \overline{C_{n}^{\mathscr{L}}} = \! -4 \sin^{2}
(\phi_{n})(g_{n})^{-1}(\delta (\varsigma_{n}))^{2} \, \me^{-2 \mi \sum_{
\genfrac{}{}{0pt}{2}{j=m+1}{j \not= n}}^{N} \phi_{j}} \prod_{\genfrac{}{}{0pt}
{2}{k=m+1}{k \not= n}}^{N} \! \left(\tfrac{\sin (\frac{1}{2}(\phi_{n}+\phi_{
k}))}{\sin (\frac{1}{2}(\phi_{n}-\phi_{k}))} \right)^{2}.
\end{equation*}
Then $\widehat{m}^{\sharp}(\zeta) \colon \mathbb{C} \setminus \sigma^{
\prime}_{\mathcal{O}^{\mathcal{D}}} \! \to \! \mathrm{M}_{2}(\mathbb{C})$ 
solves the following (augmented) {\rm RHP:}
\begin{enumerate}
\item[(i)] $\widehat{m}^{\sharp}(\zeta)$ is piecewise (sectionally) 
meromorphic $\forall \, \zeta \! \in \! \mathbb{C} \setminus \sigma_{c}^{
\prime};$
\item[(ii)] $\widehat{m}_{\pm}^{\sharp}(\zeta) \! := \! \lim_{\genfrac{}{}
{0pt}{2}{\zeta^{\prime} \, \to \, \zeta}{\zeta^{\prime} \, \in \, \pm \, 
\mathrm{side} \, \mathrm{of} \, \sigma_{\mathcal{O}^{\mathcal{D}}}^{\prime}}} 
\widehat{m}^{\sharp}(\zeta^{\prime})$ satisfy the following jump conditions,
\begin{equation*}
\widehat{m}_{+}^{\sharp}(\zeta) \! = \! \widehat{m}_{-}^{\sharp}(\zeta) 
\exp (-\mi k(\zeta)(x \! + \! 2 \lambda (\zeta)t) \mathrm{ad}(\sigma_{3})) 
\widehat{\mathcal{G}}^{\sharp}(\zeta), \quad \zeta \! \in \! \mathbb{R},
\end{equation*}
where
\begin{equation*}
\widehat{\mathcal{G}}^{\sharp}(\zeta) \! = \! 
\begin{pmatrix}
(1 \! - \! r(\zeta) \overline{r(\overline{\zeta})}) \delta_{-}(\zeta)
(\delta_{+}(\zeta))^{-1} & \, \, -\overline{r(\overline{\zeta})} \, \delta_{
-}(\zeta) \delta_{+}(\zeta) \prod_{k=m+1}^{N}(d_{k}^{+}(\zeta))^{2} \\
r(\zeta)(\delta_{-}(\zeta) \delta_{+}(\zeta))^{-1} \prod_{k=m+1}^{N}(d_{k}^{
+}(\zeta))^{-2} & \, \, (\delta_{-}(\zeta))^{-1} \delta_{+}(\zeta)
\end{pmatrix},
\end{equation*}
and
\begin{equation*}
\widehat{m}^{\sharp}_{+}(\zeta) \! = \! 
\begin{cases}
\widehat{m}^{\sharp}_{-}(\zeta) \! \left(\mathrm{I} \! + \! \tfrac{C_{n}^{
\mathscr{K}}}{(\zeta-\varsigma_{n})} \sigma_{+} \right), &\text{$\zeta \! \in 
\! \widehat{\mathscr{K}}_{n}, \quad n \! \in \! \{m \! + \! 1,m \! + \! 2,
\ldots,N\}$,} \\
\widehat{m}^{\sharp}_{-}(\zeta) \! \left(\mathrm{I} \! + \! \tfrac{C_{n}^{
\mathscr{L}}}{(\zeta-\overline{\varsigma_{n}})} \sigma_{-} \right), &\text{$
\zeta \! \in \! \widehat{\mathscr{L}}_{n}, \quad n \! \in \! \{m \! + \! 1,m 
\! + \! 2,\ldots,N\};$}
\end{cases}
\end{equation*}
\item[(iii)] $\widehat{m}^{\sharp}(\zeta)$ has simple poles in $\sigma_{d}^{
\prime}$ with
\begin{align*}
\mathrm{Res}(\widehat{m}^{\sharp}(\zeta);\varsigma_{n}) &= \! \lim_{\zeta \to 
\varsigma_{n}} \widehat{m}^{\sharp}(\zeta)g_{n}(\delta (\varsigma_{n}))^{-2} 
\! \left(\prod_{k=m+1}^{N}(d_{k}^{+}(\varsigma_{n}))^{-2} \right) \! \sigma_{
-}, & n \! &\in \! \{1,2,\ldots,m\}, \\
\mathrm{Res}(\widehat{m}^{\sharp}(\zeta);\overline{\varsigma_{n}}) &= \! 
\sigma_{1} \overline{\mathrm{Res}(\widehat{m}^{\sharp}(\zeta);\varsigma_{
n})} \, \sigma_{1}, & n \! &\in \! \{1,2,\ldots,m\};
\end{align*}
\item[(iv)] $\det (\widehat{m}^{\sharp}(\zeta)) \vert_{\zeta = \pm 1} \! = \! 
0;$
\item[(v)] $\widehat{m}^{\sharp}(\zeta) \! =_{\zeta \to 0} \! \zeta^{-1}
(\delta (0))^{\sigma_{3}} \! \left(\prod_{k=m+1}^{N}(d_{k}^{+}(0))^{\sigma_{
3}} \right) \! \sigma_{2} \! + \! \mathcal{O}(1);$
\item[(vi)] $\widehat{m}^{\sharp}(\zeta) \! =_{\genfrac{}{}{0pt}{2}{\zeta \to 
\infty}{\zeta \in \mathbb{C} \, \setminus \sigma_{\mathcal{O}^{\mathcal{
D}}}^{\prime}}} \! \mathrm{I} \! + \! \mathcal{O}(\zeta^{-1});$
\item[(vii)] $\widehat{m}^{\sharp}(\zeta) \! = \! \sigma_{1} \overline{
\widehat{m}^{\sharp}(\overline{\zeta})} \, \sigma_{1}$ and $\widehat{m}^{
\sharp}(\zeta^{-1}) \! = \! \zeta \widehat{m}^{\sharp}(\zeta)(\delta (0))^{
\sigma_{3}} \! \left(\prod_{k=m+1}^{N}(d_{k}^{+}(0))^{\sigma_{3}} \right) 
\! \sigma_{2}$.
\end{enumerate}
Let
\begin{gather}
u(x,t) \! := \! \mi \lim_{\genfrac{}{}{0pt}{2}{\zeta \to \infty}{\zeta 
\in \mathbb{C} \setminus \sigma_{\mathcal{O}^{\mathcal{D}}}^{\prime}}} \! 
\left(\zeta \! \left(\widehat{m}^{\sharp}(\zeta)(\delta (\zeta))^{\sigma_{
3}} \prod_{k=m+1}^{N}(d_{k}^{+}(\zeta))^{\sigma_{3}} \! - \! \mathrm{I} 
\right) \right)_{12},
\end{gather}
and
\begin{gather}
\int\nolimits_{+\infty}^{x}(\vert u(x^{\prime},t) \vert^{2} \! - \! 1) \, \md 
x^{\prime} \! := \! -\mi \lim_{\genfrac{}{}{0pt}{2}{\zeta \to \infty}{\zeta 
\in \mathbb{C} \setminus \sigma_{\mathcal{O}^{\mathcal{D}}}^{\prime}}} \! 
\left(\zeta \! \left(\widehat{m}^{\sharp}(\zeta)(\delta (\zeta))^{\sigma_{
3}} \prod_{k=m+1}^{N}(d_{k}^{+}(\zeta))^{\sigma_{3}} \! - \! \mathrm{I} 
\right) \right)_{11}.
\end{gather}
Then $u(x,t)$ is the solution of the Cauchy problem for the 
{\rm D${}_{f}$NLSE}.
\end{ccccc}

\emph{Proof.} {}From the definition of $\widehat{m}^{\sharp}(\zeta)$ given in 
the Lemma, one shows that, for $m \! \in \! \{1,2,\ldots,N\}$, $\widehat{m}^{
\sharp}_{+}(\zeta) \! = \! \widehat{m}^{\sharp}_{-}(\zeta) \widehat{\upsilon}^{
\sharp}_{\widehat{\mathscr{K}}_{n}}(\zeta)$, $\zeta \! \in \! \cup_{n=m+1}^{
N} \widehat{\mathscr{K}}_{n}$, and $\widehat{m}^{\sharp}_{+}(\zeta) \! = \! 
\widehat{m}^{\sharp}_{-}(\zeta) \widehat{\upsilon}^{\sharp}_{\widehat{
\mathscr{L}}_{n}}(\zeta)$, $\zeta \! \in \! \cup_{n=m+1}^{N} \widehat{
\mathscr{L}}_{n}$, where
\begin{align*}
\widehat{\upsilon}^{\sharp}_{\widehat{\mathscr{K}}_{n}}(\zeta) \! =& \left(
\prod_{k=m+1}^{N}(d_{k}^{-}(\zeta))^{\sigma_{3}} \right) \! J_{\widehat{
\mathscr{K}}_{n}}(\zeta) \! \left(\mathrm{I} \! + \! \tfrac{g_{n}(\delta 
(\varsigma_{n}))^{-2}}{(\zeta -\varsigma_{n})} \sigma_{-} \right) \! \prod_{
k=m+1}^{N}(d_{k}^{+}(\zeta))^{-\sigma_{3}}, \\
\widehat{\upsilon}^{\sharp}_{\widehat{\mathscr{L}}_{n}}(\zeta) \! =& \left(
\prod_{k=m+1}^{N}(d_{k}^{+}(\zeta))^{\sigma_{3}} \right) \! \! \left(\mathrm{
I} \! + \! \tfrac{\overline{g_{n}(\delta (\varsigma_{n}))^{-2}}}{(\zeta -
\overline{\varsigma_{n}})} \sigma_{+} \right) \! (J_{\widehat{\mathscr{
L}}_{n}}(\zeta))^{-1} \prod_{k=m+1}^{N}(d_{k}^{-}(\zeta))^{-\sigma_{3}}.
\end{align*}
Now, as in \cite{a34}, demanding that $\widehat{\upsilon}^{\sharp}_{\widehat{
\mathscr{K}}_{n}}(\zeta)$ (respectively, $\widehat{\upsilon}^{\sharp}_{
\widehat{\mathscr{L}}_{n}}(\zeta))$ have the following upper (respectively, 
lower) diagonal structure, $\widehat{\upsilon}^{\sharp}_{\widehat{\mathscr{
K}}_{n}}(\zeta) \! = \! \mathrm{I} \! + \! C_{n}^{\mathscr{K}}(\zeta \! - \! 
\varsigma_{n})^{-1} \sigma_{+}$ (respectively, $\widehat{\upsilon}^{\sharp}_{
\widehat{\mathscr{L}}_{n}}(\zeta) \! = \! \mathrm{I} \! + \! C_{n}^{\mathscr{
L}}(\zeta \! - \! \overline{\varsigma_{n}})^{-1} \sigma_{-})$, one arrives at
\begin{align*}
J_{\widehat{\mathscr{K}}_{n}}(\zeta) \! =& \! 
\begin{pmatrix}
\prod_{k=m+1}^{N} \frac{d_{k}^{+}(\zeta)}{d_{k}^{-}(\zeta)} \! - \! \frac{C_{
n}^{\mathscr{K}}g_{n}(\delta (\varsigma_{n}))^{-2}}{(\zeta -\varsigma_{n})^{
2}} \prod_{k=m+1}^{N} \frac{(d_{k}^{+}(\zeta))^{-1}}{d_{k}^{-}(\zeta)} & \, 
\, \frac{C_{n}^{\mathscr{K}}}{(\zeta -\varsigma_{n})} \prod_{k=m+1}^{N} \frac{
(d_{k}^{+}(\zeta))^{-1}}{d_{k}^{-}(\zeta)} \\
-\frac{g_{n}(\delta (\varsigma_{n}))^{-2}}{(\zeta -\varsigma_{n})} \prod_{k=m
+1}^{N} \frac{d_{k}^{-}(\zeta)}{d_{k}^{+}(\zeta)} & \, \, \prod_{k=m+1}^{N} 
\frac{d_{k}^{-}(\zeta)}{d_{k}^{+}(\zeta)}
\end{pmatrix}, \\
J_{\widehat{\mathscr{L}}_{n}}(\zeta) \! =& \! 
\begin{pmatrix}
\prod_{k=m+1}^{N} \frac{d_{k}^{+}(\zeta)}{d_{k}^{-}(\zeta)} & \, \, \frac{
\overline{g_{n}(\delta (\varsigma_{n}))^{-2}}}{(\zeta - \overline{\varsigma_{
n}})} \prod_{k=m+1}^{N} \frac{d_{k}^{+}(\zeta)}{d_{k}^{-}(\zeta)} \\
-\frac{C_{n}^{\mathscr{L}}}{(\zeta -\overline{\varsigma_{n}})} \prod_{k=m+
1}^{N} \frac{d_{k}^{-}(\zeta)}{(d_{k}^{+}(\zeta))^{-1}} & \, \, \prod_{k=m
+1}^{N} \frac{d_{k}^{-}(\zeta)}{d_{k}^{+}(\zeta)} \! - \! \frac{C_{n}^{
\mathscr{L}} \overline{g_{n}(\delta (\varsigma_{n}))^{-2}}}{(\zeta -
\overline{\varsigma_{n}})^{2}} \prod_{k=m+1}^{N} \frac{d_{k}^{-}(\zeta)}{(d_{
k}^{+}(\zeta))^{-1}}
\end{pmatrix},
\end{align*}
with $\det (J_{\widehat{\mathscr{K}}_{n}}(\zeta)) \! = \! \det (J_{\widehat{
\mathscr{L}}_{n}}(\zeta)) \! = \! 1$. Choosing $d_{n}^{\pm}(\zeta)$ as in the 
Lemma, one shows that
\begin{align*}
\mathrm{Res}(J_{\widehat{\mathscr{K}}_{n}}(\zeta);\varsigma_{n}) \! =& \! 
\begin{pmatrix}
\left. \left( \prod_{\genfrac{}{}{0pt}{2}{k=m+1}{k \not= n}}^{N} \frac{d_{k}^{
+}(\zeta)}{d_{k}^{-}(\zeta)} \! - \! \frac{C_{n}^{\mathscr{K}}g_{n}(\delta 
(\varsigma_{n}))^{-2}}{(\zeta -\overline{\varsigma_{n}})^{2}} \prod_{
\genfrac{}{}{0pt}{2}{k=m+1}{k \not= n}}^{N} \frac{(d_{k}^{+}(\zeta))^{-1}}{d_{
k}^{-}(\zeta)} \right) \! \right\vert_{\zeta = \varsigma_{n}} & 0 \\
0 & 0
\end{pmatrix}, \\
\mathrm{Res}(J_{\widehat{\mathscr{L}}_{n}}(\zeta);\overline{\varsigma_{n}}) 
\! =& \! 
\begin{pmatrix}
0 & 0 \\ 
0 & \left. \left(\prod_{\genfrac{}{}{0pt}{2}{k=m+1}{k \not= n}}^{N} \frac{d_{
k}^{-}(\zeta)}{d_{k}^{+}(\zeta)} \! - \! \frac{C_{n}^{\mathscr{L}} \overline{
g_{n}(\delta (\varsigma_{n}))^{-2}}}{(\zeta -\varsigma_{n})^{2}} \prod_{
\genfrac{}{}{0pt}{2}{k=m+1}{k \not= n}}^{N} \frac{d_{k}^{-}(\zeta)}{(d_{k}^{
+}(\zeta))^{-1}} \right) \! \right\vert_{\zeta = \overline{\varsigma_{n}}}
\end{pmatrix}:
\end{align*}
choosing $C_{n}^{\mathscr{K}}$ and $C_{n}^{\mathscr{L}}$ as in the Lemma, 
one gets that $\mathrm{Res}(J_{\widehat{\mathscr{K}}_{n}}(\zeta);\varsigma_{
n}) \! = \! \mathrm{Res}(J_{\widehat{\mathscr{L}}_{n}}(\zeta);\overline{
\varsigma_{n}}) \! = \! \mathbf{0}$; thus, $J_{\widehat{\mathscr{K}}_{n}}
(\zeta)$ (respectively, $J_{\widehat{\mathscr{L}}_{n}}(\zeta))$ is 
holomorphic in $\cup_{n=m+1}^{N} \mathrm{int}(\widehat{\mathscr{K}}_{n})$ 
(respectively, $\cup_{n=m+1}^{N} \mathrm{int}(\widehat{\mathscr{L}}_{n}))$. 
The remainder of the proof follows {}from Lemma~3.2 and the definition 
of $\widehat{m}^{\sharp}(\zeta)$ given in the Lemma via straightforward 
algebraic calculations. \hfill $\square$
\begin{eeeee}
One notes {}from the proof of Lemma~3.3 that, for $m \! \in \! \{1,2,\ldots,
N\}$, with $\eta_{n} \! := \! \sin (\phi_{n}) \! \in \! (0,1)$ and $\xi_{n} 
\! := \! \cos (\phi_{n}) \! \in \! (-1,1)$, as $t \! \to \! +\infty$ and $x 
\! \to \! -\infty$ such that $z_{o} \! := \! x/t \! < \! -2$ and $(x,t) \! 
\in \! \daleth_{m}$,
\begin{align*}
\widehat{\upsilon}^{\sharp}_{\widehat{\mathscr{K}}_{n}}(\zeta) \! =& \, 
\mathrm{I} \! + \! \tfrac{C_{n}^{\mathscr{K}}}{(\zeta -\varsigma_{n})} 
\sigma_{+} \! = \mathrm{I} \! + \! \mathcal{O} \! \left(\tfrac{\me^{-4t 
\eta_{n} \vert \xi_{n}-\xi_{m} \vert}}{(\zeta -\varsigma_{n})} \sigma_{+} 
\right), \quad \zeta \! \in \! \widehat{\mathscr{K}}_{n}, \quad n \! \in \! 
\{m \! + \! 1,m \! + \! 2,\ldots,N\}, \\
\widehat{\upsilon}^{\sharp}_{\widehat{\mathscr{L}}_{n}}(\zeta) \! =& \, 
\mathrm{I} \! + \! \tfrac{C_{n}^{\mathscr{L}}}{(\zeta -\overline{\varsigma_{
n}})} \sigma_{-} \! = \mathrm{I} \! + \! \mathcal{O} \! \left( \tfrac{
\me^{-4t \eta_{n} \vert \xi_{n}-\xi_{m} \vert}}{(\zeta -\overline{\varsigma_{
n}})} \sigma_{-} \right), \quad \zeta \! \in \! \widehat{\mathscr{L}}_{n}, 
\quad n \! \in \! \{m \! + \! 1,m \! + \! 2,\ldots,N\};
\end{align*}
hence, as $t \! \to \! +\infty$, $\widehat{\upsilon}^{\sharp}_{\widehat{
\star}_{n}}(\zeta) \! \to \! \mathrm{I}$ (uniformly), where $\star \! \in 
\! \{\mathscr{K},\mathscr{L}\}$. One also notes {}from Lemmae~3.1--3.3 that, 
for $\zeta \! \in \! \cup_{n=m+1}^{N} \mathrm{int}(\widehat{\mathscr{K}}_{
n})$,
\begin{equation*}
\widehat{m}^{\sharp}(\zeta) \! = \! \widehat{m}(\zeta) \! 
\begin{pmatrix}
\left(\frac{\zeta -\varsigma_{n}}{\zeta -\overline{\varsigma_{n}}} \right) 
\! \prod_{\genfrac{}{}{0pt}{2}{k=m+1}{k \not= n}}^{N}(d_{k}^{+}(\zeta))^{-1} 
& -\frac{C_{n}^{\mathscr{K}}}{(\zeta -\overline{\varsigma_{n}})} \prod_{
\genfrac{}{}{0pt}{2}{k=m+1}{k \not= n}}^{N}(d_{k}^{+}(\zeta))^{-1} \\
0 & \left(\frac{\zeta -\overline{\varsigma_{n}}}{\zeta -\varsigma_{n}} 
\right) \! \prod_{\genfrac{}{}{0pt}{2}{k=m+1}{k \not= n}}^{N}d_{k}^{+}
(\zeta)
\end{pmatrix},
\end{equation*}
and, for $\zeta \! \in \! \cup_{n=m+1}^{N} \mathrm{int}(\widehat{\mathscr{
L}}_{n})$,
\begin{equation*}
\widehat{m}^{\sharp}(\zeta) \! = \! \widehat{m}(\zeta) \! 
\begin{pmatrix}
\left(\frac{\zeta -\varsigma_{n}}{\zeta -\overline{\varsigma_{n}}} \right) 
\! \prod_{\genfrac{}{}{0pt}{2}{k=m+1}{k \not= n}}^{N}(d_{k}^{+}(\zeta))^{-1} 
& 0 \\
\frac{C_{n}^{\mathscr{L}}}{(\zeta -\varsigma_{n})} \prod_{\genfrac{}{}{0pt}
{2}{k=m+1}{k \not= n}}^{N}d_{k}^{+}(\zeta) & \left(\frac{\zeta -\overline{
\varsigma_{n}}}{\zeta -\varsigma_{n}} \right) \! \prod_{\genfrac{}{}{0pt}
{2}{k=m+1}{k \not= n}}^{N}d_{k}^{+}(\zeta)
\end{pmatrix};
\end{equation*}
hence, modulo singular terms like $(\zeta \! - \! \varsigma_{n})^{-1}$ and 
$(\zeta \! - \! \overline{\varsigma_{n}})^{-1}$, and recalling that (see 
above), as $t \! \to \! +\infty$, $C_{n}^{\mathscr{K}}$ and $C_{n}^{
\mathscr{L}}$ are $\mathcal{O}(\exp (-4t \eta_{n} \vert \xi_{n} \! - \! 
\xi_{m} \vert))$, one deduces that, since the RHP for $\widehat{m}(\zeta)$ 
formulated in Lemma~3.1 is asymptotically solvable \cite{a38}, there are no 
exponentially growing factors for $\widehat{m}^{\sharp}(\zeta)$ when $\zeta 
\! \in \! \cup_{n=m+1}^{N}(\mathrm{int}(\widehat{\mathscr{K}}_{n}) \cup 
\mathrm{int}(\widehat{\mathscr{L}}_{n}))$.
\end{eeeee}
By estimating the error along the trajectory of the $m$th dark soliton $(m 
\! \in \! \{1,2,\ldots,N\})$ when the jump matrices on $\{\widehat{\mathscr{
K}}_{n},\widehat{\mathscr{L}}_{n}\}_{n=m+1}^{N}$ are removed {}from the 
specification of the RHP for $\widehat{m}^{\sharp}(\zeta)$, one arrives at 
an asymptotically solvable, model RHP (see Lemma~3.5 below); however, since 
the proof of Lemma~3.5 relies substantially on the Beals-Coifman (BC) 
construction \cite{a41} for the solution of a matrix (and appropriately 
normalised) RHP on an oriented and unbounded contour, it is convenient to 
present, with some requisite preamble, a succinct and self-contained synopsis 
of it at this juncture. But first, the following result is necessary.
\begin{bbbbb}[{\rm \cite{a38}}]
The solution of the {\rm RHP} for $\widehat{m}^{\sharp}(\zeta) \colon \mathbb{
C} \setminus \sigma_{\mathcal{O}^{\mathcal{D}}}^{\prime} \! \to \! \mathrm{
M}_{2}(\mathbb{C})$ formulated in Lemma~{\rm 3.3} has the (integral equation) 
representation
\begin{equation*}
\widehat{m}^{\sharp}(\zeta) \! = \! \left(\mathrm{I} \! + \! \zeta^{-1} 
\widehat{\Delta}_{o}^{\sharp} \right) \! \widehat{\mathscr{P}}^{\sharp}
(\zeta) \! \left(\widehat{m}_{d}^{\sharp}(\zeta) \! + \! \int\nolimits_{
\sigma_{c}^{\prime}} \tfrac{\widehat{m}^{\sharp}_{-}(\mu)(\widehat{
\upsilon}^{\sharp}(\mu)-\mathrm{I})}{(\mu -\zeta)} \, \tfrac{\md \mu}{2 
\pi \mi} \right), \quad \zeta \! \in \! \mathbb{C} \setminus \sigma_{
\mathcal{O}^{\mathcal{D}}}^{\prime},
\end{equation*}
where
\begin{equation*}
\widehat{m}^{\sharp}_{d}(\zeta) \! = \! \mathrm{I} \! + \! \sum_{n=1}^{m} 
\! \left(\tfrac{\mathrm{Res}(\widehat{m}^{\sharp}(\zeta);\varsigma_{n})}{
(\zeta -\varsigma_{n})} \! + \! \tfrac{\sigma_{1} \overline{\mathrm{Res}
(\widehat{m}^{\sharp}(\zeta);\varsigma_{n})} \, \sigma_{1}}{(\zeta 
-\overline{\varsigma_{n}})} \right),
\end{equation*}
$\widehat{v}^{\sharp}(\cdot)$ is a generic notation for the jump matrices of 
$\widehat{m}^{\sharp}(\zeta)$ on $\sigma_{c}^{\prime}$ (Lemma~{\rm 3.3}, 
(ii)), and $\widehat{\Delta}_{o}^{\sharp}$ and $\widehat{\mathscr{P}}^{
\sharp}(\zeta)$ are specified below. The solution of the above (integral) 
equation can be written as the ordered factorisation
\begin{equation*}
\widehat{m}^{\sharp}(\zeta) \! = \! \left(\mathrm{I} \! + \! \zeta^{-1} 
\widehat{\Delta}_{o}^{\sharp} \right) \! \widehat{\mathscr{P}}^{\sharp}
(\zeta) \widehat{m}_{d}^{\sharp}(\zeta)m^{c}(\zeta), \quad \zeta \! \in \! 
\mathbb{C} \setminus \sigma_{\mathcal{O}^{\mathcal{D}}}^{\prime},
\end{equation*}
where $\widehat{m}^{\sharp}_{d}(\zeta) \! = \! \sigma_{1} \overline{\widehat{
m}^{\sharp}_{d}(\overline{\zeta})} \, \sigma_{1}$ $(\in \! \mathrm{SL}(2,
\mathbb{C}))$ has the representation given above, $\widehat{\mathscr{P}}^{
\sharp}(\zeta) \! = \! \sigma_{1} \overline{\widehat{\mathscr{P}}^{\sharp}
(\overline{\zeta})} \, \sigma_{1}$ is chosen so that $\widehat{\Delta}^{
\sharp}_{o}$ is idempotent, $\mathrm{I} \! + \! \zeta^{-1} \widehat{\Delta}_{
o}^{\sharp}$ $(\in \! \mathrm{M}_{2}(\mathbb{C}))$ is holomorphic in a 
punctured neighbourhood of the origin, with $\widehat{\Delta}_{o}^{\sharp} \! 
= \! \sigma_{1} \overline{\widehat{\Delta}_{o}^{\sharp}} \, \sigma_{1}$ $(\in 
\! \mathrm{GL}(2,\mathbb{C}))$ such that $\det (\mathrm{I} \! + \! \zeta^{-1} 
\widehat{\Delta}_{o}^{\sharp}) \vert_{\zeta =\pm 1} \! = \! 0$, and having 
the finite, order 2, matrix involutive structure
\begin{equation*}
\widehat{\Delta}_{o}^{\sharp} \! = \! 
\begin{pmatrix}
\widehat{\Delta}^{\sharp} \me^{\mi (k+1/2) \pi} & (1 \! + \! (\widehat{
\Delta}^{\sharp})^{2})^{1/2} \me^{-\mi \widehat{\vartheta}^{\sharp}} \\
(1 \! + \! (\widehat{\Delta}^{\sharp})^{2})^{1/2} \me^{\mi \widehat{
\vartheta}^{\sharp}} & \widehat{\Delta}^{\sharp} \me^{-\mi (k+1/2) \pi}
\end{pmatrix}, \quad k \! \in \! \mathbb{Z},
\end{equation*}
where $\widehat{\Delta}^{\sharp}$ and $\widehat{\vartheta}^{\sharp}$ are 
obtained {}from the relation $\widehat{\Delta}_{o}^{\sharp} \! = \! \widehat{
\mathscr{P}}^{\sharp}(0) \widehat{m}_{d}^{\sharp}(0)m^{c}(0)(\delta (0))^{
\sigma_{3}} \! \left(\prod_{k=m+1}^{N}(d_{k}^{+}(0))^{\sigma_{3}} \right) \! 
\sigma_{2}$, and satisfying $\mathrm{tr}(\widehat{\Delta}_{o}^{\sharp}) \! 
= \! 0$, $\det (\widehat{\Delta}_{o}^{\sharp}) \! = \! -1$, and $\widehat{
\Delta}_{o}^{\sharp} \widehat{\Delta}_{o}^{\sharp} \! = \! \mathrm{I}$, and 
$m^{c}(\zeta) \colon \mathbb{C} \setminus \sigma_{c}^{\prime} \! \to \! 
\mathrm{SL}(2,\mathbb{C})$ solves the following {\rm RHP:} {\rm (1)} $m^{c}
(\zeta)$ is piecewise (sectionally) holomorphic $\forall \, \zeta \! \in \! 
\mathbb{C} \setminus \sigma_{c}^{\prime};$ {\rm (2)} $m^{c}_{\pm}(\zeta) \! 
:= \! \lim_{\genfrac{}{}{0pt}{2}{\zeta^{\prime} \, \to \, \zeta}{\zeta^{
\prime} \, \in \, \pm \, \mathrm{side} \, \mathrm{of} \, \sigma_{c}^{\prime}}}
m^{c}(\zeta^{\prime})$ satisfy, for $\zeta \! \in \! \sigma_{c}^{\prime}$, 
the jump condition $m^{c}_{+}(\zeta) \! = \! m^{c}_{-}(\zeta) \upsilon^{c}
(\zeta)$, where $\upsilon^{c}(\zeta) \! = \! \exp (-\mi k(\zeta)(x \! + \! 
2 \lambda (\zeta)t) \mathrm{ad}(\sigma_{3})) \widehat{\mathcal{G}}^{\sharp}
(\zeta)$, $\zeta \! \in \! \mathbb{R}$, with $\widehat{\mathcal{G}}^{\sharp}
(\zeta)$ given in Lemma~{\rm 3.3}, (ii), $\upsilon^{c}(\zeta) \! = \! 
\mathrm{I} \! + \! C_{n}^{\mathscr{K}}(\zeta \! - \! \varsigma_{n})^{-1} 
\sigma_{+}$, $\zeta \! \in \! \widehat{\mathscr{K}}_{n}$, and $\upsilon^{
c}(\zeta) \! = \! \mathrm{I} \! + \! C_{n}^{\mathscr{L}}(\zeta \! - \! 
\overline{\varsigma_{n}})^{-1} \sigma_{-}$, $\zeta \! \in \! \widehat{
\mathscr{L}}_{n}$, $n \! \in \! \{m \! + \! 1,m \! + \! 2,\ldots,N\}$, with 
$C_{n}^{\mathscr{K}}$ and $C_{n}^{\mathscr{L}}$ given in Lemma~{\rm 3.3;} 
{\rm (3)} $m^{c}(\zeta) \! =_{\genfrac{}{}{0pt}{2}{\zeta \to \infty}{\zeta 
\in \mathbb{C} \setminus \sigma_{c}^{\prime}}} \! \mathrm{I} \! + \! 
\mathcal{O}(\zeta^{-1});$ and {\rm (4)} $m^{c}(\zeta) \! = \! \sigma_{1} 
\overline{m^{c}(\overline{\zeta})} \, \sigma_{1}$.
\end{bbbbb}
The BC formulation \cite{a41} now follows. One agrees to call a contour 
$\Gamma^{\sharp}$ oriented if: (1) $\mathbb{C} \setminus \Gamma^{\sharp}$ has 
finitely many open connected components; (2) $\mathbb{C} \setminus \Gamma^{
\sharp}$ is the disjoint union of two, possibly disconnected, open regions, 
denoted by $\boldsymbol{\mho}^{+}$ and $\boldsymbol{\mho}^{-}$; and (3) 
$\Gamma^{\sharp}$ may be viewed as either the positively oriented boundary 
for $\boldsymbol{\mho}^{+}$ or the negatively oriented boundary for 
$\boldsymbol{\mho}^{-}$ ($\mathbb{C} \setminus \Gamma^{\sharp}$ is coloured 
by two colours, $\pm)$. Let $\Gamma^{\sharp}$, as a closed set, be the union 
of finitely many oriented simple piecewise-smooth arcs. Denote the set of 
all self-intersections of $\Gamma^{\sharp}$ by $\widehat{\Gamma}^{\sharp}$ 
(with $\mathrm{card}(\widehat{\Gamma}^{\sharp}) \! < \! \infty$ assumed 
throughout). Set $\widetilde{\Gamma}^{\sharp} \! := \! \Gamma^{\sharp} 
\setminus \widehat{\Gamma}^{\sharp}$. The BC construction for the solution 
of a (matrix) RHP, in the absence of a discrete spectrum and spectral 
singularities \cite{a45,a53}, on an oriented contour $\Gamma^{\sharp}$ 
consists of finding an $\mathrm{M}_{2}(\mathbb{C})$-valued function 
$\mathcal{X}(\lambda)$ such that: (1) $\mathcal{X}(\lambda)$ is piecewise 
holomorphic $\forall \, \lambda \! \in \! \mathbb{C} \setminus \Gamma^{
\sharp}$; (2) $\mathcal{X}_{+}(\lambda) \! = \! \mathcal{X}_{-}(\lambda) 
\upsilon (\lambda)$, $\lambda \! \in \! \widetilde{\Gamma}^{\sharp}$, for 
some ``jump'' matrix $\upsilon (\lambda) \colon \widetilde{\Gamma}^{\sharp} 
\! \to \! \mathrm{GL}(2,\mathbb{C})$; and (3) uniformly as $\lambda \! \to 
\! \infty$, $\lambda \! \in \! \mathbb{C} \setminus \Gamma^{\sharp}$, 
$\mathcal{X}(\lambda) \! = \! \mathrm{I} \! + \! \mathcal{O}(\lambda^{-1})$. 
Let $\upsilon (\lambda) \! := \! (\mathrm{I} \! - \! w_{-}(\lambda))^{-1}
(\mathrm{I} \! + \! w_{+}(\lambda))$, $\lambda \! \in \! \widetilde{\Gamma}^{
\sharp}$, be a factorisation for $\upsilon (\lambda)$, where $w_{\pm}
(\lambda)$ are some upper/lower, or lower/upper, triangular (depending on 
the orientation of $\Gamma^{\sharp})$ nilpotent matrices, with degree of 
nilpotency 2, and $w_{\pm}(\lambda) \! \in \! \cap_{p \in \{2,\infty\}} 
\mathcal{L}^{p}_{\mathrm{M}_{2}(\mathbb{C})}(\widetilde{\Gamma}^{\sharp})$ 
(if $\widetilde{\Gamma}^{\sharp}$ is unbounded, one requires that $w_{\pm}
(\lambda) \! =_{\genfrac{}{}{0pt}{2}{\lambda \to \infty}{\lambda \in 
\widetilde{\Gamma}^{\sharp}}} \! \mathbf{0})$. Define $w(\lambda) \! := \! 
w_{+}(\lambda) \! + \! w_{-}(\lambda)$, and introduce the Cauchy operators 
on $\mathcal{L}^{2}_{\mathrm{M}_{2}(\mathbb{C})}(\Gamma^{\sharp})$, $(C_{\pm}
f)(\lambda) \! := \! \lim_{\genfrac{}{}{0pt}{2}{\lambda^{\prime} \, \to \, 
\lambda}{\lambda^{\prime} \, \in \, \pm \, \mathrm{side} \, \mathrm{of} \, 
\Gamma^{\sharp}}} \int_{\Gamma^{\sharp}} \tfrac{f(z)}{(z-\lambda^{\prime})} 
\, \tfrac{\md z}{2 \pi \mi}$, where $f(\cdot) \! \in \! \mathcal{L}^{2}_{
\mathrm{M}_{2}(\mathbb{C})}(\Gamma^{\sharp})$, with $C_{\pm} \colon \mathcal{
L}^{2}_{\mathrm{M}_{2}(\mathbb{C})}(\Gamma^{\sharp}) \! \to \! \mathcal{L}^{
2}_{\mathrm{M}_{2}(\mathbb{C})}(\Gamma^{\sharp})$ bounded in operator 
norm\footnote{$\vert \vert C_{\pm} \vert \vert_{\mathscr{N}(\Gamma^{
\sharp})} \! < \! \infty$, where $\mathscr{N}(\ast)$ denotes the space of 
all bounded linear operators acting {}from $\mathcal{L}^{2}_{\mathrm{M}_{2}
(\mathbb{C})}(\ast)$ into $\mathcal{L}^{2}_{\mathrm{M}_{2}(\mathbb{C})}
(\ast)$.}, and $\vert \vert (C_{\pm}f)(\cdot) \vert \vert_{\mathcal{L}^{2}_{
\mathrm{M}_{2}(\mathbb{C})}(\ast)} \! \leqslant \! \mathrm{const.} \vert 
\vert f(\cdot) \vert \vert_{\mathcal{L}^{2}_{\mathrm{M}_{2}(\mathbb{C})}
(\ast)}$. Introduce the BC operator:
\begin{equation*}
C_{w}f \! := \! C_{+}(fw_{-}) \! + \! C_{-}(fw_{+}), \qquad f(\cdot) \! 
\in \! \mathcal{L}^{2}_{\mathrm{M}_{2}(\mathbb{C})}(\ast);
\end{equation*}
moreover, since $\mathbb{C} \setminus \Gamma^{\sharp}$ can be coloured by two 
colours $(\pm)$, $C_{\pm}$ are complementary projections \cite{a45}, namely, 
$C_{+}^{2} \! = \! C_{+}$, $C_{-}^{2} \! = \! -C_{-}$, $C_{+}C_{-} \! = \! 
C_{-}C_{+} \! = \! \underline{\mathbf{0}}$ (the null operator), and $C_{+} 
\! - \! C_{-} \! = \! \id$ (the identity operator): in the case that $C_{+}$ 
and $-C_{-}$ are complementary, the contour $\Gamma^{\sharp}$ can always be 
oriented in such a way that the $\pm$ regions lie on the $\pm$ sides of the 
contour, respectively. Specialising the BC construction to the solution 
of the RHP for $m^{c}(\zeta)$ on $\sigma_{c}^{\prime}$ formulated in 
Proposition~3.1, and writing $\upsilon^{c}(\zeta)$ as the following (bounded) 
algebraic factorisation $\upsilon^{c}(\zeta) \! := \! (\mathrm{I} \! - \! 
w_{-}^{c}(\zeta))^{-1}(\mathrm{I} \! + \! w_{+}^{c}(\zeta))$, $\zeta \! \in 
\! \sigma_{c}^{\prime}$, the integral representation for $m^{c}(\zeta)$ is 
given by the following
\begin{ccccc}[Beals and Coifman {\rm \cite{a41}}]
Let
\begin{equation*}
\mu^{c}(\zeta) \! = \! m^{c}_{+}(\zeta)(\mathrm{I} \! + \! w_{+}^{c}
(\zeta))^{-1} \! = \! m^{c}_{-}(\zeta)(\mathrm{I} \! - \! w_{-}^{c}
(\zeta))^{-1}, \quad \zeta \! \in \! \sigma_{c}^{\prime}.
\end{equation*}
If $\mu^{c}(\zeta) \! \in \! \mathrm{I} \! + \! \mathcal{L}^{2}_{\mathrm{M}_{
2}(\mathbb{C})}(\sigma_{c}^{\prime}) \! := \! \{\mathstrut \mathrm{I} \! + \! 
h(\cdot); \, h(\cdot) \! \in \! \mathcal{L}^{2}_{\mathrm{M}_{2}(\mathbb{C})}
(\sigma_{c}^{\prime})\}$\footnote{For $f(\zeta) \! \in \! \mathrm{I} \! + \! 
\mathcal{L}^{2}_{\mathrm{M}_{2}(\mathbb{C})}(\ast)$, $\vert \vert f(\cdot) 
\vert \vert_{\mathrm{I}+\mathcal{L}^{2}_{\mathrm{M}_{2}(\mathbb{C})}(\ast)} 
\! := \! \left(\vert \vert f(\infty) \vert \vert^{2}_{\mathcal{L}^{\infty}_{
\mathrm{M}_{2}(\mathbb{C})}(\ast)} \! + \! \vert \vert f(\cdot) \! - \! 
f(\infty) \vert \vert^{2}_{\mathcal{L}^{2}_{\mathrm{M}_{2}(\mathbb{C})}
(\ast)} \right)^{1/2}$ \cite{a44}.} solves the linear singular integral 
equation
\begin{equation*}
(\id \! - \! C_{w^{c}})(\mu^{c}(\zeta) \! - \! \mathrm{I})=C_{w^{c}} 
\mathrm{I}=C_{+}(w_{-}^{c}) \! + \! C_{-}(w_{+}^{c}), \quad \zeta \! \in 
\! \sigma_{c}^{\prime},
\end{equation*}
where $\id$ is the identity operator on $\mathcal{L}^{2}_{\mathrm{M}_{2}
(\mathbb{C})}(\sigma_{c}^{\prime})$, then the solution of the {\rm RHP} for 
$m^{c}(\zeta)$ is
\begin{equation*}
m^{c}(\zeta)=\mathrm{I} \! + \! \int\nolimits_{\sigma_{c}^{\prime}} \dfrac{
\mu^{c}(z)w^{c}(z)}{(z \! - \! \zeta)} \, \dfrac{\md z}{2 \pi \mi}, \qquad 
\zeta \! \in \! \mathbb{C} \setminus \sigma_{c}^{\prime},
\end{equation*}
where $\mu^{c}(\zeta) \! = \! ((\id \! - \! C_{w^{c}})^{-1} \mathrm{I})
(\zeta)$, and $w^{c}(\zeta) \! := \! w_{+}^{c}(\zeta) \! + \! w_{-}^{c}
(\zeta)$.
\end{ccccc}
Finally, one arrives at, and is in a position to prove, the following
\begin{ccccc}
For $m \! \in \! \{1,2,\ldots,N\}$, set $\widehat{\sigma}_{d} \! := \! 
\cup_{n=1}^{m}(\{\varsigma_{n}\} \cup \{\overline{\varsigma_{n}}\})$, and 
let $\sigma_{c} \! = \! \{\mathstrut \zeta; \, \Im (\zeta) \! = \! 0\}$  
with orientation {}from $-\infty$ to $+\infty$. Let $\widehat{\chi}(\zeta) 
\colon \mathbb{C} \setminus (\widehat{\sigma}_{d} \cup \sigma_{c}) \! \to 
\! \mathrm{M}_{2}(\mathbb{C})$ solve the following {\rm RHP:}
\begin{enumerate}
\item[(i)] $\widehat{\chi}(\zeta)$ is piecewise (sectionally) meromorphic 
$\forall \, \zeta \! \in \! \mathbb{C} \setminus \sigma_{c};$
\item[(ii)] $\widehat{\chi}_{\pm}(\zeta) \! := \! \lim_{\genfrac{}{}{0pt}{2}
{\zeta^{\prime} \, \to \, \zeta}{\zeta^{\prime} \, \in \, \pm \, 
\mathrm{side} \, \mathrm{of} \, \sigma_{c}}} \widehat{\chi}(\zeta^{\prime})$ 
satisfy the jump condition
\begin{equation*}
\widehat{\chi}_{+}(\zeta) \! = \! \widehat{\chi}_{-}(\zeta) \exp (-\mi 
k(\zeta)(x \! + \! 2 \lambda (\zeta)t) \mathrm{ad}(\sigma_{3})) \widehat{
\mathcal{G}}^{\sharp}(\zeta), \quad \zeta \! \in \! \mathbb{R};
\end{equation*}
\item[(iii)] $\widehat{\chi}(\zeta)$ has simple poles in $\widehat{\sigma}_{
d}$ with
\begin{align*}
\mathrm{Res}(\widehat{\chi}(\zeta);\varsigma_{n}) &= \! \lim_{\zeta \to 
\varsigma_{n}} \widehat{\chi}(\zeta)g_{n}(\delta (\varsigma_{n}))^{-2} \! 
\left(\prod_{k=m+1}^{N}(d_{k}^{+}(\varsigma_{n}))^{-2} \right) \! \sigma_{-}, 
& n \! &\in \! \{1,2,\ldots,m\}, \\
\mathrm{Res}(\widehat{\chi}(\zeta);\overline{\varsigma_{n}}) &= \! \sigma_{
1} \overline{\mathrm{Res}(\widehat{\chi}(\zeta);\varsigma_{n})} \, \sigma_{
1}, & n \! &\in \! \{1,2,\ldots,m\};
\end{align*}
\item[(iv)] $\det (\widehat{\chi}(\zeta)) \vert_{\zeta =\pm 1} \! = \! 0;$
\item[(v)] $\widehat{\chi}(\zeta) \! =_{\zeta \to 0} \! \zeta^{-1}(\delta 
(0))^{\sigma_{3}} \! \left(\prod_{k=m+1}^{N}(d_{k}^{+}(0))^{\sigma_{3}} 
\right) \! \sigma_{2} \! + \! \mathcal{O}(1);$
\item[(vi)] $\widehat{\chi}(\zeta) \! =_{\genfrac{}{}{0pt}{2}{\zeta \to 
\infty}{\zeta \in \mathbb{C} \setminus (\widehat{\sigma}_{d} \cup \sigma_{
c})}} \! \mathrm{I} \! + \! \mathcal{O}(\zeta^{-1});$
\item[(vii)] $\widehat{\chi}(\zeta) \! = \! \sigma_{1} \overline{\widehat{
\chi}(\overline{\zeta})} \, \sigma_{1}$ and $\widehat{\chi}(\zeta^{-1}) \! = 
\! \zeta \widehat{\chi}(\zeta)(\delta (0))^{\sigma_{3}} \! \left(\prod_{k=m
+1}^{N}(d_{k}^{+}(0))^{\sigma_{3}} \right) \! \sigma_{2}$.
\end{enumerate}
Then, as $t \! \to \! +\infty$ and $x \! \to \! -\infty$ such that $z_{o} \! 
:= \! x/t \! < \! -2$ and $(x,t) \! \in \! \daleth_{m}$, $\widehat{m}^{\sharp}
(\zeta) \colon \mathbb{C} \setminus \sigma^{\prime}_{\mathcal{O}^{\mathcal{
D}}} \! \to \! \mathrm{M}_{2}(\mathbb{C})$ has the following asymptotics:
\begin{equation*}
\widehat{m}^{\sharp}(\zeta) \! = \! \left(\mathrm{I} \! + \! \mathcal{O} 
\! \left(\widehat{\mathscr{F}}(\zeta) \exp \! \left(-\widehat{\beth} \, t 
\right) \right) \right) \! \widehat{\chi}(\zeta),
\end{equation*}
where $\widehat{\beth} \! := \! 4 \min_{\genfrac{}{}{0pt}{2}{m \in \{1,2,
\ldots,N\}}{n \in \{m+1,m+2,\ldots,N\}}}\{\sin (\phi_{n}) \vert \cos (\phi_{
n}) \! - \! \cos (\phi_{m}) \vert\}$ $(> \! 0)$, and, for $i,j \! \in \! 
\{1,2\}$, $(\widehat{\mathscr{F}}(\zeta))_{ij} \! =_{\zeta \to \infty} \! 
\mathcal{O}(\vert \zeta \vert^{-1})$ and $(\widehat{\mathscr{F}}(\zeta))_{
ij} \! =_{\zeta \to 0} \! \mathcal{O}(1)$. Furthermore, let
\begin{gather}
u(x,t) \! := \! \mi \lim_{\genfrac{}{}{0pt}{2}{\zeta \to \infty}{\zeta \in 
\mathbb{C} \setminus (\widehat{\sigma}_{d} \cup \sigma_{c})}} \! \left(\zeta 
\! \left(\widehat{\chi}(\zeta)(\delta (\zeta))^{\sigma_{3}} \! \prod_{k
=m+1}^{N}(d_{k}^{+}(\zeta))^{\sigma_{3}} \! - \! \mathrm{I} \right) 
\right)_{12} \! + \! \mathcal{O} \! \left(\exp \! \left(-\widehat{\beth} \, 
t \right) \right),
\end{gather}
and
\begin{gather}
\int\nolimits_{+\infty}^{x}(\vert u(x^{\prime},t) \vert^{2} \! - \! 1) \, \md 
x^{\prime} \! := \! -\mi \lim_{\genfrac{}{}{0pt}{2}{\zeta \to \infty}{\zeta 
\in \mathbb{C} \setminus (\widehat{\sigma}_{d} \cup \sigma_{c})}} \! \left(
\zeta \! \left(\widehat{\chi}(\zeta)(\delta (\zeta))^{\sigma_{3}} \! \prod_{
k=m+1}^{N}(d_{k}^{+}(\zeta))^{\sigma_{3}} \! - \! \mathrm{I} \right) 
\right)_{11} \! + \! \mathcal{O} \! \left(\exp \! \left(-\widehat{\beth} \, 
t \right) \right).
\end{gather}
Then $u(x,t)$ is the solution of the Cauchy problem for the 
{\rm D${}_{f}$NLSE}.
\end{ccccc}
\begin{eeeee}
The solution of the (normalised at $\infty)$ RHP for $\widehat{\chi}(\zeta) 
\colon \mathbb{C} \setminus (\widehat{\sigma}_{d} \cup \sigma_{c}) \! \to 
\! \mathrm{M}_{2}(\mathbb{C})$ formulated in Lemma~3.5 has a factorised 
representation analogous to that of $\widehat{m}^{\sharp}(\zeta)$ given 
in Proposition~3.1 (with appropriate change(s) of notation).
\end{eeeee}

\emph{Proof.} Define $\mathscr{E}(\zeta) \! := \! \widehat{m}^{\sharp}
(\zeta)(\widehat{\chi}(\zeta))^{-1}$. {}From this definition, the RHPs for 
$\widehat{m}^{\sharp}(\zeta)$ and $\widehat{\chi}(\zeta)$ formulated in 
Lemmae~3.3 and~3.5, respectively, Proposition~3.1, and Remark~3.5, one shows 
that, for $m \! \in \! \{1,2,\ldots,N\}$ and $n \! \in \! \{m \! + \! 1,m 
\! + \! 2,\ldots,N\}$, $\mathscr{E}(\zeta)$ solves the following RHP: (1) 
$\mathscr{E}(\zeta)$ is piecewise (sectionally) holomorphic $\forall \, 
\zeta \! \in \! \mathbb{C} \setminus \Sigma_{\mathscr{E}}$, where $\Sigma_{
\mathscr{E}} \! = \! \cup_{n=m+1}^{N} \Sigma_{\mathscr{E}}^{n}$, with 
$\Sigma_{\mathscr{E}}^{n} \! := \! \widehat{\mathscr{K}}_{n} \cup \widehat{
\mathscr{L}}_{n}$ (with orientations preserved); (2) $\mathscr{E}_{\pm}
(\zeta) \! := \! \lim_{\genfrac{}{}{0pt}{2}{\zeta^{\prime} \to \zeta}{
\zeta^{\prime} \, \in \, \pm \, \mathrm{side} \, \mathrm{of} \, \Sigma_{
\mathscr{E}}}} \mathscr{E}(\zeta^{\prime})$ satisfy the jump condition 
$\mathscr{E}_{+}(\zeta) \! = \! \mathscr{E}_{-}(\zeta) \upsilon_{\mathscr{
E}}(\zeta)$, $\zeta \! \in \! \Sigma_{\mathscr{E}}$, where
\begin{align*}
\upsilon_{\mathscr{E}}(\zeta) \! = \! 
\begin{cases}
\mathrm{I} \! + \! \widetilde{\mathscr{W}}_{\mathscr{E}}^{\widehat{\mathscr{
K}}_{n}}(\zeta), &\text{$\zeta \! \in \! \cup_{n=m+1}^{N} \widehat{\mathscr{
K}}_{n} \, (\subset \! \Sigma_{\mathscr{E}}), \quad n \! \in \! \{m \! + \! 
1,m \! + \! 2,\ldots,N\}$,} \\
\mathrm{I} \! + \! \widetilde{\mathscr{W}}_{\mathscr{E}}^{\widehat{\mathscr{
L}}_{n}}(\zeta), &\text{$\zeta \! \in \! \cup_{n=m+1}^{N} \widehat{\mathscr{
L}}_{n} \, (\subset \! \Sigma_{\mathscr{E}}), \quad n \! \in \! \{m \! + \! 
1,m \! + \! 2,\ldots,N\}$,}
\end{cases}
\end{align*}
with $\widetilde{\mathscr{W}}_{\mathscr{E}}^{\widehat{\mathscr{K}}_{n}}(\zeta) 
\! = \! C_{n}^{\mathscr{K}}(\zeta \! - \! \varsigma_{n})^{-1} \mathcal{X}^{
\flat}(\zeta)$, $\widetilde{\mathscr{W}}_{\mathscr{E}}^{\widehat{\mathscr{
L}}_{n}}(\zeta) \! = \! C_{n}^{\mathscr{L}}(\zeta \! - \! \overline{
\varsigma_{n}})^{-1} \mathcal{X}^{\natural}(\zeta)$,
\begin{equation*}
\mathcal{X}^{\flat}(\zeta) \! = \! 
\begin{pmatrix}
-\widehat{\chi}_{11}(\zeta) \widehat{\chi}_{21}(\zeta) & \, \, (\widehat{
\chi}_{11}(\zeta))^{2} \\
-(\widehat{\chi}_{21}(\zeta))^{2} & \, \, \widehat{\chi}_{11}(\zeta) 
\widehat{\chi}_{21}(\zeta)
\end{pmatrix}, \qquad \mathcal{X}^{\natural}(\zeta) \! = \! 
\begin{pmatrix}
\widehat{\chi}_{12}(\zeta) \widehat{\chi}_{22}(\zeta) & \, \, -(\widehat{
\chi}_{12}(\zeta))^{2} \\
(\widehat{\chi}_{22}(\zeta))^{2} & \, \, -\widehat{\chi}_{12}(\zeta) 
\widehat{\chi}_{22}(\zeta)
\end{pmatrix},
\end{equation*}
and $C_{n}^{\mathscr{K}}$ and $C_{n}^{\mathscr{L}}$ given in Lemma~3.3; (3) 
$\det (\mathscr{E}(\zeta)) \vert_{\zeta =\pm 1} \! = \! 1;$ (4) $\mathscr{E}
(\zeta) \! =_{\zeta \to 0} \! \mathcal{O}(1)$ and $\mathscr{E} \! 
=_{\genfrac{}{}{0pt}{2}{\zeta \to \infty}{\zeta \in \mathbb{C} \setminus 
\Sigma_{\mathscr{E}}}} \! \mathrm{I} \! + \! \mathcal{O}(\zeta^{-1});$ and 
(5) $\mathscr{E}(\zeta) \! = \! \sigma_{1} \overline{\mathscr{E}(\overline{
\zeta})} \, \sigma_{1}$ and $\mathscr{E}(\zeta^{-1}) \! = \! \mathscr{E}
(\zeta)$. Note, in particular, that $\mathscr{E}(\zeta)$ has no jump 
discontinuity on $\mathbb{R}$, and no poles. Recall, now, the BC 
construction (see the paragraph preceding Lemma~3.4). Write the following 
(bounded) algebraic factorisation for $\upsilon_{\mathscr{E}}(\zeta)$, 
$\upsilon_{\mathscr{E}}(\zeta) \! = \! (\mathrm{I} \! - \! w^{\mathscr{
E}}_{-}(\zeta))^{-1}(\mathrm{I} \! + \! w^{\mathscr{E}}_{+}(\zeta))$, $\zeta 
\! \in \! \Sigma_{\mathscr{E}}$, and choose \cite{a46} $w^{\mathscr{E}}_{-}
(\zeta) \! = \! \mathbf{0}$; hence, $w^{\mathscr{E}}_{+}(\zeta) \! = \! 
\widetilde{\mathscr{W}}_{\mathscr{E}}^{\widehat{\mathscr{K}}_{n}}(\zeta)$, 
$\zeta \! \in \! \cup_{n=m+1}^{N} \widehat{\mathscr{K}}_{n}$, and $w^{
\mathscr{E}}_{+}(\zeta) \! = \! \widetilde{\mathscr{W}}_{\mathscr{E}}^{
\widehat{\mathscr{L}}_{n}}(\zeta)$, $\zeta \! \in \! \cup_{n=m+1}^{N} 
\widehat{\mathscr{L}}_{n}$. Let $\mu^{\mathscr{E}}(\zeta)$ be the solution 
of the BC linear singular integral equation $(\id_{\mathscr{E}} \!- \! C_{
w^{\mathscr{E}}}) \mu^{\mathscr{E}}(\zeta) \! = \! \mathrm{I}$, $\zeta \! 
\in \! \Sigma_{\mathscr{E}}$, where $\id_{\mathscr{E}}$ is the identity 
operator on $\mathcal{L}_{\mathrm{M}_{2}(\mathbb{C})}^{2}(\Sigma_{\mathscr{
E}})$, and, for $f(\cdot) \! \in \! \mathcal{L}^{2}_{\mathrm{M}_{2}(\mathbb{
C})}(\Sigma_{\mathscr{E}})$, set $C_{w^{\mathscr{E}}}f \! := \! C_{+}(fw^{
\mathscr{E}}_{-}) \! + \! C_{-}(fw^{\mathscr{E}}_{+}) \! = \! C_{-}(fw^{
\mathscr{E}}_{+})$, with $(C_{\pm}f)(\zeta) \! := \! \lim_{\genfrac{}{}{0pt}
{2}{\zeta^{\prime} \to \zeta}{\zeta^{\prime} \, \in \, \pm \, \mathrm{side} 
\, \mathrm{of} \, \Sigma_{\mathscr{E}}}} \! \int_{\Sigma_{\mathscr{E}}} 
\tfrac{f(z)}{(z-\zeta^{\prime})} \, \tfrac{\md z}{2 \pi \mi}$. It was shown 
in \cite{a38} that $\vert \vert (\id_{\mathscr{E}} \! - \! C_{w^{\mathscr{
E}}})^{-1} \vert \vert_{\mathscr{N}(\Sigma_{\mathscr{E}})} \! < \! \infty$, 
where $\mathscr{N}(\ast)$ denotes the space of bounded linear operators 
{}from $\mathcal{L}^{2}_{\mathrm{M}_{2}(\mathbb{C})}(\ast)$ to $\mathcal{
L}^{2}_{\mathrm{M}_{2}(\mathbb{C})}(\ast)$. According to the BC construction, 
the solution of the (normalised at $\infty)$ RHP for $\mathscr{E}(\zeta)$ has 
the integral representation $\mathscr{E}(\zeta) \! = \! \mathrm{I} \! + \! 
\int_{\Sigma_{\mathscr{E}}} \! \tfrac{\mu^{\mathscr{E}}(z)w^{\mathscr{E}}
(z)}{(z-\zeta)} \, \tfrac{\md z}{2 \pi \mi}$, $\zeta \! \in \! \mathbb{C} 
\setminus \Sigma_{\mathscr{E}}$, where $\mu^{\mathscr{E}}(\zeta) \! = \! 
((\id_{\mathscr{E}} \! - \! C_{w^{\mathscr{E}}})^{-1} \mathrm{I})(\zeta)$, 
and $w^{\mathscr{E}}(\zeta) \! = \! \sum_{l \in \{\pm\}} \! w^{\mathscr{
E}}_{l}(\zeta) \! = \! w^{\mathscr{E}}_{+}(\zeta)$. Since 
(cf.~Definition~3.1), for $i \! \not= \! j \! \in \! \{m \! + \! 1,m \! + 
\! 2,\ldots,N\}$, $\widehat{\mathscr{K}}_{i} \cap \widehat{\mathscr{L}}_{i} 
\! = \! \widehat{\mathscr{K}}_{i} \cap \widehat{\mathscr{K}}_{j} \! = \! 
\widehat{\mathscr{L}}_{i} \cap \widehat{\mathscr{L}}_{j} \! = \! \emptyset$, 
it follows that
\begin{equation*}
\mathscr{E}(\zeta) \! = \! \mathrm{I} \! + \! \sum_{n=m+1}^{N} \! \left( 
\int\nolimits_{\widehat{\mathscr{K}}_{n}} \tfrac{\mu^{\mathscr{E}}(z) 
\widetilde{\mathscr{W}}_{\mathscr{E}}^{\widehat{\mathscr{K}}_{n}}(z)}{(z-
\zeta)} \tfrac{\md z}{2 \pi \mi} \! + \! \int\nolimits_{\widehat{\mathscr{
L}}_{n}} \tfrac{\mu^{\mathscr{E}}(z) \widetilde{\mathscr{W}}_{\mathscr{E}}^{
\widehat{\mathscr{L}}_{n}}(z)}{(z-\zeta)} \tfrac{\md z}{2 \pi \mi} \right), 
\quad \zeta \! \in \! \mathbb{C} \setminus \Sigma_{\mathscr{E}}.
\end{equation*}
{}From the second resolvent identity and the expressions for $\widetilde{
\mathscr{W}}_{\mathscr{E}}^{\widehat{\mathscr{K}}_{n}}(\zeta)$ and 
$\widetilde{\mathscr{W}}_{\mathscr{E}}^{\widehat{\mathscr{L}}_{n}}(\zeta)$, 
one shows that
\begin{align*}
\mathscr{E}(\zeta) \! - \! \mathrm{I} \! =& \sum_{n=m+1}^{N} \! \left(\int_{
\widehat{\mathscr{K}}_{n}} \tfrac{C_{n}^{\mathscr{K}} \mathcal{X}^{\flat}(z)}
{(z-\varsigma_{n})(z-\zeta)} \tfrac{\md z}{2 \pi \mi} \! + \! \int_{\widehat{
\mathscr{L}}_{n}} \tfrac{C_{n}^{\mathscr{L}} \mathcal{X}^{\natural}(z)}{(z-
\overline{\varsigma_{n}})(z-\zeta)} \tfrac{\md z}{2 \pi \mi} \! + \! \int_{
\widehat{\mathscr{K}}_{n}} \tfrac{C_{n}^{\mathscr{K}}((\id_{\mathscr{E}}-C_{
w^{\mathscr{E}}})^{-1}C_{w^{\mathscr{E}}} \mathrm{I})(z) \mathcal{X}^{\flat}
(z)}{(z-\varsigma_{n})(z-\zeta)} \tfrac{\md z}{2 \pi \mi} \right. \\
+& \left. \int_{\widehat{\mathscr{L}}_{n}} \tfrac{C_{n}^{\mathscr{L}}((\id_{
\mathscr{E}}-C_{w^{\mathscr{E}}})^{-1}C_{w^{\mathscr{E}}} \mathrm{I})(z) 
\mathcal{X}^{\natural}(z)}{(z-\overline{\varsigma_{n}})(z-\zeta)} \tfrac{\md 
z}{2 \pi \mi} \right), \quad \zeta \! \in \! \mathbb{C} \setminus \Sigma_{
\mathscr{E}}.
\end{align*}
Using the Cauchy-Schwarz inequality for integrals, one arrives at
\begin{align*}
\vert \mathscr{E}(\zeta) \! - \! \mathrm{I} \vert &\leqslant \sum_{n=m+1}^{
N} \! \left( \tfrac{\vert C_{n}^{\mathscr{K}} \vert}{2 \pi} \vert \vert 
\mathcal{X}^{\flat}(\cdot) \vert \vert_{\mathcal{L}^{2}_{\mathrm{M}_{2}
(\mathbb{C})}(\widehat{\mathscr{K}}_{n})} \! \left\vert \left\vert \tfrac{
\mathrm{I}}{(\cdot -\varsigma_{n})(\cdot -\zeta)} \right\vert \right\vert_{
\mathcal{L}^{2}_{\mathrm{M}_{2}(\mathbb{C})}(\widehat{\mathscr{K}}_{n})} \! 
+\tfrac{\vert C_{n}^{\mathscr{L}} \vert}{2 \pi} \vert \vert \mathcal{X}^{
\natural}(\cdot) \vert \vert_{\mathcal{L}^{2}_{\mathrm{M}_{2}(\mathbb{C})}
(\widehat{\mathscr{L}}_{n})} \right. \\
&\times \left. \left\vert \left\vert \tfrac{\mathrm{I}}{(\cdot -\overline{
\varsigma_{n}})(\cdot -\zeta)} \right\vert \right\vert_{\mathcal{L}^{2}_{
\mathrm{M}_{2}(\mathbb{C})}(\widehat{\mathscr{L}}_{n})} \! +\tfrac{\vert C_{
n}^{\mathscr{K}} \vert}{2 \pi} \vert \vert (\id_{\mathscr{E}} \! - \! C_{w^{
\mathscr{E}}})^{-1} \vert \vert_{\mathscr{N}(\widehat{\mathscr{K}}_{n})} 
\vert \vert (C_{w^{\mathscr{E}}} \mathrm{I})(\cdot) \vert \vert_{\mathcal{
L}^{2}_{\mathrm{M}_{2}(\mathbb{C})}(\widehat{\mathscr{K}}_{n})} \right. \\
&\times \left. \vert \vert \mathcal{X}^{\flat}(\cdot) \vert \vert_{\mathcal{
L}^{2}_{\mathrm{M}_{2}(\mathbb{C})}(\widehat{\mathscr{K}}_{n})} \! \left\vert 
\left\vert \tfrac{\mathrm{I}}{(\cdot -\varsigma_{n})(\cdot -\zeta)} 
\right\vert \right\vert_{\mathcal{L}^{2}_{\mathrm{M}_{2}(\mathbb{C})}
(\widehat{\mathscr{K}}_{n})} \! +\tfrac{\vert C_{n}^{\mathscr{L}} \vert}{2 
\pi} \vert \vert (\id_{\mathscr{E}} \! - \! C_{w^{\mathscr{E}}})^{-1} \vert 
\vert_{\mathscr{N}(\widehat{\mathscr{L}}_{n})} \right. \\
&\times \left. \vert \vert (C_{w^{\mathscr{E}}} \mathrm{I})(\cdot) \vert 
\vert_{\mathcal{L}^{2}_{\mathrm{M}_{2}(\mathbb{C})}(\widehat{\mathscr{L}}_{
n})} \vert \vert \mathcal{X}^{\natural}(\cdot) \vert \vert_{\mathcal{L}^{
2}_{\mathrm{M}_{2}(\mathbb{C})}(\widehat{\mathscr{L}}_{n})} \! \left\vert 
\left\vert \tfrac{\mathrm{I}}{(\cdot -\overline{\varsigma_{n}})(\cdot-\zeta)} 
\right\vert \right\vert_{\mathcal{L}^{2}_{\mathrm{M}_{2}(\mathbb{C})}
(\widehat{\mathscr{L}}_{n})} \right), \quad \zeta \! \in \! \mathbb{C} 
\setminus \Sigma_{\mathscr{E}}.
\end{align*}
One shows that, for $\zeta \! \in \! \mathbb{C} \setminus \Sigma_{\mathscr{
E}}$,
\begin{align*}
\left\vert \left\vert \tfrac{\mathrm{I}}{(\cdot -\varsigma_{n})(\cdot-\zeta)} 
\right\vert \right\vert_{\mathcal{L}^{2}_{\mathrm{M}_{2}(\mathbb{C})}
(\widehat{\mathscr{K}}_{n})} \! \leqslant& \, \sqrt{\tfrac{2}{\widehat{
\varepsilon}_{n}^{\mathscr{K}}}} \! \left(\int\nolimits_{0}^{2 \pi} \tfrac{
\md \omega}{\vert \zeta -\widehat{\varepsilon}_{n}^{\mathscr{K}} \me^{-\mi 
\omega} \vert^{2}} \right)^{1/2} \! =: \! \sqrt{\tfrac{2}{\widehat{
\varepsilon}_{n}^{\mathscr{K}}}} \, \mathscr{F}_{\widehat{\mathscr{K}}_{n}}
(\zeta), \\
\left\vert \left\vert \tfrac{\mathrm{I}}{(\cdot -\overline{\varsigma_{n}})
(\cdot -\zeta)} \right\vert \right\vert_{\mathcal{L}^{2}_{\mathrm{M}_{2}
(\mathbb{C})}(\widehat{\mathscr{L}}_{n})} \! \leqslant& \, \sqrt{\tfrac{2}
{\widehat{\varepsilon}_{n}^{\mathscr{L}}}} \! \left(\int\nolimits_{0}^{2 
\pi} \tfrac{\md \omega}{\vert \zeta -\widehat{\varepsilon}_{n}^{\mathscr{L}} 
\me^{\mi \omega} \vert^{2}} \right)^{1/2} \! =: \! \sqrt{\tfrac{2}{\widehat{
\varepsilon}_{n}^{\mathscr{L}}}} \, \mathscr{F}_{\widehat{\mathscr{L}}_{n}}
(\zeta),
\end{align*}
with $\mathscr{F}_{\widehat{\star}_{n}}(\zeta) \! =_{\zeta \to \infty} \! 
\mathcal{O}(\vert \zeta \vert^{-1})$ and $\mathscr{F}_{\widehat{\star}_{n}}
(\zeta) \! =_{\zeta \to 0} \! \mathcal{O}(1)$, $\star \! \in \! \{\mathscr{
K},\mathscr{L}\}$. Again, via the Cauchy-Schwarz i\-n\-e\-q\-u\-a\-l\-i\-t\-y 
for integrals,
\begin{align*}
\vert \vert (C_{w^{\mathscr{E}}} \mathrm{I})(\cdot) \vert \vert_{\mathcal{
L}^{2}_{\mathrm{M}_{2}(\mathbb{C})}(\widehat{\mathscr{K}}_{n})} \! &\leqslant 
\vert \vert (C_{-}(\mathrm{I}w^{\mathscr{E}}_{+}))(\cdot) \vert \vert_{
\mathcal{L}^{2}_{\mathrm{M}_{2}(\mathbb{C})}(\widehat{\mathscr{K}}_{n})} \! 
\leqslant \! \vert \vert C_{-} \vert \vert_{\mathscr{N}(\widehat{\mathscr{
K}}_{n})} \vert \vert w^{\mathscr{E}}_{+}(\cdot) \vert \vert_{\mathcal{L}^{
2}_{\mathrm{M}_{2}(\mathbb{C})}(\widehat{\mathscr{K}}_{n})} \\
&\leqslant \vert \vert C_{-} \vert \vert_{\mathscr{N}(\widehat{\mathscr{
K}}_{n})} \! \left\vert \left\vert \tfrac{C_{n}^{\mathscr{K}}}{(\cdot -
\varsigma_{n})} \mathcal{X}^{\flat}(\cdot) \right\vert \right\vert_{\mathcal{
L}^{2}_{\mathrm{M}_{2}(\mathbb{C})}(\widehat{\mathscr{K}}_{n})} \\
&\leqslant \vert \vert C_{-} \vert \vert_{\mathscr{N}(\widehat{\mathscr{K}}_{
n})} \vert C_{n}^{\mathscr{K}} \vert \vert \vert \mathcal{X}^{\flat}(\cdot) 
\vert \vert_{\mathcal{L}^{2}_{\mathrm{M}_{2}(\mathbb{C})}(\widehat{\mathscr{
K}}_{n})} \! \left\vert \left\vert \tfrac{\mathrm{I}}{(\cdot -\varsigma_{n})} 
\right\vert \right\vert_{\mathcal{L}^{2}_{\mathrm{M}_{2}(\mathbb{C})}
(\widehat{\mathscr{K}}_{n})} \\
&\leqslant 2 \sqrt{\tfrac{\pi}{\widehat{\varepsilon}_{n}^{\mathscr{K}}}} \, 
\vert C_{n}^{\mathscr{K}} \vert \vert \vert C_{-} \vert \vert_{\mathscr{N}
(\widehat{\mathscr{K}}_{n})} \vert \vert \mathcal{X}^{\flat}(\cdot) \vert 
\vert_{\mathcal{L}^{2}_{\mathrm{M}_{2}(\mathbb{C})}(\widehat{\mathscr{K}}_{
n})},
\end{align*}
with an analogous estimate for $\vert \vert (C_{w^{\mathscr{E}}} \mathrm{I})
(\cdot) \vert \vert_{\mathcal{L}^{2}_{\mathrm{M}_{2}(\mathbb{C})}(\widehat{
\mathscr{L}}_{n})}$:
\begin{equation*}
\vert \vert (C_{w^{\mathscr{E}}} \mathrm{I})(\cdot) \vert \vert_{\mathcal{
L}^{2}_{\mathrm{M}_{2}(\mathbb{C})}(\widehat{\mathscr{L}}_{n})} \! \leqslant 
2 \sqrt{\tfrac{\pi}{\widehat{\varepsilon}_{n}^{\mathscr{L}}}} \, \vert C_{
n}^{\mathscr{L}} \vert \vert \vert C_{-} \vert \vert_{\mathscr{N}(\widehat{
\mathscr{L}}_{n})} \vert \vert \mathcal{X}^{\natural}(\cdot) \vert \vert_{
\mathcal{L}^{2}_{\mathrm{M}_{2}(\mathbb{C})}(\widehat{\mathscr{L}}_{n})}.
\end{equation*}
Hence, for $\zeta \! \in \! \mathbb{C} \setminus \Sigma_{\mathscr{E}}$, 
\begin{align*}
\vert \mathscr{E}(\zeta) \! - \! \mathrm{I} \vert &\leqslant \sum_{n=m+1}^{N} 
\! \left(\tfrac{\vert C_{n}^{\mathscr{K}} \vert \mathscr{F}_{\widehat{
\mathscr{K}}_{n}}(\zeta)}{\pi \sqrt{2 \widehat{\varepsilon}_{n}^{\mathscr{
K}}}} \vert \vert \mathcal{X}^{\flat}(\cdot) \vert \vert_{\mathcal{L}^{2}_{
\mathrm{M}_{2}(\mathbb{C})}(\widehat{\mathscr{K}}_{n})} \! +\tfrac{\vert C_{
n}^{\mathscr{L}} \vert \mathscr{F}_{\widehat{\mathscr{L}}_{n}}(\zeta)}{\pi 
\sqrt{2 \widehat{\varepsilon}_{n}^{\mathscr{L}}}} \vert \vert \mathcal{X}^{
\natural}(\cdot) \vert \vert_{\mathcal{L}^{2}_{\mathrm{M}_{2}(\mathbb{C})}
(\widehat{\mathscr{L}}_{n})} \right. \\
&+ \left. \tfrac{\sqrt{2} \, \vert C_{n}^{\mathscr{K}} \vert^{2} \mathscr{
F}_{\widehat{\mathscr{K}}_{n}}(\zeta)}{\sqrt{\pi} \, \widehat{\varepsilon}_{
n}^{\mathscr{K}}} \vert \vert (\id_{\mathscr{E}} \! - \! C_{w^{\mathscr{
E}}})^{-1} \vert \vert_{\mathscr{N}(\widehat{\mathscr{K}}_{n})} \vert \vert 
C_{-} \vert \vert_{\mathscr{N}(\widehat{\mathscr{K}}_{n})} \vert \vert 
\mathcal{X}^{\flat}(\cdot) \vert \vert_{\mathcal{L}^{2}_{\mathrm{M}_{2}
(\mathbb{C})}(\widehat{\mathscr{K}}_{n})}^{2} \right. \\
&+ \left. \tfrac{\sqrt{2} \, \vert C_{n}^{\mathscr{L}} \vert^{2} \mathscr{
F}_{\widehat{\mathscr{L}}_{n}}(\zeta)}{\sqrt{\pi} \, \widehat{\varepsilon}_{
n}^{\mathscr{L}}} \vert \vert (\id_{\mathscr{E}} \! - \! C_{w^{\mathscr{
E}}})^{-1} \vert \vert_{\mathscr{N}(\widehat{\mathscr{L}}_{n})} \vert \vert 
C_{-} \vert \vert_{\mathscr{N}(\widehat{\mathscr{L}}_{n})} \vert \vert 
\mathcal{X}^{\natural}(\cdot) \vert \vert_{\mathcal{L}^{2}_{\mathrm{M}_{2}
(\mathbb{C})}(\widehat{\mathscr{L}}_{n})}^{2} \right).
\end{align*}
It is shown, \emph{a posteriori}, in Section~4 that the RHP for $\widehat{
\chi}(\zeta)$ formulated in the Lemma is asymptotically solvable, whence 
$\vert \vert \mathcal{X}^{\flat}(\cdot) \vert \vert_{\mathcal{L}^{2}_{
\mathrm{M}_{2}(\mathbb{C})}(\widehat{\mathscr{K}}_{n})}^{2} \! \leqslant 
\! \mathrm{const.} \! = \! \underline{c}$ and $\vert \vert \mathcal{X}^{
\natural}(\cdot) \vert \vert_{\mathcal{L}^{2}_{\mathrm{M}_{2}(\mathbb{C})}
(\widehat{\mathscr{L}}_{n})}^{2} \! \leqslant \! \mathrm{const.} \! = \! 
\underline{c}$. Furthermore \cite{a38}, $\vert \vert (\id_{\mathscr{E}} \! 
- \! C_{w^{\mathscr{E}}})^{-1} \vert \vert_{\mathscr{N}(\widehat{\star}_{n})} 
\! \leqslant \! \mathrm{const.} \, \vert \vert (\id_{\mathscr{E}} \! - \! 
C_{w^{\mathscr{E}}})^{-1} \vert \vert_{\mathscr{N}(\Sigma_{\mathscr{E}})} 
\! \leqslant \! \underline{c}$ (see above), $\star \! \in \! \{\mathscr{K},
\mathscr{L}\}$. Recalling the expressions for $C_{n}^{\mathscr{K}}$ and 
$C_{n}^{\mathscr{L}}$ given in Lemma~3.3, that as $t \! \to \! +\infty$ and 
$x \! \to \! -\infty$ such that $z_{o} \! := \! x/t \! < \! -2$ and $(x,t) 
\! \in \! \daleth_{m}$, $(g_{n})^{-1} \! = \! \mathcal{O}(\exp (-4t \sin 
(\phi_{n}) \vert \cos (\phi_{n}) \! - \! \cos (\phi_{m}) \vert))$, and the 
definition $\vert \vert \mathscr{E}(\cdot) \! - \! \mathrm{I} \vert \vert_{
\mathcal{L}^{2}_{\mathrm{M}_{2}(\mathbb{C})}(\mathbb{C} \, \setminus 
\Sigma_{\mathscr{E}})} \! := \! \max_{i,j \in \{1,2\}} \sup_{\zeta \in 
\mathbb{C} \, \setminus \Sigma_{\mathscr{E}}} \vert (\mathscr{E}(\zeta) \! 
- \! \mathrm{I})_{ij} \vert$, assembling the above, one arrives at
\begin{equation*}
\vert \vert \mathscr{E}(\cdot) \! - \! \mathrm{I} \vert \vert_{\mathcal{L}^{
2}_{\mathrm{M}_{2}(\mathbb{C})}(\mathbb{C} \, \setminus \Sigma_{\mathscr{E}})} 
\! \leqslant \! \mathcal{O} \! \left(\mathscr{F}_{\mathscr{E}}(\zeta) \exp \! 
\left(-4t \min_{\genfrac{}{}{0pt}{2}{m \in \{1,2,\ldots,N\}}{n \in \{m+1,m+2,
\ldots,N\}}}\{\sin (\phi_{n}) \vert \cos (\phi_{n}) \! - \! \cos (\phi_{m}) 
\vert\} \right) \right),
\end{equation*}
where $\mathscr{F}_{\mathscr{E}}(\zeta) \! =_{\zeta \to \infty} \! \mathcal{O}
(\vert \zeta \vert^{-1})$ and $\mathscr{F}_{\mathscr{E}}(\zeta) \! =_{\zeta 
\to 0} \! \mathcal{O}(1)$; hence, the asymptotic estimate for $\widehat{m}^{
\sharp}(\zeta)$ stated in the Lemma. Finally, {}from the asymptotics for 
$\mathscr{E}(\zeta) \! - \! \mathrm{I}$ derived above, the ordered 
factorisation for $\widehat{m}^{\sharp}(\zeta)$ given in Proposition~3.1, 
and Eqs.~(61) and~(62), the large-$\zeta$ asymptotics lead one to Eqs.~(63) 
and~(64). \hfill $\square$
\section{Asymptotic Solution of the Model RHP}
In this section, the model (normalised at $\infty)$ RHP for $\widehat{\chi}
(\zeta)$ formulated in Lemma~3.5 is solved asymptotically as $t \! \to \! 
+\infty$ and $x \! \to \! -\infty$ such that $z_{o} \! := \! x/t \! < \! -2$ 
and $(x,t) \! \in \! \daleth_{m}$, $m \! \in \! \{1,2,\ldots,N\}$, and the 
corresponding (asymptotic) results for $u(x,t)$, the solution of the Cauchy 
problem for the D${}_{f}$NLSE, and $\int_{\pm \infty}^{x}(\vert u(x^{\prime},
t) \vert^{2} \! - \! 1) \, \md x^{\prime}$ stated in Theorem~2.2.1 (for the 
upper sign) are derived.
\begin{ccccc}
The solution of the {\rm RHP} for $\widehat{\chi}(\zeta) \colon \mathbb{C} 
\setminus (\widehat{\sigma}_{d} \cup \sigma_{c}) \! \to \! \mathrm{M}_{2}
(\mathbb{C})$ formulated in Lemma~{\rm 3.5} is given by the following ordered 
factorisation,
\begin{equation*}
\widehat{\chi}(\zeta) \! = \! \left(\mathrm{I} \! + \! \zeta^{-1} \widehat{
\Delta}_{o} \right) \! \widehat{\mathscr{P}}(\zeta) \widehat{m}_{d}(\zeta) 
\chi^{c}(\zeta), \quad \zeta \! \in \! \mathbb{C} \setminus (\widehat{
\sigma}_{d} \cup \sigma_{c}),
\end{equation*}
where $\widehat{m}_{d}(\zeta) \! = \! \sigma_{1} \overline{\widehat{m}_{d}
(\overline{\zeta})} \, \sigma_{1}$ $(\in \! \mathrm{SL}(2,\mathbb{C}))$ has 
the representation
\begin{equation*}
\widehat{m}_{d}(\zeta) \! = \! \mathrm{I} \! + \! \sum_{n=1}^{m} \! \left(
\dfrac{\mathrm{Res}(\widehat{\chi}(\zeta);\varsigma_{n})}{(\zeta \! - \! 
\varsigma_{n})} \! + \! \dfrac{\sigma_{1} \overline{\mathrm{Res}(\widehat{
\chi}(\zeta);\varsigma_{n})} \, \sigma_{1}}{(\zeta \! - \! \overline{
\varsigma_{n}})} \right),
\end{equation*}
$\widehat{\mathscr{P}}(\zeta) \! = \! \sigma_{1} \overline{\widehat{\mathscr{
P}}(\overline{\zeta})} \, \sigma_{1}$ is chosen (see Lemma~{\rm 4.3} below) 
so that $\widehat{\Delta}_{o}$ is idempotent, $\mathrm{I} \! + \! \zeta^{-1} 
\widehat{\Delta}_{o}$ is holomorphic in a punctured neighbourhood of the 
origin, with $\widehat{\Delta}_{o} \! = \! \sigma_{1} \overline{\widehat{
\Delta}_{o}} \, \sigma_{1}$ $(\in \! \mathrm{GL}(2,\mathbb{C}))$ and $\det 
(\mathrm{I} \! + \! \zeta^{-1} \widehat{\Delta}_{o}) \vert_{\zeta =\pm 1} 
\! = \! 0$, and determined by the relation
\begin{equation*}
\widehat{\Delta}_{o} \! = \! \widehat{\mathscr{P}}(0) \widehat{m}_{d}(0) 
\chi^{c}(0)(\delta (0))^{\sigma_{3}} \! \left(\prod_{k=m+1}^{N}(d_{k}^{+}
(0))^{\sigma_{3}} \right) \! \sigma_{2},
\end{equation*}
and satisfying $\mathrm{tr}(\widehat{\Delta}_{o}) \! = \! 0$, $\det 
(\widehat{\Delta}_{o}) \! = \! -1$, and $\widehat{\Delta}_{o} \widehat{
\Delta}_{o} \! = \! \mathrm{I}$, and $\chi^{c}(\zeta) \colon \mathbb{C} 
\setminus \sigma_{c} \! \to \! \mathrm{SL}(2,\mathbb{C})$ solves the 
following {\rm RHP:} {\rm (1)} $\chi^{c}(\zeta)$ is piecewise (sectionally) 
holomorphic $\forall \, \zeta \! \in \! \mathbb{C} \setminus \sigma_{c};$ 
{\rm (2)} $\chi^{c}_{\pm}(\zeta) \! := \! \lim_{\genfrac{}{}{0pt}{2}{\zeta^{
\prime} \to \zeta}{\pm \Im (\zeta^{\prime})>0}} \chi^{c}(\zeta^{\prime})$ 
satisfy, for $\zeta \! \in \! \mathbb{R}$, the jump condition
\begin{equation*}
\chi^{c}_{+}(\zeta) \! = \! \chi^{c}_{-}(\zeta) \me^{-\mi k(\zeta)(x+2\lambda 
(\zeta)t) \mathrm{ad}(\sigma_{3})} \! 
\left(
\begin{smallmatrix}
(1-r(\zeta) \overline{r(\overline{\zeta})}) \delta_{-}(\zeta)/\delta_{+}
(\zeta) & -\frac{\overline{r(\overline{\zeta})}}{(\delta_{-}(\zeta) \delta_{
+}(\zeta))^{-1}} \prod_{k=m+1}^{N}(d_{k}^{+}(\zeta))^{2} \\
\frac{r(\zeta)}{\delta_{-}(\zeta) \delta_{+}(\zeta)} \prod_{k=m+1}^{N}
(d_{k}^{+}(\zeta))^{-2} & \delta_{+}(\zeta)/\delta_{-}(\zeta)
\end{smallmatrix}
\right);
\end{equation*}
{\rm (3)} $\chi^{c}(\zeta) \! =_{\genfrac{}{}{0pt}{2}{\zeta \to \infty}{
\zeta \in \mathbb{C} \setminus \sigma_{c}}} \! \mathrm{I} \! + \! \mathcal{
O}(\zeta^{-1});$ and {\rm (4)} $\chi^{c}(\zeta) \! = \! \sigma_{1} \overline{
\chi^{c}(\overline{\zeta})} \, \sigma_{1}$.
\end{ccccc}

\emph{Proof.} One verifies that, modulo the explicit determination of 
$\widehat{\Delta}_{o}$, $\widehat{\mathscr{P}}(\zeta)$, $\widehat{m}_{d}
(\zeta)$, and $\chi^{c}(\zeta)$, the ordered factorisation for $\widehat{\chi}
(\zeta)$ stated in the Lemma, with the conditions on $\widehat{\Delta}_{o}$, 
$\widehat{\mathscr{P}}(\zeta)$, $\widehat{m}_{d}(\zeta)$, and $\chi^{c}
(\zeta)$ stated therein, solves the RHP for $\widehat{\chi}(\zeta)$ stated 
in Lemma~3.5. \hfill $\square$

The determination of the asymptotics for the solution of the RHP for $\chi^{
c}(\zeta) \colon \mathbb{C} \setminus \sigma_{c} \! \to \! \mathrm{SL}(2,
\mathbb{C})$ stated in Lemma~4.1 was the (principal) subject of study in 
\cite{a38}, and is given by the following
\begin{ccccc}
Let $\varepsilon$ be an arbitrarily fixed, sufficiently small positive real 
number, and, for $z\! \in \! \{\lambda_{1},\lambda_{2}\}$, with $\lambda_{1}$ 
and $\lambda_{2}$ given in Theorem~{\rm 2.2.1}, Eq.~{\rm (10)}, set  $\mathbb{
U}(z;\varepsilon) \! := \! \{\mathstrut \zeta; \, \vert \zeta \! - \! z \vert 
\! < \! \varepsilon\}$. Then, as $t \! \to \! +\infty$ and $x \! \to \! 
-\infty$ such that $z_{o} \! := \! x/t \! < \! -2$, for $\zeta \! \in \! 
\mathbb{C} \setminus \cup_{z \in \{\lambda_{1},\lambda_{2}\}} \mathbb{U}(z;
\varepsilon)$, $\chi^{c}(\zeta)$ has the following asymptotics:
\begin{align*}
\chi^{c}_{11}(\zeta) \! &= \! 1 \! + \! \mathcal{O} \! \left( \! \left(\dfrac{
c^{\mathcal{S}}(\lambda_{1}) \underline{c}(\lambda_{2},\lambda_{3},\overline{
\lambda_{3}})}{\sqrt{\lambda_{2}(z_{o}^{2} \! + \! 32)} \, (\zeta \! - \! 
\lambda_{1})} \! + \! \dfrac{c^{\mathcal{S}}(\lambda_{2}) \underline{c}
(\lambda_{1},\lambda_{3},\overline{\lambda_{3}})}{\sqrt{\lambda_{1}(z_{o}^{
2} \! + \! 32)} \, (\zeta \! - \! \lambda_{2})} \right) \! \dfrac{\ln t}{
(\lambda_{1} \! - \! \lambda_{2}) t} \right), \\
\chi^{c}_{12}(\zeta) \! &= \! \me^{\frac{\mi \Xi^{+}(0)}{2}} \! \left(\dfrac{
\sqrt{\nu (\lambda_{1})} \, \lambda_{1}^{2 \mi \nu (\lambda_{1})}}{\sqrt{t
(\lambda_{1} \! - \! \lambda_{2})} \, (z_{o}^{2} \! + \! 32)^{1/4}} \! \left( 
\dfrac{\lambda_{1} \me^{-\mi (\Theta^{+}(z_{o},t)+\frac{\pi}{4})}}{(\zeta \! 
- \! \lambda_{1})} \! + \! \dfrac{\lambda_{2} \me^{\mi (\Theta^{+}(z_{o},t)+
\frac{\pi}{4})}}{(\zeta \! - \! \lambda_{2})} \! \right) \right. \\
&+ \left. \! \mathcal{O} \! \left( \! \left(\dfrac{c^{\mathcal{S}}(\lambda_{
1}) \underline{c}(\lambda_{2},\lambda_{3},\overline{\lambda_{3}})}{\sqrt{
\lambda_{2}(z_{o}^{2} \! + \! 32)} \, (\zeta \! - \! \lambda_{1})} \! + \! 
\dfrac{c^{\mathcal{S}}(\lambda_{2}) \underline{c}(\lambda_{1},\lambda_{3},
\overline{\lambda_{3}})}{\sqrt{\lambda_{1}(z_{o}^{2} \! + \! 32)} \, (\zeta 
\! - \! \lambda_{2})} \right) \! \dfrac{\ln t}{(\lambda_{1} \! - \! \lambda_{
2}) t} \right) \! \right), \\
\chi^{c}_{21}(\zeta) \! &= \! \me^{-\frac{\mi \Xi^{+}(0)}{2}} \! \left(\dfrac{
\sqrt{\nu (\lambda_{1})} \, \lambda_{1}^{-2 \mi \nu (\lambda_{1})}}{\sqrt{t
(\lambda_{1} \! - \! \lambda_{2})} \, (z_{o}^{2} \! + \! 32)^{1/4}} \! \left(
\dfrac{\lambda_{1} \me^{\mi (\Theta^{+}(z_{o},t)+\frac{\pi}{4})}}{(\zeta \! - 
\! \lambda_{1})} \! + \! \dfrac{\lambda_{2} \me^{-\mi (\Theta^{+}(z_{o},t)+
\frac{\pi}{4})}}{(\zeta \! - \! \lambda_{2})} \! \right) \right. \\
&+ \left. \! \mathcal{O} \! \left( \! \left(\dfrac{c^{\mathcal{S}}(\lambda_{
1}) \underline{c}(\lambda_{2},\lambda_{3},\overline{\lambda_{3}})}{\sqrt{
\lambda_{2}(z_{o}^{2} \! + \! 32)} \, (\zeta \! - \! \lambda_{1})} \! + \! 
\dfrac{c^{\mathcal{S}}(\lambda_{2}) \underline{c}(\lambda_{1},\lambda_{3},
\overline{\lambda_{3}})}{\sqrt{\lambda_{1}(z_{o}^{2} \! + \! 32)} \, (\zeta 
\! - \! \lambda_{2})} \right) \! \dfrac{\ln t}{(\lambda_{1} \! - \! \lambda_{
2}) t} \right) \! \right), \\
\chi^{c}_{22}(\zeta) \! &= \! 1 \! + \! \mathcal{O} \! \left( \! \left(\dfrac{
c^{\mathcal{S}}(\lambda_{1}) \underline{c}(\lambda_{2},\lambda_{3},\overline{
\lambda_{3}})}{\sqrt{\lambda_{2}(z_{o}^{2} \! + \! 32)} \, (\zeta \! - \! 
\lambda_{1})} \! + \! \dfrac{c^{\mathcal{S}}(\lambda_{2}) \underline{c}
(\lambda_{1},\lambda_{3},\overline{\lambda_{3}})}{\sqrt{\lambda_{1}(z_{o}^{
2} \! + \! 32)} \, (\zeta \! - \! \lambda_{2})} \right) \! \dfrac{\ln t}{
(\lambda_{1} \! - \! \lambda_{2}) t} \right),
\end{align*}
where $\lambda_{3}$, $\nu (\cdot)$, $\Theta^{+}(z_{o},t)$, and $\Xi^{+}
(\cdot)$, respectively, are given in Theorem~{\rm 2.2.1}, Eqs.~{\rm (10)}, 
{\rm (11)}, {\rm (17)}, and~{\rm (18)}, $\vert \vert (\cdot \! - \! \lambda_{
k})^{-1} \vert \vert_{\mathcal{L}^{\infty}(\mathbb{C} \, \setminus \cup_{z 
\in \{\lambda_{1},\lambda_{2}\}} \mathbb{U}(z;\varepsilon))} \! < \! \infty$, 
$k \! \in \! \{1,2\}$, $\chi^{c}(\zeta) \! = \! \sigma_{1} \overline{\chi^{c}
(\overline{\zeta})} \, \sigma_{1}$, and $(\chi^{c}(0) \sigma_{2})^{2} \! = \! 
\mathrm{I}$ $(+ \, \mathcal{O}(t^{-1} \ln t))$.
\end{ccccc}

\emph{Sketch of Proof.} Proceeding as in the proof of Lemma~6.1 in \cite{a38} 
and particularising it to the case of the RHP for $\chi^{c}(\zeta)$ stated in 
Lemma~4.1, one arrives at
\begin{align*}
\chi^{c}_{11}(\zeta) \! &= \! 1 \! - \! \tfrac{\widehat{r}(\lambda_{1})
(\delta_{B}^{0})^{-2} \me^{\frac{\pi \nu}{2}} \me^{\frac{\mi \pi}{4}}}{2 \pi 
\mi (\zeta -\lambda_{1}) \beta^{\Sigma_{B^{0}}}_{21} \mathcal{X}_{B} \sqrt{
t}} \int\nolimits_{0}^{+\infty}(\me^{-\frac{\mi \pi}{4}} \partial_{z} \mathbf{
D}_{-\mi \nu}(z) \! - \! \tfrac{\mi}{2} \me^{\frac{\mi \pi}{4}}z \mathbf{D}_{
-\mi \nu}(z))z^{-\mi \nu} \me^{-\frac{z^{2}}{4}} \, \md z \\
&+ \tfrac{\widehat{r}(\lambda_{1})(1-\vert \widehat{r}(\lambda_{1}) \vert^{
2})^{-1}(\delta_{B}^{0})^{-2} \me^{-\frac{3 \pi \mi}{4}}}{2 \pi \mi (\zeta 
-\lambda_{1}) \beta^{\Sigma_{B^{0}}}_{21} \me^{\frac{3 \pi \nu}{2}} \mathcal{
X}_{B} \sqrt{t}} \int\nolimits_{0}^{+\infty}(\me^{\frac{3 \pi \mi}{4}} 
\partial_{z} \mathbf{D}_{-\mi \nu}(z) \! - \! \tfrac{\mi}{2} \me^{-\frac{3 
\pi \mi}{4}}z \mathbf{D}_{-\mi \nu}(z))z^{-\mi \nu} \me^{-\frac{z^{2}}{4}} \, 
\md z \\
&- \tfrac{\overline{\widehat{r}(\lambda_{1})}(\delta_{A}^{0})^{-2} \me^{-
\frac{\pi \nu}{2}}(-1)^{\mi \nu} \me^{\frac{3 \pi \mi}{4}}}{2 \pi \mi 
(\zeta -\lambda_{2}) \beta^{\Sigma_{A^{0}}}_{21} \mathcal{X}_{A} \sqrt{t}} 
\int\nolimits_{0}^{+\infty}(\me^{-\frac{3 \pi \mi}{4}} \partial_{z} \mathbf{
D}_{\mi \nu}(z) \! + \! \tfrac{\mi}{2} \me^{\frac{3 \pi \mi}{4}}z \mathbf{
D}_{\mi \nu}(z))z^{\mi \nu} \me^{-\frac{z^{2}}{4}} \, \md z \\
&+ \tfrac{\overline{\widehat{r}(\lambda_{1})}(1 -\vert \widehat{r}(\lambda_{
1}) \vert^{2})^{-1}(\delta_{A}^{0})^{-2}(-1)^{\mi \nu} \me^{-\frac{\mi \pi}{
4}}}{2 \pi \mi (\zeta -\lambda_{2}) \beta^{\Sigma_{A^{0}}}_{21} \me^{\frac{
\pi \nu}{2}} \mathcal{X}_{A} \sqrt{t}} \int\nolimits_{0}^{+\infty}(\me^{
\frac{\mi \pi}{4}} \partial_{z} \mathbf{D}_{\mi \nu}(z) \! + \! \tfrac{\mi}
{2} \me^{-\frac{\mi \pi}{4}}z \mathbf{D}_{\mi \nu}(z)) z^{\mi \nu} \me^{-
\frac{z^{2}}{4}} \, \md z \\
&+ \mathcal{O} \! \left( \! \left(\tfrac{c^{\mathcal{S}}(\lambda_{1}) 
\underline{c}(\lambda_{2},\lambda_{3},\overline{\lambda_{3}})(\delta_{B}^{
0})^{-2}}{(\zeta -\lambda_{1}) \vert \lambda_{1}-\lambda_{3} \vert \sqrt{
(\lambda_{1}-\lambda_{2})} \, \, \mathcal{X}_{B}} \! + \! \tfrac{c^{\mathcal{
S}}(\lambda_{2}) \underline{c}(\lambda_{1},\lambda_{3},\overline{\lambda_{
3}})(\delta_{A}^{0})^{-2}}{(\zeta -\lambda_{2}) \vert \lambda_{2}-\lambda_{
3} \vert \sqrt{(\lambda_{1}-\lambda_{2})} \, \, \mathcal{X}_{A}} \right) \! 
\tfrac{\ln t}{t} \right), \\
\chi^{c}_{12}(\zeta) \! &= \! \left(\tfrac{\overline{\widehat{r}(\lambda_{
1})}(\delta_{B}^{0})^{2} \me^{\frac{\pi \nu}{2}} \me^{-\frac{\mi \pi}{4}}}
{2 \pi \mi (\zeta -\lambda_{1}) \mathcal{X}_{B} \sqrt{t}} \! - \! \tfrac{
\overline{\widehat{r}(\lambda_{1})}(1-\vert \widehat{r}(\lambda_{1}) \vert^{
2})^{-1}(\delta_{B}^{0})^{2} \me^{\frac{3 \pi \mi}{4}}}{2 \pi \mi (\zeta 
-\lambda_{1}) \me^{\frac{3 \pi \nu}{2}} \mathcal{X}_{B} \sqrt{t}} \right) \! 
\int\nolimits_{0}^{+\infty} \mathbf{D}_{\mi \nu}(z)z^{\mi \nu} \me^{-\frac{
z^{2}}{4}} \, \md z \\
&+ \left(\tfrac{\widehat{r}(\lambda_{1})(\delta_{A}^{0})^{2} \me^{-\frac{\pi 
\nu}{2}} \me^{-\frac{3 \pi \mi}{4}}}{2 \pi \mi (\zeta -\lambda_{2})(-1)^{\mi 
\nu} \mathcal{X}_{A} \sqrt{t}} \! - \! \tfrac{\widehat{r}(\lambda_{1})(1-
\vert \widehat{r}(\lambda_{1}) \vert^{2})^{-1}(\delta_{A}^{0})^{2} \me^{\frac{
\mi \pi}{4}}}{2 \pi \mi (\zeta -\lambda_{2}) \me^{\frac{\pi \nu}{2}}(-1)^{\mi 
\nu} \mathcal{X}_{A} \sqrt{t}} \right) \! \int\nolimits_{0}^{+\infty} \mathbf{
D}_{-\mi \nu}(z)z^{-\mi \nu} \me^{-\frac{z^{2}}{4}} \, \md z \\
&+ \mathcal{O} \! \left( \! \left(\tfrac{c^{\mathcal{S}}(\lambda_{1}) 
\underline{c}(\lambda_{2},\lambda_{3},\overline{\lambda_{3}})(\delta_{B}^{
0})^{2}}{(\zeta -\lambda_{1}) \vert \lambda_{1}-\lambda_{3} \vert \sqrt{
(\lambda_{1}-\lambda_{2})} \, \, \mathcal{X}_{B}} \! + \! \tfrac{c^{\mathcal{
S}}(\lambda_{2}) \underline{c}(\lambda_{1},\lambda_{3},\overline{\lambda_{
3}})(\delta_{A}^{0})^{2}}{(\zeta -\lambda_{2}) \vert \lambda_{2}-\lambda_{3} 
\vert \sqrt{(\lambda_{1}-\lambda_{2})} \, \, \mathcal{X}_{A}} \right) \! 
\tfrac{\ln t}{t} \right), \\
\chi^{c}_{21}(\zeta) \! &= \! -\left(\tfrac{\widehat{r}(\lambda_{1})(\delta_{
B}^{0})^{-2} \me^{\frac{\pi \nu}{2}} \me^{\frac{\mi \pi}{4}}}{2 \pi \mi 
(\zeta -\lambda_{1}) \mathcal{X}_{B} \sqrt{t}} \! - \! \tfrac{\widehat{r}
(\lambda_{1})(1-\vert \widehat{r}(\lambda_{1}) \vert^{2})^{-1}(\delta_{B}^{
0})^{-2} \me^{-\frac{3 \pi \mi}{4}}}{2 \pi \mi (\zeta -\lambda_{1}) \me^{
\frac{3 \pi \nu}{2}} \mathcal{X}_{B} \sqrt{t}} \right) \! \int\nolimits_{
0}^{+\infty} \mathbf{D}_{-\mi \nu}(z)z^{-\mi \nu} \me^{-\frac{z^{2}}{4}} \, 
\md z \\
&-\left(\tfrac{\overline{\widehat{r}(\lambda_{1})}(\delta_{A}^{0})^{-2} \me^{
-\frac{\pi \nu}{2}} \me^{\frac{3 \pi \mi}{4}}}{2 \pi \mi (\zeta -\lambda_{2})
(-1)^{-\mi \nu} \mathcal{X}_{A} \sqrt{t}} \! - \! \tfrac{\overline{\widehat{
r}(\lambda_{1})}(1-\vert \widehat{r}(\lambda_{1}) \vert^{2})^{-1}(\delta_{
A}^{0})^{-2} \me^{-\frac{\mi \pi}{4}}}{2 \pi \mi (\zeta -\lambda_{2}) \me^{
\frac{\pi \nu}{2}}(-1)^{-\mi \nu} \mathcal{X}_{A} \sqrt{t}} \right) \! 
\int\nolimits_{0}^{+\infty} \mathbf{D}_{\mi \nu}(z) z^{\mi \nu} \me^{-\frac{
z^{2}}{4}} \, \md z \\
&+\mathcal{O} \! \left( \! \left(\tfrac{c^{\mathcal{S}}(\lambda_{1}) 
\underline{c}(\lambda_{2},\lambda_{3},\overline{\lambda_{3}})(\delta_{B}^{
0})^{-2}}{(\zeta -\lambda_{1}) \vert \lambda_{1}-\lambda_{3} \vert \sqrt{
(\lambda_{1}-\lambda_{2})} \, \, \mathcal{X}_{B}} \! + \! \tfrac{c^{\mathcal{
S}}(\lambda_{2}) \underline{c}(\lambda_{1},\lambda_{3},\overline{\lambda_{
3}})(\delta_{A}^{0})^{-2}}{(\zeta -\lambda_{2}) \vert \lambda_{2}-\lambda_{
3} \vert \sqrt{(\lambda_{1}-\lambda_{2})} \, \, \mathcal{X}_{A}} \right) \! 
\tfrac{\ln t}{t} \right), \\
\chi^{c}_{22}(\zeta) \! &= \! 1 \! + \! \tfrac{\overline{\widehat{r}(\lambda_{
1})}(\delta_{B}^{0})^{2} \me^{\frac{\pi \nu}{2}} \me^{-\frac{\mi \pi}{4}}}{2 
\pi \mi (\zeta -\lambda_{1}) \beta^{\Sigma_{B^{0}}}_{12} \mathcal{X}_{B} 
\sqrt{t}} \int\nolimits_{0}^{+\infty}(\me^{\frac{\mi \pi}{4}} \partial_{z} 
\mathbf{D}_{\mi \nu}(z) \! + \! \tfrac{\mi}{2} \me^{-\frac{\mi \pi}{4}}z 
\mathbf{D}_{\mi \nu}(z))z^{\mi \nu} \me^{-\frac{z^{2}}{4}} \, \md z \\
&-\tfrac{\overline{\widehat{r}(\lambda_{1})}(1-\vert \widehat{r}(\lambda_{1}) 
\vert^{2})^{-1}(\delta_{B}^{0})^{2} \me^{\frac{3 \pi \mi}{4}}}{2 \pi \mi 
(\zeta -\lambda_{1}) \beta^{\Sigma_{B^{0}}}_{12} \me^{\frac{3 \pi \nu}{2}} 
\mathcal{X}_{B} \sqrt{t}} \int\nolimits_{0}^{+\infty}(\me^{-\frac{3 \pi \mi}
{4}} \partial_{z} \mathbf{D}_{\mi \nu}(z) \! + \! \tfrac{\mi}{2} \me^{\frac{
3 \pi \mi}{4}}z \mathbf{D}_{\mi \nu}(z))z^{\mi \nu} \me^{-\frac{z^{2}}{4}} 
\, \md z
\end{align*}
\begin{align*}
&+ \tfrac{\widehat{r}(\lambda_{1})(\delta_{A}^{0})^{2} \me^{-\frac{\pi \nu}
{2}} \me^{-\frac{3 \pi \mi}{4}}}{2 \pi \mi (\zeta -\lambda_{2}) \beta^{
\Sigma_{A^{0}}}_{12}(-1)^{\mi \nu} \mathcal{X}_{A} \sqrt{t}} \int\nolimits_{
0}^{+\infty}(\me^{\frac{3 \pi \mi}{4}} \partial_{z} \mathbf{D}_{-\mi \nu}(z) 
\! - \! \tfrac{\mi}{2} \me^{-\frac{3 \pi \mi}{4}}z \mathbf{D}_{-\mi \nu}(z))
z^{-\mi \nu} \me^{-\frac{z^{2}}{4}} \, \md z \\
&- \tfrac{\widehat{r}(\lambda_{1})(1-\vert \widehat{r}(\lambda_{1}) \vert^{
2})^{-1}(\delta_{A}^{0})^{2} \me^{\frac{\mi \pi}{4}}}{2 \pi \mi (\zeta 
-\lambda_{2}) \beta^{\Sigma_{A^{0}}}_{12} \me^{\frac{\pi \nu}{2}}(-1)^{\mi 
\nu} \mathcal{X}_{A} \sqrt{t}} \int\nolimits_{0}^{+\infty}(\me^{-\frac{\mi 
\pi}{4}} \partial_{z} \mathbf{D}_{-\mi \nu}(z) \! - \! \tfrac{\mi}{2} \me^{
\frac{\mi \pi}{4}}z \mathbf{D}_{-\mi \nu}(z))z^{-\mi \nu} \me^{-\frac{z^{2}}
{4}} \, \md z \\
&+ \mathcal{O} \! \left( \! \left(\tfrac{c^{\mathcal{S}}(\lambda_{1}) 
\underline{c}(\lambda_{2},\lambda_{3},\overline{\lambda_{3}})(\delta_{B}^{
0})^{2}}{(\zeta -\lambda_{1}) \vert \lambda_{1}-\lambda_{3} \vert \sqrt{
(\lambda_{1}-\lambda_{2})} \, \, \mathcal{X}_{B}} \! + \! \tfrac{c^{\mathcal{
S}}(\lambda_{2}) \underline{c}(\lambda_{1},\lambda_{3},\overline{\lambda_{
3}})(\delta_{A}^{0})^{2}}{(\zeta -\lambda_{2}) \vert \lambda_{2}-\lambda_{3} 
\vert \sqrt{(\lambda_{1}-\lambda_{2})} \, \, \mathcal{X}_{A}} \right) \! 
\tfrac{\ln t}{t} \right),
\end{align*}
where $\widehat{r}(\zeta) \! = \! r(\zeta) \prod_{k=m+1}^{N}(d_{k}^{+}
(\zeta))^{-2}$ $\left(\vert \widehat{r}(\lambda_{1}) \vert \! = \! \vert 
r(\lambda_{1}) \vert \right)$, $\nu \! = \! \nu (\lambda_{1})$,
\begin{gather*}
\delta_{B}^{0} \! = \! \vert \lambda_{1} \! - \! \lambda_{3} \vert^{-\mi \nu} 
\! \left(2t(\lambda_{1} \! - \! \lambda_{2})^{3} \lambda_{1}^{-3} \right)^{-
\frac{\mi \nu}{2}} \me^{\mathscr{Z}(\lambda_{1})} \exp \! \left(-\tfrac{\mi 
t}{2}(\lambda_{1} \! - \! \lambda_{2})(z_{o} \! + \! \lambda_{1} \! + \! 
\lambda_{2}) \right), \\
\delta_{A}^{0} \! = \! \vert \lambda_{2} \! - \! \lambda_{3} \vert^{\mi \nu} 
\! \left(2t(\lambda_{1} \! - \! \lambda_{2})^{3} \lambda_{2}^{-3} \right)^{
\frac{\mi \nu}{2}} \me^{\mathscr{Z}(\lambda_{2})} \exp \! \left(\tfrac{\mi 
t}{2}(\lambda_{1} \! - \! \lambda_{2})(z_{o} \! + \! \lambda_{1} \! + \! 
\lambda_{2}) \right), \\
\mathscr{Z}(\lambda_{1}) \! = \! \dfrac{\mi}{2 \pi} \int_{-\infty}^{0} \ln 
\vert \mu \! - \! \lambda_{1} \vert \md \ln (1 \! - \! \vert r(\mu) \vert^{
2}) \! + \! \dfrac{\mi}{2 \pi} \int_{\lambda_{2}}^{\lambda_{1}} \ln \vert 
\mu \! - \! \lambda_{1} \vert \md \ln (1 \! - \! \vert r(\mu) \vert^{2}), \\
\mathscr{Z}(\lambda_{2}) \! = \! -\mathscr{Z}(\lambda_{1}) \! + \! \dfrac{
\mi}{2 \pi} \int_{-\infty}^{0} \ln \vert \mu \vert \md \ln (1 \! - \! \vert 
r(\mu) \vert^{2}) \! + \! \dfrac{\mi}{2 \pi} \int_{\lambda_{2}}^{\lambda_{1}} 
\ln \vert \mu \vert \md \ln (1 \! - \! \vert r(\mu) \vert^{2}), \\
\mathcal{X}_{B} \! = \! \tfrac{\vert \lambda_{1}-\lambda_{3} \vert}{\lambda_{
1}} \sqrt{\tfrac{2(\lambda_{1}-\lambda_{2})}{\lambda_{1}}} \,, \qquad \quad 
\mathcal{X}_{A} \! = \! \tfrac{\vert \lambda_{2}-\lambda_{3} \vert}{\lambda_{
2}} \sqrt{\tfrac{2(\lambda_{1}-\lambda_{2})}{\lambda_{2}}} \,, \\
\beta^{\Sigma_{B^{0}}}_{12} \! = \, \overline{\beta^{\Sigma_{B^{0}}}_{21}}
=\tfrac{\sqrt{2 \pi} \, \me^{-\frac{\pi \nu}{2}} \me^{\frac{\mi \pi}{4}}}
{\widehat{r}(\lambda_{1}) \, \overline{\Gamma (\mi \nu)}}, \qquad \quad 
\beta^{\Sigma_{A^{0}}}_{12} \! = \, \overline{\beta^{\Sigma_{A^{0}}}_{21}}
=\tfrac{\sqrt{2 \pi} \, \me^{-\frac{\pi \nu}{2}} \me^{-\frac{\mi \pi}{4}}}
{\overline{\widehat{r}(\lambda_{1})} \, \Gamma (\mi \nu)},
\end{gather*}
$\Gamma (\cdot)$ is the gamma function \cite{a51}, and $\mathbf{D}_{\ast}
(\cdot)$ is the parabolic cylinder function \cite{a51}. Using Eq.~(10), one 
shows that $\vert \lambda_{k} \! - \! \lambda_{3} \vert \lambda_{k}^{-1} \! 
= \! (2 \lambda_{k})^{-1/2}(z_{o}^{2} \! + \! 32)^{1/4}$, $k \! \in \! \{1,
2\}$. Using the identities \cite{a51} $\partial_{z} \mathbf{D}_{z_{1}}(z) \! 
= \! \tfrac{1}{2}(z_{1} \mathbf{D}_{z_{1}-1}(z) \! - \! \mathbf{D}_{z_{1}+1}
(z))$, $z \mathbf{D}_{z_{1}}(z) \! = \! \mathbf{D}_{z_{1}+1}(z) \! + \! z_{
1} \mathbf{D}_{z_{1}-1}(z)$, and $\vert \Gamma (\mi \nu) \vert^{2} \! = \! 
\tfrac{\pi}{\nu \sinh (\pi \nu)}$, and the integral \cite{a51} $\int_{0}^{
+\infty} \mathbf{D}_{-z_{1}}(z)z^{z_{2}-1} \me^{-z^{2}/4} \, \md z \! = \! 
\tfrac{\sqrt{\pi} \, \exp \left(-\frac{1}{2}(z_{1}+z_{2}) \ln 2 \right) 
\Gamma (z_{2})}{\Gamma (\frac{1}{2}(z_{1}+z_{2})+\frac{1}{2})}$, $\Re (z_{
2}) \! > \! 0$, {}from the above expressions for $\chi^{c}_{ij}(\zeta)$, $i,
j \! \in \! \{1,2\}$, and repeated application of the relation $\vert r
(\lambda_{1}) \vert \vert \Gamma (\mi \nu) \vert \nu \me^{\frac{\pi \nu}{2}} 
\! = \! (2 \pi \nu)^{1/2}$, one obtains the result stated in the Lemma. 
Furthermore, one shows that the symmetry reduction $\chi^{c}(\zeta) \! = \! 
\sigma_{1} \overline{\chi^{c}(\overline{\zeta})} \, \sigma_{1}$ is satisfied, 
and verifies that $(\chi^{c}(0) \sigma_{2})^{2} \! = \! \mathrm{I} \! + \! 
\mathcal{O}(t^{-1} \ln t)$. \hfill $\square$
\begin{bbbbb}
For $m \! \in \! \{1,2,\ldots,N\}$, set $\mathrm{Res}(\widehat{\chi}(\zeta);
\varsigma_{n}) \! := \! 
\left(
\begin{smallmatrix}
a_{n} & b_{n} \\
c_{n} & d_{n}
\end{smallmatrix}
\right)$, $n \! \in \! \{1,2,\ldots,m\}$. Then $b_{n} \! = \! -a_{n} \chi^{
c}_{12}(\varsigma_{n})/\chi^{c}_{22}(\varsigma_{n})$, $d_{n} \! = \! -c_{n} 
\chi^{c}_{12}(\varsigma_{n})/\chi^{c}_{22}(\varsigma_{n})$, and $\{a_{n},
\overline{c_{n}}\}_{n=1}^{m}$ satisfy the following (non-singular) system of 
$2m$ linear inhomogeneous algebraic 
equations,
\begin{eqnarray*}
\left[\begin{array}{cccccc}
\left. \begin{array}{ccc} \cline{1-3}
\multicolumn{1}{|c}{} &   & \multicolumn{1}{c|}{} \\
\multicolumn{1}{|c}{} & \widehat{\mathscr{A}} & 
\multicolumn{1}{c|}{} \\
\multicolumn{1}{|c}{} &   & \multicolumn{1}{c|}{} \\ 
\cline{1-3}   
\end{array} \right. & \left. \begin{array}{ccc} 
\cline{1-3}
\multicolumn{1}{|c}{} &   & \multicolumn{1}{c|}{} \\
\multicolumn{1}{|c}{} & \widehat{\mathscr{B}} & 
\multicolumn{1}{c|}{} \\
\multicolumn{1}{|c}{} &   & \multicolumn{1}{c|}{} \\ 
\cline{1-3}   
\end{array} \right. \\
& \\
\left. \begin{array}{ccc} \cline{1-3}
\multicolumn{1}{|c}{} &   & \multicolumn{1}{c|}{} \\
\multicolumn{1}{|c}{} & \overline{\widehat{\mathscr{B}}} & 
\multicolumn{1}{c|}{} \\
\multicolumn{1}{|c}{} &   & \multicolumn{1}{c|}{} \\ 
\cline{1-3}   
\end{array} \right. & \left. \begin{array}{ccc} 
\cline{1-3}
\multicolumn{1}{|c}{} &   & \multicolumn{1}{c|}{} \\
\multicolumn{1}{|c}{} & \overline{\widehat{\mathscr{A}} \,} & 
\multicolumn{1}{c|}{} \\
\multicolumn{1}{|c}{} &   & \multicolumn{1}{c|}{} \\ 
\cline{1-3}   
\end{array} \right. 
\end{array} \right] 
\left[\begin{array}{c}
          a_{1} \\
          a_{2} \\
          \vdots \\
          a_{m} \\
          \overline{c_{1}} \\
          \overline{c_{2}} \\
          \vdots \\
          \overline{c_{m}}
      \end{array} \right] = 
\left[\begin{array}{c}
          g^{\ast}_{1} \chi^{c}_{12}(\varsigma_{1}) \\
          g^{\ast}_{2} \chi^{c}_{12}(\varsigma_{2}) \\
          \vdots \\
          g^{\ast}_{m} \chi^{c}_{12}(\varsigma_{m}) \\
          \overline{g^{\ast}_{1} \chi^{c}_{22}(\varsigma_{1})} \\
          \overline{g^{\ast}_{2} \chi^{c}_{22}(\varsigma_{2})} \\
          \vdots \\
          \overline{g^{\ast}_{m} \chi^{c}_{22}(\varsigma_{m})}
       \end{array} \right],
\end{eqnarray*}
where
\begin{align*}
\widehat{\mathscr{A}}_{ij} \! :=& \! 
\begin{cases}
\dfrac{\det (\chi^{c}(\varsigma_{i})) \! + \! g_{i}^{\ast} \mathrm{W}(\chi^{
c}_{12}(\varsigma_{i}),\chi^{c}_{22}(\varsigma_{i}))}{\chi^{c}_{22}
(\varsigma_{i})}, &\text{$i \! = \! j \! \in \! \{1,2,\ldots,m\}$,} \\
-\dfrac{g_{i}^{\ast}(\chi^{c}_{12}(\varsigma_{i}) \chi^{c}_{22}(\varsigma_{
j}) \! - \! \chi^{c}_{22}(\varsigma_{i}) \chi^{c}_{12}(\varsigma_{j}))}{
(\varsigma_{i} \! - \! \varsigma_{j}) \chi^{c}_{22}(\varsigma_{j})}, 
&\text{$i \! \not= \! j \! \in \! \{1,2,\ldots,m\}$,}
\end{cases} \\
\widehat{\mathscr{B}}_{ij} \! :=& \! -\dfrac{g_{i}^{\ast}(\chi^{c}_{22}
(\varsigma_{i}) \overline{\chi^{c}_{22}(\varsigma_{j})} \! - \! \chi^{c}_{12}
(\varsigma_{i}) \overline{\chi^{c}_{12}(\varsigma_{j})})}{(\varsigma_{i} \! 
- \! \overline{\varsigma_{j}}) \overline{\chi^{c}_{22}(\varsigma_{j})}}, 
\quad \, \, i,j \! \in \! \{1,2,\ldots,m\},
\end{align*}
\begin{equation*}
g_{j}^{\ast} \! = \! \vert g_{j} \vert \me^{\mi \theta_{g_{j}}} \exp \! 
\left(2 \mi k(\varsigma_{j})(x \! + \! 2 \lambda (\varsigma_{j})t) \right) 
\! (\delta (\varsigma_{j}))^{-2} \! \prod_{k=m+1}^{N}(d_{k}^{+}(\varsigma_{
j}))^{-2}, \quad j \! \in \! \{1,2,\ldots,m\},
\end{equation*}
with $\vert g_{j} \vert$ and $\theta_{g_{j}}$ given in Lemma~{\rm 3.1}, 
(iii), and $\mathrm{W}(\chi^{c}_{12}(z),\chi^{c}_{22}(z)) \! = \! 
\left\vert
\begin{smallmatrix}
\chi^{c}_{12}(z) & \chi^{c}_{22}(z) \\
\partial_{z} \chi^{c}_{12}(z) & \partial_{z} \chi^{c}_{22}(z)
\end{smallmatrix}
\right\vert$.
\end{bbbbb}

\emph{Proof.} Recall {}from Lemma~4.1 that $\widehat{\chi}(\zeta) \colon 
\mathbb{C} \setminus (\widehat{\sigma}_{d} \cup \sigma_{c}) \! \to \! \mathrm{
M}_{2}(\mathbb{C})$ has the factorised representation $\widehat{\chi}(\zeta) 
\! = \! \left(\mathrm{I} \! + \! \zeta^{-1} \widehat{\Delta}_{o} \right) \! 
\widehat{\mathscr{P}}(\zeta) \! \left(\mathrm{I} \! + \! \sum_{n=1}^{m} 
\! \left(\tfrac{\mathrm{Res}(\widehat{\chi}(\zeta);\varsigma_{n})}{(\zeta 
-\varsigma_{n})} \! + \! \tfrac{\sigma_{1} \overline{\mathrm{Res}(\widehat{
\chi}(\zeta);\varsigma_{n})} \, \sigma_{1}}{(\zeta -\overline{\varsigma_{
n}})} \right) \right) \! \chi^{c}(\zeta)$, where $\chi^{c}(\zeta)$ is given 
in Lemma 4.2. For $m \! \in \! \{1,2,\ldots,N\}$, set $\mathrm{Res}(\widehat{
\chi}(\zeta);\varsigma_{n}) \! := \! 
\left(
\begin{smallmatrix}
a_{n} & b_{n} \\
c_{n} & d_{n}
\end{smallmatrix}
\right)$, whence $\sigma_{1} \overline{\mathrm{Res}(\widehat{\chi}(\zeta);
\varsigma_{n})} \, \sigma_{1} \! = \! 
\left(
\begin{smallmatrix}
\overline{d_{n}} & \, \, \overline{c_{n}} \\
\overline{b_{n}} & \, \, \overline{a_{n}}
\end{smallmatrix}
\right)$; thus,
\begin{align}
\widehat{\chi}(\zeta) \! =& \! \left(\mathrm{I} \! + \! \tfrac{1}{\zeta} 
\widehat{\Delta}_{o} \right) \! \widehat{\mathscr{P}}(\zeta) \! 
\begin{pmatrix}
1+\frac{a_{n}}{\zeta-\varsigma_{n}}+\sum_{\genfrac{}{}{0pt}{2}{k=1}{k \not= 
n}}^{m} \! \frac{a_{k}}{\zeta-\varsigma_{k}}+\sum_{k=1}^{m} \! \frac{
\overline{d_{k}}}{\zeta-\overline{\varsigma_{k}}} & \, \frac{b_{n}}{\zeta-
\varsigma_{n}}+\sum_{\genfrac{}{}{0pt}{2}{k=1}{k \not= n}}^{m} \! \frac{b_{
k}}{\zeta-\varsigma_{k}}+\sum_{k=1}^{m} \! \frac{\overline{c_{k}}}{\zeta-
\overline{\varsigma_{k}}} \\
\frac{c_{n}}{\zeta-\varsigma_{n}}+\sum_{\genfrac{}{}{0pt}{2}{k=1}{k \not= 
n}}^{m} \! \frac{c_{k}}{\zeta-\varsigma_{k}}+\sum_{k=1}^{m} \! \frac{
\overline{b_{k}}}{\zeta-\overline{\varsigma_{k}}} & \, 1+\frac{d_{n}}{\zeta
-\varsigma_{n}}+\sum_{\genfrac{}{}{0pt}{2}{k=1}{k \not= n}}^{m} \! \frac{d_{
k}}{\zeta-\varsigma_{k}}+\sum_{k=1}^{m} \! \frac{\overline{a_{k}}}{\zeta-
\overline{\varsigma_{k}}}
\end{pmatrix} \nonumber \\
\times& 
\begin{pmatrix}
\chi^{c}_{11}(\zeta) & \chi^{c}_{12}(\zeta) \\
\chi^{c}_{21}(\zeta) & \chi^{c}_{22}(\zeta)
\end{pmatrix}.
\end{align}
As in the BC construction \cite{a41}, one now Taylor expands $\chi^{c}
(\zeta)$ about $\{\varsigma_{n}\}_{n=1}^{m}$: $\chi^{c}_{ij}(\zeta) \! = 
\! \chi^{c}_{ij}(\varsigma_{n}) \! + \! (\partial_{\zeta} \chi^{c}_{ij}
(\varsigma_{n}))(\zeta \! - \! \varsigma_{n}) \! + \! \mathcal{O}((\zeta \! - 
\! \varsigma_{n})^{2})$, $i,j \! \in \! \{1,2\}$, where $\partial_{\zeta} 
\chi^{c}_{ij}(\varsigma_{n}) \! = \! \partial_{\zeta} \chi^{c}_{ij}(\zeta) 
\vert_{\zeta =\varsigma_{n}}$. Recalling {}from Lemma~3.5, \emph{(iii)}, that 
$\widehat{\chi}(\zeta)$ satisfies the polar (residue) conditions $\mathrm{Res}
(\widehat{\chi}(\zeta);\varsigma_{n})=\lim_{\zeta \to \varsigma_{n}} \widehat{
\chi}(\zeta)g_{n}^{\ast} \sigma_{-}$ and $\mathrm{Res}(\widehat{\chi}(\zeta);
\linebreak[4]
\overline{\varsigma_{n}}) \! = \! \sigma_{1} \overline{\mathrm{Res}(\widehat{
\chi}(\zeta);\varsigma_{n})} \, \sigma_{1}$, $n \! \in \! \{1,2,\ldots,m\}$, 
with $g_{n}^{\ast}$ given in the Proposition, assembling the above, one shows 
that the only non-trivial conditions are
\begin{gather*}
a_{n} \chi^{c}_{12}(\varsigma_{n}) \! + \! b_{n} \chi^{c}_{22}(\varsigma_{n}) 
\! = \! 0, \\
c_{n} \chi^{c}_{12}(\varsigma_{n}) \! + \! d_{n} \chi^{c}_{22}(\varsigma_{n}) 
\! = \! 0, \\
a_{n} \chi^{c}_{11}(\varsigma_{n}) \! + \! b_{n} \chi^{c}_{21}(\varsigma_{n}) 
\! = \! a_{n}g_{n}^{\ast} \partial_{\zeta} \chi^{c}_{12}(\varsigma_{n}) \! + 
\! \left(1 \! + \! \sum_{\genfrac{}{}{0pt}{2}{k=1}{k \not= n}}^{m} \dfrac{
a_{k}}{\varsigma_{n} \! - \! \varsigma_{k}} \! + \! \sum_{k=1}^{m} \dfrac{
\overline{d_{k}}}{\varsigma_{n} \! - \! \overline{\varsigma_{k}}} \right) \! 
g_{n}^{\ast} \chi^{c}_{12}(\varsigma_{n}) \\
+ \, b_{n}g_{n}^{\ast} \partial_{\zeta} \chi^{c}_{22}(\varsigma_{n}) \! + \! 
\left(\sum_{\genfrac{}{}{0pt}{2}{k=1}{k \not= n}}^{m} \dfrac{b_{k}}{\varsigma_{
n} \! - \! \varsigma_{k}} \! + \! \sum_{k=1}^{m} \dfrac{\overline{c_{k}}}{
\varsigma_{n} \! - \! \overline{\varsigma_{k}}} \right) \! g_{n}^{\ast} \chi^{
c}_{22}(\varsigma_{n}) \! + \! \lim_{\zeta \to \varsigma_{n}}(\, \underbrace{
a_{n} \chi^{c}_{12}(\varsigma_{n}) \! + \! b_{n} \chi^{c}_{22}(\varsigma_{n}) 
\,}_{0}) \dfrac{g_{n}^{\ast}}{\zeta \! - \! \varsigma_{n}}, \\
c_{n} \chi^{c}_{11}(\varsigma_{n}) \! + \! d_{n} \chi^{c}_{21}(\varsigma_{
n}) \! = \! c_{n}g_{n}^{\ast} \partial_{\zeta} \chi^{c}_{12}(\varsigma_{n}) 
\! + \! \left(\sum_{\genfrac{}{}{0pt}{2}{k=1}{k \not= n}}^{m} \dfrac{c_{k}}
{\varsigma_{n} \! - \! \varsigma_{k}} \! + \! \sum_{k=1}^{m} \dfrac{\overline{
b_{k}}}{\varsigma_{n} \! - \! \overline{\varsigma_{k}}} \right) \! g_{n}^{
\ast} \chi^{c}_{12}(\varsigma_{n}) \\
+ \, d_{n}g_{n}^{\ast} \partial_{\zeta} \chi^{c}_{22}(\varsigma_{n}) \! + \! 
\left(1 \! + \! \sum_{\genfrac{}{}{0pt}{2}{k=1}{k \not= n}}^{m} \dfrac{d_{k}}
{\varsigma_{n} \! - \! \varsigma_{k}} \! + \! \sum_{k=1}^{m} \dfrac{\overline{
a_{k}}}{\varsigma_{n} \! - \! \overline{\varsigma_{k}}} \right) \! g_{n}^{
\ast} \chi^{c}_{22}(\varsigma_{n}) \! + \! \lim_{\zeta \to \varsigma_{n}}(\, 
\underbrace{c_{n} \chi^{c}_{12}(\varsigma_{n}) \! + \! d_{n} \chi^{c}_{22}
(\varsigma_{n}) \,}_{0}) \dfrac{g_{n}^{\ast}}{\zeta \! - \! \varsigma_{n}}.
\end{gather*}
{}From the first two equations of the above system, one gets that $b_{n} \! = 
\! -a_{n} \chi^{c}_{12}(\varsigma_{n})/\chi^{c}_{22}(\varsigma_{n})$ and $d_{
n} \! = \! -c_{n} \chi^{c}_{12}(\varsigma_{n})/\chi^{c}_{22}(\varsigma_{n})$ 
(whence, $\det \! 
\left(
\begin{smallmatrix}
a_{n} & b_{n} \\
c_{n} & d_{n}
\end{smallmatrix}
\right) \! = \! 0)$: using the latter (two) relations, it follows {}from the 
last two equations of the above system that, for $n \! \in \! \{1,2,\ldots,
m\}$,
\begin{equation*}
a_{n} \mathcal{A}_{n} \! = \! \sum_{\genfrac{}{}{0pt}{2}{k=1}{k \not= n}}^{m} 
\dfrac{a_{k}g_{n}^{\ast} \mathcal{B}_{nk}}{\varsigma_{n} \! - \! \varsigma_{
k}} \! + \! \sum_{k=1}^{m} \dfrac{\overline{c_{k}} \, g_{n}^{\ast} \mathcal{
D}_{nk}}{\varsigma_{n} \! - \! \overline{\varsigma_{k}}} \! + \! g_{n}^{\ast} 
\chi^{c}_{12}(\varsigma_{n}), \quad \, \overline{c_{n} \mathcal{A}_{n}} \! = 
\! \sum_{\genfrac{}{}{0pt}{2}{k=1}{k \not= n}}^{m} \dfrac{\overline{c_{k}g_{
n}^{\ast} \mathcal{B}_{nk}}}{\overline{\varsigma_{n}} \! - \! \overline{
\varsigma_{k}}} \! + \! \sum_{k=1}^{m} \dfrac{a_{k} \overline{g_{n}^{\ast} 
\mathcal{D}_{nk}}}{\overline{\varsigma_{n}} \! - \! \varsigma_{k}} \! + \! 
\overline{g_{n}^{\ast} \chi^{c}_{22}(\varsigma_{n})},
\end{equation*}
where
\begin{gather*}
\mathcal{A}_{n} \! = \! \dfrac{\det (\chi^{c}(\varsigma_{n})) \! + \! g_{
n}^{\ast} \mathrm{W}(\chi^{c}_{12}(\varsigma_{n}),\chi^{c}_{22}(\varsigma_{
n}))}{\chi^{c}_{22}(\varsigma_{n})}, \qquad \, \, \mathcal{B}_{nk} \! = \! 
\dfrac{\chi^{c}_{12}(\varsigma_{n}) \chi^{c}_{22}(\varsigma_{k}) \! - \! 
\chi^{c}_{12}(\varsigma_{k}) \chi^{c}_{22}(\varsigma_{n})}{\chi^{c}_{22}
(\varsigma_{k})}, \\
\mathcal{D}_{nk} \! = \! \dfrac{\chi^{c}_{22}(\varsigma_{n}) \overline{\chi^{
c}_{22}(\varsigma_{k})}- \! \chi^{c}_{12}(\varsigma_{n}) \overline{\chi^{c}_{
12}(\varsigma_{k})}}{\overline{\chi^{c}_{22}(\varsigma_{k})}};
\end{gather*}
thus, the (rank $2m)$ linear inhomogeneous algebraic system for $\{a_{n},
\overline{c_{n}}\}_{n=1}^{m}$ stated in the Proposition. The non-degeneracy 
of the $(2m \times 2m)$ coefficient matrix is a consequence of the asymptotic 
solvability of the original RHP formulated in Lemma~2.1.2 \cite{a38} (see, 
also, Eq.~(66) below). \hfill $\square$
\begin{bbbbb}
As $t \! \to \! +\infty$ and $x \! \to \! -\infty$ such that $z_{o} \! := \! 
x/t \! < \! -2$ and $(x,t) \! \in \! \daleth_{m}$, $m \! \in \! \{1,2,\ldots,
N\}$, for $n \! \in \! \{1,2,\ldots,m \! - \! 1\}$,
\begin{gather*}
a_{n} \! = \! \mathcal{O} \! \left(\me^{-\gimel^{+}t} \right), \qquad \quad 
b_{n} \! = \! \mathcal{O} \! \left(t^{-1/2}(z_{o}^{2} \! + \! 32)^{-1/4} 
\me^{-\gimel^{+}t} \right), \\
c_{n} \! = \! \mathcal{O} \! \left(\me^{-\gimel^{+}t} \right), \qquad \quad 
d_{n} \! = \! \mathcal{O} \! \left(t^{-1/2}(z_{o}^{2} \! + \! 32)^{-1/4} 
\me^{-\gimel^{+}t} \right),
\end{gather*}
where $\gimel^{+} \! := \! 4 \min_{\genfrac{}{}{0pt}{2}{m \in \{1,2,\ldots,
N\}}{n \in \{1,2,\ldots,m-1\}}}\{\sin (\phi_{n}) \vert \cos (\phi_{n}) \! - 
\! \cos (\phi_{m}) \vert\}$ $(> \! 0)$, and
\begin{align*}
a_{m} \! &= \, a_{m}^{0} \! + \! \tfrac{1}{\sqrt{t}}a_{m}^{1} \! + \! 
\mathcal{O} \! \left(\tfrac{c^{\mathcal{S}}(z_{o})}{(z_{o}^{2}+32)^{1/2}} 
\tfrac{\ln t}{t} \right) \\
&=: \tfrac{g_{m}^{\ast} \overline{g_{m}^{\ast}} \, (\varsigma_{m}-\overline{
\varsigma_{m}})^{-1}}{(1+g_{m}^{\ast} \overline{g_{m}^{\ast}} \, (\varsigma_{
m}-\overline{\varsigma_{m}})^{-2})} \! + \! \tfrac{1}{\sqrt{t}} \! \left(
\tfrac{g_{m}^{\ast} \overline{g_{m}^{\ast}} \, (\varsigma_{m}-\overline{
\varsigma_{m}})^{-1}(g_{m}^{\ast} \partial_{\zeta} \widetilde{\chi}^{c}_{12}
(\varsigma_{m})+\overline{g_{m}^{\ast} \partial_{\zeta} \widetilde{\chi}^{c}_{
12}(\varsigma_{m})})}{(1+g_{m}^{\ast} \overline{g_{m}^{\ast}} \, (\varsigma_{
m}-\overline{\varsigma_{m}})^{-2})^{2}} \! + \! \tfrac{g_{m}^{\ast} 
\widetilde{\chi}^{c}_{12}(\varsigma_{m})}{(1+g_{m}^{\ast} \overline{g_{m}^{
\ast}} \, (\varsigma_{m}-\overline{\varsigma_{m}})^{-2})} \right) \\
&+\, \mathcal{O} \! \left(\tfrac{c^{\mathcal{S}}(z_{o})}{(z_{o}^{2}+32)^{
1/2}} \tfrac{\ln t}{t} \right), \\
b_{m} \! &= \, \tfrac{1}{\sqrt{t}}b_{m}^{1} \! + \! \mathcal{O} \! \left(
\tfrac{c^{\mathcal{S}}(z_{o})}{(z_{o}^{2}+32)^{1/2}} \tfrac{\ln t}{t} \right) 
\! =: \! -\tfrac{1}{\sqrt{t}} \tfrac{g_{m}^{\ast} \overline{g_{m}^{\ast}} \, 
(\varsigma_{m}-\overline{\varsigma_{m}})^{-1} \widetilde{\chi}^{c}_{12}
(\varsigma_{m})}{(1+g_{m}^{\ast} \overline{g_{m}^{\ast}} \, (\varsigma_{m}-
\overline{\varsigma_{m}})^{-2})} \! + \! \mathcal{O} \! \left(\tfrac{c^{
\mathcal{S}}(z_{o})}{(z_{o}^{2}+32)^{1/2}} \tfrac{\ln t}{t} \right), \\
c_{m} \! &= \, c_{m}^{0} \! + \! \tfrac{1}{\sqrt{t}}c_{m}^{1} \! + \! 
\mathcal{O} \! \left(\tfrac{c^{\mathcal{S}}(z_{o})}{(z_{o}^{2}+32)^{1/2}} 
\tfrac{\ln t}{t} \right) \\
&=: \, \tfrac{g_{m}^{\ast}}{(1+g_{m}^{\ast} \overline{g_{m}^{\ast}} \, 
(\varsigma_{m}-\overline{\varsigma_{m}})^{-2})} \! + \! \tfrac{1}{\sqrt{t}} 
\! \left(\tfrac{g_{m}^{\ast} \overline{g_{m}^{\ast}} \, (\varsigma_{m}-
\overline{\varsigma_{m}})^{-1} \overline{\widetilde{\chi}^{c}_{12}
(\varsigma_{m})}-g_{m}^{\ast} \overline{g_{m}^{\ast} \partial_{\zeta} 
\widetilde{\chi}^{c}_{12}(\varsigma_{m})}}{(1+g_{m}^{\ast} \overline{g_{m}^{
\ast}} \, (\varsigma_{m}-\overline{\varsigma_{m}})^{-2})} \! + \! \tfrac{g_{
m}^{\ast}(g_{m}^{\ast} \partial_{\zeta} \widetilde{\chi}^{c}_{12}(\varsigma_{
m})+\overline{g_{m}^{\ast} \partial_{\zeta} \widetilde{\chi}^{c}_{12}
(\varsigma_{m})})}{(1+g_{m}^{\ast} \overline{g_{m}^{\ast}} \, (\varsigma_{m}
-\overline{\varsigma_{m}})^{-2})^{2}} \right) \\
&+\, \mathcal{O} \! \left(\tfrac{c^{\mathcal{S}}(z_{o})}{(z_{o}^{2}+32)^{
1/2}} \tfrac{\ln t}{t} \right), \\
d_{m} \! &= \, \tfrac{1}{\sqrt{t}}d_{m}^{1} \! + \! \mathcal{O} \! \left(
\tfrac{c^{\mathcal{S}}(z_{o})}{(z_{o}^{2}+32)^{1/2}} \tfrac{\ln t}{t} \right) 
\! =: \! -\tfrac{1}{\sqrt{t}} \tfrac{g_{m}^{\ast} \widetilde{\chi}^{c}_{12} \, 
(\varsigma_{m})}{(1+g_{m}^{\ast} \overline{g_{m}^{\ast}} \, (\varsigma_{m}-
\overline{\varsigma_{m}})^{-2})} \! + \! \mathcal{O} \! \left(\tfrac{c^{
\mathcal{S}}(z_{o})}{(z_{o}^{2}+32)^{1/2}} \tfrac{\ln t}{t} \right),
\end{align*}
where
\begin{equation*}
\widetilde{\chi}^{c}_{12}(\zeta) \! = \! \dfrac{\sqrt{\nu (\lambda_{1})} \, 
\me^{\frac{\mi \Xi^{+}(0)}{2}} \lambda_{1}^{2 \mi \nu (\lambda_{1})}}{\sqrt{
(\lambda_{1} \! - \! \lambda_{2})} \, (z_{o}^{2} \! + \! 32)^{1/4}} \! \left( 
\dfrac{\lambda_{1} \me^{-\mi (\Theta^{+}(z_{o},t)+\frac{\pi}{4})}}{(\zeta \! 
- \! \lambda_{1})} \! + \! \dfrac{\lambda_{2} \me^{\mi (\Theta^{+}(z_{o},t)+
\frac{\pi}{4})}}{(\zeta \! - \! \lambda_{2})} \! \right),
\end{equation*}
with $\nu (\cdot)$, $\lambda_{1}$, $\lambda_{2}$, $\lambda_{3}$, $\Xi^{+}
(\cdot)$, and $\Theta^{+}(z_{o},t)$ specified in Lemma~{\rm 4.2}, and $c^{
\mathcal{S}}(z_{o}) \! = \! \tfrac{c^{\mathcal{S}}(\lambda_{1}) \underline{
c}(\lambda_{2},\lambda_{3},\overline{\lambda_{3}})}{\sqrt{\lambda_{1}} \, 
(\lambda_{1}-\lambda_{2})} \! + \! \tfrac{c^{\mathcal{S}}(\lambda_{2}) 
\underline{c}(\lambda_{1},\lambda_{3},\overline{\lambda_{3}})}{\sqrt{
\lambda_{2}} \, (\lambda_{1}-\lambda_{2})}$.
\end{bbbbb}

\emph{Proof.} Noting that, as $t \! \to \! +\infty$ and $x \! \to \! -\infty$ 
such that $z_{o} \! < \! -2$ and $(x,t) \! \in \! \daleth_{m}$, $g_{n}^{\ast} 
\! \! \upharpoonright_{\daleth_{m}} \! = \! \mathcal{O}(1)$, $n \! = \! m$, 
and $g_{n}^{\ast} \! \! \upharpoonright_{\daleth_{m}} \! = \! \mathcal{O}
(\exp (-4t \sin (\phi_{n}) \vert \cos (\phi_{n}) \! - \! \cos (\phi_{m}) 
\vert))$, $n \! \in \! \{1,2,\ldots,m \! - \! 1\}$, one deduces {}from 
Proposition~4.1 that $\{a_{n},\overline{c_{n}}\}_{n=1}^{m}$ solve
\begin{eqnarray*}
\, \, \, \left[\begin{array}{cccccccccccc}
\left. \begin{array}{cccccc} \cline{1-6}
\multicolumn{1}{|c}{} & \raisebox{-0.45ex}{$\mathcal{A}_{1}$} & \raisebox{
-0.45ex}{$o(1)$} & \dots & \raisebox{-0.45ex}{$o(1)$} & 
\multicolumn{1}{c|}{} \\
\multicolumn{1}{|c}{} & o(1) & \mathcal{A}_{2} & \ddots & \vdots & 
\multicolumn{1}{c|}{} \\
\multicolumn{1}{|c}{} & \vdots & \ddots & \ddots & o(1) & 
\multicolumn{1}{c|}{} \\
\multicolumn{1}{|c}{} & \raisebox{0.55ex}{$-\frac{g_{m}^{\ast} \mathcal{B}_{
m1}}{\varsigma_{m}-\varsigma_{1}}$} & \cdots & \raisebox{0.55ex}{$-\frac{g_{
m}^{\ast} \mathcal{B}_{mm-1}}{\varsigma_{m}-\varsigma_{m-1}}$} & \raisebox{
0.45ex}{$\mathcal{A}_{m}$} & \multicolumn{1}{c|}{} \\
\cline{1-6} 
\end{array} \right. & \left. \begin{array}{cccccc} 
\cline{1-6}
\multicolumn{1}{|c}{} & \raisebox{-0.45ex}{$o(1)$} & \dots & \dots & 
\raisebox{-0.45ex}{$o(1)$} & \multicolumn{1}{c|}{} \\
\multicolumn{1}{|c}{} & \vdots & \ddots & \ddots & \vdots & 
\multicolumn{1}{c|}{} \\
\multicolumn{1}{|c}{} & \vdots & \ddots & \ddots & \vdots & 
\multicolumn{1}{c|}{} \\
\multicolumn{1}{|c}{} & \raisebox{0.55ex}{$-\frac{g_{m}^{\ast} \mathcal{D}_{
m1}}{\varsigma_{m}-\overline{\varsigma_{1}}}$} & \cdots & \cdots & \raisebox{
0.55ex}{$-\frac{g_{m}^{\ast} \mathcal{D}_{mm}}{\varsigma_{m}-\overline{
\varsigma_{m}}}$} & \multicolumn{1}{c|}{} \\
\cline{1-6}
\end{array} \right. \\
& \\
\left. \begin{array}{cccccc} \cline{1-6}
\multicolumn{1}{|c}{} & \raisebox{-0.45ex}{$o(1)$} & \dots & \dots & 
\raisebox{-0.45ex}{$o(1)$} & \multicolumn{1}{c|}{} \\
\multicolumn{1}{|c}{} & \vdots & \ddots & \ddots & \vdots & 
\multicolumn{1}{c|}{} \\
\multicolumn{1}{|c}{} & \vdots & \ddots & \ddots & \vdots & 
\multicolumn{1}{c|}{} \\
\multicolumn{1}{|c}{} & \raisebox{0.55ex}{$-\frac{\overline{g_{m}^{\ast} 
\mathcal{D}_{m1}}}{\overline{\varsigma_{m}}-\varsigma_{1}}$} & \cdots & 
\cdots & \raisebox{0.55ex}{$-\frac{\overline{g_{m}^{\ast} \mathcal{D}_{mm}}}
{\overline{\varsigma_{m}}-\varsigma_{m}}$} & 
\multicolumn{1}{c|}{} \\
\cline{1-6} 
\end{array} \right. & \left. \begin{array}{cccccc} 
\cline{1-6}
\multicolumn{1}{|c}{} & \overline{\raisebox{-0.45ex}{$\mathcal{A}_{1}$}} & 
\raisebox{-0.45ex}{$o(1)$} & \dots & \raisebox{-0.45ex}{$o(1)$} & 
\multicolumn{1}{c|}{} \\
\multicolumn{1}{|c}{} & \vdots & \overline{\mathcal{A}_{2}} & \ddots & \vdots 
& \multicolumn{1}{c|}{} \\
\multicolumn{1}{|c}{} & \vdots & \ddots & \ddots & o(1) & 
\multicolumn{1}{c|}{} \\
\multicolumn{1}{|c}{} & \raisebox{0.55ex}{$-\frac{\overline{g_{m}^{\ast} 
\mathcal{B}_{m1}}}{\overline{\varsigma_{m}}-\overline{\varsigma_{1}}}$} & 
\cdots & \raisebox{0.55ex}{$-\frac{\overline{g_{m}^{\ast} \mathcal{B}_{mm-
1}}}{\overline{\varsigma_{m}}-\overline{\varsigma_{m-1}}}$} & \raisebox{
0.55ex}{$\overline{\mathcal{A}_{m}}$} & \multicolumn{1}{c|}{} \\
\cline{1-6} 
\end{array} \right. 
\end{array} \right] 
\left[\begin{array}{c}
          a_{1} \\
          a_{2} \\
          \vdots \\
          \vdots \\  
          a_{m} \\
          \overline{c_{1}} \\
          \overline{c_{2}} \\
          \vdots \\
          \vdots \\   
          \overline{c_{m}}
      \end{array} \right]
\end{eqnarray*}
\begin{equation*}
= \! [\underbrace{o(1),\ldots,o(1),g^{\ast}_{m} \chi^{c}_{12}(\varsigma_{
m})}_{m},\underbrace{o(1),\ldots,o(1),\overline{g^{\ast}_{m} \chi^{c}_{22}
(\varsigma_{m})}}_{m} \,]^{\mathrm{T}},
\end{equation*}
where ${}^{\mathrm{T}}$ denotes transposition, $\mathcal{A}_{n}$, $\mathcal{
B}_{nk}$, and $\mathcal{D}_{nk}$ are given in the proof of Proposition~4.1, 
$o(1) \! := \! \mathcal{O}(\exp (-4t \min_{\genfrac{}{}{0pt}{2}{m \in \{1,2,
\ldots,N\}}{n \in \{1,2,\ldots,m-1\}}}\{\sin (\phi_{n}) \vert \cos (\phi_{n}) 
\! - \! \cos (\phi_{m}) \vert\}))$, and $\chi^{c}_{12}(\cdot)$ and $\chi^{c}_{
22}(\cdot)$ are given in Lemma 4.2. Solving the above system for $\{a_{n},
\overline{c_{n}}\}_{n=1}^{m}$ via the Cauchy-Binet formula, or Cramer's Rule, 
recalling the expressions for $\chi^{c}_{ij}(\zeta)$, $i,j \! \in \! \{1,
2\}$, given in Lemma~4.2, setting $\widetilde{\chi}^{c}_{12}(\zeta)$ and $c^{
\mathcal{S}}(z_{o})$ as in the Proposition, and recalling {}from 
Proposition~4.1 that $b_{n} \! = \! -a_{n} \chi^{c}_{12}(\varsigma_{n})
/\chi^{c}_{22}(\varsigma_{n})$ and $d_{n} \! = \! -c_{n} \chi^{c}_{12}
(\varsigma_{n})/\chi^{c}_{22}(\varsigma_{n})$, one gets the estimates for 
$\{a_{n},b_{n},c_{n},d_{n}\}_{n=1}^{m-1}$ and the explicit---asymptotic 
expansion---formulae for $\{a_{m},b_{m},c_{m},d_{m}\}$ stated in the 
Proposition. Furthermore, setting $\mathcal{Y} \! := \! 
\left(
\begin{smallmatrix}
\boxed{\widehat{\mathscr{A}}} & \boxed{\widehat{\mathscr{B}}} \\
\boxed{\overline{\widehat{\mathscr{B}}}} & \boxed{\overline{\widehat{
\mathscr{A}} \,}}
\end{smallmatrix}
\right)$, with $\widehat{\mathscr{A}}$ and $\widehat{\mathscr{B}}$ defined 
in Proposition~4.1, {}from the asymptotic estimates above for $\{a_{n},b_{n},
c_{n},d_{n}\}_{n=1}^{m}$, and recalling that, as a consequence of the 
asymptotic solvability of the original RHP formulated in Lemma~2.1.2, $\det 
(\mathcal{Y}) \! \not\equiv \! 0$, an application of Hadamard's Inequality 
$(\vert \det (\mathcal{Y}) \vert^{2} \! \leqslant \! \prod_{j=1}^{2m} \sum_{
i=1}^{2m} \vert \mathcal{Y}_{ij} \vert^{2}$, where $\mathcal{Y}_{ij}$ denotes 
the $(i \, j)$-element of $\mathcal{Y})$ shows that
\begin{equation}
0 \! < \! \vert \det (\mathcal{Y}) \vert^{2} \! \leqslant \! \prod_{j=1}^{m} 
\! \left(1 \! + \! \tfrac{\sin^{2}(\phi_{m}) \vert \gamma_{m} \vert^{2}P^{2}
(\phi_{m},\phi_{k})Q^{2}(\phi_{m})}{\sin^{2}(\frac{1}{2}(\phi_{m}+\phi_{j}))} 
\me^{2 \phi (x,t)} \right)^{2} \! +\mathcal{O} \! \left( \tfrac{c^{\mathcal{
S}}(z_{o})}{(z_{o}^{2}+32)^{1/2}} \tfrac{\ln t}{t} \right),
\end{equation}
where
\begin{gather}
\phi (x,t) \! := \! -2 \sin (\phi_{m})(x \! + \! 2t \cos \phi_{m}), \\
P(\phi_{m},\phi_{k}) \! := \! \left(\prod_{k=1}^{m-1} \dfrac{\sin (\frac{1}
{2}(\phi_{m} \! + \! \phi_{k}))}{\sin (\frac{1}{2}(\phi_{m} \! - \! \phi_{k}
))} \right) \! \! \left(\prod_{k=m+1}^{N} \dfrac{\sin (\frac{1}{2}(\phi_{m} 
\! + \! \phi_{k}))}{\sin (\frac{1}{2}(\phi_{m} \! - \! \phi_{k}))} \right)^{
-1}, \\
Q(\phi_{m}) \! := \! \exp \! \left( \! \left(\int_{0}^{\lambda_{2}} \! + \! 
\int_{\lambda_{1}}^{+\infty} \! - \! \int_{-\infty}^{0} \! - \! \int_{
\lambda_{2}}^{\lambda_{1}} \right) \! \dfrac{\sin (\phi_{m}) \ln (1 \! - \! 
\vert r(\mu) \vert^{2})}{(\mu^{2} \! - \! 2 \mu \cos (\phi_{m}) \! + \! 1)} 
\, \dfrac{\md \mu}{2 \pi} \right),
\end{gather}
and $c^{\mathcal{S}}(z_{o})$ is given in the Proposition. \hfill $\square$

The following Lemma is proved via the higher-order generalisation \cite{a54} 
of the Deift-Zhou (DZ) non-linear steepest descent method \cite{a55} (see, 
also, \cite{a56}), but its proof is far beyond the scope of the present work 
(it shall be presented elsewhere).
\begin{eeeee}
Even though in Lemma~4.3 below, in the \emph{sensus strictu} of asymptotic 
analysis, the exponentially small terms should be neglected, and thus not 
written out explicitly, in lieu of the $t^{-p/2}(\ln t)^{q}$ corrections, 
$p \! \geqslant \! 1$, $q \! \in \! \{0,1,\ldots,p \! - \! 1\}$, they are 
written there, and there only (see, also, Appendix~A, Lemma~A.1.7), in order 
to bring to the reader's attention the fact that there are additional, 
albeit exponentially small, terms that are due to the remaining solitons: 
thereafter, exponentially small terms are neglected.
\end{eeeee}
\begin{ccccc}
As $t \! \to \! +\infty$ and $x \! \to \! -\infty$ such that $z_{o} \! := \! 
x/t \! < \! -2$ and $(x,t) \! \in \! \daleth_{m}$, $m \! \in \! \{1,2,\ldots,
N\}$,
\begin{equation*}
\widehat{\mathscr{P}}(\zeta) \! = \! 
\begin{pmatrix}
\frac{\zeta +\widehat{a}_{1}^{+}}{\zeta +\widehat{a}_{2}^{+}} & \frac{
\widehat{a}_{3}^{+}}{\zeta +\widehat{a}_{4}^{+}} \\
\frac{\overline{\widehat{a}_{3}^{+}}}{\zeta +\overline{\widehat{a}_{4}^{+}}} 
& \frac{\zeta +\overline{\widehat{a}_{1}^{+}}}{\zeta +\overline{\widehat{a}_{
2}^{+}}}
\end{pmatrix},
\end{equation*}
where
\begin{align*}
\widehat{a}_{1}^{+} \! =& \, \overline{\widehat{a}_{2}^{+}}= \! 1 \! + \! 
\sum_{p=1}^{\infty} \sum_{q=0}^{p-1} \dfrac{\widehat{a}_{pq}^{1}(z_{o})(\ln 
t)^{q}}{t^{p/2}} \! + \! \mathcal{O} \! \left( \me^{-4t \min_{\genfrac{}{}
{0pt}{2}{m \in \{1,2,\ldots,N\}}{n \in \{1,2,\ldots,m-1\}}}\{\sin (\phi_{n}) 
\vert \cos (\phi_{n})-\cos (\phi_{m}) \vert\}} \right), \\
\widehat{a}_{3}^{+} \! =& \, \sum_{p=1}^{\infty} \sum_{q=0}^{p-1} \dfrac{
\widehat{a}_{pq}^{3}(z_{o})(\ln t)^{q}}{t^{p/2}} \! + \! \mathcal{O} \! 
\left(\me^{-4t \min_{\genfrac{}{}{0pt}{2}{m \in \{1,2,\ldots,N\}}{n \in \{1,
2,\ldots,m-1\}}}\{\sin (\phi_{n}) \vert \cos (\phi_{n})-\cos (\phi_{m}) 
\vert\}} \right), \\
\widehat{a}_{4}^{+} \! =& \, 1 \! + \! \sum_{p=1}^{\infty} \sum_{q=0}^{p-1} 
\dfrac{\widehat{a}_{pq}^{4}(z_{o})(\ln t)^{q}}{t^{p/2}} \! + \! \mathcal{O} 
\! \left( \me^{-4t \min_{\genfrac{}{}{0pt}{2}{m \in \{1,2,\ldots,N\}}{n \in 
\{1,2,\ldots,m-1\}}}\{\sin (\phi_{n}) \vert \cos (\phi_{n})-\cos (\phi_{m}) 
\vert\}} \right),
\end{align*}
$\widehat{a}_{pq}^{k}(z_{o}) \! \in \! c^{\mathcal{S}}(z_{o})$, $k \! \in \! 
\{1,3,4\}$, and $\widehat{\mathscr{P}}(\zeta) \! = \! \sigma_{1} \overline{
\widehat{\mathscr{P}}(\overline{\zeta})} \, \sigma_{1}$.
\end{ccccc}
\begin{eeeee}
Even though Lemma~4.3 is not proven in this paper, it will be shown that (see 
the proof of Proposition~4.6 below), up to the leading-order terms retained 
in this work, namely, terms that are $\mathcal{O}(\tfrac{c^{\mathcal{S}}
(z_{o})}{(z_{o}^{2}+32)^{1/2}} \tfrac{\ln t}{t})$, $\widehat{a}^{3}_{10}
(z_{o}) \! = \! 0$; thus, actually, $\widehat{a}_{3}^{+} \! = \! \sum_{p=2}^{
\infty} \! \sum_{q=0}^{p-1} \! \tfrac{\widehat{a}_{pq}^{3}(z_{o})(\ln t)^{q}}
{t^{p/2}} \! + \! \mathcal{O}(\exp (-4t \min_{\genfrac{}{}{0pt}{2}{m \in \{1,
2,\ldots,N\}}{n \in \{1,2,\ldots,m-1\}}} \linebreak[4]
\{\sin (\phi_{n}) \vert \cos (\phi_{n}) \! - \! \cos (\phi_{m}) \vert\})) \! 
= \! \mathcal{O}(\tfrac{c^{\mathcal{S}}(z_{o})}{(z_{o}^{2}+32)^{1/2}} \frac{
\ln t}{t})$. Furthermore, to $\mathcal{O}(\tfrac{c^{\mathcal{S}}(z_{o})}{(z_{
o}^{2}+32)^{1/2}} \frac{\ln t}{t})$, the asymptotic expansion for $\widehat{
a}_{4}^{+}$ plays, in fact, no role in the final formulae of this paper. 
As a possible prelude to a motivation of why $\widehat{a}_{i}^{+}$, $i 
\! \in \! \{1,2,3,4\}$, have, modulo exponentially small terms, the asymptotic 
expansions stated in Lemma~4.3, one can apply the higher-order generalisation 
of the DZ method \cite{a54} to the proof of Lemma~6.1 in \cite{a38} to show 
that $\chi^{c}_{ij}(\zeta)$, $i,j \! \in \! \{1,2\}$, have the asymptotic 
expansion $\chi^{c}_{ij}(\zeta) \! = \! \delta_{ij} \! + \! \sum_{p=1}^{
\infty} \sum_{q=0}^{p-1} \tfrac{(\chi^{c}_{ij}(z_{o}))_{pq}(f_{ij}(\zeta))_{
pq}(\ln t)^{q}}{t^{p/2}}$, where $\delta_{ij}$ is the Kronecker delta, 
$(\chi^{c}_{11}(\cdot))_{10} \! = \! (\chi^{c}_{22}(\cdot))_{10} \! = \! 
0$, and $\vert \vert (f_{ij}(\cdot))_{pq} \vert \vert_{\mathcal{L}^{\infty}
(\mathbb{C} \, \setminus \cup_{z \in \{\lambda_{1},\lambda_{2}\}} \mathbb{U}
(z;\varepsilon))} \! < \! \infty$; however, as stated heretofore, these 
details are omitted in this paper (it is the author's conjecture that 
$\widehat{a}_{3}^{+} \! = \! \widehat{a}_{4}^{+} \! = \! 0$, namely, 
$\widehat{\mathscr{P}}(\zeta)$ is diagonal).
\end{eeeee}
\begin{bbbbb}
Set $\widehat{a}_{10}^{1}(z_{o}) \! =: \! \widehat{a}_{1}$, $\widehat{a}^{
2}_{10}(z_{o}) \! =: \! \widehat{a}_{2}$, $\widehat{a}^{3}_{10}(z_{o}) \! =: 
\! \widehat{a}_{3}$, and $\widehat{a}^{4}_{10}(z_{o}) \! =: \! \widehat{a}_{
4}$. Then as $t \! \to \! +\infty$ and $x \! \to \! -\infty$ such that $z_{o} 
\! := \! x/t \! < \! -2$ and $(x,t) \! \in \! \daleth_{m}$, $m \! \in \! \{1,
2,\ldots,N\}$,
\begin{align*}
(\widehat{\Delta}_{o})_{11} \! =& \, -\dfrac{\overline{c_{m}^{0}}}{\overline{
\varsigma_{m}}} \mi \delta^{-1}(0) \me^{2 \mi \sum_{k=m+1}^{N} \phi_{k}} \! + 
\! \dfrac{\mi \delta^{-1}(0) \me^{2 \mi \sum_{k=m+1}^{N} \phi_{k}}}{\sqrt{t}} 
\! \left(-(\widehat{a}_{1} \! - \! \widehat{a}_{2}) \dfrac{\overline{c_{m}^{
0}}}{\overline{\varsigma_{m}}} \! - \! \left(\dfrac{b_{m}^{1}}{\varsigma_{m}} 
\! + \! \dfrac{\overline{c_{m}^{1}}}{\overline{\varsigma_{m}}} \right) \right. 
\\
+& \left. \, \widehat{a}_{3} \! \left(1 \! - \! \dfrac{\overline{a_{m}^{0}}}
{\overline{\varsigma_{m}}} \right) \! - \! \left(1 \! - \! \dfrac{a_{m}^{0}}
{\varsigma_{m}} \right) \! \dfrac{2 \delta (0) \sqrt{\nu (\lambda_{1})} \, 
\cos (\Theta^{+}(z_{o},t) \! + \! \frac{\pi}{4})}{\sqrt{(\lambda_{1} \! - \! 
\lambda_{2})} \, (z_{o}^{2} \! + \! 32)^{1/4}} \right) \! + \! \mathcal{O} \! 
\left(\dfrac{c^{\mathcal{S}}(z_{o})}{(z_{o}^{2} \! + \! 32)^{1/2}} \dfrac{\ln 
t}{t} \right), \\
(\widehat{\Delta}_{o})_{12} \! =& \, -\left(1 \! - \! \dfrac{a_{m}^{0}}{
\varsigma_{m}} \right) \! \mi \delta (0) \me^{-2 \mi \sum_{k=m+1}^{N} \phi_{
k}} \! + \! \dfrac{\mi \delta (0) \me^{-2 \mi \sum_{k=m+1}^{N} \phi_{k}}}{
\sqrt{t}} \! \left(-(\widehat{a}_{1} \! - \! \widehat{a}_{2}) \! \left(1 \! 
- \! \dfrac{a_{m}^{0}}{\varsigma_{m}} \right) \! + \! \left(\dfrac{a_{m}^{1}}
{\varsigma_{m}} \! + \! \dfrac{\overline{d_{m}^{1}}}{\overline{\varsigma_{
m}}} \right) \right. \\
+& \left. \, \widehat{a}_{3} \dfrac{c_{m}^{0}}{\varsigma_{m}} \! - \! \dfrac{
\overline{c_{m}^{0}}}{\overline{\varsigma_{m}}} \dfrac{2 \delta^{-1}(0) \sqrt{
\nu (\lambda_{1})} \, \cos (\Theta^{+}(z_{o},t) \! + \! \frac{\pi}{4})}{\sqrt{
(\lambda_{1} \! - \! \lambda_{2})} \, (z_{o}^{2} \! + \! 32)^{1/4}} \right) 
\! + \! \mathcal{O} \! \left(\dfrac{c^{\mathcal{S}}(z_{o})}{(z_{o}^{2} \! + 
\! 32)^{1/2}} \dfrac{\ln t}{t} \right), \\
(\widehat{\Delta}_{o})_{21} \! =& \, \left(1 \! - \! \dfrac{\overline{a_{m}^{
0}}}{\overline{\varsigma_{m}}} \right) \! \mi \delta^{-1}(0) \me^{2 \mi \sum_{
k=m+1}^{N} \phi_{k}} \! + \! \dfrac{\mi \delta^{-1}(0) \me^{2 \mi \sum_{k=m+
1}^{N} \phi_{k}}}{\sqrt{t}} \! \left((\overline{\widehat{a}_{1}} \! - \! 
\overline{\widehat{a}_{2}}) \! \left(1 \! - \! \dfrac{\overline{a_{m}^{0}}}{
\overline{\varsigma_{m}}} \right) \! - \! \left(\dfrac{\overline{a_{m}^{1}}}
{\overline{\varsigma_{m}}} \! + \! \dfrac{d_{m}^{1}}{\varsigma_{m}} \right) 
\right. \\
-& \left. \, \overline{\widehat{a}_{3}} \, \dfrac{\overline{c_{m}^{0}}}{
\overline{\varsigma_{m}}} \! + \! \dfrac{c_{m}^{0}}{\varsigma_{m}} \dfrac{2 
\delta (0) \sqrt{\nu (\lambda_{1})} \, \cos (\Theta^{+}(z_{o},t) \! + \! 
\frac{\pi}{4})}{\sqrt{(\lambda_{1} \! - \! \lambda_{2})} \, (z_{o}^{2} \! + 
\! 32)^{1/4}} \right) \! + \! \mathcal{O} \! \left(\dfrac{c^{\mathcal{S}}
(z_{o})}{(z_{o}^{2} \! + \! 32)^{1/2}} \dfrac{\ln t}{t} \right), \\
(\widehat{\Delta}_{o})_{22} \! =& \, \dfrac{c_{m}^{0}}{\varsigma_{m}} \mi 
\delta (0) \me^{-2 \mi \sum_{k=m+1}^{N} \phi_{k}} \! + \! \dfrac{\mi \delta 
(0) \me^{-2 \mi \sum_{k=m+1}^{N} \phi_{k}}}{\sqrt{t}} \! \left((\overline{
\widehat{a}_{1}} \! - \! \overline{\widehat{a}_{2}}) \dfrac{c_{m}^{0}}{
\varsigma_{m}} \! + \! \left(\dfrac{\overline{b_{m}^{1}}}{\overline{
\varsigma_{m}}} \! + \! \dfrac{c_{m}^{1}}{\varsigma_{m}} \right) \right. \\
-& \left. \, \overline{\widehat{a}_{3}} \! \left(1 \! - \! \dfrac{a_{m}^{0}}
{\varsigma_{m}} \right) \! + \! \left(1 \! - \! \dfrac{\overline{a_{m}^{0}}}
{\overline{\varsigma_{m}}} \right) \! \dfrac{2 \delta^{-1}(0) \sqrt{\nu 
(\lambda_{1})} \, \cos (\Theta^{+}(z_{o},t) \! + \! \frac{\pi}{4})}{\sqrt{
(\lambda_{1} \! - \! \lambda_{2})} \, (z_{o}^{2} \! + \! 32)^{1/4}} \right) 
\! + \! \mathcal{O} \! \left(\dfrac{c^{\mathcal{S}}(z_{o})}{(z_{o}^{2} \! + 
\! 32)^{1/2}} \dfrac{\ln t}{t} \right).
\end{align*}
\end{bbbbb}

\emph{Proof.} Recall {}from Lemma~4.1 that $\widehat{\Delta}_{o} \! = \! 
\widehat{\mathscr{P}}(0) \widehat{m}_{d}(0) \chi^{c}(0)(\delta (0))^{\sigma_{
3}} \! \left(\prod_{k=m+1}^{N}(d_{k}^{+}(0))^{\sigma_{3}} \right) \! \sigma_{
2}$. Collect, now, the following facts: (1) {}from Lemma~4.3, $\widehat{
\mathscr{P}}(0) \! = \! 
\left(
\begin{smallmatrix}
\widehat{a}_{1}^{+}/\widehat{a}_{2}^{+} & \, \, \, \widehat{a}_{3}^{+}/
\widehat{a}_{4}^{+} \\
\overline{\widehat{a}_{3}^{+}}/\overline{\widehat{a}_{4}^{+}} & \, \, \, 
\overline{\widehat{a}_{1}^{+}}/\overline{\widehat{a}_{2}^{+}}
\end{smallmatrix}
\right)$; (2) {}from the expression for $\widehat{m}_{d}(\zeta)$ given in 
Lemma~4.1, the definition (cf.~Proposition~4.1) $\mathrm{Res}(\widehat{\chi}
(\zeta);\varsigma_{n}) \! = \! 
\left(
\begin{smallmatrix}
a_{n} & b_{n} \\
c_{n} & d_{n}
\end{smallmatrix}
\right)$, $n \! \in \! \{1,2,\ldots,m\}$, and the asymptotics for $\{a_{n},
b_{n},c_{n},d_{n}\}_{n=1}^{m}$ given in Proposition~4.2, one shows that
\begin{equation*}
\widehat{m}_{d}(0) \! = \! 
\begin{pmatrix}
1-\frac{a_{m}}{\varsigma_{m}}-\frac{\overline{d_{m}}}{\overline{\varsigma_{
m}}} & \, \, -\frac{b_{m}}{\varsigma_{m}}-\frac{\overline{c_{m}}}{\overline{
\varsigma_{m}}} \\
-\frac{\overline{b_{m}}}{\overline{\varsigma_{m}}}-\frac{c_{m}}{\varsigma_{
m}} & \, \, 1-\frac{\overline{a_{m}}}{\overline{\varsigma_{m}}}-\frac{d_{m}}
{\varsigma_{m}}
\end{pmatrix} \! + \! \mathcal{O}_{2 \times 2} \! \left(\me^{-4t \min_{
\genfrac{}{}{0pt}{2}{m \in \{1,2,\ldots,N\}}{n \in \{1,2,\ldots,m-1\}}}\{
\sin (\phi_{n}) \vert \cos (\phi_{n})-\cos (\phi_{m}) \vert\}} \right),
\end{equation*}
where $\mathcal{O}_{2 \times 2}(\blacklozenge)$ denotes a $2 \times 2$ matrix 
each of whose entries are $\mathcal{O}(\blacklozenge)$; (3) {}from Lemma~4.2 
and the formula for $\widetilde{\chi}^{c}_{12}(\zeta)$ $(= \! \overline{
\widetilde{\chi}^{c}_{21}(\overline{\zeta})})$ given in Proposition~4.2, 
one shows that $\chi^{c}(0) \! = \! 
\left(
\begin{smallmatrix}
1 & \, \, \frac{1}{\sqrt{t}} \widetilde{\chi}^{c}_{12}(0) \\
\frac{1}{\sqrt{t}} \widetilde{\chi}^{c}_{21}(0) & \, \, 1
\end{smallmatrix}
\right) \! + \! \mathcal{O}_{2 \times 2}(\tfrac{c^{\mathcal{S}}(z_{o})}{(z_{
o}^{2}+32)^{1/2}} \tfrac{\ln t}{t})$; and (4) $(\delta (0))^{\sigma_{3}} \! 
\left(\prod_{k=m+1}^{N}(d_{k}^{+}(0))^{\sigma_{3}} \right) \! \sigma_{2}=
\mi \delta^{-1}(0) \! \left(\prod_{k=m+1}^{N}(d_{k}^{+}(0))^{-1} \right) \! 
\sigma_{-}-\mi \delta (0) \linebreak[4]
\cdot \left(\prod_{k=m+1}^{N}d_{k}^{+}(0) \right) \! \sigma_{+}$. Using the 
results of (1)--(4), and recalling the expression for $\widehat{\Delta}_{o}$ 
given above, one arrives at
\begin{align*}
(\widehat{\Delta}_{o})_{11} \! =& \, \tfrac{\widehat{a}_{1}^{+}}{\widehat{
a}_{2}^{+}} \mi \delta^{-1}(0) \! \left(\tfrac{\widetilde{\chi}^{c}_{12}(0)}
{\sqrt{t}}(1 \! - \! \tfrac{a_{m}}{\varsigma_{m}}-\tfrac{\overline{d_{m}}}
{\overline{\varsigma_{m}}}) \! - \! (\tfrac{b_{m}}{\varsigma_{m}} \! + \! 
\tfrac{\overline{c_{m}}}{\overline{\varsigma_{m}}}) \right) \! \prod_{k=m+
1}^{N}(d_{k}^{+}(0))^{-1} \\
+& \, \tfrac{\widehat{a}_{3}^{+}}{\widehat{a}_{4}^{+}} \mi \delta^{-1}(0) \! 
\left((1 \! - \! \tfrac{\overline{a_{m}}}{\overline{\varsigma_{m}}} \! - \! 
\tfrac{d_{m}}{\varsigma_{m}}) \! - \! \tfrac{\widetilde{\chi}_{12}^{c}(0)}
{\sqrt{t}}(\tfrac{\overline{b_{m}}}{\overline{\varsigma_{m}}} \! + \! \tfrac{
c_{m}}{\varsigma_{m}}) \right) \! \prod_{k=m+1}^{N}(d_{k}^{+}(0))^{-1}, \\
(\widehat{\Delta}_{o})_{12} \! =& \, \tfrac{\widehat{a}_{1}^{+}}{\widehat{
a}_{2}^{+}} \mi \delta (0) \! \left(-(1 \! - \! \tfrac{a_{m}}{\varsigma_{m}}
-\tfrac{\overline{d_{m}}}{\overline{\varsigma_{m}}}) \! + \! \tfrac{
\widetilde{\chi}^{c}_{21}(0)}{\sqrt{t}}(\tfrac{b_{m}}{\varsigma_{m}} \! + \! 
\tfrac{\overline{c_{m}}}{\overline{\varsigma_{m}}}) \right) \! \prod_{k=m
+1}^{N}d_{k}^{+}(0) \\
+& \, \tfrac{\widehat{a}_{3}^{+}}{\widehat{a}_{4}^{+}} \mi \delta (0) \! 
\left(-\tfrac{\widetilde{\chi}^{c}_{21}(0)}{\sqrt{t}}(1 \! - \! \tfrac{
\overline{a_{m}}}{\overline{\varsigma_{m}}} \! - \! \tfrac{d_{m}}{\varsigma_{
m}}) \! + \! (\tfrac{\overline{b_{m}}}{\overline{\varsigma_{m}}} \! + \! 
\tfrac{c_{m}}{\varsigma_{m}}) \right) \! \prod_{k=m+1}^{N}d_{k}^{+}(0),
\end{align*}
\begin{align*}
(\widehat{\Delta}_{o})_{21} \! =& \, \tfrac{\overline{\widehat{a}_{1}^{+}}}{
\overline{\widehat{a}_{2}^{+}}} \mi \delta^{-1}(0) \! \left((1 \! - \! 
\tfrac{\overline{a_{m}}}{\overline{\varsigma_{m}}}-\tfrac{d_{m}}{\varsigma_{
m}}) \! - \! \tfrac{\widetilde{\chi}^{c}_{12}(0)}{\sqrt{t}}(\tfrac{\overline{
b_{m}}}{\overline{\varsigma_{m}}} \! + \! \tfrac{c_{m}}{\varsigma_{m}}) 
\right) \! \prod_{k=m+1}^{N}(d_{k}^{+}(0))^{-1} \\
+& \, \tfrac{\overline{\widehat{a}_{3}^{+}}}{\overline{\widehat{a}_{4}^{+}}} 
\mi \delta^{-1} (0) \! \left(\tfrac{\widetilde{\chi}^{c}_{12}(0)}{\sqrt{t}}
(1 \! - \! \tfrac{a_{m}}{\varsigma_{m}} \! - \! \tfrac{\overline{d_{m}}}{
\overline{\varsigma_{m}}}) \! - \! (\tfrac{b_{m}}{\varsigma_{m}} \! + \! 
\tfrac{\overline{c_{m}}}{\overline{\varsigma_{m}}}) \right) \! \prod_{k=m
+1}^{N}(d_{k}^{+}(0))^{-1}, \\
(\widehat{\Delta}_{o})_{22} \! =& \, \tfrac{\overline{\widehat{a}_{1}^{+}}}{
\overline{\widehat{a}_{2}^{+}}} \mi \delta (0) \! \left(-\tfrac{\widetilde{
\chi}^{c}_{21}(0)}{\sqrt{t}}(1 \! - \! \tfrac{\overline{a_{m}}}{\overline{
\varsigma_{m}}}-\tfrac{d_{m}}{\varsigma_{m}}) \! + \! (\tfrac{\overline{b_{
m}}}{\overline{\varsigma_{m}}} \! + \! \tfrac{c_{m}}{\varsigma_{m}}) \right) 
\! \prod_{k=m+1}^{N}d_{k}^{+}(0) \\
+& \, \tfrac{\overline{\widehat{a}_{3}^{+}}}{\overline{\widehat{a}_{4}^{+}}} 
\mi \delta (0) \! \left(-(1 \! - \! \tfrac{a_{m}}{\varsigma_{m}} \! - \! 
\tfrac{\overline{d_{m}}}{\overline{\varsigma_{m}}}) \! + \! \tfrac{\widetilde{
\chi}_{21}^{c}(0)}{\sqrt{t}}(\tfrac{b_{m}}{\varsigma_{m}} \! + \! \tfrac{
\overline{c_{m}}}{\overline{\varsigma_{m}}}) \right) \! \prod_{k=m+1}^{N}
d_{k}^{+}(0).
\end{align*}
Using the asymptotic expansions for $\{a_{m},b_{m},c_{m},d_{m}\}$ 
(respectively, $\{\widehat{a}_{i}^{+}\}_{i=1}^{4})$ given in Proposition~4.2 
(respectively, Lemma~4.3), one arrives at the leading-order results stated 
in the Proposition. \hfill $\square$
\begin{eeeee}
In Propositions~4.4 and~4.6 below, one should keep, everywhere, the upper 
(respectively, lower) signs for $\theta_{\gamma_{m}} \! = \! +\pi/2$ 
(respectively, $\theta_{\gamma_{m}} \! = \! -\pi/2)$.
\end{eeeee}
\begin{bbbbb}
Let $\phi (x,t)$, $P(\phi_{m},\phi_{k})$, and $Q(\phi_{m})$ be defined by 
Eqs.~{\rm (67)}, {\rm (68)}, and~{\rm (69)}, respectively. Then, for 
$\theta_{\gamma_{m}} \! = \! \pm \pi/2$, as $t \! \to \! +\infty$ and $x 
\! \to \! -\infty$ such that $z_{o} \! := \! x/t \! < \! -2$ and $(x,t) \! 
\in \! \daleth_{m}$, $m \! \in \! \{1,2,\ldots,N\}$,
\begin{align*}
a_{m}^{0} \! =& -\dfrac{2 \mi \sin (\phi_{m}) \vert \gamma_{m} \vert^{2}P^{2}
(\phi_{m},\phi_{k})Q^{2}(\phi_{m}) \me^{2 \phi (x,t)}}{(1 \! - \! \vert 
\gamma_{m} \vert^{2}P^{2}(\phi_{m},\phi_{k})Q^{2}(\phi_{m}) \me^{2 \phi 
(x,t)})}, \\
c_{m}^{0} \! =& \mp \dfrac{2 \sin (\phi_{m}) \vert \gamma_{m} \vert \delta^{
-1}(0) \me^{\mi (\phi_{m}+s^{+})+\phi (x,t)}P(\phi_{m},\phi_{k})Q(\phi_{m})
}{(1 \! - \! \vert \gamma_{m} \vert^{2}P^{2}(\phi_{m},\phi_{k})Q^{2}(\phi_{
m}) \me^{2 \phi (x,t)})}, \\
a_{m}^{1} \! =& \mp \dfrac{16 \mi \lambda_{1}^{2} \sin^{2}(\phi_{m}) \vert 
\gamma_{m} \vert^{3} \sqrt{\nu (\lambda_{1})} \, P^{3}(\phi_{m},\phi_{k})Q^{
3}(\phi_{m}) \cos (s^{+}) \me^{3 \phi (x,t)}}{(1 \! - \! \vert \gamma_{m} 
\vert^{2}P^{2}(\phi_{m},\phi_{k})Q^{2}(\phi_{m}) \me^{2 \phi (x,t)})^{2}
(\lambda_{1}^{2} \! - \! 2 \lambda_{1} \cos (\phi_{m}) \! + \! 1)^{2} \sqrt{
(\lambda_{1} \! - \! \lambda_{2})} \, (z_{o}^{2} \! + \! 32)^{1/4}} \\
\times& \left(((\lambda_{1} \! + \! \lambda_{2}) \cos (\phi_{m}) \! - \! 2) 
\cos (\Theta^{+}(z_{o},t) \! + \! \tfrac{\pi}{4}) \! + \! (\lambda_{1} \! - 
\! \lambda_{2}) \sin (\phi_{m}) \sin (\Theta^{+}(z_{o},t) \! + \! \tfrac{\pi}
{4}) \right) \\
\mp& \, \dfrac{2 \lambda_{1} \sin (\phi_{m}) \vert \gamma_{m} \vert \sqrt{\nu 
(\lambda_{1})} \, P(\phi_{m},\phi_{k})Q(\phi_{m}) \me^{\phi (x,t)}}{(1 \! - 
\! \vert \gamma_{m} \vert^{2}P^{2}(\phi_{m},\phi_{k})Q^{2}(\phi_{m}) \me^{2 
\phi (x,t)})(\lambda_{1}^{2} \! - \! 2 \lambda_{1} \cos (\phi_{m}) \! + \! 1) 
\sqrt{(\lambda_{1} \! - \! \lambda_{2})} \, (z_{o}^{2} \! + \! 32)^{1/4}} \\
\times& \left(2 \cos (s^{+}) \cos (\Theta^{+}(z_{o},t) \! + \! \tfrac{\pi}
{4}) \! - \! (\lambda_{1} \! + \! \lambda_{2}) \cos (\phi_{m} \! + \! s^{+}) 
\cos (\Theta^{+}(z_{o},t) \! + \! \tfrac{\pi}{4}) \right. \\
-& \left. (\lambda_{1} \! - \! \lambda_{2}) \sin (\phi_{m} \! + \! s^{+}) 
\sin (\Theta^{+}(z_{o},t) \! + \! \tfrac{\pi}{4}) \! + \! 2 \mi \sin (s^{+}) 
\cos (\Theta^{+}(z_{o},t) \! + \! \tfrac{\pi}{4}) \right. \\
-& \left. \mi (\lambda_{1} \! + \! \lambda_{2}) \sin (\phi_{m} \! + \! s^{+}) 
\cos (\Theta^{+}(z_{o},t) \! + \! \tfrac{\pi}{4}) \! + \! \mi (\lambda_{1} \! 
- \! \lambda_{2}) \cos (\phi_{m} \! + \! s^{+}) \sin (\Theta^{+}(z_{o},t) \! 
+ \! \tfrac{\pi}{4}) \right), \\
b_{m}^{1} \! =& \, \dfrac{2 \mi \lambda_{1} \sin (\phi_{m}) \vert \gamma_{m} 
\vert^{2} \sqrt{\nu (\lambda_{1})} \, \delta (0) \me^{-\mi (\phi_{m}+s^{+})
+2 \phi (x,t)}P^{2}(\phi_{m},\phi_{k})Q^{2}(\phi_{m})}{(1 \! - \! \vert 
\gamma_{m} \vert^{2}P^{2}(\phi_{m},\phi_{k})Q^{2}(\phi_{m}) \me^{2 \phi (x,
t)})(\lambda_{1}^{2} \! - \! 2 \lambda_{1} \cos (\phi_{m}) \! + \! 1) \sqrt{
(\lambda_{1} \! - \! \lambda_{2})} \, (z_{o}^{2} \! + \! 32)^{1/4}} \\
\times& \left(2 \cos (s^{+}) \cos (\Theta^{+}(z_{o},t) \! + \! \tfrac{\pi}
{4}) \! - \! (\lambda_{1} \! + \! \lambda_{2}) \cos (\phi_{m} \! + \! s^{+}) 
\cos (\Theta^{+}(z_{o},t) \! + \! \tfrac{\pi}{4}) \right. \\
-& \left. (\lambda_{1} \! - \! \lambda_{2}) \sin (\phi_{m} \! + \! s^{+}) \sin 
(\Theta^{+}(z_{o},t) \! + \! \tfrac{\pi}{4}) \! + \! 2 \mi \sin (s^{+}) \cos 
(\Theta^{+}(z_{o},t) \! + \! \tfrac{\pi}{4}) \right. \\
-& \left. \mi (\lambda_{1} \! + \! \lambda_{2}) \sin (\phi_{m} \! + \! s^{+}) 
\cos (\Theta^{+}(z_{o},t) \! + \! \tfrac{\pi}{4}) \! + \! \mi (\lambda_{1} 
\! - \! \lambda_{2}) \cos (\phi_{m} \! + \! s^{+}) \sin (\Theta^{+}(z_{o},t) 
\! + \! \tfrac{\pi}{4}) \right), \\
c_{m}^{1} \! =& -\dfrac{16 \lambda_{1}^{2} \sin^{2}(\phi_{m}) \vert \gamma_{
m} \vert^{2} \sqrt{\nu (\lambda_{1})} \, \delta^{-1}(0) \me^{\mi (\phi_{m}+
s^{+})+2 \phi (x,t)}P^{2}(\phi_{m},\phi_{k})Q^{2}(\phi_{m}) \cos (s^{+})}{(1 
\! - \! \vert \gamma_{m} \vert^{2}P^{2}(\phi_{m},\phi_{k})Q^{2}(\phi_{m}) 
\me^{2 \phi (x,t)})^{2}(\lambda_{1}^{2} \! - \! 2 \lambda_{1} \cos (\phi_{m}) 
\! + \! 1)^{2} \sqrt{(\lambda_{1} \! - \! \lambda_{2})} \, (z_{o}^{2} \! + \! 
32)^{1/4}} \\
\times& \left(((\lambda_{1} \! + \! \lambda_{2}) \cos (\phi_{m}) \! - \! 2) 
\cos (\Theta^{+}(z_{o},t) \! + \! \tfrac{\pi}{4}) \! + \! (\lambda_{1} \! - 
\! \lambda_{2}) \sin (\phi_{m}) \sin (\Theta^{+}(z_{o},t) \! + \! \tfrac{\pi}
{4}) \right) \\
-& \, \dfrac{2 \mi \lambda_{1} \sin (\phi_{m}) \vert \gamma_{m} \vert^{2} 
\sqrt{\nu (\lambda_{1})} \, \delta^{-1}(0) \me^{\mi (\phi_{m}+s^{+})+2 \phi 
(x,t)}P^{2}(\phi_{m},\phi_{k})Q^{2}(\phi_{m})}{(1 \! - \! \vert \gamma_{m} 
\vert^{2}P^{2}(\phi_{m},\phi_{k})Q^{2}(\phi_{m}) \me^{2 \phi (x,t)})
(\lambda_{1}^{2} \! - \! 2 \lambda_{1} \cos (\phi_{m}) \! + \! 1) \sqrt{
(\lambda_{1} \! - \! \lambda_{2})} \, (z_{o}^{2} \! + \! 32)^{1/4}} \\
\times& \left(2 \cos (s^{+}) \cos (\Theta^{+}(z_{o},t) \! + \! \tfrac{\pi}{
4}) \! - \! (\lambda_{1} \! + \! \lambda_{2}) \cos (\phi_{m} \! + \! s^{+}) 
\cos (\Theta^{+}(z_{o},t) \! + \! \tfrac{\pi}{4}) \right. \\
-& \left. (\lambda_{1} \! - \! \lambda_{2}) \sin (\phi_{m} \! + \! s^{+}) \sin 
(\Theta^{+}(z_{o},t) \! + \! \tfrac{\pi}{4}) \! - \! 2 \mi \sin (s^{+}) \cos 
(\Theta^{+}(z_{o},t) \! + \! \tfrac{\pi}{4}) \right. \\
+& \left. \mi (\lambda_{1} \! + \! \lambda_{2}) \sin (\phi_{m} \! + \! s^{+}) 
\cos (\Theta^{+}(z_{o},t) \! + \! \tfrac{\pi}{4}) \! - \! \mi (\lambda_{1} \! 
- \! \lambda_{2}) \cos (\phi_{m} \! + \! s^{+}) \sin (\Theta^{+}(z_{o},t) \! 
+ \! \tfrac{\pi}{4}) \right) \\
+& \, \dfrac{8 \lambda_{1}^{2} \sin^{2}(\phi_{m}) \vert \gamma_{m} \vert^{2} 
\sqrt{\nu (\lambda_{1})} \, \delta^{-1}(0) \me^{\mi (\phi_{m}+s^{+})+2 \phi 
(x,t)}P^{2}(\phi_{m},\phi_{k})Q^{2}(\phi_{m})}{(1 \! - \! \vert \gamma_{m} 
\vert^{2}P^{2}(\phi_{m},\phi_{k})Q^{2}(\phi_{m}) \me^{2 \phi (x,t)})
(\lambda_{1}^{2} \! - \! 2 \lambda_{1} \cos (\phi_{m}) \! + \! 1)^{2} 
\sqrt{(\lambda_{1} \! - \! \lambda_{2})} \, (z_{o}^{2} \! + \! 32)^{1/4}}
\end{align*}
\begin{align*}
\times& \left(\left(((\lambda_{1} \! + \! \lambda_{2}) \cos (\phi_{m}) \! - 
\! 2) \cos (\Theta^{+}(z_{o},t) \! + \! \tfrac{\pi}{4}) \! + \! (\lambda_{1} 
\! - \! \lambda_{2}) \sin (\phi_{m}) \sin (\Theta^{+}(z_{o},t) \! + \! 
\tfrac{\pi}{4}) \right) \cos (s^{+}) \right. \\
-& \left. \mi \! \left(((\lambda_{1} \! + \! \lambda_{2}) \cos (\phi_{m}) \! 
- \! 2) \cos (\Theta^{+}(z_{o},t) \! + \! \tfrac{\pi}{4}) \! + \! (\lambda_{
1} \! - \! \lambda_{2}) \sin (\phi_{m}) \sin (\Theta^{+}(z_{o},t) \! + \! 
\tfrac{\pi}{4}) \right) \sin (s^{+}) \right), \\
d_{m}^{1} \! =& \pm \dfrac{2 \lambda_{1} \sin (\phi_{m}) \vert \gamma_{m} 
\vert \sqrt{\nu (\lambda_{1})} \, P(\phi_{m},\phi_{k})Q(\phi_{m}) \me^{\phi 
(x,t)}}{(1 \! - \! \vert \gamma_{m} \vert^{2}P^{2}(\phi_{m},\phi_{k})Q^{2}
(\phi_{m}) \me^{2 \phi (x,t)})(\lambda_{1}^{2} \! - \! 2 \lambda_{1} \cos 
(\phi_{m}) \! + \! 1) \sqrt{(\lambda_{1} \! - \! \lambda_{2})} \, (z_{o}^{2} 
\! + \! 32)^{1/4}} \\
\times& \left(2 \cos (s^{+}) \cos (\Theta^{+}(z_{o},t) \! + \! \tfrac{\pi}
{4}) \! - \! (\lambda_{1} \! + \! \lambda_{2}) \cos (\phi_{m} \! + \! s^{+}) 
\cos (\Theta^{+}(z_{o},t) \! + \! \tfrac{\pi}{4}) \right. \\
-& \left. (\lambda_{1} \! - \! \lambda_{2}) \sin (\phi_{m} \! + \! s^{+}) \sin 
(\Theta^{+}(z_{o},t) \! + \! \tfrac{\pi}{4}) \! + \! 2 \mi \sin (s^{+}) \cos 
(\Theta^{+}(z_{o},t) \! + \! \tfrac{\pi}{4}) \right. \\
-& \left. \mi (\lambda_{1} \! + \! \lambda_{2}) \sin (\phi_{m} \! + \! s^{+}) 
\cos (\Theta^{+}(z_{o},t) \! + \! \tfrac{\pi}{4}) \! + \! \mi (\lambda_{1} \! 
- \! \lambda_{2}) \cos (\phi_{m} \! + \! s^{+}) \sin (\Theta^{+}(z_{o},t) \! 
+ \! \tfrac{\pi}{4}) \right),
\end{align*}
where $s^{+}$ is given in Theorem~{\rm 2.2.1}, Eq.~{\rm (11)}.
\end{bbbbb}

\emph{Proof.} Recalling the definitions of $\{a_{m}^{0},a_{m}^{1},b_{m}^{1},
c_{m}^{0},c_{m}^{1},d_{m}^{1}\}$ given in Proposition~4.2, substituting 
into them the expressions for $g_{m}^{\ast}$ and $\widetilde{\chi}^{c}_{12}
(\zeta)$ given in Propositions~4.1 and~4.2, respectively, using standard 
trigonometric identities, and defining $\phi (x,t)$, $P(\phi_{m},\phi_{k})$, 
and $Q(\phi_{m})$ as in Eqs.~(67), (68), and~(69), respectively, one obtains, 
after tedious, but otherwise straightforward calculations, the result stated 
in the Proposition. \hfill $\square$
\begin{bbbbb}
As $t \! \to \! +\infty$ and $x \! \to \! -\infty$ such that $z_{o} \! := \! 
x/t \! < \! -2$ and $(x,t) \! \in \! \daleth_{m}$, $m \! \in \! \{1,2,\ldots,
N\}$,
\begin{align}
u(x,t) \! =& \, \mi \! \left((\widehat{\Delta}_{o})_{12} \! + \! \widehat{
a}_{3}^{+} \! + \! b_{m} \! + \! \overline{c_{m}} \! + \! \tfrac{\sqrt{\nu 
(\lambda_{1})} \, \me^{\frac{\mi \Xi^{+}(0)}{2}} \lambda_{1}^{2 \mi \nu 
(\lambda_{1})}}{\sqrt{t(\lambda_{1}-\lambda_{2})} \, (z_{o}^{2}+32)^{1/4}} 
\! \left(\lambda_{1} \me^{-\mi (\Theta^{+}(z_{o},t)+\frac{\pi}{4})} \! + \! 
\lambda_{2} \me^{\mi (\Theta^{+}(z_{o},t)+\frac{\pi}{4})} \right) \right) 
\nonumber \\
+& \, \mathcal{O} \! \left(\tfrac{c^{\mathcal{S}}(z_{o})}{(z_{o}^{2}+32)^{
1/2}} \tfrac{\ln t}{t} \right),
\end{align}
\begin{align}
\int_{+\infty}^{x}(\vert u(x^{\prime},t) \vert^{2} \! - \! 1) \, \md x^{
\prime} \! =& \, -\mi \! \left((\widehat{\Delta}_{o})_{11} \! + \! \widehat{
a}_{1}^{+} \! - \! \widehat{a}_{2}^{+} \! + \! a_{m} \! + \! \overline{d_{
m}} \! + \! 2 \mi \sum_{k=m+1}^{N} \sin (\phi_{k}) \right. \nonumber \\
+& \left. \mi \! \left(\int\nolimits_{-\infty}^{0} \! + \! \int\nolimits_{
\lambda_{2}}^{\lambda_{1}} \right) \! \ln (1 \! - \! \vert r(\mu) \vert^{2}) 
\, \tfrac{\md \mu}{2 \pi} \right) \! + \! \mathcal{O} \! \left(\tfrac{c^{
\mathcal{S}}(z_{o})}{(z_{o}^{2}+32)^{1/2}} \tfrac{\ln t}{t} \right),
\end{align}
\begin{align}
\int_{-\infty}^{x}(&\vert u(x^{\prime},t) \vert^{2} \! - \! 1) \, \md x^{
\prime} \! = \! \int_{+\infty}^{x}(\vert u(x^{\prime},t) \vert^{2} \! - 
\! 1) \, \md x^{\prime} \! - \! 2 \sum_{n=1}^{N} \sin (\phi_{n}) \! - \! 
\int\nolimits_{-\infty}^{+\infty} \ln (1 \! - \! \vert r(\mu) \vert^{2}) 
\, \tfrac{\md \mu}{2 \pi}.
\end{align}
\end{bbbbb}

\emph{Proof.} Recall Eqs.~(63), (64), and~(65) for $u(x,t)$, $\int_{+\infty}^{
x}(\vert u(x^{\prime},t) \vert^{2} \! - \! 1) \, \md x^{\prime}$, and 
$\widehat{\chi}(\zeta)$, respectively. Using the result for $\widehat{
\mathscr{P}}(\zeta)$ (respectively, $\chi^{c}_{ij}(\zeta)$, $i,j \! \in \! 
\{1,2\})$ stated in Lemma~4.3 (respectively, Lemma~4.2), noting that $(\delta 
(\zeta))^{\pm 1} \! =_{\zeta \to \infty} \! 1 \! \pm \! \mi \! \left( \! 
\left(\int_{-\infty}^{0} \! + \! \int_{\lambda_{2}}^{\lambda_{1}} \right) \! 
\ln (1 \! - \! \vert r(\mu) \vert^{2}) \tfrac{\md \mu}{2 \pi} \right) \! 
\zeta^{-1} \! + \! \mathcal{O}(\zeta^{-2})$ and $\prod_{k=m+1}^{N}(d_{k}^{+}
(\zeta))^{\pm 1} \! =_{\zeta \to \infty} \! 1 \! \pm \! \left(\sum_{k=m+1}^{
N}(\varsigma_{k} \! - \! \overline{\varsigma_{k}}) \right) \! \zeta^{-1} \! + 
\! \mathcal{O}(\zeta^{-2})$, and using the asymptotic estimates for $\{a_{n},
b_{n},c_{n},d_{n}\}_{n=1}^{m-1}$ given in Proposition~4.2, one forms the 
large-$\zeta$ asymptotics for $\widehat{\chi}(\zeta)$ given in Eq.~(65) to 
show that
\begin{align*}
(\zeta (\widehat{\chi}(\zeta)(\delta (\zeta))^{\sigma_{3}} \! &\prod_{k
=m+1}^{N}(d_{k}^{+}(\zeta))^{\sigma_{3}} \! - \! \mathrm{I}))_{11} \! 
\underset{\genfrac{}{}{0pt}{2}{\zeta \to \infty}{\zeta \in \mathbb{C} 
\setminus (\widehat{\sigma}_{d} \cup \sigma_{c})}}{=} \! (\widehat{\Delta}_{
o})_{11} \! + \! \widehat{a}_{1}^{+} \! - \! \widehat{a}_{2}^{+} \! + \! 
a_{m} \! + \! \overline{d_{m}} \! + \! \sum_{k=m+1}^{N}(\varsigma_{k} \! - 
\! \overline{\varsigma_{k}}) \\
+& \, \mi \! \left(\int\nolimits_{-\infty}^{0} \! + \! \int\nolimits_{\lambda_{
2}}^{\lambda_{1}} \right) \! \ln (1 \! - \! \vert r(\mu) \vert^{2}) \tfrac{
\md \mu}{2 \pi} \! + \! \mathcal{O} \! \left(\tfrac{c^{\mathcal{S}}(z_{o})}{
(z_{o}^{2}+32)^{1/2}} \tfrac{\ln t}{t} \right) \! + \! \mathcal{O} \! \left(
\me^{-\varrho t} \right), \\
(\zeta (\widehat{\chi}(\zeta)(\delta (\zeta))^{\sigma_{3}} \! &\prod_{k=
m+1}^{N}(d_{k}^{+}(\zeta))^{\sigma_{3}} \! - \! \mathrm{I}))_{12} \! 
\underset{\genfrac{}{}{0pt}{2}{\zeta \to \infty}{\zeta \in \mathbb{C} 
\setminus (\widehat{\sigma}_{d} \cup \sigma_{c})}}{=} \! (\widehat{\Delta}_{
o})_{12} \! + \! \widehat{a}_{3}^{+} \! + \! b_{m} \! + \! \overline{c_{m}} 
\! + \! \tfrac{\sqrt{\nu (\lambda_{1})} \, \me^{\frac{\mi \Xi^{+}(0)}{2}} 
\lambda_{1}^{2 \mi \nu (\lambda_{1})}}{\sqrt{t(\lambda_{1}-\lambda_{2})} \, 
(z_{o}^{2}+32)^{1/4}} \\
\times& \! \left(\lambda_{1} \me^{-\mi (\Theta^{+}(z_{o},t)+\frac{\pi}{4})} 
\! + \! \lambda_{2} \me^{\mi (\Theta^{+}(z_{o},t)+\frac{\pi}{4})} \right) \! 
+ \! \mathcal{O} \! \left(\tfrac{c^{\mathcal{S}}(z_{o})}{(z_{o}^{2}+32)^{
1/2}} 
\tfrac{\ln t}{t} \right) \! + \! \mathcal{O} \! \left(\me^{-\varrho t} 
\right),
\end{align*}
where $\varrho \! := \! 4 \min_{\genfrac{}{}{0pt}{2}{m \in \{1,2,\ldots,N\}}
{n \not= m \in \{1,2,\ldots,N\}}}\{\sin (\phi_{n}) \vert \cos (\phi_{n}) \! 
- \! \cos (\phi_{m}) \vert\}$ $(> \! 0)$. Neglecting exponentially small 
terms (cf.~Remark~4.1), {}from the expressions for $u(x,t)$ and $\int_{+
\infty}^{x}(\vert u(x^{\prime},t) \vert^{2} \! - \! 1) \, \md x^{\prime}$ 
given, respectively, in Eqs.~(63) and~(64), and the trace identity 
(cf.~Eq.~(4)) $\int_{-\infty}^{+\infty}(\vert u(x^{\prime},t) \vert^{2} \! 
- \! 1) \, \md x^{\prime} \! = \! (\int_{-\infty}^{x} \! + \! \int_{x}^{
+\infty})(\vert u(x^{\prime},t) \vert^{2} \! - \! 1) \, \md x^{\prime} \! = 
\! -2 \sum_{n=1}^{N} \sin (\phi_{n}) \! - \! \int_{-\infty}^{+\infty} \ln 
(1 \! - \! \vert r(\mu) \vert^{2}) \, \tfrac{\md \mu}{2 \pi}$, one obtains 
the results stated in the Proposition. \hfill $\square$
\begin{bbbbb}
As $t \! \to \! +\infty$ and $x \! \to \! -\infty$ such that $z_{o} \! := \! 
x/t \! < \! -2$ and $(x,t) \! \in \! \daleth_{m}$, $m \! \in \! \{1,2,\ldots,
N\}$, for $\theta_{\gamma_{m}} \! = \! \pm \pi/2$,
\begin{align*}
(\widehat{\Delta}_{o})_{11} =& \, \mi \! \left(\pm \dfrac{2 \sin (\phi_{m}) 
\vert \gamma_{m} \vert P(\phi_{m},\phi_{k})Q(\phi_{m}) \me^{\phi (x,t)}}{(1 
\! - \! \vert \gamma_{m} \vert^{2}P^{2}(\phi_{m},\phi_{k})Q^{2}(\phi_{m}) 
\me^{2 \phi (x,t)})} \! + \! \dfrac{\sqrt{\nu (\lambda_{1})}}{\sqrt{t
(\lambda_{1} \! - \! \lambda_{2})} \, (z_{o}^{2} \! + \! 32)^{1/4}} \! 
\left(-2 \cos (\Theta^{+}(z_{o},t) \right. \right. \\
+& \left. \left. \! \tfrac{\pi}{4}) \cos (s^{+}) \! + \! \dfrac{4 \sin (\phi_{
m}) \vert \gamma_{m} \vert^{2}P^{2}(\phi_{m},\phi_{k})Q^{2}(\phi_{m}) \sin 
(s^{+} \! - \! \phi_{m}) \cos (\Theta^{+}(z_{o},t) \! + \! \tfrac{\pi}{4}) 
\me^{2 \phi (x,t)}}{(1 \! - \! \vert \gamma_{m} \vert^{2}P^{2}(\phi_{m},
\phi_{k})Q^{2}(\phi_{m}) \me^{2 \phi (x,t)})} \right. \right. \\
+& \left. \left. \! \dfrac{8 \lambda_{1}^{2} \sin^{2}(\phi_{m}) \vert \gamma_{
m} \vert^{2}P^{2}(\phi_{m},\phi_{k})Q^{2}(\phi_{m})(1 \! + \! \vert \gamma_{
m} \vert^{2}P^{2}(\phi_{m},\phi_{k})Q^{2}(\phi_{m}) \me^{2 \phi (x,t)}) \me^{
2 \phi (x,t)}}{(1 \! - \! \vert \gamma_{m} \vert^{2}P^{2}(\phi_{m},\phi_{k})
Q^{2}(\phi_{m}) \me^{2 \phi (x,t)})^{2}(\lambda_{1}^{2} \! - \! 2 \lambda_{
1} \cos (\phi_{m}) \! + \! 1)^{2}} \right. \right. \\
\times& \left. \left. \! \left(((\lambda_{1} \! + \! \lambda_{2}) \cos 
(\phi_{m}) \! - \! 2) \cos (\Theta^{+}(z_{o},t) \! + \! \tfrac{\pi}{4}) \! + 
\! (\lambda_{1} \! - \! \lambda_{2}) \sin (\phi_{m}) \sin (\Theta^{+}(z_{o},
t) \! + \! \tfrac{\pi}{4}) \right) \cos (s^{+}) \right. \right. \\
+& \left. \left. \! \dfrac{4 \lambda_{1} \sin (\phi_{m}) \cos (\phi_{m}) 
\vert \gamma_{m} \vert^{2}P^{2}(\phi_{m},\phi_{k})Q^{2}(\phi_{m}) \me^{2 
\phi (x,t)}}{(1 \! - \! \vert \gamma_{m} \vert^{2}P^{2}(\phi_{m},\phi_{k})
Q^{2}(\phi_{m}) \me^{2 \phi (x,t)})(\lambda_{1}^{2} \! - \! 2 \lambda_{1} 
\cos (\phi_{m}) \! + \! 1)} \! \left(2 \sin (s^{+} \! - \! \phi_{m}) \cos 
(\Theta^{+}(z_{o},t) \right. \right. \right. \\
+& \left. \left. \left. \tfrac{\pi}{4}) \! - \! (\lambda_{1} \! + \! 
\lambda_{2}) \sin (s^{+}) \cos (\Theta^{+}(z_{o},t) \! + \! \tfrac{\pi}{4}) 
\! + \! (\lambda_{1} \! - \! \lambda_{2}) \cos (s^{+}) \sin (\Theta^{+}(z_{
o},t) \! + \! \tfrac{\pi}{4}) \right) \right) \right) \\
+& \, \mathcal{O} \! \left(\tfrac{c^{\mathcal{S}}(z_{o})}{(z_{o}^{2}+32)^{
1/2}} \tfrac{\ln t}{t} \right), \\
(\widehat{\Delta}_{o})_{12} =& -\mi \me^{-\mi (\theta^{+}(1)+s^{+})} \! + \! 
\dfrac{2 \sin (\phi_{m}) \vert \gamma_{m} \vert^{2}P^{2}(\phi_{m},\phi_{k})
Q^{2}(\phi_{m}) \me^{-\mi (\theta^{+}(1)+\phi_{m}+s^{+})+2 \phi (x,t)}}{(1 
\! - \! \vert \gamma_{m} \vert^{2}P^{2}(\phi_{m},\phi_{k})Q^{2}(\phi_{m}) 
\me^{2 \phi (x,t)})} \\
+& \, \dfrac{1}{\sqrt{t}} \! \left(\dfrac{2 \mi \Im (\widehat{a}_{1} \! - \! 
\widehat{a}_{2}) \sin (\phi_{m}) \vert \gamma_{m} \vert^{2}P^{2}(\phi_{m},
\phi_{k})Q^{2}(\phi_{m}) \me^{-\mi (\theta^{+}(1)+\phi_{m}+s^{+})+2 \phi 
(x,t)}}{(1 \! - \! \vert \gamma_{m} \vert^{2}P^{2}(\phi_{m},\phi_{k})Q^{2}
(\phi_{m}) \me^{2 \phi (x,t)})} \right. \\
\pm& \left. \dfrac{4 \mi \sin (\phi_{m}) \vert \gamma_{m} \vert P(\phi_{m},
\phi_{k})Q(\phi_{m}) \sqrt{\nu (\lambda_{1})} \, \me^{-\mi (\theta^{+}(1)+2
s^{+})+\phi (x,t)} \cos (\Theta^{+}(z_{o},t) \! + \! \tfrac{\pi}{4})}{(1 \! 
- \! \vert \gamma_{m} \vert^{2}P^{2}(\phi_{m},\phi_{k})Q^{2}(\phi_{m}) \me^{
2 \phi (x,t)}) \sqrt{(\lambda_{1} \! - \! \lambda_{2})} \, (z_{o}^{2} \! + 
\! 32)^{1/4}} \right. \\
\mp& \left. \dfrac{16 \mi \lambda_{1}^{2} \sin^{2}(\phi_{m}) \vert \gamma_{
m} \vert^{3}P^{3}(\phi_{m},\phi_{k})Q^{3}(\phi_{m}) \sqrt{\nu (\lambda_{1})} 
\, \me^{-\mi (\theta^{+}(1)+s^{+})+3\phi (x,t)} \cos (s^{+})}{(1 \! - \! 
\vert \gamma_{m} \vert^{2}P^{2}(\phi_{m},\phi_{k})Q^{2}(\phi_{m}) \me^{2 \phi 
(x,t)})^{2}(\lambda_{1}^{2} \! - \! 2 \lambda_{1} \cos (\phi_{m}) \! + \! 
1)^{2} \sqrt{(\lambda_{1} \! - \! \lambda_{2})} \, (z_{o}^{2} \! + \! 32)^{
1/4}} \right. \\
\times& \left. \left(((\lambda_{1} \! + \! \lambda_{2}) \cos (\phi_{m}) \! - 
\! 2) \sin (\phi_{m}) \cos (\Theta^{+}(z_{o},t) \! + \! \tfrac{\pi}{4}) \! + 
\! (\lambda_{1} \! - \! \lambda_{2}) \sin^{2}(\phi_{m}) \sin (\Theta^{+}(z_{
o},t) \! + \! \tfrac{\pi}{4}) \right. \right. \\
+& \left. \left. \mi (((\lambda_{1} \! + \! \lambda_{2}) \cos (\phi_{m}) \! 
- \! 2) \cos (\phi_{m}) \cos (\Theta^{+}(z_{o},t) \! + \! \tfrac{\pi}{4}) \! 
+ \! (\lambda_{1} \! - \! \lambda_{2}) \sin (\phi_{m}) \cos (\phi_{m}) \right. 
\right. \\
\times& \left. \left. \sin (\Theta^{+}(z_{o},t) \! + \! \tfrac{\pi}{4}) 
\right) \! + \! \Im (\widehat{a}_{1} \! - \! \widehat{a}_{2}) \me^{-\mi 
(\theta^{+}(1)+s^{+})} \right. \\
-& \left. \dfrac{4 \lambda_{1} \sin (\phi_{m}) \vert \gamma_{m} \vert P(\phi_{
m},\phi_{k})Q(\phi_{m}) \sqrt{\nu (\lambda_{1})} \, \me^{-\mi (\theta^{+}(1)
+s^{+})+\phi (x,t)}}{(1 \! - \! \vert \gamma_{m} \vert^{2}P^{2}(\phi_{m},
\phi_{k})Q^{2}(\phi_{m}) \me^{2 \phi (x,t)})(\lambda_{1}^{2} \! - \! 2 
\lambda_{1} \cos (\phi_{m}) \! + \! 1) \sqrt{(\lambda_{1} \! - \! \lambda_{
2})} \, (z_{o}^{2} \! + \! 32)^{1/4}} \right. \\
\times& \left. \left(\mp 2 \sin (s^{+} \! - \! \phi_{m}) \cos (\Theta^{+}
(z_{o},t) \! + \! \tfrac{\pi}{4}) \! \pm \! (\lambda_{1} \! + \! \lambda_{
2}) \sin (s^{+}) \cos (\Theta^{+}(z_{o},t) \! + \! \tfrac{\pi}{4}) \right. 
\right. \\
\mp& \left. \left. (\lambda_{1} \! - \! \lambda_{2}) \cos (s^{+}) \sin 
(\Theta^{+}(z_{o},t) \! + \! \tfrac{\pi}{4}) \right) \right) \! + \! 
\mathcal{O} \! \left(\tfrac{c^{\mathcal{S}}(z_{o})}{(z_{o}^{2}+32)^{1/2}} 
\tfrac{\ln t}{t} \right),
\end{align*}
\begin{align*}
\Im (\widehat{a}_{1} \! - \! \widehat{a}_{2}) =& \pm \dfrac{\sqrt{\nu 
(\lambda_{1})} \sin (s^{+}) \cos (\Theta^{+}(z_{o},t) \! + \! \tfrac{\pi}{4})
(1 \! - \! \vert \gamma_{m} \vert^{2}P^{2}(\phi_{m},\phi_{k})Q^{2}(\phi_{m}) 
\me^{2 \phi (x,t)})}{\sqrt{(\lambda_{1} \! - \! \lambda_{2})} \, (z_{o}^{2} 
\! + \! 32)^{1/4} \sin (\phi_{m}) \vert \gamma_{m} \vert P(\phi_{m},\phi_{k})
Q(\phi_{m}) \me^{\phi (x,t)}} \\
\pm& \, \dfrac{4 \lambda_{1}^{2} \sqrt{\nu (\lambda_{1})} \sin (\phi_{m}) 
\vert \gamma_{m} \vert P(\phi_{m},\phi_{k})Q(\phi_{m}) \sin (s^{+}) \me^{\phi 
(x,t)}}{(\lambda_{1}^{2} \! - \! 2 \lambda_{1} \cos (\phi_{m}) \! + \! 1)^{2} 
\sqrt{(\lambda_{1} \! - \! \lambda_{2})} \, (z_{o}^{2} \! + \! 32)^{1/4}} \\
\times& \left(((\lambda_{1} \! + \! \lambda_{2}) \cos (\phi_{m}) \! - \! 2) 
\cos (\Theta^{+}(z_{o},t) \! + \! \tfrac{\pi}{4}) \! + \! (\lambda_{1} \! - 
\! \lambda_{2}) \sin (\phi_{m}) \sin (\Theta^{+}(z_{o},t) \! + \! \tfrac{\pi}
{4}) \right) \\
\pm& \, \dfrac{2 \lambda_{1} \sqrt{\nu (\lambda_{1})} \cos (\phi_{m}) \vert 
\gamma_{m} \vert P(\phi_{m},\phi_{k})Q(\phi_{m}) \me^{\phi (x,t)}}{(\lambda_{
1}^{2} \! - \! 2 \lambda_{1} \cos (\phi_{m}) \! + \! 1) \sqrt{(\lambda_{1} \! 
- \! \lambda_{2})} \, (z_{o}^{2} \! + \! 32)^{1/4}} \! \left(2 \cos (s^{+} \! 
- \! \phi_{m}) \cos (\Theta^{+}(z_{o},t) \! + \! \tfrac{\pi}{4}) \right. \\
-& \left. (\lambda_{1} \! + \! \lambda_{2}) \cos (s^{+}) \cos (\Theta^{+}(z_{
o},t) \! + \! \tfrac{\pi}{4}) \! - \! (\lambda_{1} \! - \! \lambda_{2}) \sin 
(s^{+}) \sin (\Theta^{+}(z_{o},t) \! + \! \tfrac{\pi}{4}) \right) \\
\pm& \, \dfrac{2 \sqrt{\nu (\lambda_{1})} \vert \gamma_{m} \vert P(\phi_{m},
\phi_{k})Q(\phi_{m}) \cos (s^{+} \! - \! \phi_{m}) \cos (\Theta^{+}(z_{o},t) 
\! + \! \tfrac{\pi}{4}) \me^{\phi (x,t)}}{\sqrt{(\lambda_{1} \! - \! \lambda_{
2})} \, (z_{o}^{2} \! + \! 32)^{1/4}} \\
+& \, \mathcal{O} \! \left(\tfrac{c^{\mathcal{S}}(z_{o})}{(z_{o}^{2}+32)^{
1/2}} \tfrac{\ln t}{t} \right),
\end{align*}
\begin{equation*}
\Re (\widehat{a}_{1} \! - \! \widehat{a}_{2}) \! = \! \Re (\widehat{a}_{3}) 
\! = \! \Im (\widehat{a}_{3}) \! = \! 0,
\end{equation*}
where $\theta^{+}(\cdot)$ is given in Theorem~{\rm 2.2.1}, Eq.~{\rm (8)}.
\end{bbbbb}

\emph{Proof.} Recall {}from Lemma~4.1 that: (1) $\widehat{\Delta}_{o} \! = \! 
\widehat{\mathscr{P}}(0) \widehat{m}_{d}(0) \chi^{c}(0)(\delta (0))^{\sigma_{
3}} \! \left(\prod_{k=m+1}^{N}(d_{k}^{+}(0))^{\sigma_{3}} \right) \! \sigma_{
2}$; (2) $\mathrm{tr}(\widehat{\Delta}_{o}) \! = \! 0$; (3) $\det (\widehat{
\Delta}_{o}) \! = \! -1$; and (4) $\widehat{\Delta}_{o} \widehat{\Delta}_{
o} \! = \! \mathrm{I}$. Taking the determinant of both sides of the above 
expression for $\widehat{\Delta}_{o}$ and using the fact that $\det 
(\widehat{\Delta}_{o}) \! = \! -1$, it follows that, modulo terms that 
are $\mathcal{O}(\tfrac{c^{\mathcal{S}}(z_{o})}{(z_{o}^{2}+32)^{1/2}} \tfrac{
\ln t}{t})$, and always ignoring exponentially small terms, $\det (\widehat{
\mathscr{P}}(0)) \! = \! (\det (\widehat{m}_{d}(0)))^{-1}$. Before proceeding 
further, this will be verified; in particular, since $\widehat{m}_{d}(\zeta) 
\! \in \! \mathrm{SL}(2,\mathbb{C})$, it must be the case that, modulo terms 
that are $\mathcal{O}(\tfrac{c^{\mathcal{S}}(z_{o})}{(z_{o}^{2}+32)^{1/2}} 
\tfrac{\ln t}{t})$, $\det (\widehat{m}_{d}(0)) \! = \! 1$. {}From Lemma~4.3, 
keeping only leading-order terms, one shows that $\det (\widehat{\mathscr{P}}
(0)) \! = \! 1 \! + \! \tfrac{2 \Re (\widehat{a}_{1}-\widehat{a}_{2})}{\sqrt{
t}} \! + \! \mathcal{O}(\tfrac{c^{\mathcal{S}}(z_{o})}{(z_{o}^{2}+32)^{1/2}} 
\tfrac{\ln t}{t})$, and, {}from the proof of Proposition~4.1, the estimates 
of Proposition~4.2, and noting that $(1 \! - \! \tfrac{a_{m}^{0}}{\varsigma_{
m}})(1 \! - \! \tfrac{\overline{a_{m}^{0}}}{\overline{\varsigma_{m}}}) \! - 
\! \tfrac{c_{m}^{0}}{\varsigma_{m}} \tfrac{\overline{c_{m}^{0}}}{\overline{
\varsigma_{m}}} \! = \! 1$ and $\int_{-\infty}^{+\infty} \tfrac{(1-\mu^{2}) 
\ln (1-\vert r(\mu) \vert^{2})}{(\mu^{2}-2 \mu \cos (\phi_{m})+1)} \tfrac{
\md \mu}{\mu} \! = \! 0$ (which is proven using the symmetry reduction 
$r(\zeta^{-1}) \! = \! -\overline{r(\overline{\zeta})})$, one shows that 
$(\det (\widehat{m}_{d}(0)))^{-1} \! = \! 1 \! + \! \tfrac{2}{\sqrt{t}} \Re 
\! \left((1 \! - \! \tfrac{\overline{a_{m}^{0}}}{\overline{\varsigma_{m}}})
(\tfrac{a_{m}^{1}}{\varsigma_{m}} \! + \! \tfrac{\overline{d_{m}^{1}}}{
\overline{\varsigma_{m}}}) \! + \! \tfrac{c_{m}^{0}}{\varsigma_{m}}(\tfrac{
b_{m}^{1}}{\varsigma_{m}} \! + \! \tfrac{\overline{c_{m}^{1}}}{\overline{
\varsigma_{m}}}) \right) \! + \! \mathcal{O}(\tfrac{c^{\mathcal{S}}(z_{o})}
{(z_{o}^{2}+32)^{1/2}} \tfrac{\ln t}{t})$; thus, {}from the---yet to be 
verified---identity $\det (\widehat{\mathscr{P}}(0)) \! = \! (\det (\widehat{
m}_{d}(0)))^{-1}$, and the above, it follows that $\Re (\widehat{a}_{1} \! 
- \! \widehat{a}_{2}) \! = \! \Re \! \left((1 \! - \! \tfrac{\overline{a_{
m}^{0}}}{\overline{\varsigma_{m}}})(\tfrac{a_{m}^{1}}{\varsigma_{m}} \! + 
\! \tfrac{\overline{d_{m}^{1}}}{\overline{\varsigma_{m}}}) \! + \! \tfrac{
c_{m}^{0}}{\varsigma_{m}}(\tfrac{b_{m}^{1}}{\varsigma_{m}} \! + \! \tfrac{
\overline{c_{m}^{1}}}{\overline{\varsigma_{m}}}) \right)$. If the formulae 
presented thus far are correct, then one must be able to show {}from them 
that the right-hand side of the latter relation equals zero. {}From 
Proposition~4.2, and repeated application of standard trigonometric 
identities, one shows, after a very lengthy and tedious algebraic 
calculation, that $(\theta_{\gamma_{m}} \! = \! \pm \pi/2)$
\begin{align*}
\Re \! \left(\tfrac{c_{m}^{0}}{\varsigma_{m}}(\tfrac{b_{m}^{1}}{\varsigma_{
m}} \! + \! \tfrac{\overline{c_{m}^{1}}}{\overline{\varsigma_{m}}}) \right) 
&= \pm \tfrac{16 \lambda_{1}^{2} \sin^{3}(\phi_{m}) \vert \gamma_{m} \vert^{
3}P^{3}(\phi_{m},\phi_{k})Q^{3}(\phi_{m}) \sqrt{\nu (\lambda_{1})} \, (1+
\vert \gamma_{m} \vert^{2}P^{2}(\phi_{m},\phi_{k})Q^{2}(\phi_{m}) \me^{2 \phi 
(x,t)}) \me^{3 \phi (x,t)}}{(\lambda_{1}^{2}-2 \lambda_{1} \cos (\phi_{m})
+1)^{2} \sqrt{(\lambda_{1}-\lambda_{2})} \, (z_{o}^{2}+32)^{1/4}(1-\vert 
\gamma_{m} \vert^{2}P^{2}(\phi_{m},\phi_{k})Q^{2}(\phi_{m}) \me^{2 \phi (x,
t)})^{3}} \\
\times& \left(((\lambda_{1} \! + \! \lambda_{2}) \cos (\phi_{m}) \! - \! 2) 
\cos (\Theta^{+}(z_{o},t) \! + \! \tfrac{\pi}{4}) \! + \! (\lambda_{1} \! - 
\! \lambda_{2}) \sin (\phi_{m}) \sin (\Theta^{+}(z_{o},t) \! + \! \tfrac{\pi}
{4}) \right) \\
\times& \, \cos (s^{+}) \! \pm \! \tfrac{8 \lambda_{1} \sin^{2}(\phi_{m}) \cos 
(\phi_{m}) \vert \gamma_{m} \vert^{3}P^{3}(\phi_{m},\phi_{k})Q^{3}(\phi_{m}) 
\sqrt{\nu (\lambda_{1})} \, \me^{3 \phi (x,t)}}{(1-\vert \gamma_{m} \vert^{2}
P^{2}(\phi_{m},\phi_{k})Q^{2}(\phi_{m}) \me^{2 \phi (x,t)})^{2}(\lambda_{1}^{
2}-2 \lambda_{1} \cos (\phi_{m})+1) \sqrt{(\lambda_{1}-\lambda_{2})} \, (z_{
o}^{2}+32)^{1/4}} \\
\times& \left(2 \sin (s^{+} \! - \! \phi_{m}) \cos (\Theta^{+}(z_{o},t) \! + 
\! \tfrac{\pi}{4}) \! - \! (\lambda_{1} \! + \! \lambda_{2}) \sin (s^{+}) 
\cos (\Theta^{+}(z_{o},t) \! + \! \tfrac{\pi}{4}) \right. \\
+& \left. (\lambda_{1} \! - \! \lambda_{2}) \cos (s^{+}) \sin (\Theta^{+}(z_{
o},t) \! + \! \tfrac{\pi}{4}) \right),
\end{align*}
and
\begin{align*}
\Re \! \left((1 \! - \! \tfrac{\overline{a_{m}^{0}}}{\overline{\varsigma_{
m}}})(\tfrac{a_{m}^{1}}{\varsigma_{m}} \! + \! \tfrac{\overline{d_{m}^{1}}}
{\overline{\varsigma_{m}}}) \right) &= \mp \tfrac{16 \lambda_{1}^{2} \sin^{3}
(\phi_{m}) \vert \gamma_{m} \vert^{3}P^{3}(\phi_{m},\phi_{k})Q^{3}(\phi_{m}) 
\sqrt{\nu (\lambda_{1})} \, (1+\vert \gamma_{m} \vert^{2}P^{2}(\phi_{m},\phi_{
k})Q^{2}(\phi_{m}) \me^{2 \phi (x,t)}) \me^{3 \phi (x,t)}}{(\lambda_{1}^{2}-
2 \lambda_{1} \cos (\phi_{m})+1)^{2} \sqrt{(\lambda_{1}-\lambda_{2})} \, (z_{
o}^{2}+32)^{1/4}(1-\vert \gamma_{m} \vert^{2}P^{2}(\phi_{m},\phi_{k})Q^{2}
(\phi_{m}) \me^{2 \phi (x,t)})^{3}} \\
\times& \left(((\lambda_{1} \! + \! \lambda_{2}) \cos (\phi_{m}) \! - \! 2) 
\cos (\Theta^{+}(z_{o},t) \! + \! \tfrac{\pi}{4}) \! + \! (\lambda_{1} \! - 
\! \lambda_{2}) \sin (\phi_{m}) \sin (\Theta^{+}(z_{o},t) \! + \! \tfrac{\pi}
{4}) \right) \\
\times& \cos (s^{+}) \! \mp \! \tfrac{8 \lambda_{1} \sin^{2}(\phi_{m}) \cos 
(\phi_{m}) \vert \gamma_{m} \vert^{3}P^{3}(\phi_{m},\phi_{k})Q^{3}(\phi_{m}) 
\sqrt{\nu (\lambda_{1})} \, \me^{3 \phi (x,t)}}{(1-\vert \gamma_{m} \vert^{2}
P^{2}(\phi_{m},\phi_{k})Q^{2}(\phi_{m}) \me^{2 \phi (x,t)})^{2}(\lambda_{1}^{
2}-2 \lambda_{1} \cos (\phi_{m})+1) \sqrt{(\lambda_{1}-\lambda_{2})} \, (z_{
o}^{2}+32)^{1/4}} \\
\times& \left(2 \sin (s^{+} \! - \! \phi_{m}) \cos (\Theta^{+}(z_{o},t) \! + 
\! \tfrac{\pi}{4}) \! - \! (\lambda_{1} \! + \! \lambda_{2}) \sin (s^{+}) 
\cos (\Theta^{+}(z_{o},t) \! + \! \tfrac{\pi}{4}) \right. \\
+& \left. (\lambda_{1} \! - \! \lambda_{2}) \cos (s^{+}) \sin (\Theta^{+}(z_{
o},t) \! + \! \tfrac{\pi}{4}) \right);
\end{align*}
thus, adding, $\Re \! \left((1 \! - \! \tfrac{\overline{a_{m}^{0}}}{\overline{
\varsigma_{m}}})(\tfrac{a_{m}^{1}}{\varsigma_{m}} \! + \! \tfrac{\overline{d_{
m}^{1}}}{\overline{\varsigma_{m}}}) \! + \! \tfrac{c_{m}^{0}}{\varsigma_{m}}
(\tfrac{b_{m}^{1}}{\varsigma_{m}} \! + \! \tfrac{\overline{c_{m}^{1}}}{
\overline{\varsigma_{m}}}) \right) \! = \! 0$, whence $\Re (\widehat{a}_{1} 
\! - \! \widehat{a}_{2}) \! = \! 0$. Recalling the expression for $(\widehat{
\Delta}_{o})_{11}$ given in Proposition~4.3, the estimates and expansions of 
Proposition~4.2, and the fact---just established---that $\Re (\widehat{a}_{1} 
\! - \! \widehat{a}_{2}) \! = \! 0$, one shows that
\begin{equation*}
(\widehat{\Delta}_{o})_{11} \! = \! \tfrac{1}{\sqrt{t}}(\widehat{\Delta}_{
o})_{11}^{\alpha} \! + \! \mi \! \left((\widehat{\Delta}_{o})_{11}^{\beta} 
\! + \! \tfrac{1}{\sqrt{t}}(\widehat{\Delta}_{o})_{11}^{\gamma} \right) \! 
+ \! \mathcal{O} \! \left( \tfrac{c^{\mathcal{S}}(z_{o})}{(z_{o}^{2}+32)^{
1/2}} \tfrac{\ln t}{t} \right),
\end{equation*}
where
\begin{align*}
(\widehat{\Delta}_{o})_{11}^{\alpha} :=& \mp \tfrac{2 \Im (\widehat{a}_{1}-
\widehat{a}_{2}) \sin (\phi_{m}) \vert \gamma_{m} \vert P(\phi_{m},\phi_{k})
Q(\phi_{m}) \me^{\phi (x,t)}}{(1-\vert \gamma_{m} \vert^{2}P^{2}(\phi_{m},
\phi_{k})Q^{2}(\phi_{m}) \me^{2 \phi (x,t)})} \\
+& \tfrac{2 \sqrt{\nu (\lambda_{1})} \cos (\Theta^{+}(z_{o},t)+\frac{\pi}
{4}) \sin (s^{+})}{\sqrt{(\lambda_{1}-\lambda_{2})} \, (z_{o}^{2}+32)^{1/4}} 
\! - \! \Re (\widehat{a}_{3}) \sin (\theta^{+}(1) \! + \! s^{+}) \! - \! \Im 
(\widehat{a}_{3}) \cos (\theta^{+}(1) \! + \! s^{+}) \\
+& \tfrac{4 \sin (\phi_{m}) \vert \gamma_{m} \vert^{2}P^{2}(\phi_{m},\phi_{k})
Q^{2}(\phi_{m}) \sqrt{\nu (\lambda_{1})} \cos (\Theta^{+}(z_{o},t)+\frac{\pi}
{4}) \cos (s^{+}-\phi_{m}) \me^{2 \phi (x,t)}}{(1-\vert \gamma_{m} \vert^{2}
P^{2}(\phi_{m},\phi_{k})Q^{2}(\phi_{m}) \me^{2 \phi (x,t)}) \sqrt{(\lambda_{
1}-\lambda_{2})} \, (z_{o}^{2}+32)^{1/4}} \\
+& \tfrac{2 \Re (\widehat{a}_{3}) \sin (\phi_{m}) \vert \gamma_{m} \vert^{2}
P^{2}(\phi_{m},\phi_{k})Q^{2}(\phi_{m}) \cos (s^{+}+\phi_{m}+\theta^{+}(1)) 
\me^{2 \phi (x,t)}}{(1- \vert \gamma_{m} \vert^{2}P^{2}(\phi_{m},\phi_{k})
Q^{2}(\phi_{m}) \me^{2 \phi (x,t)})} \\
-& \tfrac{2 \Im (\widehat{a}_{3}) \sin (\phi_{m}) \vert \gamma_{m} \vert^{2}
P^{2}(\phi_{m},\phi_{k})Q^{2}(\phi_{m}) \sin (s^{+}+\phi_{m}+\theta^{+}(1)) 
\me^{2 \phi (x,t)}}{(1- \vert \gamma_{m} \vert^{2}P^{2}(\phi_{m},\phi_{k})
Q^{2}(\phi_{m}) \me^{2 \phi (x,t)})} \\
+& \tfrac{8 \lambda_{1}^{2} \sin^{2}(\phi_{m}) \vert \gamma_{m} \vert^{2}P^{2}
(\phi_{m},\phi_{k})Q^{2}(\phi_{m}) \sqrt{\nu (\lambda_{1})} \sin (s^{+}) \me^{
2 \phi (x,t)}}{(1- \vert \gamma_{m} \vert^{2}P^{2}(\phi_{m},\phi_{k})Q^{2}
(\phi_{m}) \me^{2 \phi (x,t)})(\lambda_{1}^{2}-2 \lambda_{1} \cos (\phi_{m})
+1)^{2} \sqrt{(\lambda_{1}-\lambda_{2})} \, (z_{o}^{2}+32)^{1/4}}
\end{align*}
\begin{align*}
\times& \left(((\lambda_{1} \! + \! \lambda_{2}) \cos (\phi_{m}) \! - \! 2) 
\cos (\Theta^{+}(z_{o},t)+\tfrac{\pi}{4}) \! + \! (\lambda_{1} \! - \! 
\lambda_{2}) \sin (\phi_{m}) \sin (\Theta^{+}(z_{o},t)+\tfrac{\pi}{4}) 
\right) \\
+& \tfrac{4 \lambda_{1} \sin (\phi_{m}) \cos (\phi_{m}) \vert \gamma_{m} 
\vert^{2}P^{2}(\phi_{m},\phi_{k})Q^{2}(\phi_{m}) \sqrt{\nu (\lambda_{1})} \, 
\me^{2 \phi (x,t)}}{(1- \vert \gamma_{m} \vert^{2}P^{2}(\phi_{m},\phi_{k})
Q^{2}(\phi_{m}) \me^{2 \phi (x,t)})(\lambda_{1}^{2}-2 \lambda_{1} \cos 
(\phi_{m})+1) \sqrt{(\lambda_{1}-\lambda_{2})} \, (z_{o}^{2}+32)^{1/4}} \\
\times& \left(2 \cos (s^{+} \! - \! \phi_{m}) \cos (\Theta^{+}(z_{o},t) \! + 
\! \tfrac{\pi}{4}) \! - \! (\lambda_{1} \! + \! \lambda_{2}) \cos (s^{+}) 
\cos (\Theta^{+}(z_{o},t) \! + \! \tfrac{\pi}{4}) \right. \\
-& \left. (\lambda_{1} \! - \! \lambda_{2}) \sin (s^{+}) \sin (\Theta^{+}
(z_{o},t) \! + \! \tfrac{\pi}{4}) \right),
\end{align*}
\begin{equation*}
(\widehat{\Delta}_{o})_{11}^{\beta} := \pm \tfrac{2 \sin (\phi_{m}) \vert 
\gamma_{m} \vert P(\phi_{m},\phi_{k})Q(\phi_{m}) \me^{\phi (x,t)}}{(1- \vert 
\gamma_{m} \vert^{2}P^{2}(\phi_{m},\phi_{k})Q^{2}(\phi_{m}) \me^{2 \phi (x,
t)})},
\end{equation*}
\begin{align*}
(\widehat{\Delta}_{o})_{11}^{\gamma} :=& -\tfrac{2 \sqrt{\nu (\lambda_{1})} 
\cos (\Theta^{+}(z_{o},t)+\frac{\pi}{4}) \cos (s^{+})}{\sqrt{(\lambda_{1}-
\lambda_{2})} \, (z_{o}^{2}+32)^{1/4}} \! + \! \Re (\widehat{a}_{3}) \cos 
(\theta^{+}(1) \! + \! s^{+}) \! - \! \Im (\widehat{a}_{3}) \sin (\theta^{
+}(1) \! + \! s^{+}) \\
+& \tfrac{4 \sin (\phi_{m}) \vert \gamma_{m} \vert^{2}P^{2}(\phi_{m},\phi_{k})
Q^{2}(\phi_{m}) \sqrt{\nu (\lambda_{1})} \cos (\Theta^{+}(z_{o},t)+\frac{\pi}
{4}) \sin (s^{+}-\phi_{m}) \me^{2 \phi (x,t)}}{(1- \vert \gamma_{m} \vert^{2}
P^{2}(\phi_{m},\phi_{k})Q^{2}(\phi_{m}) \me^{2 \phi (x,t)}) \sqrt{(\lambda_{1}
-\lambda_{2})} \, (z_{o}^{2}+32)^{1/4}} \\
+& \tfrac{2 \Re (\widehat{a}_{3}) \sin (\phi_{m}) \vert \gamma_{m} \vert^{2}
P^{2}(\phi_{m},\phi_{k})Q^{2}(\phi_{m}) \sin (s^{+}+\phi_{m}+\theta^{+}(1)) 
\me^{2 \phi (x,t)}}{(1-\vert \gamma_{m} \vert^{2}P^{2}(\phi_{m},\phi_{k})Q^{2}
(\phi_{m}) \me^{2 \phi (x,t)})} \\
+& \tfrac{2 \Im (\widehat{a}_{3}) \sin (\phi_{m}) \vert \gamma_{m} \vert^{2}
P^{2}(\phi_{m},\phi_{k})Q^{2}(\phi_{m}) \cos (s^{+}+\phi_{m}+\theta^{+}(1)) 
\me^{2 \phi (x,t)}}{(1- \vert \gamma_{m} \vert^{2}P^{2}(\phi_{m},\phi_{k})
Q^{2}(\phi_{m}) \me^{2 \phi (x,t)})} \\
+& \tfrac{8 \lambda_{1}^{2} \sin^{2}(\phi_{m}) \vert \gamma_{m} \vert^{2}P^{
2}(\phi_{m},\phi_{k})Q^{2}(\phi_{m}) \sqrt{\nu (\lambda_{1})} \cos (s^{+})(1+
\vert \gamma_{m} \vert^{2}P^{2}(\phi_{m},\phi_{k})Q^{2}(\phi_{m}) \me^{2 \phi 
(x,t)}) \me^{2 \phi (x,t)}}{(\lambda_{1}^{2}-2 \lambda_{1} \cos (\phi_{m})
+1)^{2} \sqrt{(\lambda_{1}-\lambda_{2})} \, (z_{o}^{2}+32)^{1/4}(1- \vert 
\gamma \vert^{2}P^{2}(\phi_{m},\phi_{k})Q^{2}(\phi_{m}) \me^{2 \phi (x,
t)})^{2}} \\
\times& \left(((\lambda_{1} \! + \! \lambda_{2}) \cos (\phi_{m}) \! - \! 2) 
\cos (\Theta^{+}(z_{o},t)+\tfrac{\pi}{4}) \! + \! (\lambda_{1} \! - \! 
\lambda_{2}) \sin (\phi_{m}) \sin (\Theta^{+}(z_{o},t)+\tfrac{\pi}{4}) 
\right) \\
+& \tfrac{4 \lambda_{1} \sin (\phi_{m}) \cos (\phi_{m}) \vert \gamma_{m} 
\vert^{2}P^{2}(\phi_{m},\phi_{k})Q^{2}(\phi_{m}) \sqrt{\nu (\lambda_{1})} \, 
\me^{2 \phi (x,t)}}{(1- \vert \gamma_{m} \vert^{2}P^{2}(\phi_{m},\phi_{k})
Q^{2}(\phi_{m}) \me^{2 \phi (x,t)})(\lambda_{1}^{2}-2 \lambda_{1} \cos 
(\phi_{m})+1) \sqrt{(\lambda_{1}-\lambda_{2})} \, (z_{o}^{2}+32)^{1/4}} \\
\times& \left(2 \sin (s^{+} \! - \! \phi_{m}) \cos (\Theta^{+}(z_{o},t) \! + 
\! \tfrac{\pi}{4}) \! - \! (\lambda_{1} \! + \! \lambda_{2}) \sin (s^{+}) 
\cos (\Theta^{+}(z_{o},t) \! + \! \tfrac{\pi}{4}) \right. \\
+& \left. (\lambda_{1} \! - \! \lambda_{2}) \cos (s^{+}) \sin (\Theta^{+}
(z_{o},t) \! + \! \tfrac{\pi}{4}) \right),
\end{align*}
and $\theta^{+}(\cdot)$ is specified in the Proposition. Recalling that 
$\mathrm{tr}(\widehat{\Delta}_{o}) \! = \! 0$, it follows that 
$\Re ((\widehat{\Delta}_{o})_{11}) \! = \! 0$; thus, $(\widehat{\Delta}_{
o})_{11}^{\alpha} \! = \! 0$, which gives a relation for $\Im (\widehat{a}_{
1} \! - \! \widehat{a}_{2})$, but, since $\Re (\widehat{a}_{3})$ and $\Im 
(\widehat{a}_{3})$ are as yet undetermined, this is not enough. Towards this 
end, one uses the condition $\det (\widehat{\Delta}_{o}) \! = \! (\widehat{
\Delta}_{o})_{11} \overline{(\widehat{\Delta}_{o})_{11}} \! - \! (\widehat{
\Delta}_{o})_{12} \overline{(\widehat{\Delta}_{o})_{12}} \! = \! -1$ (Note: 
if the conditions $\mathrm{tr}(\widehat{\Delta}_{o}) \! = \! 0$ and $\det 
(\widehat{\Delta}_{o}) \! = \! -1$ are satisfied, then it follows that 
$\widehat{\Delta}_{o} \overline{\widehat{\Delta}_{o}} \! = \! \mathrm{I}$ is 
also satisfied, so it is enough to use the condition $\det (\widehat{\Delta}_{
o}) \! = \! -1)$. {}From the formula for $(\widehat{\Delta}_{o})_{11}$ 
given above, and the expression for $(\widehat{\Delta}_{o})_{12}$ given in 
Proposition~4.3, one shows that
\begin{align*}
(\widehat{\Delta}_{o})_{12} \overline{(\widehat{\Delta}_{o})_{12}} =& \, (1 
\! - \! \tfrac{a_{m}^{0}}{\varsigma_{m}})(1 \! - \! \tfrac{\overline{a_{m}^{
0}}}{\overline{\varsigma_{m}}}) \! + \! \tfrac{2}{\sqrt{t}} \! \left(\Re \! 
\left((\widehat{a}_{1} \! - \! \widehat{a}_{2})(1 \! - \! \tfrac{a_{m}^{0}}{
\varsigma_{m}})(1 \! - \! \tfrac{\overline{a_{m}^{0}}}{\overline{\varsigma_{
m}}}) \right) \right. \\
-& \left. \Re \! \left((1 \! - \! \tfrac{\overline{a_{m}^{0}}}{\overline{
\varsigma_{m}}})(\tfrac{a_{m}^{1}}{\varsigma_{m}} \! + \! \tfrac{\overline{
d_{m}^{1}}}{\overline{\varsigma_{m}}}) \right) \! + \! \Re \! \left((1 \! - 
\! \tfrac{a_{m}^{0}}{\varsigma_{m}}) \tfrac{2c_{m}^{0} \sqrt{\nu (\lambda_{
1})} \, \delta (0) \cos (\Theta^{+}(z_{o},t)+\frac{\pi}{4})}{\varsigma_{m} 
\sqrt{(\lambda_{1}-\lambda_{2})} \, (z_{o}^{2}+32)^{1/4}} \right) \right. \\
-& \left. \Re \! \left((1 \! - \! \tfrac{\overline{a_{m}^{0}}}{\overline{
\varsigma_{m}}}) \tfrac{c_{m}^{0} \widehat{a}_{3}}{\varsigma_{m}} \right) 
\right) \! + \! \mathcal{O} \! \left( \tfrac{c^{\mathcal{S}}(z_{o})}{(z_{
o}^{2}+32)^{1/2}} \tfrac{\ln t}{t} \right).
\end{align*}
Using the estimates given in Proposition~4.2, and recalling that $\Re 
(\widehat{a}_{1} \! - \! \widehat{a}_{2}) \! = \! 0$, one gets that
\begin{align*}
(\widehat{\Delta}_{o})_{12} &\overline{(\widehat{\Delta}_{o})_{12}} = \, 1 \! 
+ \! \tfrac{4 \sin^{2}(\phi_{m}) \vert \gamma_{m} \vert^{2}P^{2}(\phi_{m},
\phi_{k})Q^{2}(\phi_{m}) \me^{2 \phi (x,t)}}{(1- \vert \gamma_{m} \vert^{2}
P^{2}(\phi_{m},\phi_{k})Q^{2}(\phi_{m}) \me^{2 \phi (x,t)})} \! + \! \tfrac{
4 \sin^{2}(\phi_{m}) \vert \gamma_{m} \vert^{4}P^{4}(\phi_{m},\phi_{k})Q^{4}
(\phi_{m}) \me^{4 \phi (x,t)}}{(1- \vert \gamma_{m} \vert^{2}P^{2}(\phi_{m},
\phi_{k})Q^{2}(\phi_{m}) \me^{2 \phi (x,t)})^{2}} \\
+& \, \tfrac{2}{\sqrt{t}} \! \left(\mp \tfrac{4 \sin (\phi_{m}) \vert \gamma_{
m} \vert P(\phi_{m},\phi_{k})Q(\phi_{m}) \sqrt{\nu (\lambda_{1})} \, \cos (s^{
+}) \cos (\Theta^{+}(z_{o},t)+\frac{\pi}{4}) \me^{\phi (x,t)}}{(1- \vert 
\gamma_{m} \vert^{2}P^{2}(\phi_{m},\phi_{k})Q^{2}(\phi_{m}) \me^{2 \phi (x,
t)}) \sqrt{(\lambda_{1}-\lambda_{2})} \, (z_{o}^{2}+32)^{1/4}} \right. \\
\pm& \left. \tfrac{8 \sin^{2}(\phi_{m}) \vert \gamma_{m} \vert^{3}P^{3}
(\phi_{m},\phi_{k})Q^{3}(\phi_{m}) \sqrt{\nu (\lambda_{1})} \, \sin (s^{+}-
\phi_{m}) \cos (\Theta^{+}(z_{o},t)+\frac{\pi}{4}) \me^{3 \phi (x,t)}}{(1- 
\vert \gamma_{m} \vert^{2}P^{2}(\phi_{m},\phi_{k})Q^{2}(\phi_{m}) \me^{2 
\phi (x,t)})^{2} \sqrt{(\lambda_{1}-\lambda_{2})} \, (z_{o}^{2}+32)^{1/4}} 
\right. \\
\pm& \left. \tfrac{2 \sin (\phi_{m}) \vert \gamma_{m} \vert P(\phi_{m},\phi_{
k})Q(\phi_{m}) \me^{\phi (x,t)}}{(1- \vert \gamma_{m} \vert^{2}P^{2}(\phi_{m},
\phi_{k})Q^{2}(\phi_{m}) \me^{2 \phi (x,t)})} \! \left(\Re (\widehat{a}_{3}) 
\! \cos (s^{+} \! + \! \theta^{+}(1)) \! - \! \Im (\widehat{a}_{3}) \! \sin 
(s^{+} \! + \! \theta^{+}(1)) \right) \right. \\
\pm& \left. \tfrac{4 \sin^{2}(\phi_{m}) \vert \gamma_{m} \vert^{3}P^{3}(\phi_{
m},\phi_{k})Q^{3}(\phi_{m}) \me^{3 \phi (x,t)}}{(1- \vert \gamma_{m} \vert^{2}
P^{2}(\phi_{m},\phi_{k})Q^{2}(\phi_{m}) \me^{2 \phi (x,t)})^{2}} \! \left(\Re 
(\widehat{a}_{3}) \! \sin (s^{+} \! + \! \theta^{+}(1) \! + \! \phi_{m}) \! + 
\! \Im (\widehat{a}_{3}) \! \cos (s^{+} \! + \! \theta^{+}(1) \! + \! \phi_{
m}) \right) \right. \\
\pm& \left. \tfrac{16 \lambda_{1}^{2} \sin^{3}(\phi_{m}) \vert \gamma_{m} 
\vert^{3}P^{3}(\phi_{m},\phi_{k})Q^{3}(\phi_{m}) \sqrt{\nu (\lambda_{1})} \, 
\cos (s^{+})(1+\vert \gamma_{m} \vert^{2}P^{2}(\phi_{m},\phi_{k})Q^{2}(\phi_{
m}) \me^{2 \phi (x,t)}) \me^{3 \phi (x,t)}}{(\lambda_{1}^{2}-2 \lambda_{1} 
\cos (\phi_{m})+1)^{2} \sqrt{(\lambda_{1}-\lambda_{2})} \, (z_{o}^{2}+32)^{
1/4}(1- \vert \gamma_{m} \vert^{2}P^{2}(\phi_{m},\phi_{k})Q^{2}(\phi_{m}) 
\me^{2 \phi (x,t)})^{3}} \right. \\
\times& \left. \left(((\lambda_{1} \! + \! \lambda_{2}) \cos (\phi_{m}) \! - 
\! 2) \cos (\Theta^{+}(z_{o},t) \! + \! \tfrac{\pi}{4}) \! + \! (\lambda_{1} 
\! - \! \lambda_{2}) \sin (\phi_{m}) \sin (\Theta^{+}(z_{o},t) \! + \! 
\tfrac{\pi}{4}) \right) \right.
\end{align*}
\begin{align*}
\pm& \left. \tfrac{8 \lambda_{1} \sin^{2}(\phi_{m}) \cos (\phi_{m}) \vert 
\gamma_{m} \vert^{3}P^{3}(\phi_{m},\phi_{k})Q^{3}(\phi_{m}) \sqrt{\nu 
(\lambda_{1})} \, \me^{3 \phi (x,t)}}{(1- \vert \gamma_{m} \vert^{2}P^{2}
(\phi_{m},\phi_{k})Q^{2}(\phi_{m}) \me^{2 \phi (x,t)})^{2}(\lambda_{1}^{2}-2 
\lambda_{1} \cos (\phi_{m})+1) \sqrt{(\lambda_{1}-\lambda_{2})} \, (z_{o}^{
2}+32)^{1/4}} \right. \\
\times& \left. \left(2 \sin (s^{+} \! - \! \phi_{m}) \cos (\Theta^{+}(z_{o},
t) \! + \! \tfrac{\pi}{4}) \! - \! (\lambda_{1} \! + \! \lambda_{2}) \sin 
(s^{+}) \cos (\Theta^{+}(z_{o},t) \! + \! \tfrac{\pi}{4}) \right. \right. \\
+& \left. \left. (\lambda_{1} \! - \! \lambda_{2}) \cos (s^{+}) \sin (\Theta^{
+}(z_{o},t) \! + \! \tfrac{\pi}{4}) \right) \right) \! + \! \mathcal{O} \! 
\left( \tfrac{c^{\mathcal{S}}(z_{o})}{(z_{o}^{2}+32)^{1/2}} \tfrac{\ln t}{t} 
\right).
\end{align*}
{}From the expression for $(\widehat{\Delta}_{o})_{11}$ given above, and using 
the fact that $(\widehat{\Delta}_{o})_{11}^{\alpha} \! = \! 0$, one shows 
that
\begin{align*}
(\widehat{\Delta}_{o})_{11} \overline{(\widehat{\Delta}_{o})_{11}} =& \, 
\tfrac{4 \sin^{2}(\phi_{m}) \vert \gamma_{m} \vert^{2}P^{2}(\phi_{m},\phi_{k})
Q^{2}(\phi_{m}) \me^{2 \phi (x,t)}}{(1-\vert \gamma_{m} \vert^{2}P^{2}(\phi_{
m},\phi_{k})Q^{2}(\phi_{m}) \me^{2 \phi (x,t)})^{2}} \\
+& \, \tfrac{2}{\sqrt{t}} \! \left(\mp \tfrac{4 \sin (\phi_{m}) \vert \gamma_{
m} \vert P(\phi_{m},\phi_{k})Q(\phi_{m}) \sqrt{\nu (\lambda_{1})} \, \cos 
(s^{+}) \cos (\Theta^{+}(z_{o},t)+\frac{\pi}{4}) \me^{\phi (x,t)}}{(1- \vert 
\gamma_{m} \vert^{2}P^{2}(\phi_{m},\phi_{k})Q^{2}(\phi_{m}) \me^{2 \phi (x,
t)}) \sqrt{(\lambda_{1}-\lambda_{2})} \, (z_{o}^{2}+32)^{1/4}} \right. \\
\pm& \left. \tfrac{8 \sin^{2}(\phi_{m}) \vert \gamma_{m} \vert^{3}P^{3}(\phi_{
m},\phi_{k})Q^{3}(\phi_{m}) \sqrt{\nu (\lambda_{1})} \, \sin (s^{+}-\phi_{m}) 
\cos (\Theta^{+}(z_{o},t)+\frac{\pi}{4}) \me^{3 \phi (x,t)}}{(1- \vert 
\gamma_{m} \vert^{2}P^{2}(\phi_{m},\phi_{k})Q^{2}(\phi_{m}) \me^{2 \phi (x,
t)})^{2} \sqrt{(\lambda_{1}-\lambda_{2})} \, (z_{o}^{2}+32)^{1/4}} \right. \\
\pm& \left. \tfrac{2 \Re (\widehat{a}_{3}) \sin (\phi_{m}) \vert \gamma_{m} 
\vert P(\phi_{m},\phi_{k})Q(\phi_{m}) \me^{\phi (x,t)}}{(1-\vert \gamma_{m} 
\vert^{2}P^{2}(\phi_{m},\phi_{k})Q^{2}(\phi_{m}) \me^{2 \phi (x,t)})} \! 
\left(\cos (s^{+} \! + \! \theta^{+}(1)) \right. \right. \\
+& \left. \left. \tfrac{2 \sin (\phi_{m}) \vert \gamma_{m} \vert^{2}P^{2}
(\phi_{m},\phi_{k})Q^{2}(\phi_{m}) \sin (s^{+}+\phi_{m}+\theta^{+}(1)) \me^{
2 \phi (x,t)}}{(1-\vert \gamma_{m} \vert^{2}P^{2}(\phi_{m},\phi_{k})Q^{2}
(\phi_{m}) \me^{2 \phi (x,t)})} \right) \right. \\
\pm& \left. \tfrac{2 \Im (\widehat{a}_{3}) \sin (\phi_{m}) \vert \gamma_{m} 
\vert P(\phi_{m},\phi_{k})Q(\phi_{m}) \me^{\phi (x,t)}}{(1-\vert \gamma_{m} 
\vert^{2}P^{2}(\phi_{m},\phi_{k})Q^{2}(\phi_{m}) \me^{2 \phi (x,t)})} \! 
\left(-\sin (s^{+} \! + \! \theta^{+}(1)) \right. \right. \\
+& \left. \left. \tfrac{2 \sin (\phi_{m}) \vert \gamma_{m} \vert^{2}P^{2}
(\phi_{m},\phi_{k})Q^{2}(\phi_{m}) \cos (s^{+}+\phi_{m}+\theta^{+}(1)) \me^{
2 \phi (x,t)}}{(1-\vert \gamma_{m} \vert^{2}P^{2}(\phi_{m},\phi_{k})Q^{2}
(\phi_{m}) \me^{2 \phi (x,t)})} \right) \right. \\
\pm& \left. \tfrac{16 \lambda_{1}^{2} \sin^{3}(\phi_{m}) \vert \gamma_{m} 
\vert^{3}P^{3}(\phi_{m},\phi_{k})Q^{3}(\phi_{m}) \sqrt{\nu (\lambda_{1})} \, 
\cos (s^{+})(1+\vert \gamma_{m} \vert^{2}P^{2}(\phi_{m},\phi_{k})Q^{2}(\phi_{
m}) \me^{2 \phi (x,t)}) \me^{3 \phi (x,t)}}{(\lambda_{1}^{2}-2 \lambda_{1} 
\cos (\phi_{m})+1)^{2} \sqrt{(\lambda_{1}-\lambda_{2})} \, (z_{o}^{2}+32)^{
1/4}(1- \vert \gamma_{m} \vert^{2}P^{2}(\phi_{m},\phi_{k})Q^{2}(\phi_{m}) 
\me^{2 \phi (x,t)})^{3}} \right. \\
\times& \left. \left(((\lambda_{1} \! + \! \lambda_{2}) \cos (\phi_{m}) \! - 
\! 2) \cos (\Theta^{+}(z_{o},t) \! + \! \tfrac{\pi}{4}) \! + \! (\lambda_{1} 
\! - \! \lambda_{2}) \sin (\phi_{m}) \sin (\Theta^{+}(z_{o},t) \! + \! 
\tfrac{\pi}{4}) \right) \right. \\
\pm& \left. \tfrac{8 \lambda_{1} \sin^{2}(\phi_{m}) \cos (\phi_{m}) \vert 
\gamma_{m} \vert^{3}P^{3}(\phi_{m},\phi_{k})Q^{3}(\phi_{m}) \sqrt{\nu 
(\lambda_{1})} \, \me^{3 \phi (x,t)}}{(1- \vert \gamma_{m} \vert^{2}P^{2}
(\phi_{m},\phi_{k})Q^{2}(\phi_{m}) \me^{2 \phi (x,t)})^{2}(\lambda_{1}^{2}-2 
\lambda_{1} \cos (\phi_{m})+1) \sqrt{(\lambda_{1}-\lambda_{2})} \, (z_{o}^{2}
+32)^{1/4}} \right. \\
\times& \left. \left(2 \sin (s^{+} \! - \! \phi_{m}) \cos (\Theta^{+}(z_{o},
t) \! + \! \tfrac{\pi}{4}) \! - \! (\lambda_{1} \! + \! \lambda_{2}) \sin 
(s^{+}) \cos (\Theta^{+}(z_{o},t) \! + \! \tfrac{\pi}{4}) \right. \right. \\
+& \left. \left. (\lambda_{1} \! - \! \lambda_{2}) \cos (s^{+}) \sin (\Theta^{
+}(z_{o},t) \! + \! \tfrac{\pi}{4}) \right) \right) \! + \! \mathcal{O} \! 
\left( \tfrac{c^{\mathcal{S}}(z_{o})}{(z_{o}^{2}+32)^{1/2}} \tfrac{\ln t}{t} 
\right).
\end{align*}
Now, taking note of the relation $(\widehat{\Delta}_{o})_{11}^{\beta}
(\widehat{\Delta}_{o})_{11}^{\beta} \! - \! (1 \! - \! \tfrac{a_{m}^{0}}
{\varsigma_{m}})(1 \! - \! \tfrac{\overline{a_{m}^{0}}}{\overline{\varsigma_{
m}}}) \! = \! -1$, one substitutes the above-derived formulae for $\vert 
(\widehat{\Delta}_{o})_{11} \vert^{2}$ and $\vert (\widehat{\Delta}_{o})_{12} 
\vert^{2}$ into $\vert (\widehat{\Delta}_{o})_{11} \vert^{2} \! - \! \vert 
(\widehat{\Delta}_{o})_{12} \vert^{2} \! = \! -1$, and, modulo terms that are 
$\mathcal{O}(\tfrac{c^{\mathcal{S}}(z_{o})}{(z_{o}^{2}+32)^{1/2}} \tfrac{\ln 
t}{t})$, gets \textbf{exact} cancellation at $\mathcal{O}(1)$ and $\mathcal{
O}(t^{-1/2})$; thus, one concludes that $\Re (\widehat{a}_{3}) \! = \! \Im 
(\widehat{a}_{3}) \! = \! 0$. Recalling that $(\widehat{\Delta}_{o})_{11}^{
\alpha} \! = \! 0$, and using the fact that $\Re (\widehat{a}_{3}) \! = \! \Im 
(\widehat{a}_{3}) \! = \! 0$, {}from the expression for $(\widehat{\Delta}_{
0})_{11}$ given above, one obtains, after some straightforward algebra, 
the expressions for $\Im (\widehat{a}_{1} \! - \! \widehat{a}_{2})$ and 
$(\widehat{\Delta}_{o})_{11}$ stated in the Proposition. {}From 
Proposition~4.2, and the fact that $\Re (\widehat{a}_{3}) \! = \! \Im 
(\widehat{a}_{3}) \! = \! \Re (\widehat{a}_{1} \! - \! \widehat{a}_{2}) \! 
= \! 0$, one obtains, upon recalling the expression for $(\widehat{\Delta}_{
o})_{12}$ given in Proposition~4.3, the formula for $(\widehat{\Delta}_{
o})_{12}$ given in the Proposition. \hfill $\square$
\begin{ccccc}
As $t \! \to \! +\infty$ and $x \! \to \! -\infty$ such that $z_{o} \! := 
\! x/t \! < \! -2$ and $(x,t) \! \in \! \daleth_{m}$, $m \! \in \! \{1,2,
\ldots,N\}$, $u(x,t)$, the solution of the Cauchy problem for the 
{\rm D${}_{f}$NLSE}, and $\int_{\pm \infty}^{x}(\vert u(x^{\prime},t) 
\vert^{2} \! - \! 1) \, \md x^{\prime}$ have the leading-order asymptotic 
expansions (for the upper sign) stated in Theorem~{\rm 2.2.1}, 
Eqs.~{\rm (7)--(20)}.
\end{ccccc}

\emph{Proof.} The asymptotic expansions for $u(x,t)$ and $\int_{\pm 
\infty}^{x}(\vert u(x^{\prime},t) \vert^{2} \! - \! 1) \, \md x^{\prime}$ 
follow {}from Proposition~4.2, Proposition~4.4, Eqs.~(70)--(72), and 
Proposition~4.6 after tedious, but otherwise straightforward algebraic 
calculations. \hfill $\square$
\clearpage
\section*{Appendix A. Asymptotic Analysis as $t \! \to \! -\infty$}
\setcounter{section}{1}
\setcounter{z0}{1}
\setcounter{z1}{1}
\setcounter{z2}{1}
\setcounter{z3}{1}
In this appendix, a silhouette of the asymptotic analysis for $u(x,t)$ and 
$\int_{\pm \infty}^{x}(\vert u(x^{\prime},t) \vert^{2} \! - \! 1) \, \md 
x^{\prime}$ as $t \! \to \! -\infty$ and $x \! \to \! +\infty$ such that 
$z_{o} \! := \! x/t \! < \! -2$ and $(x,t) \! \in \! \daleth_{m}$, $m \! \in 
\! \{1,2,\ldots,N\}$, is presented. Since the calculations are analogous 
to those of Sections~3 and~4, only final results/statements, with in one 
instance a sketch of a proof, are given: one mimics the scheme of the 
calculation in Sections~3 and~4 to arrive at the corresponding asymptotic 
results.

The analogue of Lemma~3.1 is
\begin{ay}
For $r(\zeta) \! \in \! \mathcal{S}_{\mathbb{C}}^{1}(\mathbb{R})$, let 
$m(\zeta) \colon \mathbb{C} \setminus (\sigma_{d} \cup \sigma_{c}) \! \to \! 
\mathrm{M}_{2}(\mathbb{C})$ be the solution of the {\rm RHP} formulated in 
Lemma~{\rm 2.1.2}. Set $\widetilde{m}(\zeta) \! := \! m(\zeta)(\widetilde{
\delta}(\zeta))^{-\sigma_{3}}$, where $\widetilde{\delta}(\zeta) \! = \! \exp 
\! \left( \! \left( \int_{0}^{\lambda_{2}} \! + \! \int_{\lambda_{1}}^{+
\infty} \right) \! \tfrac{\ln (1-\vert r(\mu) \vert^{2})}{(\mu -\zeta)} 
\tfrac{\md \mu}{2 \pi \mi} \right)$, with $\lambda_{1}$ and $\lambda_{2}$ 
given in Theorem~{\rm 2.2.1}, Eq.~{\rm (10)}, $\widetilde{\delta}(\zeta) 
\overline{\widetilde{\delta}(\overline{\zeta})} \! = \! 1$, $\widetilde{
\delta}(\zeta) \widetilde{\delta}(\zeta^{-1}) \! = \! \widetilde{\delta}(0)$, 
and $\vert \vert (\widetilde{\delta}(\cdot))^{\pm 1} \vert \vert_{\mathcal{
L}^{\infty}(\mathbb{C})} \! \! := \! \sup_{\zeta \in \mathbb{C}} \vert 
(\widetilde{\delta}(\zeta))^{\pm 1} \vert \! < \! \infty$. Then $\widetilde{
m}(\zeta) \colon \mathbb{C} \setminus (\sigma_{d} \cup \sigma_{c}) \! \to 
\! \mathrm{M}_{2}(\mathbb{C})$ solves the following {\rm RHP:}
\begin{enumerate}
\item[(i)] $\widetilde{m}(\zeta)$ is piecewise (sectionally) meromorphic 
$\forall \, \zeta \! \in \! \mathbb{C} \setminus \sigma_{c};$
\item[(ii)] $\widetilde{m}_{\pm}(\zeta) \! := \! \lim_{\genfrac{}{}{0pt}{2}
{\zeta^{\prime} \, \to \, \zeta}{\pm \Im (\zeta^{\prime})>0}} \widetilde{m}
(\zeta^{\prime})$ satisfy the jump condition
\begin{equation*}
\widetilde{m}_{+}(\zeta) \! = \! \widetilde{m}_{-}(\zeta) \exp (-\mi k(\zeta)
(x \! + \! 2 \lambda (\zeta)t) \mathrm{ad}(\sigma_{3})) \widetilde{\mathcal{
G}}(\zeta), \quad \zeta \! \in \! \mathbb{R},
\end{equation*}
where
\begin{equation*}
\widetilde{\mathcal{G}}(\zeta) \! = \! 
\begin{pmatrix}
(1 \! - \! r(\zeta) \overline{r(\overline{\zeta})}) \widetilde{\delta}_{-}
(\zeta)(\widetilde{\delta}_{+}(\zeta))^{-1} & \, \, -\overline{r(\overline{
\zeta})} \, \widetilde{\delta}_{-}(\zeta) \widetilde{\delta}_{+}(\zeta) \\
r(\zeta)(\widetilde{\delta}_{-}(\zeta) \widetilde{\delta}_{+}(\zeta))^{-1} & 
\, \, (\widetilde{\delta}_{-}(\zeta))^{-1} \widetilde{\delta}_{+}(\zeta)
\end{pmatrix};
\end{equation*}
\item[(iii)] $\widetilde{m}(\zeta)$ has simple poles in $\sigma_{d} \! = \! 
\cup_{n=1}^{N}(\{\varsigma_{n}\} \cup \{\overline{\varsigma_{n}}\})$ with
\begin{align*}
\mathrm{Res}(\widetilde{m}(\zeta);\varsigma_{n}) &= \! \lim_{\zeta \to 
\varsigma_{n}} \widetilde{m}(\zeta)g_{n}(\widetilde{\delta}(\varsigma_{n}))^{
-2} \sigma_{-}, & n \! &\in \! \{1,2,\ldots,N\}, \\
\mathrm{Res}(\widetilde{m}(\zeta);\overline{\varsigma_{n}}) &= \! \sigma_{1} 
\overline{\mathrm{Res}(\widetilde{m}(\zeta);\varsigma_{n})} \, \sigma_{1}, & 
n \! &\in \! \{1,2,\ldots,N\},
\end{align*}
where $g_{n}$ is defined in Lemma~{\rm 3.1}, (iii);
\item[(iv)] $\det (\widetilde{m}(\zeta)) \vert_{\zeta = \pm 1} \! = \! 0;$
\item[(v)] $\widetilde{m}(\zeta) \! =_{\zeta \to 0} \! \zeta^{-1}(\widetilde{
\delta}(0))^{\sigma_{3}} \sigma_{2} \! + \! \mathcal{O}(1);$
\item[(vi)] $\widetilde{m}(\zeta) \! =_{\genfrac{}{}{0pt}{2}{\zeta \to \infty}
{\zeta \in \mathbb{C} \setminus (\sigma_{d} \cup \sigma_{c})}} \! \mathrm{I} 
\! + \! \mathcal{O}(\zeta^{-1});$
\item[(vii)] $\widetilde{m}(\zeta) \! = \! \sigma_{1} \overline{\widetilde{
m}(\overline{\zeta})} \, \sigma_{1}$ and $\widetilde{m}(\zeta^{-1}) \! = \! 
\zeta \widetilde{m}(\zeta)(\widetilde{\delta}(0))^{\sigma_{3}} \sigma_{2}$.
\end{enumerate}
Let
\begin{gather}
u(x,t) \! := \! \mi \lim_{\genfrac{}{}{0pt}{2}{\zeta \to \infty}{\zeta \in 
\mathbb{C} \setminus (\sigma_{d} \cup \sigma_{c})}}(\zeta (\widetilde{m}
(\zeta)(\widetilde{\delta}(\zeta))^{\sigma_{3}} \! - \! \mathrm{I}))_{12},
\end{gather}
and
\begin{gather}
\int\nolimits_{+\infty}^{x}(\vert u(x^{\prime},t) \vert^{2} \! - \! 1) \, \md 
x^{\prime} \! := \! -\mi \lim_{\genfrac{}{}{0pt}{2}{\zeta \to \infty}{\zeta 
\in \mathbb{C} \setminus (\sigma_{d} \cup \sigma_{c})}}(\zeta (\widetilde{m}
(\zeta)(\widetilde{\delta}(\zeta))^{\sigma_{3}} \! - \! \mathrm{I}))_{11}.
\end{gather}
Then $u(x,t)$ is the solution of the Cauchy problem for the 
{\rm D${}_{f}$NLSE}.
\end{ay}

The analogue of Definition~3.1 is
\begin{cy}
For $m \! \in \! \{1,2,\ldots,N\}$ and $\{\varsigma_{n}\}_{n=1}^{m-1} \subset 
\mathbb{C}_{+}$ (respectively, $\{\overline{\varsigma_{n}}\}_{n=1}^{m-1} 
\subset \mathbb{C}_{-})$, define the clockwise (respectively, 
counter-clockwise) oriented circles $\widetilde{\mathscr{K}}_{n} \! := \! 
\{\mathstrut \zeta; \, \vert \zeta \! - \! \varsigma_{n} \vert \! = \! 
\widetilde{\varepsilon}_{n}^{\mathscr{K}}\}$ (respectively, $\widetilde{
\mathscr{L}}_{n} \! := \! \{\mathstrut \zeta; \, \vert \zeta \! - \! 
\overline{\varsigma_{n}} \vert \! = \! \widetilde{\varepsilon}_{n}^{
\mathscr{L}}\})$, with $\widetilde{\varepsilon}_{n}^{\mathscr{K}}$ 
(respectively, $\widetilde{\varepsilon}_{n}^{\mathscr{L}})$ chosen 
sufficiently small such that $\widetilde{\mathscr{K}}_{n} \cap \widetilde{
\mathscr{K}}_{n^{\prime}} \! = \! \widetilde{\mathscr{L}}_{n} \cap 
\widetilde{\mathscr{L}}_{n^{\prime}} \! = \! \widetilde{\mathscr{K}}_{n} 
\cap \widetilde{\mathscr{L}}_{n} \! = \! \widetilde{\mathscr{K}}_{n} \cap 
\sigma_{c} \! = \! \widetilde{\mathscr{L}}_{n} \cap \sigma_{c} \! = \! 
\emptyset$ $\forall \, \, n \! \not= \! n^{\prime} \! \in \! \{1,2,\ldots,
m \! - \! 1\}$.
\end{cy}

The analogue of Lemma~3.2 is
\setcounter{z2}{1}
\setcounter{z3}{2}
\begin{ay}
For $r(\zeta) \! \in \! \mathcal{S}_{\mathbb{C}}^{1}(\mathbb{R})$, let 
$\widetilde{m}(\zeta) \colon \mathbb{C} \setminus (\sigma_{d} \cup \sigma_{
c}) \! \to \! \mathrm{M}_{2}(\mathbb{C})$ be the solution of the {\rm RHP} 
formulated in Lemma~{\rm A.1.1}. Set
\begin{equation*}
\widetilde{m}^{\flat}(\zeta) \! := \! 
\begin{cases}
\widetilde{m}(\zeta), &\text{$\zeta \! \in \! \mathbb{C} \setminus (\sigma_{
c} \cup (\cup_{n=1}^{m-1}(\widetilde{\mathscr{K}}_{n} \cup \mathrm{int}
(\widetilde{\mathscr{K}}_{n}) \cup \widetilde{\mathscr{L}}_{n} \cup 
\mathrm{int}(\widetilde{\mathscr{L}}_{n}))))$,} \\
\widetilde{m}(\zeta) \! \left(\mathrm{I} \! - \! \tfrac{g_{n}(\widetilde{
\delta}(\varsigma_{n}))^{-2}}{(\zeta -\varsigma_{n})} \sigma_{-} \right), 
&\text{$\zeta \! \in \! \mathrm{int}(\widetilde{\mathscr{K}}_{n}), \quad n 
\! \in \! \{1,2,\ldots,m \! - \! 1\}$,} \\
\widetilde{m}(\zeta) \! \left(\mathrm{I} \! + \! \tfrac{\overline{g_{n}
(\widetilde{\delta}(\varsigma_{n}))^{-2}}}{(\zeta -\overline{\varsigma_{n}})} 
\sigma_{+} \right), &\text{$\zeta \! \in \! \mathrm{int}(\widetilde{\mathscr{
L}}_{n}), \quad n \! \in \! \{1,2,\ldots,m \! - \! 1\}$.}
\end{cases}
\end{equation*}
Then $\widetilde{m}^{\flat}(\zeta) \colon \mathbb{C} \setminus ((\sigma_{d} 
\setminus \cup_{n=1}^{m-1}(\{\varsigma_{n}\} \cup \{\overline{\varsigma_{n}}
\})) \cup (\sigma_{c} \cup (\cup_{n=1}^{m-1}(\widetilde{\mathscr{K}}_{n} \cup 
\widetilde{\mathscr{L}}_{n})))) \! \to \! \mathrm{M}_{2}(\mathbb{C})$ solves 
the following {\rm RHP:}
\begin{enumerate}
\item[(i)] $\widetilde{m}^{\flat}(\zeta)$ is piecewise (sectionally) 
meromorphic $\forall \, \zeta \! \in \! \mathbb{C} \setminus (\sigma_{c} \cup 
(\cup_{n=1}^{m-1}(\widetilde{\mathscr{K}}_{n} \cup \widetilde{\mathscr{L}}_{
n})));$
\item[(ii)] $\widetilde{m}_{\pm}^{\flat}(\zeta) \! := \! \lim_{\genfrac{}{}
{0pt}{2}{\zeta^{\prime} \, \to \, \zeta}{\zeta^{\prime} \, \in \, \pm \, 
\mathrm{side} \, \mathrm{of} \, \sigma_{c} \cup (\cup_{n=1}^{m-1}(\widetilde{
\mathscr{K}}_{n} \cup \widetilde{\mathscr{L}}_{n}))}} \widetilde{m}^{\flat}
(\zeta^{\prime})$ satisfy the jump condition
\begin{equation*}
\widetilde{m}_{+}^{\flat}(\zeta) \! = \! \widetilde{m}_{-}^{\flat}(\zeta) 
\widetilde{\upsilon}^{\flat}(\zeta), \quad \zeta \! \in \! \sigma_{c} \cup 
(\cup_{n=1}^{m-1}(\widetilde{\mathscr{K}}_{n} \cup \widetilde{\mathscr{
L}}_{n})),
\end{equation*}
where
\begin{equation*}
\widetilde{\upsilon}^{\flat}(\zeta) \! = \! 
\begin{cases}
\exp (-\mi k(\zeta)(x \! + \! 2 \lambda (\zeta)t) \mathrm{ad}(\sigma_{3})) 
\widetilde{\mathcal{G}}(\zeta), &\text{$\zeta \! \in \! \mathbb{R}$,} \\
\mathrm{I} \! + \! \tfrac{g_{n}(\widetilde{\delta}(\varsigma_{n}))^{-2}}
{(\zeta -\varsigma_{n})} \sigma_{-}, &\text{$\zeta \! \in \! \widetilde{
\mathscr{K}}_{n}, \quad n \! \in \! \{1,2,\ldots,m \! - \! 1\}$,} \\
\mathrm{I} \! + \! \tfrac{\overline{g_{n}(\widetilde{\delta}(\varsigma_{
n}))^{-2}}}{(\zeta -\overline{\varsigma_{n}})} \sigma_{+}, &\text{$\zeta \! 
\in \! \widetilde{\mathscr{L}}_{n}, \quad n \! \in \! \{1,2,\ldots,m \! - 
\! 1\}$,}
\end{cases}
\end{equation*}
with $\widetilde{\mathcal{G}}(\zeta)$ given in Lemma~{\rm A.1.1}, (ii);
\item[(iii)] $\widetilde{m}^{\flat}(\zeta)$ has simple poles in $\sigma_{d} 
\setminus \cup_{n=1}^{m-1}(\{\varsigma_{n}\} \cup \{\overline{\varsigma_{n}}
\})$ with
\begin{align*}
\mathrm{Res}(\widetilde{m}^{\flat}(\zeta);\varsigma_{n}) &= \! \lim_{\zeta 
\to \varsigma_{n}} \widetilde{m}^{\flat}(\zeta)g_{n}(\widetilde{\delta}
(\varsigma_{n}))^{-2} \sigma_{-}, & n \! &\in \! \{m,m \! + \! 1,\ldots,
N\}, \\
\mathrm{Res}(\widetilde{m}^{\flat}(\zeta);\overline{\varsigma_{n}}) &= \! 
\sigma_{1} \overline{\mathrm{Res}(\widetilde{m}^{\flat}(\zeta);\varsigma_{
n})} \, \sigma_{1}, & n \! &\in \! \{m,m \! + \! 1,\ldots,N\};
\end{align*}
\item[(iv)] $\det (\widetilde{m}^{\flat}(\zeta)) \vert_{\zeta = \pm 1} \! = 
\! 0;$
\item[(v)] $\widetilde{m}^{\flat}(\zeta) \! =_{\zeta \to 0} \! \zeta^{-1}
(\widetilde{\delta}(0))^{\sigma_{3}} \sigma_{2} \! + \! \mathcal{O}(1);$
\item[(vi)] as $\zeta \! \to \! \infty$, $\zeta \! \in \! \mathbb{C} 
\setminus ((\sigma_{d} \setminus \cup_{n=1}^{m-1}(\{\varsigma_{n}\} \cup 
\{\overline{\varsigma_{n}}\})) \cup (\sigma_{c} \cup (\cup_{n=1}^{m-1}
(\widetilde{\mathscr{K}}_{n} \cup \widetilde{\mathscr{L}}_{n}))))$, 
$\widetilde{m}^{\flat}(\zeta) \! = \! \mathrm{I} \! + \! \mathcal{O}
(\zeta^{-1});$
\item[(vii)] $\widetilde{m}^{\flat}(\zeta) \! = \! \sigma_{1} \overline{
\widetilde{m}^{\flat}(\overline{\zeta})} \, \sigma_{1}$ and $\widetilde{m}^{
\flat}(\zeta^{-1}) \! = \! \zeta \widetilde{m}^{\flat}(\zeta)(\widetilde{
\delta}(0))^{\sigma_{3}} \sigma_{2}$.
\end{enumerate}
For $\zeta \! \in \! \mathbb{C} \setminus ((\sigma_{d} \setminus \cup_{
n=1}^{m-1}(\{\varsigma_{n}\} \cup \{\overline{\varsigma_{n}}\})) \cup 
(\sigma_{c} \cup (\cup_{n=1}^{m-1}(\widetilde{\mathscr{K}}_{n} \cup 
\widetilde{\mathscr{L}}_{n}))))$, let
\begin{gather}
u(x,t) \! := \! \mi \lim_{\zeta \to \infty}(\zeta (\widetilde{m}^{\flat}
(\zeta)(\widetilde{\delta}(\zeta))^{\sigma_{3}} \! - \! \mathrm{I}))_{12},
\end{gather}
and
\begin{gather}
\int\nolimits_{+\infty}^{x}(\vert u(x^{\prime},t) \vert^{2} \! - \! 1) \, \md 
x^{\prime} \! := \! -\mi \lim_{\zeta \to \infty}(\zeta (\widetilde{m}(\zeta)
(\widetilde{\delta}(\zeta))^{\sigma_{3}} \! - \! \mathrm{I}))_{11}.
\end{gather}
Then $u(x,t)$ is the solution of the Cauchy problem for the 
{\rm D${}_{f}$NLSE}.
\end{ay}

The analogue of Lemma~3.3 is
\setcounter{z2}{1}
\setcounter{z3}{3}
\begin{ay}
For $m \! \in \! \{1,2,\ldots,N\}$, let $\sigma^{\prime \prime}_{d} \! := \! 
\sigma_{d} \setminus \cup_{n=1}^{m-1}(\{\varsigma_{n}\} \cup \{\overline{
\varsigma_{n}}\})$, $\sigma^{\prime \prime}_{c} \! := \! \sigma_{c} \cup 
(\cup_{n=1}^{m-1}(\widetilde{\mathscr{K}}_{n} \cup \widetilde{\mathscr{L}}_{
n}))$, where $\widetilde{\mathscr{K}}_{n}$ and $\widetilde{\mathscr{L}}_{n}$ 
are given in Definition~{\rm A.1.1}, and $\sigma^{\prime \prime}_{\mathcal{
O}^{\mathcal{D}}} \! := \! \sigma^{\prime \prime}_{d} \cup \sigma^{\prime 
\prime}_{c}$ $(\sigma_{d}^{\prime \prime} \cap \sigma_{c}^{\prime \prime} \! 
= \! \emptyset)$. Set
\begin{equation*}
\widetilde{m}^{\sharp}(\zeta) \! := \! 
\begin{cases}
\widetilde{m}^{\flat}(\zeta) \prod_{k=1}^{m-1}(d_{k}^{+}(\zeta))^{-\sigma_{
3}}, &\text{
$\zeta \! \in \! \mathbb{C} \setminus (\sigma_{c}^{\prime \prime} 
\cup (\cup_{n=1}^{m-1}(\mathrm{int}(\widetilde{\mathscr{K}}_{n}) \cup 
\mathrm{int}(\widetilde{\mathscr{L}}_{n}))))$,} \\
\widetilde{m}^{\flat}(\zeta)(\widetilde{J}_{\widetilde{\mathscr{K}}_{n}}
(\zeta))^{-1} \prod_{k=1}^{m-1}(d_{k}^{-}(\zeta))^{-\sigma_{3}}, &\text{
$\zeta \! \in \! \mathrm{int}(\widetilde{\mathscr{K}}_{n}), \quad n \! \in 
\! \{1,2,\ldots,m \! - \! 1\}$,} \\
\widetilde{m}^{\flat}(\zeta)(\widetilde{J}_{\widetilde{\mathscr{L}}_{n}}
(\zeta))^{-1} \prod_{k=1}^{m-1}(d_{k}^{-}(\zeta))^{-\sigma_{3}}, &\text{
$\zeta \! \in \! \mathrm{int}(\widetilde{\mathscr{L}}_{n}), \quad n \! \in 
\! \{1,2,\ldots,m \! - \! 1\}$,}
\end{cases}
\end{equation*}
where $d_{n}^{\pm}(\zeta)$ are given in Lemma~{\rm 3.3}, $\widetilde{J}_{
\widetilde{\mathscr{K}}_{n}}(\zeta)$ $(\in \! \mathrm{SL}(2,\mathbb{C}))$ and 
$\widetilde{J}_{\widetilde{\mathscr{L}}_{n}}(\zeta)$ $(\in \! \mathrm{SL}(2,
\mathbb{C}))$, respectively, are holomorphic in $\cup_{k=1}^{m-1} \mathrm{int}
(\widetilde{\mathscr{K}}_{k})$ and $\cup_{l=1}^{m-1} \mathrm{int}(\widetilde{
\mathscr{L}}_{l})$, with
\begin{align*}
\widetilde{J}_{\widetilde{\mathscr{K}}_{n}}(\zeta) \! =& \! 
\begin{pmatrix}
\frac{\prod_{\genfrac{}{}{0pt}{2}{k=1}{k \not= n}}^{m-1} \frac{d_{k}^{+}
(\zeta)}{d_{k}^{-}(\zeta)}-\frac{\widetilde{C}_{n}^{\mathscr{K}}g_{n}
(\widetilde{\delta}(\varsigma_{n}))^{-2}}{(\zeta -\overline{\varsigma_{n}})^{
2}} \prod_{\genfrac{}{}{0pt}{2}{k=1}{k \not= n}}^{m-1} \frac{(d_{k}^{+}
(\zeta))^{-1}}{d_{k}^{-}(\zeta)}}{(\zeta -\varsigma_{n})} & \frac{\widetilde{
C}_{n}^{\mathscr{K}}}{(\zeta -\overline{\varsigma_{n}})^{2}} \prod_{\genfrac{}
{}{0pt}{2}{k=1}{k \not= n}}^{m-1} \frac{(d_{k}^{+}(\zeta))^{-1}}{d_{k}^{-}
(\zeta)} \\
-g_{n}(\widetilde{\delta}(\varsigma_{n}))^{-2} \prod_{\genfrac{}{}{0pt}{2}
{k=1}{k \not= n}}^{m-1} \frac{d_{k}^{-}(\zeta)}{d_{k}^{+}(\zeta)} & (\zeta \! 
- \! \varsigma_{n}) \prod_{\genfrac{}{}{0pt}{2}{k=1}{k \not= n}}^{m-1} \frac{
d_{k}^{-}(\zeta)}{d_{k}^{+}(\zeta)}
\end{pmatrix}, \\
\widetilde{J}_{\widetilde{\mathscr{L}}_{n}}(\zeta) \! =& \! 
\begin{pmatrix}
(\zeta \! - \! \overline{\varsigma_{n}}) \prod_{\genfrac{}{}{0pt}{2}{k=1}
{k \not= n}}^{m-1} \frac{d_{k}^{+}(\zeta)}{d_{k}^{-}(\zeta)} & \overline{g_{n}
(\widetilde{\delta}(\varsigma_{n}))^{-2}} \prod_{\genfrac{}{}{0pt}{2}{k=1}
{k \not= n}}^{m-1} \frac{d_{k}^{+}(\zeta)}{d_{k}^{-}(\zeta)} \\
-\frac{\widetilde{C}_{n}^{\mathscr{L}}}{(\zeta -\varsigma_{n})^{2}} \prod_{
\genfrac{}{}{0pt}{2}{k=1}{k \not= n}}^{m-1} \frac{d_{k}^{-}(\zeta)}{(d_{k}^{
+}(\zeta))^{-1}} & \frac{\prod_{\genfrac{}{}{0pt}{2}{k=1}{k \not= n}}^{m-1} 
\frac{d_{k}^{-}(\zeta)}{d_{k}^{+}(\zeta)}-\frac{\widetilde{C}_{n}^{\mathscr{
L}} \overline{g_{n}(\widetilde{\delta}(\varsigma_{n}))^{-2}}}{(\zeta 
-\varsigma_{n})^{2}} \prod_{\genfrac{}{}{0pt}{2}{k=1}{k \not= n}}^{m-1} 
\frac{d_{k}^{-}(\zeta)}{(d_{k}^{+}(\zeta))^{-1}}}{(\zeta -\overline{
\varsigma_{n}})}
\end{pmatrix},
\end{align*}
and
\begin{equation*}
\widetilde{C}_{n}^{\mathscr{K}} \! = \overline{\widetilde{C}_{n}^{\mathscr{
L}}} = \! -4 \sin^{2}(\phi_{n})(g_{n})^{-1}(\widetilde{\delta}(\varsigma_{
n}))^{2} \, \me^{-2 \mi \sum_{\genfrac{}{}{0pt}{2}{j=1}{j \not= n}}^{m-1} 
\phi_{j}} \prod_{\genfrac{}{}{0pt}{2}{k=1}{k \not= n}}^{m-1} \! \left(\tfrac{
\sin (\frac{1}{2}(\phi_{n}+\phi_{k}))}{\sin (\frac{1}{2}(\phi_{n}-\phi_{k}))} 
\right)^{2}.
\end{equation*}
Then $\widetilde{m}^{\sharp}(\zeta) \colon \mathbb{C} \setminus \sigma^{
\prime \prime}_{\mathcal{O}^{\mathcal{D}}} \! \to \! \mathrm{M}_{2}(\mathbb{
C})$ solves the following (augmented) {\rm RHP:}
\begin{enumerate}
\item[(i)] $\widetilde{m}^{\sharp}(\zeta)$ is piecewise (sectionally) 
meromorphic $\forall \, \zeta \! \in \! \mathbb{C} \setminus \sigma_{c}^{
\prime \prime};$
\item[(ii)] $\widetilde{m}_{\pm}^{\sharp}(\zeta) \! := \! \lim_{\genfrac{}{}
{0pt}{2}{\zeta^{\prime} \, \to \, \zeta}{\zeta^{\prime} \, \in \, \pm \, 
\mathrm{side} \, \mathrm{of} \, \sigma_{\mathcal{O}^{\mathcal{D}}}^{\prime 
\prime}}} \widetilde{m}^{\sharp}(\zeta^{\prime})$ satisfy the following jump 
conditions,
\begin{equation*}
\widetilde{m}_{+}^{\sharp}(\zeta) \! = \! \widetilde{m}_{-}^{\sharp}(\zeta) 
\exp (-\mi k(\zeta)(x \! + \! 2 \lambda (\zeta)t) \mathrm{ad}(\sigma_{3})) 
\widetilde{\mathcal{G}}^{\sharp}(\zeta), \quad \zeta \! \in \! \mathbb{R},
\end{equation*}
where
\begin{equation*}
\widetilde{\mathcal{G}}^{\sharp}(\zeta) \! = \! 
\begin{pmatrix}
(1 \! - \! r(\zeta) \overline{r(\overline{\zeta})}) \widetilde{\delta}_{-}
(\zeta)(\widetilde{\delta}_{+}(\zeta))^{-1} & \, \, -\overline{r(\overline{
\zeta})} \, \widetilde{\delta}_{-}(\zeta) \widetilde{\delta}_{+}(\zeta) 
\prod_{k=1}^{m-1}(d_{k}^{+}(\zeta))^{2} \\
r(\zeta)(\widetilde{\delta}_{-}(\zeta) \widetilde{\delta}_{+}(\zeta))^{-1} 
\prod_{k=1}^{m-1}(d_{k}^{+}(\zeta))^{-2} & \, \, (\widetilde{\delta}_{-}
(\zeta))^{-1} \widetilde{\delta}_{+}(\zeta)
\end{pmatrix},
\end{equation*}
and
\begin{equation*}
\widetilde{m}^{\sharp}_{+}(\zeta) \! = \! 
\begin{cases}
\widetilde{m}^{\sharp}_{-}(\zeta) \! \left(\mathrm{I} \! + \! \tfrac{
\widetilde{C}_{n}^{\mathscr{K}}}{(\zeta-\varsigma_{n})} \sigma_{+} \right), 
&\text{$\zeta \! \in \! \widetilde{\mathscr{K}}_{n}, \quad n \! \in \! \{1,2,
\ldots,m \! - \! 1\}$,} \\
\widetilde{m}^{\sharp}_{-}(\zeta) \! \left(\mathrm{I} \! + \! \tfrac{
\widetilde{C}_{n}^{\mathscr{L}}}{(\zeta-\overline{\varsigma_{n}})} \sigma_{-} 
\right), &\text{$\zeta \! \in \! \widetilde{\mathscr{L}}_{n}, \quad n \! \in 
\! \{1,2,\ldots,m \! - \! 1\};$}
\end{cases}
\end{equation*}
\item[(iii)] $\widetilde{m}^{\sharp}(\zeta)$ has simple poles in $\sigma_{
d}^{\prime \prime}$ with
\begin{align*}
\mathrm{Res}(\widetilde{m}^{\sharp}(\zeta);\varsigma_{n}) &= \! \lim_{\zeta 
\to \varsigma_{n}} \widetilde{m}^{\sharp}(\zeta)g_{n}(\widetilde{\delta}
(\varsigma_{n}))^{-2} \! \left(\prod_{k=1}^{m-1}(d_{k}^{+}(\varsigma_{n}))^{
-2} \right) \! \sigma_{-}, & n \! &\in \! \{m,m \! + \! 1,\ldots,N\}, \\
\mathrm{Res}(\widetilde{m}^{\sharp}(\zeta);\overline{\varsigma_{n}}) &= \! 
\sigma_{1} \overline{\mathrm{Res}(\widetilde{m}^{\sharp}(\zeta);\varsigma_{
n})} \, \sigma_{1}, & n \! &\in \! \{m,m \! + \! 1,\ldots,N\};
\end{align*}
\item[(iv)] $\det (\widetilde{m}^{\sharp}(\zeta)) \vert_{\zeta = \pm 1} \! 
= \! 0;$
\item[(v)] $\widetilde{m}^{\sharp}(\zeta) \! =_{\zeta \to 0} \! \zeta^{-1}
(\widetilde{\delta}(0))^{\sigma_{3}} \! \left(\prod_{k=1}^{m-1}(d_{k}^{+}
(0))^{\sigma_{3}} \right) \! \sigma_{2} \! + \! \mathcal{O}(1);$
\item[(vi)] $\widetilde{m}^{\sharp}(\zeta) \! =_{\genfrac{}{}{0pt}{2}{\zeta 
\to \infty}{\zeta \in \mathbb{C} \setminus \sigma_{\mathcal{O}^{\mathcal{
D}}}^{\prime \prime}}} \! \mathrm{I} \! + \! \mathcal{O}(\zeta^{-1});$
\item[(vii)] $\widetilde{m}^{\sharp}(\zeta) \! = \! \sigma_{1} \overline{
\widetilde{m}^{\sharp}(\overline{\zeta})} \, \sigma_{1}$ and $\widetilde{m}^{
\sharp}(\zeta^{-1}) \! = \! \zeta \widetilde{m}^{\sharp}(\zeta)(\widetilde{
\delta}(0))^{\sigma_{3}} \! \left(\prod_{k=1}^{m-1}(d_{k}^{+}(0))^{\sigma_{
3}} \right) \! \sigma_{2}$.
\end{enumerate}
Let
\begin{gather}
u(x,t) \! := \! \mi \lim_{\genfrac{}{}{0pt}{2}{\zeta \to \infty}{\zeta \in 
\mathbb{C} \setminus \sigma_{\mathcal{O}^{\mathcal{D}}}^{\prime \prime}}} 
\! \left(\zeta \! \left(\widetilde{m}^{\sharp}(\zeta)(\widetilde{\delta}
(\zeta))^{\sigma_{3}} \prod_{k=1}^{m-1}(d_{k}^{+}(\zeta))^{\sigma_{3}} \! - 
\! \mathrm{I} \right) \right)_{12},
\end{gather}
and
\begin{gather}
\int\nolimits_{+\infty}^{x}(\vert u(x^{\prime},t) \vert^{2} \! - \! 1) \, \md 
x^{\prime} \! := \! -\mi \lim_{\genfrac{}{}{0pt}{2}{\zeta \to \infty}{\zeta 
\in \mathbb{C} \setminus \sigma_{\mathcal{O}^{\mathcal{D}}}^{\prime \prime}}} 
\! \left(\zeta \! \left(\widetilde{m}^{\sharp}(\zeta)(\widetilde{\delta}
(\zeta))^{\sigma_{3}} \prod_{k=1}^{m-1}(d_{k}^{+}(\zeta))^{\sigma_{3}} \! - 
\! \mathrm{I} \right) \right)_{11}.
\end{gather}
Then $u(x,t)$ is the solution of the Cauchy problem for the 
{\rm D${}_{f}$NLSE}.
\end{ay}

The analogue of Proposition~3.1 is
\setcounter{z0}{1}
\setcounter{z1}{1}
\begin{by}[{\rm \cite{a38}}]
The solution of the {\rm RHP} for $\widetilde{m}^{\sharp}(\zeta) \colon 
\mathbb{C} \setminus \sigma_{\mathcal{O}^{\mathcal{D}}}^{\prime \prime} \! 
\to \! \mathrm{M}_{2}(\mathbb{C})$ formulated in Lemma~{\rm A.1.3} has the 
(integral equation) representation
\begin{equation*}
\widetilde{m}^{\sharp}(\zeta) \! = \! \left(\mathrm{I} \! + \! \zeta^{-1} 
\widetilde{\Delta}_{o}^{\sharp} \right) \! \widetilde{\mathscr{P}}^{\sharp}
(\zeta) \! \left(\widetilde{m}_{d}^{\sharp}(\zeta) \! + \! \int\nolimits_{
\sigma_{c}^{\prime \prime}} \tfrac{\widetilde{m}^{\sharp}_{-}(\mu)
(\widetilde{\upsilon}^{\sharp}(\mu)-\mathrm{I})}{(\mu -\zeta)} \, \tfrac{\md 
\mu}{2 \pi \mi} \right), \quad \zeta \! \in \! \mathbb{C} \setminus \sigma_{
\mathcal{O}^{\mathcal{D}}}^{\prime \prime},
\end{equation*}
where
\begin{equation*}
\widetilde{m}^{\sharp}_{d}(\zeta) \! = \! \mathrm{I} \! + \! \sum_{n=m}^{N} 
\! \left(\tfrac{\mathrm{Res}(\widetilde{m}^{\sharp}(\zeta);\varsigma_{n})}
{(\zeta -\varsigma_{n})} \! + \! \tfrac{\sigma_{1} \overline{\mathrm{Res}
(\widetilde{m}^{\sharp}(\zeta);\varsigma_{n})} \, \sigma_{1}}{(\zeta -
\overline{\varsigma_{n}})} \right),
\end{equation*}
$\widetilde{v}^{\sharp}(\cdot)$ is a generic notation for the jump matrices 
of $\widetilde{m}^{\sharp}(\zeta)$ on $\sigma_{c}^{\prime \prime}$ 
(Lemma~{\rm A.1.3}, (ii)), and $\widetilde{\Delta}_{o}^{\sharp}$ and 
$\widetilde{\mathscr{P}}^{\sharp}(\zeta)$ are specified below. The solution 
of the above (integral) equation can be written as the ordered factorisation
\begin{equation*}
\widetilde{m}^{\sharp}(\zeta) \! = \! \left(\mathrm{I} \! + \! \zeta^{-1} 
\widetilde{\Delta}_{o}^{\sharp} \right) \! \widetilde{\mathscr{P}}^{\sharp}
(\zeta) \widetilde{m}_{d}^{\sharp}(\zeta) \widetilde{m}^{c}(\zeta), \quad 
\zeta \! \in \! \mathbb{C} \setminus \sigma_{\mathcal{O}^{\mathcal{D}}}^{
\prime \prime},
\end{equation*}
where $\widetilde{m}^{\sharp}_{d}(\zeta) \! = \! \sigma_{1} \overline{
\widetilde{m}^{\sharp}_{d}(\overline{\zeta})} \, \sigma_{1}$ $(\in \! 
\mathrm{SL}(2,\mathbb{C}))$ has the representation given above, $\widetilde{
\mathscr{P}}^{\sharp}(\zeta) \! = \! \sigma_{1} \overline{\widetilde{
\mathscr{P}}^{\sharp}(\overline{\zeta})} \, \sigma_{1}$ is chosen so that 
$\widetilde{\Delta}^{\sharp}_{o}$ is idempotent, $\mathrm{I} \! + \! \zeta^{
-1} \widetilde{\Delta}_{o}^{\sharp}$ $(\in \! \mathrm{M}_{2}(\mathbb{C}))$ 
is holomorphic in a punctured neighbourhood of the origin, with $\widetilde{
\Delta}_{o}^{\sharp} \! = \! \sigma_{1} \overline{\widetilde{\Delta}_{o}^{
\sharp}} \, \sigma_{1}$ $(\in \! \mathrm{GL}(2,\mathbb{C}))$ such that $\det 
(\mathrm{I} \! + \! \zeta^{-1} \widetilde{\Delta}_{o}^{\sharp}) \vert_{\zeta 
=\pm 1} \! = \! 0$, and having the finite, order 2, matrix involutive 
structure $\widetilde{\Delta}_{o}^{\sharp} \! = \! 
\left(
\begin{smallmatrix}
\widetilde{\Delta}^{\sharp} \me^{\mi (k+1/2) \pi} & (1+(\widetilde{\Delta}^{
\sharp})^{2})^{1/2} \me^{-\mi \widetilde{\vartheta}^{\sharp}} \\
(1+(\widetilde{\Delta}^{\sharp})^{2})^{1/2} \me^{\mi \widetilde{\vartheta}^{
\sharp}} & \widetilde{\Delta}^{\sharp} \me^{-\mi (k+1/2) \pi}
\end{smallmatrix}
\right)$, $k \! \in \! \mathbb{Z}$, where $\widetilde{\Delta}^{\sharp}$ and 
$\widetilde{\vartheta}^{\sharp}$ are obtained {}from the relation 
$\widetilde{\Delta}_{o}^{\sharp} \! = \! \widetilde{\mathscr{P}}^{\sharp}(0) 
\widetilde{m}_{d}^{\sharp}(0) \widetilde{m}^{c}(0)(\widetilde{\delta}(0))^{
\sigma_{3}} \! \left(\prod_{k=1}^{m-1}(d_{k}^{+}(0))^{\sigma_{3}} \right) \! 
\sigma_{2}$, and satisfying $\mathrm{tr}(\widetilde{\Delta}_{o}^{\sharp}) 
\! = \! 0$, $\det (\widetilde{\Delta}_{o}^{\sharp}) \! = \! -1$, and 
$\widetilde{\Delta}_{o}^{\sharp} \widetilde{\Delta}_{o}^{\sharp} \! = \! 
\mathrm{I}$, and $\widetilde{m}^{c}(\zeta) \colon \mathbb{C} \setminus 
\sigma_{c}^{\prime \prime} \! \to \! \mathrm{SL}(2,\mathbb{C})$ solves the 
following {\rm RHP:} {\rm (1)} $\widetilde{m}^{c}(\zeta)$ is piecewise 
(sectionally) holomorphic $\forall \, \zeta \! \in \! \mathbb{C} \setminus 
\sigma_{c}^{\prime \prime};$ {\rm (2)} $\widetilde{m}^{c}_{\pm}(\zeta) \! 
:= \! \lim_{\genfrac{}{}{0pt}{2}{\zeta^{\prime} \, \to \, \zeta}{\zeta^{
\prime} \, \in \, \pm \, \mathrm{side} \, \mathrm{of} \, \sigma_{c}^{\prime 
\prime}}} \widetilde{m}^{c}(\zeta^{\prime})$ satisfy the jump condition 
$\widetilde{m}^{c}_{+}(\zeta) \! = \! \widetilde{m}^{c}_{-}(\zeta) 
\widetilde{\upsilon}^{c}(\zeta)$, $\zeta \! \in \! \sigma_{c}^{\prime 
\prime}$, where $\widetilde{\upsilon}^{c}(\zeta) \! = \! \exp (-\mi k(\zeta)
(x \! + \! 2 \lambda (\zeta)t) \mathrm{ad}(\sigma_{3})) \widetilde{\mathcal{
G}}^{\sharp}(\zeta)$, $\zeta \! \in \! \mathbb{R}$, with $\widetilde{\mathcal{
G}}^{\sharp}(\zeta)$ given in Lemma~{\rm A.1.3}, (ii), $\widetilde{\upsilon}^{
c}(\zeta) \! = \! \mathrm{I} \! + \! \widetilde{C}_{n}^{\mathscr{K}}(\zeta \! 
- \! \varsigma_{n})^{-1} \sigma_{+}$, $\zeta \! \in \! \widetilde{\mathscr{
K}}_{n}$, and $\widetilde{\upsilon}^{c}(\zeta) \! = \! \mathrm{I} \! + \! 
\widetilde{C}_{n}^{\mathscr{L}}(\zeta \! - \! \overline{\varsigma_{n}})^{-1} 
\sigma_{-}$, $\zeta \! \in \! \widetilde{\mathscr{L}}_{n}$, $n \! \in \! 
\{1,2,\ldots,m \! - \! 1\}$, with $\widetilde{C}_{n}^{\mathscr{K}}$ and 
$\widetilde{C}_{n}^{\mathscr{L}}$ given in Lemma~{\rm A.1.3;} {\rm (3)} 
$\widetilde{m}^{c}(\zeta) \! =_{\genfrac{}{}{0pt}{2}{\zeta \to \infty}{\zeta 
\in \mathbb{C} \setminus \sigma_{c}^{\prime \prime}}} \! \mathrm{I} \! + \! 
\mathcal{O}(\zeta^{-1});$ and {\rm (4)} $\widetilde{m}^{c}(\zeta) \! = \! 
\sigma_{1} \overline{\widetilde{m}^{c}(\overline{\zeta})} \, \sigma_{1}$.
\end{by}

The analogue of Lemma~3.5 is
\setcounter{z2}{1}
\setcounter{z3}{4}
\begin{ay}
For $m \! \in \! \{1,2,\ldots,N\}$, set $\widetilde{\sigma}_{d} \! := \! 
\cup_{n=m}^{N}(\{\varsigma_{n}\} \cup \{\overline{\varsigma_{n}}\})$, and 
let $\sigma_{c} \! = \! \{\mathstrut \zeta; \, \Im (\zeta) \! = \! 0\}$ with 
orientation {}from $-\infty$ to $+\infty$. Let $\mathcal{X}(\zeta) \colon 
\mathbb{C} \setminus (\widetilde{\sigma}_{d} \cup \sigma_{c}) \! \to \! 
\mathrm{M}_{2}(\mathbb{C})$ solve the following {\rm RHP:}
\begin{enumerate}
\item[(i)] $\mathcal{X}(\zeta)$ is piecewise (sectionally) meromorphic 
$\forall \, \zeta \! \in \! \mathbb{C} \setminus \sigma_{c};$
\item[(ii)] $\mathcal{X}_{\pm}(\zeta) \! := \! \lim_{\genfrac{}{}{0pt}{2}
{\zeta^{\prime} \, \to \, \zeta}{\zeta^{\prime} \, \in \, \pm \, 
\mathrm{side} \, \mathrm{of} \, \sigma_{c}}} \mathcal{X}(\zeta^{\prime})$ 
satisfy the jump condition
\begin{equation*}
\mathcal{X}_{+}(\zeta) \! = \! \mathcal{X}_{-}(\zeta) \exp (-\mi k(\zeta)(x 
\! + \! 2 \lambda (\zeta)t) \mathrm{ad}(\sigma_{3})) \widetilde{\mathcal{
G}}^{\sharp}(\zeta), \quad \zeta \! \in \! \mathbb{R};
\end{equation*}
\item[(iii)] $\mathcal{X}(\zeta)$ has simple poles in $\widetilde{\sigma}_{
d}$ with
\begin{align*}
\mathrm{Res}(\mathcal{X}(\zeta);\varsigma_{n}) &= \! \lim_{\zeta \to 
\varsigma_{n}} \mathcal{X}(\zeta)g_{n}(\widetilde{\delta}(\varsigma_{n}))^{
-2} \! \left(\prod_{k=1}^{m-1}(d_{k}^{+}(\varsigma_{n}))^{-2} \right) \! 
\sigma_{-}, & n \! &\in \! \{m,m \! + \! 1,\ldots,N\}, \\
\mathrm{Res}(\mathcal{X}(\zeta);\overline{\varsigma_{n}}) &= \! \sigma_{1} 
\overline{\mathrm{Res}(\mathcal{X}(\zeta);\varsigma_{n})} \, \sigma_{1}, & n 
\! &\in \! \{m,m \! + \! 1,\ldots,N\};
\end{align*}
\item[(iv)] $\det (\mathcal{X}(\zeta)) \vert_{\zeta =\pm 1} \! = \! 0;$
\item[(v)] $\mathcal{X}(\zeta) \! =_{\zeta \to 0} \! \zeta^{-1}(\widetilde{
\delta}(0))^{\sigma_{3}} \! \left(\prod_{k=1}^{m-1}(d_{k}^{+}(0))^{\sigma_{
3}} \right) \! \sigma_{2} \! + \! \mathcal{O}(1);$
\item[(vi)] $\mathcal{X}(\zeta) \! =_{\genfrac{}{}{0pt}{2}{\zeta \to \infty}
{\zeta \in \mathbb{C} \setminus (\widetilde{\sigma}_{d} \cup \sigma_{c})}} 
\! \mathrm{I} \! + \! \mathcal{O}(\zeta^{-1});$
\item[(vii)] $\mathcal{X}(\zeta) \! = \! \sigma_{1} \overline{\mathcal{X}
(\overline{\zeta})} \, \sigma_{1}$ and $\mathcal{X}(\zeta^{-1}) \! = \! 
\zeta \mathcal{X}(\zeta)(\widetilde{\delta}(0))^{\sigma_{3}} \! \left(\prod_{
k=1}^{m-1}(d_{k}^{+}(0))^{\sigma_{3}} \right) \! \sigma_{2}$.
\end{enumerate}
Then, as $t \! \to \! -\infty$ and $x \! \to \! +\infty$ such that $z_{o} 
\! := \! x/t \! < \! -2$ and $(x,t) \! \in \! \daleth_{m}$, $\widetilde{m}^{
\sharp}(\zeta) \colon \mathbb{C} \setminus \sigma^{\prime \prime}_{\mathcal{
O}^{\mathcal{D}}} \! \to \! \mathrm{M}_{2}(\mathbb{C})$ has the following 
asymptotics:
\begin{equation*}
\widetilde{m}^{\sharp}(\zeta) \! = \! \left(\mathrm{I} \! + \! \mathcal{O} \! 
\left(\widetilde{\mathscr{F}}(\zeta) \exp \! \left(-\widetilde{\beth} \vert t 
\vert \right) \right) \right) \! \mathcal{X}(\zeta),
\end{equation*}
where $\widetilde{\beth} \! := \! 4 \min_{\genfrac{}{}{0pt}{2}{m \in \{1,2,
\ldots,N\}}{n \in \{1,2,\ldots,m-1\}}}\{\sin (\phi_{n}) \vert \cos (\phi_{
n}) \! - \! \cos (\phi_{m}) \vert\}$ $(> \! 0)$, and, for $i,j \! \in \! 
\{1,2\}$, $(\widetilde{\mathscr{F}}(\zeta))_{ij} \! =_{\zeta \to \infty} \! 
\mathcal{O}(\vert \zeta \vert^{-1})$ and $(\widetilde{\mathscr{F}}(\zeta))_{
ij} \! =_{\zeta \to 0} \! \mathcal{O}(1)$. Furthermore, let
\begin{gather}
u(x,t) \! := \! \mi \lim_{\genfrac{}{}{0pt}{2}{\zeta \to \infty}{\zeta 
\in \mathbb{C} \setminus (\widetilde{\sigma}_{d} \cup \sigma_{c})}} \! 
\left(\zeta \! \left(\mathcal{X}(\zeta)(\widetilde{\delta}(\zeta))^{\sigma_{
3}} \! \prod_{k=1}^{m-1}(d_{k}^{+}(\zeta))^{\sigma_{3}} \! - \! \mathrm{I} 
\right) \! \right)_{12} \! + \! \mathcal{O} \! \left(\exp \! \left(-
\widetilde{\beth} \vert t \vert \right) \right),
\end{gather}
and
\begin{gather}
\int\nolimits_{+\infty}^{x}(\vert u(x^{\prime},t) \vert^{2} \! - \! 1) \, 
\md x^{\prime} \! := \! -\mi \lim_{\genfrac{}{}{0pt}{2}{\zeta \to \infty}
{\zeta \in \mathbb{C} \setminus (\widetilde{\sigma}_{d} \cup \sigma_{c})}} 
\! \left(\zeta \! \left(\mathcal{X}(\zeta)(\widetilde{\delta}(\zeta))^{
\sigma_{3}} \! \prod_{k=1}^{m-1}(d_{k}^{+}(\zeta))^{\sigma_{3}} \! - \! 
\mathrm{I} \right) \! \right)_{11} \! + \! \mathcal{O} \! \left(\exp \! 
\left(-\widetilde{\beth} \vert t \vert \right) \right).
\end{gather}
Then $u(x,t)$ is the solution of the Cauchy problem for the 
{\rm D${}_{f}$NLSE}.
\end{ay}

The analogue of Lemma~4.1 is
\setcounter{z2}{1}
\setcounter{z3}{5}
\begin{ay}
The solution of the {\rm RHP} for $\mathcal{X}(\zeta) \colon \mathbb{C} 
\setminus (\widetilde{\sigma}_{d} \cup \sigma_{c}) \! \to \! \mathrm{M}_{2}
(\mathbb{C})$ formulated in Lemma~{\rm A.1.4} is given by the following 
ordered factorisation,
\begin{equation*}
\mathcal{X}(\zeta) \! = \! \left(\mathrm{I} \! + \! \zeta^{-1} \widetilde{
\Delta}_{o} \right) \! \widetilde{\mathscr{P}}(\zeta) \widetilde{m}_{d}
(\zeta) \mathscr{M}^{c}(\zeta), \quad \zeta \! \in \! \mathbb{C} \setminus 
(\widetilde{\sigma}_{d} \cup \sigma_{c}),
\end{equation*}
where $\widetilde{m}_{d}(\zeta) \! = \! \sigma_{1} \overline{\widetilde{m}_{
d}(\overline{\zeta})} \, \sigma_{1}$ $(\in \! \mathrm{SL}(2,\mathbb{C}))$ 
has the (series) representation $\widetilde{m}_{d}(\zeta) \! = \! \mathrm{I} 
\! + \! \sum_{n=m}^{N}(\tfrac{\mathrm{Res}(\mathcal{X}(\zeta);\varsigma_{
n})}{\zeta -\varsigma_{n}} \! + \linebreak[4]
\tfrac{\sigma_{1} \overline{\mathrm{Res}(\mathcal{X}(\zeta);\varsigma_{n})} 
\sigma_{1}}{\zeta -\overline{\varsigma_{n}}})$, $\widetilde{\mathscr{P}}
(\zeta) \! = \! \sigma_{1} \overline{\widetilde{\mathscr{P}}(\overline{
\zeta})} \, \sigma_{1}$ is chosen (see Lemma~{\rm A.1.7} below) so that 
$\widetilde{\Delta}_{o}$ is idempotent, $\mathrm{I} \! + \! \zeta^{-1} 
\widetilde{\Delta}_{o}$ is holomorphic in a punctured neighbourhood of the 
origin, with $\widetilde{\Delta}_{o} \! = \! \sigma_{1} \overline{\widetilde{
\Delta}_{o}} \, \sigma_{1}$ $(\in \! \mathrm{GL}(2,\mathbb{C}))$ and $\det 
(\mathrm{I} \! + \! \zeta^{-1} \widetilde{\Delta}_{o}) \vert_{\zeta 
=\pm 1} \! = \! 0$, and determined by $\widetilde{\Delta}_{o} 
\! = \! \widetilde{\mathscr{P}}(0) \widetilde{m}_{d}(0) \mathscr{M}^{c}(0)
(\widetilde{\delta}(0))^{\sigma_{3}} \! \left(\prod_{k=1}^{m-1}(d_{k}^{+}
(0))^{\sigma_{3}} \right) \! \sigma_{2}$, and satisfying $\mathrm{tr}
(\widetilde{\Delta}_{o}) \! = \! 0$, $\det (\widetilde{\Delta}_{o}) \! = \! 
-1$, and $\widetilde{\Delta}_{o} \widetilde{\Delta}_{o} \! = \! \mathrm{I}$, 
and $\mathscr{M}^{c}(\zeta) \colon \mathbb{C} \setminus \sigma_{c} \! \to 
\! \mathrm{SL}(2,\mathbb{C})$ solves the following {\rm RHP:} {\rm (1)} 
$\mathscr{M}^{c}(\zeta)$ is piecewise (sectionally) holomorphic $\forall \, 
\zeta \! \in \! \mathbb{C} \setminus \sigma_{c};$ {\rm (2)} $\mathscr{M}^{
c}_{\pm}(\zeta) \! := \! \lim_{\genfrac{}{}{0pt}{2}{\zeta^{\prime} \to 
\zeta}{\pm \Im (\zeta^{\prime})>0}} \mathscr{M}^{c}(\zeta^{\prime})$ satisfy, 
for $\zeta \! \in \! \mathbb{R}$, the jump condition
\begin{equation*}
\mathscr{M}^{c}_{+}(\zeta) \! = \! \mathscr{M}^{c}_{-}(\zeta) \me^{-\mi 
k(\zeta)(x+2\lambda (\zeta)t) \mathrm{ad}(\sigma_{3})} \! 
\left(
\begin{smallmatrix}
(1-r(\zeta) \overline{r(\overline{\zeta})}) \widetilde{\delta}_{-}(\zeta)/
\widetilde{\delta}_{+}(\zeta) & \, \, -\frac{\overline{r(\overline{\zeta})}}
{(\widetilde{\delta}_{-}(\zeta) \widetilde{\delta}_{+}(\zeta))^{-1}} \prod_{
k=1}^{m-1}(d_{k}^{+}(\zeta))^{2} \\
\frac{r(\zeta)}{\widetilde{\delta}_{-}(\zeta) \widetilde{\delta}_{+}(\zeta)} 
\prod_{k=1}^{m-1}(d_{k}^{+}(\zeta))^{-2} & \, \, \widetilde{\delta}_{+}(\zeta)
/\widetilde{\delta}_{-}(\zeta)
\end{smallmatrix}
\right);
\end{equation*}
{\rm (3)} $\mathscr{M}^{c}(\zeta) \! =_{\genfrac{}{}{0pt}{2}{\zeta \to \infty}
{\zeta \in \mathbb{C} \setminus \sigma_{c}}} \! \mathrm{I} \! + \! \mathcal{
O}(\zeta^{-1});$ and {\rm (4)} $\mathscr{M}^{c}(\zeta) \! = \! \sigma_{1} 
\overline{\mathscr{M}^{c}(\overline{\zeta})} \, \sigma_{1}$.
\end{ay}

The analogue of Lemma~4.2 is
\setcounter{z2}{1}
\setcounter{z3}{6}
\begin{ay}
Let $\varepsilon$ be an arbitrarily fixed, sufficiently small positive real 
number, and, for $z\! \in \! \{\lambda_{1},\lambda_{2}\}$, with $\lambda_{1}$ 
and $\lambda_{2}$ given in Theorem~{\rm 2.2.1}, Eq.~{\rm (10)}, set  $\mathbb{
U}(z;\varepsilon) \! := \! \{\mathstrut \zeta; \, \vert \zeta \! - \! z \vert 
\! < \! \varepsilon\}$. Then, as $t \! \to \! -\infty$ and $x \! \to \! 
+\infty$ such that $z_{o} \! := \! x/t \! < \! -2$, for $\zeta \! \in \! 
\mathbb{C} \setminus \cup_{z \in \{\lambda_{1},\lambda_{2}\}} \mathbb{U}(z;
\varepsilon)$, $\mathscr{M}^{c}(\zeta)$ has the following asymptotics:
\begin{align*}
\mathscr{M}^{c}_{11}(\zeta) \! &= \! 1 \! + \! \mathcal{O} \! \left( \! 
\left(\dfrac{c^{\mathcal{S}}(\lambda_{1}) \underline{c}(\lambda_{2},\lambda_{
3},\overline{\lambda_{3}})}{\sqrt{\lambda_{2}(z_{o}^{2} \! + \! 32)} \, 
(\zeta \! - \! \lambda_{1})} \! + \! \dfrac{c^{\mathcal{S}}(\lambda_{2}) 
\underline{c}(\lambda_{1},\lambda_{3},\overline{\lambda_{3}})}{\sqrt{\lambda_{
1}(z_{o}^{2} \! + \! 32)} \, (\zeta \! - \! \lambda_{2})} \right) \! \dfrac{
\ln \vert t \vert}{(\lambda_{1} \! - \! \lambda_{2}) t} \right), \\
\mathscr{M}^{c}_{12}(\zeta) \! &= \! \me^{\frac{\mi \Xi^{-}(0)}{2}} \! \left(
\dfrac{\sqrt{\nu (\lambda_{1})} \, \lambda_{1}^{-2 \mi \nu (\lambda_{1})}}{
\sqrt{\vert t \vert (\lambda_{1} \! - \! \lambda_{2})} \, (z_{o}^{2} \! + \! 
32)^{1/4}} \! \left( \dfrac{\lambda_{1} \me^{\mi (\Theta^{-}(z_{o},t)-\frac{
3 \pi}{4})}}{(\zeta \! - \! \lambda_{1})} \! + \! \dfrac{\lambda_{2} \me^{
-\mi (\Theta^{-}(z_{o},t)-\frac{3 \pi}{4})}}{(\zeta \! - \! \lambda_{2})} \! 
\right) \right. \\
&+ \left. \! \mathcal{O} \! \left( \! \left(\dfrac{c^{\mathcal{S}}(\lambda_{
1}) \underline{c}(\lambda_{2},\lambda_{3},\overline{\lambda_{3}})}{\sqrt{
\lambda_{2}(z_{o}^{2} \! + \! 32)} \, (\zeta \! - \! \lambda_{1})} \! + \! 
\dfrac{c^{\mathcal{S}}(\lambda_{2}) \underline{c}(\lambda_{1},\lambda_{3},
\overline{\lambda_{3}})}{\sqrt{\lambda_{1}(z_{o}^{2} \! + \! 32)} \, (\zeta 
\! - \! \lambda_{2})} \right) \! \dfrac{\ln \vert t \vert}{(\lambda_{1} \! - 
\! \lambda_{2}) t} \right) \! \right),
\end{align*}
\begin{align*}
\mathscr{M}^{c}_{21}(\zeta) \! &= \! \me^{-\frac{\mi \Xi^{-}(0)}{2}} \! 
\left(\dfrac{\sqrt{\nu (\lambda_{1})} \, \lambda_{1}^{2 \mi \nu (\lambda_{
1})}}{\sqrt{\vert t \vert (\lambda_{1} \! - \! \lambda_{2})} \, (z_{o}^{2} \! 
+ \! 32)^{1/4}} \! \left(\dfrac{\lambda_{1} \me^{-\mi (\Theta^{-}(z_{o},t)-
\frac{3 \pi}{4})}}{(\zeta \! - \! \lambda_{1})} \! + \! \dfrac{\lambda_{2} 
\me^{\mi (\Theta^{-}(z_{o},t)-\frac{3 \pi}{4})}}{(\zeta \! - \! \lambda_{2})} 
\! \right) \right. \\
&+ \left. \! \mathcal{O} \! \left( \! \left(\dfrac{c^{\mathcal{S}}(\lambda_{
1}) \underline{c}(\lambda_{2},\lambda_{3},\overline{\lambda_{3}})}{\sqrt{
\lambda_{2}(z_{o}^{2} \! + \! 32)} \, (\zeta \! - \! \lambda_{1})} \! + \! 
\dfrac{c^{\mathcal{S}}(\lambda_{2}) \underline{c}(\lambda_{1},\lambda_{3},
\overline{\lambda_{3}})}{\sqrt{\lambda_{1}(z_{o}^{2} \! + \! 32)} \, (\zeta 
\! - \! \lambda_{2})} \right) \! \dfrac{\ln \vert t \vert}{(\lambda_{1} \! - 
\! \lambda_{2}) t} \right) \! \right), \\
\mathscr{M}^{c}_{22}(\zeta) \! &= \! 1 \! + \! \mathcal{O} \! \left( \! \left(
\dfrac{c^{\mathcal{S}}(\lambda_{1}) \underline{c}(\lambda_{2},\lambda_{3},
\overline{\lambda_{3}})}{\sqrt{\lambda_{2}(z_{o}^{2} \! + \! 32)} \, (\zeta 
\! - \! \lambda_{1})} \! + \! \dfrac{c^{\mathcal{S}}(\lambda_{2}) \underline{
c}(\lambda_{1},\lambda_{3},\overline{\lambda_{3}})}{\sqrt{\lambda_{1}(z_{o}^{
2} \! + \! 32)} \, (\zeta \! - \! \lambda_{2})} \right) \! \dfrac{\ln \vert t 
\vert}{(\lambda_{1} \! - \! \lambda_{2}) t} \right),
\end{align*}
where $\lambda_{3}$, $\nu (\cdot)$, $\Theta^{-}(z_{o},t)$, and $\Xi^{-}
(\cdot)$, respectively, are given in Theorem~{\rm 2.2.1}, Eqs.~{\rm (10)}, 
{\rm (11)}, {\rm (17)}, and~{\rm (19)}, $\vert \vert (\cdot \! - \! 
\lambda_{k})^{-1} \vert \vert_{\mathcal{L}^{\infty}(\mathbb{C} \, \setminus 
\cup_{z \in \{\lambda_{1},\lambda_{2}\}} \mathbb{U}(z;\varepsilon))} \! < \! 
\infty$, $k \! \in \! \{1,2\}$, $\mathscr{M}^{c}(\zeta) \! = \! \sigma_{1} 
\overline{\mathscr{M}^{c}(\overline{\zeta})} \, \sigma_{1}$, and $(\mathscr{
M}^{c}(0) \sigma_{2})^{2} \! = \! \mathrm{I}$ $(+ \, \mathcal{O}(t^{-1} \ln 
\vert t \vert))$.
\end{ay}

\emph{Sketch of Proof.} Proceeding as in the proof of Lemma~6.1 in \cite{a38} 
and particularising it to the case of the RHP for $\mathscr{M}^{c}(\zeta)$ 
stated in Lemma~A.1.5, one arrives at
\begin{align*}
\mathscr{M}^{c}_{11}(\zeta) \! &= \! 1 \! + \! \tfrac{\widetilde{r}(\lambda_{
1})(\widetilde{\delta}_{B}^{0})^{-2} \me^{-\frac{3 \pi \nu}{2}} \me^{\frac{3 
\pi \mi}{4}}}{2 \pi \mi (\zeta -\lambda_{1}) \widetilde{\beta}^{\widetilde{
\Sigma}_{B^{0}}}_{21} \widetilde{\mathcal{X}}_{B} \sqrt{\vert t \vert}} 
\int\nolimits_{0}^{+\infty}(\me^{-\frac{3 \pi \mi}{4}} \partial_{z} \mathbf{
D}_{\mi \nu}(z) \! + \! \tfrac{\mi}{2} \me^{\frac{3 \pi \mi}{4}}z \mathbf{
D}_{\mi \nu}(z))z^{\mi \nu} \me^{-\frac{z^{2}}{4}} \, \md z \\
&- \tfrac{\widetilde{r}(\lambda_{1})(1-\vert \widetilde{r}(\lambda_{1}) 
\vert^{2})^{-1}(\widetilde{\delta}_{B}^{0})^{-2} \me^{-\frac{\mi \pi}{4}}}{2 
\pi \mi (\zeta -\lambda_{1}) \widetilde{\beta}^{\widetilde{\Sigma}_{B^{0}}}_{
21} \me^{-\frac{\pi \nu}{2}} \widetilde{\mathcal{X}}_{B} \sqrt{\vert t \vert}
} \int\nolimits_{0}^{+\infty}(\me^{\frac{\mi \pi}{4}} \partial_{z} \mathbf{
D}_{\mi \nu}(z) \! + \! \tfrac{\mi}{2} \me^{-\frac{\mi \pi}{4}}z \mathbf{D}_{
\mi \nu}(z))z^{\mi \nu} \me^{-\frac{z^{2}}{4}} \, \md z \\
&+ \tfrac{\overline{\widetilde{r}(\lambda_{1})}(\widetilde{\delta}_{A}^{0})^{
-2} \me^{-\frac{\pi \nu}{2}}(-1)^{-\mi \nu} \me^{\frac{\mi \pi}{4}}}{2 \pi 
\mi (\zeta -\lambda_{2}) \widetilde{\beta}^{\widetilde{\Sigma}_{A^{0}}}_{21} 
\widetilde{\mathcal{X}}_{A} \sqrt{\vert t \vert}} \int\nolimits_{0}^{+\infty}
(\me^{-\frac{\mi \pi}{4}} \partial_{z} \mathbf{D}_{-\mi \nu}(z) \! - \! 
\tfrac{\mi}{2} \me^{\frac{\mi \pi}{4}}z \mathbf{D}_{-\mi \nu}(z))z^{-\mi 
\nu} \me^{-\frac{z^{2}}{4}} \, \md z \\
&- \tfrac{\overline{\widetilde{r}(\lambda_{1})}(1 -\vert \widetilde{r}
(\lambda_{1}) \vert^{2})^{-1}(\widetilde{\delta}_{A}^{0})^{-2} \me^{-\frac{3 
\pi \mi}{4}}}{2 \pi \mi (\zeta -\lambda_{2}) \widetilde{\beta}^{\widetilde{
\Sigma}_{A^{0}}}_{21} \me^{\frac{\pi \nu}{2}}(-1)^{\mi \nu} \widetilde{
\mathcal{X}}_{A} \sqrt{\vert t \vert}} \int\nolimits_{0}^{+\infty}(\me^{
\frac{3 \pi \mi}{4}} \partial_{z} \mathbf{D}_{-\mi \nu}(z) \! - \! \tfrac{
\mi}{2} \me^{-\frac{3 \pi \mi}{4}}z \mathbf{D}_{-\mi \nu}(z))z^{-\mi \nu} 
\me^{-\frac{z^{2}}{4}} \, \md z \\
&+ \mathcal{O} \! \left( \! \left( \tfrac{c^{\mathcal{S}}(\lambda_{1}) 
\underline{c}(\lambda_{2},\lambda_{3},\overline{\lambda_{3}})(\widetilde{
\delta}_{B}^{0})^{-2}}{(\zeta -\lambda_{1}) \vert \lambda_{1}-\lambda_{3} 
\vert \sqrt{(\lambda_{1}-\lambda_{2})} \, \, \widetilde{\mathcal{X}}_{B}} \! 
+ \! \tfrac{c^{\mathcal{S}}(\lambda_{2}) \underline{c}(\lambda_{1},\lambda_{
3},\overline{\lambda_{3}})(\widetilde{\delta}_{A}^{0})^{-2}}{(\zeta -\lambda_{
2}) \vert \lambda_{2}-\lambda_{3} \vert \sqrt{(\lambda_{1}-\lambda_{2})} \, 
\, \widetilde{\mathcal{X}}_{A}} \right) \! \tfrac{\ln \vert t \vert}{t} 
\right), \\
\mathscr{M}^{c}_{12}(\zeta) \! &= \! \left( \tfrac{\overline{\widetilde{r}
(\lambda_{1})}(1-\vert \widetilde{r}(\lambda_{1}) \vert^{2})^{-1}(\widetilde{
\delta}_{B}^{0})^{2} \me^{\frac{\mi \pi}{4}}}{2 \pi \mi (\zeta -\lambda_{1}) 
\me^{-\frac{\pi \nu}{2}} \widetilde{\mathcal{X}}_{B} \sqrt{\vert t \vert}} \! 
- \! \tfrac{\overline{\widetilde{r}(\lambda_{1})}(\widetilde{\delta}_{B}^{
0})^{2} \me^{-\frac{3 \pi \nu}{2}} \me^{-\frac{3 \pi \mi}{4}}}{2 \pi \mi 
(\zeta -\lambda_{1}) \widetilde{\mathcal{X}}_{B} \sqrt{\vert t \vert}} 
\right) \! \int\nolimits_{0}^{+\infty} \mathbf{D}_{-\mi \nu}(z)z^{-\mi \nu} 
\me^{-\frac{z^{2}}{4}} \, \md z \\
&+ \left( \tfrac{\widetilde{r}(\lambda_{1})(1-\vert \widetilde{r}(\lambda_{
1}) \vert^{2})^{-1}(\widetilde{\delta}_{A}^{0})^{2} \me^{\frac{3 \pi \mi}{
4}}}{2 \pi \mi (\zeta -\lambda_{2}) \me^{\frac{\pi \nu}{2}}(-1)^{-\mi \nu} 
\widetilde{\mathcal{X}}_{A} \sqrt{\vert t \vert}} \! - \! \tfrac{\widetilde{
r}(\lambda_{1})(\widetilde{\delta}_{A}^{0})^{2} \me^{-\frac{\pi \nu}{2}} 
\me^{-\frac{\mi \pi}{4}}}{2 \pi \mi (\zeta -\lambda_{2})(-1)^{-\mi \nu} 
\widetilde{\mathcal{X}}_{A} \sqrt{\vert t \vert}} \right) \! \int\nolimits_{
0}^{+\infty} \mathbf{D}_{\mi \nu}(z)z^{\mi \nu} \me^{-\frac{z^{2}}{4}} \, 
\md z \\
&+ \mathcal{O} \! \left( \! \left( \tfrac{c^{\mathcal{S}}(\lambda_{1}) 
\underline{c}(\lambda_{2},\lambda_{3},\overline{\lambda_{3}})(\widetilde{
\delta}_{B}^{0})^{2}}{(\zeta -\lambda_{1}) \vert \lambda_{1}-\lambda_{3} 
\vert \sqrt{(\lambda_{1}-\lambda_{2})} \, \, \widetilde{\mathcal{X}}_{B}} \! 
+ \! \tfrac{c^{\mathcal{S}}(\lambda_{2}) \underline{c}(\lambda_{1},\lambda_{
3},\overline{\lambda_{3}})(\widetilde{\delta}_{A}^{0})^{2}}{(\zeta -\lambda_{
2}) \vert \lambda_{2}-\lambda_{3} \vert \sqrt{(\lambda_{1}-\lambda_{2})} \, 
\, \widetilde{\mathcal{X}}_{A}} \right) \! \tfrac{\ln \vert t \vert}{t} 
\right), \\
\mathscr{M}^{c}_{21}(\zeta) \! &= \! -\left(\tfrac{\widetilde{r}(\lambda_{1})
(1-\vert \widetilde{r}(\lambda_{1}) \vert^{2})^{-1}(\widetilde{\delta}_{B}^{
0})^{-2} \me^{-\frac{\mi \pi}{4}}}{2 \pi \mi (\zeta -\lambda_{1}) \me^{-
\frac{\pi \nu}{2}} \widetilde{\mathcal{X}}_{B} \sqrt{\vert t \vert}} \! - \! 
\tfrac{\widetilde{r}(\lambda_{1})(\widetilde{\delta}_{B}^{0})^{-2} \me^{-
\frac{3 \pi \nu}{2}} \me^{\frac{3 \pi \mi}{4}}}{2 \pi \mi (\zeta -\lambda_{
1}) \widetilde{\mathcal{X}}_{B} \sqrt{\vert t \vert}} \right) \! 
\int\nolimits_{0}^{+\infty} \mathbf{D}_{\mi \nu}(z)z^{\mi \nu} \me^{-\frac{
z^{2}}{4}} \, \md z \\
&- \left( \tfrac{\overline{\widetilde{r}(\lambda_{1})}(1-\vert \widetilde{r}
(\lambda_{1}) \vert^{2})^{-1}(\widetilde{\delta}_{A}^{0})^{-2} \me^{-\frac{3 
\pi \mi}{4}}}{2 \pi \mi (\zeta -\lambda_{2}) \me^{\frac{\pi \nu}{2}}(-1)^{
\mi \nu} \widetilde{\mathcal{X}}_{A} \sqrt{\vert t \vert}} \! - \! \tfrac{
\overline{\widetilde{r}(\lambda_{1})}(\widetilde{\delta}_{A}^{0})^{-2} \me^{
-\frac{\pi \nu}{2}} \me^{\frac{\mi \pi}{4}}}{2 \pi \mi (\zeta -\lambda_{2})
(-1)^{\mi \nu} \widetilde{\mathcal{X}}_{A} \sqrt{\vert t \vert}} \right) \! 
\int\nolimits_{0}^{+\infty} \mathbf{D}_{-\mi \nu}(z)z^{-\mi \nu} \me^{-\frac{
z^{2}}{4}} \, \md z \\
&+ \mathcal{O} \! \left( \! \left( \tfrac{c^{\mathcal{S}}(\lambda_{1}) 
\underline{c}(\lambda_{2},\lambda_{3},\overline{\lambda_{3}})(\widetilde{
\delta}_{B}^{0})^{-2}}{(\zeta -\lambda_{1}) \vert \lambda_{1}-\lambda_{3} 
\vert \sqrt{(\lambda_{1}-\lambda_{2})} \, \, \widetilde{\mathcal{X}}_{B}} \! 
+ \! \tfrac{c^{\mathcal{S}}(\lambda_{2}) \underline{c}(\lambda_{1},\lambda_{
3},\overline{\lambda_{3}})(\widetilde{\delta}_{A}^{0})^{-2}}{(\zeta -\lambda_{
2}) \vert \lambda_{2}-\lambda_{3} \vert \sqrt{(\lambda_{1}-\lambda_{2})} \, 
\, \widetilde{\mathcal{X}}_{A}} \right) \! \tfrac{\ln \vert t \vert}{t} 
\right), \\
\mathscr{M}^{c}_{22}(\zeta) \! &= \! 1 \! - \! \tfrac{\overline{\widetilde{r}
(\lambda_{1})}(\widetilde{\delta}_{B}^{0})^{2} \me^{-\frac{3 \pi \nu}{2}} 
\me^{-\frac{3 \pi \mi}{4}}}{2 \pi \mi (\zeta -\lambda_{1}) \widetilde{
\beta}^{\widetilde{\Sigma}_{B^{0}}}_{12} \widetilde{\mathcal{X}}_{B} \sqrt{
\vert t \vert}} \int\nolimits_{0}^{+\infty}(\me^{\frac{3 \pi \mi}{4}} 
\partial_{z} \mathbf{D}_{-\mi \nu}(z) \! - \! \tfrac{\mi}{2} \me^{-\frac{3 
\pi \mi}{4}}z \mathbf{D}_{-\mi \nu}(z))z^{-\mi \nu} \me^{-\frac{z^{2}}{4}} 
\, \md z \\
&+ \tfrac{\overline{\widetilde{r}(\lambda_{1})}(1-\vert \widetilde{r}
(\lambda_{1}) \vert^{2})^{-1}(\widetilde{\delta}_{B}^{0})^{2} \me^{\frac{\mi 
\pi}{4}}}{2 \pi \mi (\zeta -\lambda_{1}) \widetilde{\beta}^{\widetilde{
\Sigma}_{B^{0}}}_{12} \me^{-\frac{\pi \nu}{2}} \widetilde{\mathcal{X}}_{B} 
\sqrt{\vert t \vert}} \int\nolimits_{0}^{+\infty}(\me^{-\frac{\mi \pi}{4}} 
\partial_{z} \mathbf{D}_{-\mi \nu}(z) \! - \! \tfrac{\mi}{2} \me^{\frac{\mi 
\pi}{4}}z \mathbf{D}_{-\mi \nu}(z))z^{-\mi \nu} \me^{-\frac{z^{2}}{4}} \, 
\md z \\
&- \tfrac{\widetilde{r}(\lambda_{1})(\widetilde{\delta}_{A}^{0})^{2} \me^{
-\frac{\pi \nu}{2}}(-1)^{\mi \nu} \me^{-\frac{\mi \pi}{4}}}{2 \pi \mi (\zeta 
-\lambda_{2}) \widetilde{\beta}^{\widetilde{\Sigma}_{A^{0}}}_{12} \widetilde{
\mathcal{X}}_{A} \sqrt{\vert t \vert}} \int\nolimits_{0}^{+\infty}(\me^{
\frac{\mi \pi}{4}} \partial_{z} \mathbf{D}_{\mi \nu}(z) \! + \! \tfrac{\mi}
{2} \me^{-\frac{\mi \pi}{4}}z \mathbf{D}_{\mi \nu}(z))z^{\mi \nu} \me^{-
\frac{z^{2}}{4}} \, \md z \\
&+ \tfrac{\widetilde{r}(\lambda_{1})(1-\vert \widetilde{r}(\lambda_{1}) 
\vert^{2})^{-1}(\widetilde{\delta}_{A}^{0})^{2}(-1)^{\mi \nu} \me^{\frac{3 
\pi \mi}{4}}}{2 \pi \mi (\zeta -\lambda_{2}) \widetilde{\beta}^{\widetilde{
\Sigma}_{A^{0}}}_{12} \me^{\frac{\pi \nu}{2}} \widetilde{\mathcal{X}}_{A} 
\sqrt{\vert t \vert}} \int\nolimits_{0}^{+\infty}(\me^{-\frac{3 \pi \mi}{4}} 
\partial_{z} \mathbf{D}_{\mi \nu}(z) \! + \! \tfrac{\mi}{2} \me^{\frac{3 \pi 
\mi}{4}}z \mathbf{D}_{\mi \nu}(z))z^{\mi \nu} \me^{-\frac{z^{2}}{4}} \, \md 
z \\
&+ \mathcal{O} \! \left( \! \left( \tfrac{c^{\mathcal{S}}(\lambda_{1}) 
\underline{c}(\lambda_{2},\lambda_{3},\overline{\lambda_{3}})(\widetilde{
\delta}_{B}^{0})^{2}}{(\zeta -\lambda_{1}) \vert \lambda_{1}-\lambda_{3} 
\vert \sqrt{(\lambda_{1}-\lambda_{2})} \, \, \widetilde{\mathcal{X}}_{B}} \! 
+ \! \tfrac{c^{\mathcal{S}}(\lambda_{2}) \underline{c}(\lambda_{1},\lambda_{
3},\overline{\lambda_{3}})(\widetilde{\delta}_{A}^{0})^{2}}{(\zeta -\lambda_{
2}) \vert \lambda_{2}-\lambda_{3} \vert \sqrt{(\lambda_{1}-\lambda_{2})} 
\, \, \widetilde{\mathcal{X}}_{A}} \right) \! \tfrac{\ln \vert t \vert}{t} 
\right),
\end{align*}
where $\widetilde{r}(\zeta) \! = \! r(\zeta) \prod_{k=1}^{m-1}(d_{k}^{+}
(\zeta))^{-2}$ $\left(\vert \widetilde{r}(\lambda_{1}) \vert \! = \! \vert 
r(\lambda_{1}) \vert \right)$, $\nu \! = \! \nu (\lambda_{1})$,
\begin{gather*}
\widetilde{\delta}_{B}^{0} \! = \! \vert \lambda_{1} \! - \! \lambda_{3} 
\vert^{\mi \nu} \! \left(2 \vert t \vert (\lambda_{1} \! - \! \lambda_{2})^{
3} \lambda_{1}^{-3} \right)^{\frac{\mi \nu}{2}} \me^{\mathcal{Y}(\lambda_{
1})} \exp \! \left(-\tfrac{\mi t}{2}(\lambda_{1} \! - \! \lambda_{2})(z_{o} 
\! + \! \lambda_{1} \! + \! \lambda_{2}) \right), \\
\widetilde{\delta}_{A}^{0} \! = \! \vert \lambda_{2} \! - \! \lambda_{3} 
\vert^{-\mi \nu} \! \left(2 \vert t \vert (\lambda_{1} \! - \! \lambda_{2})^{
3} \lambda_{2}^{-3} \right)^{-\frac{\mi \nu}{2}} \me^{\mathcal{Y}(\lambda_{
2})} \exp \! \left(\tfrac{\mi t}{2}(\lambda_{1} \! - \! \lambda_{2})(z_{o} \! 
+ \! \lambda_{1} \! + \! \lambda_{2}) \right), \\
\mathcal{Y}(\lambda_{1}) \! = \! \dfrac{\mi}{2 \pi} \int_{0}^{\lambda_{2}} 
\ln \vert \mu \! - \! \lambda_{1} \vert \md \ln (1 \! - \! \vert r(\mu) 
\vert^{2}) \! + \! \dfrac{\mi}{2 \pi} \int_{\lambda_{1}}^{+\infty} \ln \vert 
\mu \! - \! \lambda_{1} \vert \md \ln (1 \! - \! \vert r(\mu) \vert^{2}), \\
\mathcal{Y}(\lambda_{2}) \! = \! -\mathcal{Y}(\lambda_{1}) \! + \! \dfrac{
\mi}{2 \pi} \int_{0}^{\lambda_{2}} \ln \vert \mu \vert \md \ln (1 \! - \! 
\vert r(\mu) \vert^{2}) \! + \! \dfrac{\mi}{2 \pi} \int_{\lambda_{1}}^{
+\infty} \ln \vert \mu \vert \md \ln (1 \! - \! \vert r(\mu) \vert^{2}), \\
\widetilde{\mathcal{X}}_{B} \! = \! \mathcal{X}_{B}, \quad \widetilde{
\mathcal{X}}_{A} \! = \! \mathcal{X}_{A}, \quad \widetilde{\beta}^{
\widetilde{\Sigma}_{B^{0}}}_{12} \! = \, \overline{\widetilde{\beta}^{
\widetilde{\Sigma}_{B^{0}}}_{21}} = \! \tfrac{\sqrt{2 \pi} \, \me^{-\frac{
\pi \nu}{2}} \me^{\frac{3 \pi \mi}{4}}}{\widetilde{r}(\lambda_{1}) \Gamma 
(\mi \nu)}, \quad \quad \widetilde{\beta}^{\widetilde{\Sigma}_{A^{0}}}_{12} 
\! = \, \overline{\widetilde{\beta}^{\widetilde{\Sigma}_{A^{0}}}_{21}} = \! 
\tfrac{\sqrt{2 \pi} \, \me^{-\frac{\pi \nu}{2}} \me^{-\frac{3 \pi \mi}{4}}}
{\overline{\widetilde{r}(\lambda_{1})} \, \overline{\Gamma (\mi \nu)}},
\end{gather*}
$\Gamma (\cdot)$ is the gamma function \cite{a51}, and $\mathbf{D}_{\ast}
(\cdot)$ is the parabolic cylinder function \cite{a51}. Proceeding, now, as 
at the end of the sketch of the proof of Lemma~4.2, one obtains the result 
stated in the Lemma. Furthermore, one shows that the symmetry reduction 
$\mathscr{M}^{c}(\zeta) \! = \! \sigma_{1} \overline{\mathscr{M}^{c}
(\overline{\zeta})} \, \sigma_{1}$ is satisfied, and verifies that 
$(\mathscr{M}^{c}(0) \sigma_{2})^{2} \! = \! \mathrm{I} \! + \! \mathcal{O}
(t^{-1} \ln \vert t \vert)$. \hfill $\square$

The analogue of Proposition~4.1 is
\setcounter{z0}{1}
\setcounter{z1}{2}
\begin{by}
For $m \! \in \! \{1,2,\ldots,N\}$, set $\mathrm{Res}(\mathcal{X}(\zeta);
\varsigma_{n}) \! := \! 
\left(
\begin{smallmatrix}
\mathfrak{a}_{n} & \mathfrak{b}_{n} \\
\mathfrak{c}_{n} & \mathfrak{d}_{n}
\end{smallmatrix}
\right)$, $n \! \in \! \{m,m \! + \! 1,\ldots,N\}$. Then $\mathfrak{b}_{n} \! 
= \! -\mathfrak{a}_{n} \mathscr{M}^{c}_{12}(\varsigma_{n})/\mathscr{M}^{c}_{
22}(\varsigma_{n})$, $\mathfrak{d}_{n} \! = \! -\mathfrak{c}_{n} \mathscr{
M}^{c}_{12}(\varsigma_{n})/\mathscr{M}^{c}_{22}(\varsigma_{n})$, and 
$\{\mathfrak{a}_{n},\overline{\mathfrak{c}_{n}}\}_{n=m}^{N}$ satisfy the 
following (non-singular) system of $2(N \! - \! m \! + \! 1)$ linear 
inhomogeneous algebraic equations,
\begin{eqnarray*}
\left[\begin{array}{cccccc}
\left. \begin{array}{ccc} \cline{1-3}
\multicolumn{1}{|c}{} &   & \multicolumn{1}{c|}{} \\
\multicolumn{1}{|c}{} & \widetilde{\mathscr{A}} & 
\multicolumn{1}{c|}{} \\
\multicolumn{1}{|c}{} &   & \multicolumn{1}{c|}{} \\ 
\cline{1-3}   
\end{array} \right. & \left. \begin{array}{ccc} 
\cline{1-3}
\multicolumn{1}{|c}{} &   & \multicolumn{1}{c|}{} \\
\multicolumn{1}{|c}{} & \widetilde{\mathscr{B}} & 
\multicolumn{1}{c|}{} \\
\multicolumn{1}{|c}{} &   & \multicolumn{1}{c|}{} \\ 
\cline{1-3}   
\end{array} \right. \\
& \\
\left. \begin{array}{ccc} \cline{1-3}
\multicolumn{1}{|c}{} &   & \multicolumn{1}{c|}{} \\
\multicolumn{1}{|c}{} & \overline{\widetilde{\mathscr{B}}} & 
\multicolumn{1}{c|}{} \\
\multicolumn{1}{|c}{} &   & \multicolumn{1}{c|}{} \\ 
\cline{1-3}   
\end{array} \right. & \left. \begin{array}{ccc} 
\cline{1-3}
\multicolumn{1}{|c}{} &   & \multicolumn{1}{c|}{} \\
\multicolumn{1}{|c}{} & \overline{\widetilde{\mathscr{A}} \,} & 
\multicolumn{1}{c|}{} \\
\multicolumn{1}{|c}{} &   & \multicolumn{1}{c|}{} \\ 
\cline{1-3}   
\end{array} \right. 
\end{array} \right] 
\left[\begin{array}{c}
          \mathfrak{a}_{m} \\
          \mathfrak{a}_{m+1} \\
          \vdots \\
          \mathfrak{a}_{N} \\
          \overline{\mathfrak{c}_{m}} \\
          \overline{\mathfrak{c}_{m+1}} \\
          \vdots \\
          \overline{\mathfrak{c}_{N}}
      \end{array} \right] = 
\left[\begin{array}{c}
          g^{\star}_{m} \mathscr{M}^{c}_{12}(\varsigma_{m}) \\
          g^{\star}_{m+1} \mathscr{M}^{c}_{12}(\varsigma_{m+1}) \\
          \vdots \\
          g^{\star}_{N} \mathscr{M}^{c}_{12}(\varsigma_{N}) \\
          \overline{g^{\star}_{m} \mathscr{M}^{c}_{22}(\varsigma_{m})} \\
          \overline{g^{\star}_{m+1} \mathscr{M}^{c}_{22}(\varsigma_{m+1})} \\
          \vdots \\
          \overline{g^{\star}_{N} \mathscr{M}^{c}_{22}(\varsigma_{N})}
       \end{array} \right],
\end{eqnarray*}
where
\begin{align*}
\widetilde{\mathscr{A}}_{ij} \! :=& \! 
\begin{cases}
\dfrac{\det (\mathscr{M}^{c}(\varsigma_{i})) \! + \! g_{i}^{\star} \mathrm{W}
(\mathscr{M}^{c}_{12}(\varsigma_{i}),\mathscr{M}^{c}_{22}(\varsigma_{i}))}{
\mathscr{M}^{c}_{22}(\varsigma_{i})}, &\text{$i \! = \! j \! \in \! \{m,m \! 
+ \! 1,\ldots,N\}$,} \\
-\dfrac{g_{i}^{\star}(\mathscr{M}^{c}_{12}(\varsigma_{i}) \mathscr{M}^{c}_{
22}(\varsigma_{j}) \! - \! \mathscr{M}^{c}_{22}(\varsigma_{i}) \mathscr{M}^{
c}_{12}(\varsigma_{j}))}{(\varsigma_{i} \! - \! \varsigma_{j}) \mathscr{M}^{
c}_{22}(\varsigma_{j})}, 
&\text{$i \! \not= \! j \! \in \! \{m,m \! + \! 1,\ldots,N\}$,}
\end{cases} \\
\widetilde{\mathscr{B}}_{ij} \! :=& \! -\dfrac{g_{i}^{\star}(\mathscr{M}^{
c}_{22}(\varsigma_{i}) \overline{\mathscr{M}^{c}_{22}(\varsigma_{j})} \! - 
\! \mathscr{M}^{c}_{12}(\varsigma_{i}) \overline{\mathscr{M}^{c}_{12}
(\varsigma_{j})})}{(\varsigma_{i} \! - \! \overline{\varsigma_{j}}) 
\overline{\mathscr{M}^{c}_{22}(\varsigma_{j})}}, \quad \, \, i,j \! \in \! 
\{m,m \! + \! 1,\ldots,N\},
\end{align*}
\begin{equation*}
g_{j}^{\star} \! = \! \vert g_{j} \vert \me^{\mi \theta_{g_{j}}} \exp (2 \mi 
k(\varsigma_{j})(x \! + \! 2 \lambda (\varsigma_{j})t))(\widetilde{\delta}
(\varsigma_{j}))^{-2} \! \prod_{k=1}^{m-1}(d_{k}^{+}(\varsigma_{j}))^{-2}, 
\quad j \! \in \! \{m,m \! + \! 1,\ldots,N\},
\end{equation*}
with $\vert g_{j} \vert$ and $\theta_{g_{j}}$ given in Lemma~{\rm 3.1}, 
(iii), and $\mathrm{W}(\mathscr{M}^{c}_{12}(z),\mathscr{M}^{c}_{22}(z)) \! 
= \! 
\left\vert
\begin{smallmatrix}
\mathscr{M}^{c}_{12}(z) & \mathscr{M}^{c}_{22}(z) \\
\partial_{z} \mathscr{M}^{c}_{12}(z) & \partial_{z} \mathscr{M}^{c}_{22}(z)
\end{smallmatrix}
\right\vert$.
\end{by}

The analogue of Proposition~4.2 is
\setcounter{z0}{1}
\setcounter{z1}{3}
\begin{by}
As $t \! \to \! -\infty$ and $x \! \to \! +\infty$ such that $z_{o} \! := \! 
x/t \! < \! -2$ and $(x,t) \! \in \! \daleth_{m}$, $m \! \in \! \{1,2,\ldots,
m\}$, for $n \! \in \! \{m \! + \! 1,m \! + \! 2,\ldots,N\}$,
\begin{gather*}
\mathfrak{a}_{n} \! = \! \mathcal{O} \! \left(\me^{-\gimel^{-} \vert t 
\vert} \right), \quad \qquad \mathfrak{b}_{n} \! = \! \mathcal{O} \! \left(
t^{-1/2}(z_{o}^{2} \! + \! 32)^{-1/4} \me^{-\gimel^{-} \vert t \vert} 
\right), \\
\mathfrak{c}_{n} \! = \! \mathcal{O} \! \left(\me^{-\gimel^{-} \vert t 
\vert} \right), \quad \qquad \mathfrak{d}_{n} \! = \! \mathcal{O} \! \left(
t^{-1/2}(z_{o}^{2} \! + \! 32)^{-1/4} \me^{-\gimel^{-} \vert t \vert} \right),
\end{gather*}
where $\gimel^{-} \! := \! 4 \min_{\genfrac{}{}{0pt}{2}{m \in \{1,2,\ldots,
N\}}{n \in \{m+1,m+2,\ldots,N\}}}\{\sin (\phi_{n}) \vert \cos (\phi_{n}) \! - 
\! \cos (\phi_{m}) \vert\}$ $(> \! 0)$, and
\begin{align*}
\mathfrak{a}_{m} \! &= \, \mathfrak{a}_{m}^{0} \! + \! \tfrac{1}{\sqrt{\vert 
t \vert}} \mathfrak{a}_{m}^{1} \! + \! \mathcal{O} \! \left(\tfrac{c^{
\mathcal{S}}(z_{o})}{(z_{o}^{2}+32)^{1/2}} \tfrac{\ln \vert t \vert}{t} 
\right) \\
&=: \, \tfrac{g_{m}^{\star} \overline{g_{m}^{\star}}(\varsigma_{m}-\overline{
\varsigma_{m}})^{-1}}{(1+g_{m}^{\star} \overline{g_{m}^{\star}}(\varsigma_{
m}-\overline{\varsigma_{m}})^{-2})} \! + \! \tfrac{1}{\sqrt{\vert t \vert}} 
\! \left(\tfrac{g_{m}^{\star} \overline{g_{m}^{\star}}(\varsigma_{m}-
\overline{\varsigma_{m}})^{-1}(g_{m}^{\star} \partial_{\zeta} \widetilde{
\mathscr{M}}^{c}_{12}(\varsigma_{m})+\overline{g_{m}^{\star} \partial_{\zeta} 
\widetilde{\mathscr{M}}^{c}_{12}(\varsigma_{m})})}{(1+g_{m}^{\star} \overline{
g_{m}^{\star}}(\varsigma_{m}-\overline{\varsigma_{m}})^{-2})^{2}} \! + \! 
\tfrac{g_{m}^{\star} \widetilde{\mathscr{M}}^{c}_{12}(\varsigma_{m})}{(1+g_{
m}^{\star} \overline{g_{m}^{\star}}(\varsigma_{m}-\overline{\varsigma_{m}})^{
-2})} \right) \\
&+ \, \mathcal{O} \! \left(\tfrac{c^{\mathcal{S}}(z_{o})}{(z_{o}^{2}+32)^{
1/2}} \tfrac{\ln \vert t \vert}{t} \right), \\
\mathfrak{b}_{m} \! &= \, \tfrac{1}{\sqrt{\vert t \vert}} \mathfrak{b}_{m}^{
1} \! + \! \mathcal{O} \! \left(\tfrac{c^{\mathcal{S}}(z_{o})}{(z_{o}^{2}+
32)^{1/2}} \tfrac{\ln \vert t \vert}{t} \right) \! =: \! -\tfrac{1}{\sqrt{
\vert t \vert}} \tfrac{g_{m}^{\star} \overline{g_{m}^{\star}} \, (\varsigma_{
m}-\overline{\varsigma_{m}})^{-1} \widetilde{\mathscr{M}}^{c}_{12}(\varsigma_{
m})}{(1+g_{m}^{\star} \overline{g_{m}^{\star}}(\varsigma_{m}-\overline{
\varsigma_{m}})^{-2})} \! + \! \mathcal{O} \! \left(\tfrac{c^{\mathcal{S}}
(z_{o})}{(z_{o}^{2}+32)^{1/2}} \tfrac{\ln \vert t \vert}{t} \right), \\
\mathfrak{c}_{m} \! &= \, \mathfrak{c}_{m}^{0} \! + \! \tfrac{1}{\sqrt{\vert 
t \vert}} \mathfrak{c}_{m}^{1} \! + \! \mathcal{O} \! \left(\tfrac{c^{
\mathcal{S}}(z_{o})}{(z_{o}^{2}+32)^{1/2}} \tfrac{\ln \vert t \vert}{t} 
\right) \\
&=: \tfrac{g_{m}^{\star}}{(1+g_{m}^{\star} \overline{g_{m}^{\star}}(\varsigma_{
m}-\overline{\varsigma_{m}})^{-2})} \! + \! \tfrac{1}{\sqrt{\vert t \vert}} \! 
\! \left( \! \tfrac{g_{m}^{\star} \overline{g_{m}^{\star}}(\varsigma_{m}-
\overline{\varsigma_{m}})^{-1} \overline{\widetilde{\mathscr{M}}^{c}_{12}
(\varsigma_{m})}-g_{m}^{\star} \overline{g_{m}^{\star} \partial_{\zeta} 
\widetilde{\mathscr{M}}^{c}_{12}(\varsigma_{m})}}{(1+g_{m}^{\star} \overline{
g_{m}^{\star}}(\varsigma_{m}-\overline{\varsigma_{m}})^{-2})} \! + \! \tfrac{
g_{m}^{\star}(g_{m}^{\star} \partial_{\zeta} \widetilde{\mathscr{M}}^{c}_{12}
(\varsigma_{m})+\overline{g_{m}^{\star} \partial_{\zeta} \widetilde{\mathscr{
M}}^{c}_{12}(\varsigma_{m})})}{(1+g_{m}^{\star} \overline{g_{m}^{\star}}
(\varsigma_{m}-\overline{\varsigma_{m}})^{-2})^{2}} \! \right) \\
&+ \, \mathcal{O} \! \left(\tfrac{c^{\mathcal{S}}(z_{o})}{(z_{o}^{2}+32)^{
1/2}} \tfrac{\ln \vert t \vert}{t} \right), \\
\mathfrak{d}_{m} \! &= \, \tfrac{1}{\sqrt{\vert t \vert}} \mathfrak{d}_{m}^{
1} \! + \! \mathcal{O} \! \left(\tfrac{c^{\mathcal{S}}(z_{o})}{(z_{o}^{2}+
32)^{1/2}} \tfrac{\ln \vert t \vert}{t} \right) \! =: \! -\tfrac{1}{\sqrt{
\vert t \vert}} \tfrac{g_{m}^{\star} \widetilde{\mathscr{M}}^{c}_{12}
(\varsigma_{m})}{(1+g_{m}^{\star} \overline{g_{m}^{\star}}(\varsigma_{m}-
\overline{\varsigma_{m}})^{-2})} \! + \! \mathcal{O} \! \left(\tfrac{c^{
\mathcal{S}}(z_{o})}{(z_{o}^{2}+32)^{1/2}} \tfrac{\ln \vert t \vert}{t} 
\right),
\end{align*}
where
\begin{equation*}
\widetilde{\mathscr{M}}^{c}_{12}(\zeta) \! = \! \dfrac{\sqrt{\nu (\lambda_{
1})} \, \me^{\frac{\mi \Xi^{-}(0)}{2}} \lambda_{1}^{-2 \mi \nu (\lambda_{
1})}}{\sqrt{(\lambda_{1} \! - \! \lambda_{2})} \, (z_{o}^{2} \! + \! 32)^{
1/4}} \! \left(\dfrac{\lambda_{1} \me^{\mi (\Theta^{-}(z_{o},t)-\frac{3 \pi}
{4})}}{(\zeta \! - \! \lambda_{1})} \! + \! \dfrac{\lambda_{2} \me^{-\mi 
(\Theta^{-}(z_{o},t)-\frac{3 \pi}{4})}}{(\zeta \! - \! \lambda_{2})} \! 
\right),
\end{equation*}
with $\nu (\cdot)$, $\lambda_{1}$, $\lambda_{2}$, $\lambda_{3}$, $\Xi^{-}
(\cdot)$, and $\Theta^{-}(z_{o},t)$ specified in Lemma~{\rm A.1.6}, and 
$c^{\mathcal{S}}(z_{o})$ given in Proposition~{\rm 4.2}. Furthermore, setting 
$\widetilde{\mathcal{Y}} \! := \! 
\left(
\begin{smallmatrix}
\boxed{\widetilde{\mathscr{A}}} & \boxed{\widetilde{\mathscr{B}}} \\
\boxed{\overline{\widetilde{\mathscr{B}}}} & \boxed{\overline{\widetilde{
\mathscr{A}} \,}}
\end{smallmatrix}
\right)$,
\begin{equation}
0 \! < \! \vert \det (\widetilde{\mathcal{Y}}) \vert^{2} \! \leqslant \! 
\prod_{j=m}^{N} \! \left(1 \! + \! \tfrac{\sin^{2}(\phi_{m}) \vert \gamma_{
m} \vert^{2} P^{-2}(\phi_{m},\phi_{k})Q^{-2}(\phi_{m})}{\sin^{2}(\frac{1}{2}
(\phi_{m}+\phi_{j}))} \me^{2 \phi (x,t)} \right)^{2} \! +\mathcal{O} \! \left(
\tfrac{c^{\mathcal{S}}(z_{o})}{(z_{o}^{2}+32)^{1/2}} \tfrac{\ln \vert t \vert}
{t} \right),
\end{equation}
where $\phi (x,t)$, $P(\phi_{m},\phi_{k})$, and $Q(\phi_{m})$ are defined in 
Eqs.~{\rm (67)}, {\rm (68)}, and {\rm (69)}, respectively.
\end{by}

The analogue of Lemma~4.3 is (see, also, Remark~4.1)
\setcounter{z2}{1}
\setcounter{z3}{7}
\begin{ay}
As $t \! \to \! -\infty$ and $x \! \to \! +\infty$ such that $z_{o} \! := \! 
x/t \! < \! -2$ and $(x,t) \! \in \! \daleth_{m}$, $m \! \in \! \{1,2,\ldots,
N\}$,
\begin{equation*}
\widetilde{\mathscr{P}}(\zeta) \! = \! 
\begin{pmatrix}
\frac{\zeta +\widetilde{a}_{1}^{-}}{\zeta +\widetilde{a}_{2}^{-}} & \frac{
\widetilde{a}_{3}^{-}}{\zeta +\widetilde{a}_{4}^{-}} \\
\frac{\overline{\widetilde{a}_{3}^{-}}}{\zeta +\overline{\widetilde{a}_{4}^{
-}}} & \frac{\zeta +\overline{\widetilde{a}_{1}^{-}}}{\zeta +\overline{
\widetilde{a}_{2}^{-}}}
\end{pmatrix},
\end{equation*}
where
\begin{align*}
\widetilde{a}_{1}^{-} \! =& \, \overline{\widetilde{a}_{2}^{-}} \! =1 \! + 
\! \sum_{p=1}^{\infty} \sum_{q=0}^{p-1} \dfrac{\widetilde{a}_{pq}^{1}(z_{o})
(\ln \vert t \vert)^{q}}{\vert t \vert^{p/2}} \! + \! \mathcal{O} \! \left(
\me^{-4 \vert t \vert \min_{\genfrac{}{}{0pt}{2}{m \in \{1,2,\ldots,N\}}
{n \in \{m+1,m+2,\ldots,N\}}}\{\sin (\phi_{n}) \vert \cos (\phi_{n})-\cos 
(\phi_{m}) \vert\}} \right), \\
\widetilde{a}_{3}^{-} \! =& \, \sum_{p=1}^{\infty} \sum_{q=0}^{p-1} \dfrac{
\widetilde{a}_{pq}^{3}(z_{o})(\ln \vert t \vert)^{q}}{\vert t \vert^{p/2}} \! 
+ \! \mathcal{O} \! \left(\me^{-4 \vert t \vert \min_{\genfrac{}{}{0pt}{2}
{m \in \{1,2,\ldots,N\}}{n \in \{m+1,m+2,\ldots,N\}}}\{\sin (\phi_{n}) \vert 
\cos (\phi_{n})-\cos (\phi_{m}) \vert\}} \right), \\
\widetilde{a}_{4}^{-} \! =& \, 1 \! + \! \sum_{p=1}^{\infty} \sum_{q=0}^{p-1} 
\dfrac{\widetilde{a}_{pq}^{4}(z_{o})(\ln \vert t \vert)^{q}}{\vert t \vert^{
p/2}} \! + \! \mathcal{O} \! \left(\me^{-4 \vert t \vert \min_{\genfrac{}{}
{0pt}{2}{m \in \{1,2,\ldots,N\}}{n \in \{m+1,m+2,\ldots,N\}}}\{\sin (\phi_{
n}) \vert \cos (\phi_{n})-\cos (\phi_{m}) \vert\}} \right),
\end{align*}
$\widetilde{a}_{pq}^{k}(z_{o}) \! \in \! c^{\mathcal{S}}(z_{o})$, $k \! 
\in \! \{1,3,4\}$, and $\widetilde{\mathscr{P}}(\zeta) \! = \! \sigma_{1} 
\overline{\widetilde{\mathscr{P}}(\overline{\zeta})} \, \sigma_{1}$.
\end{ay}

The analogue of Proposition~4.3 is
\setcounter{z0}{1}
\setcounter{z1}{4}
\begin{by}
Set $\widetilde{a}_{10}^{1}(z_{o}) \! =: \! \widetilde{a}_{1}$, $\widetilde{
a}^{2}_{10}(z_{o}) \! =: \! \widetilde{a}_{2}$, $\widetilde{a}^{3}_{10}(z_{
o}) \! =: \! \widetilde{a}_{3}$, and $\widetilde{a}^{4}_{10}(z_{o}) \! =: \! 
\widetilde{a}_{4}$. Then as $t \! \to \! -\infty$ and $x \! \to \! +\infty$ 
such that $z_{o} \! := \! x/t \! < \! -2$ and $(x,t) \! \in \! \daleth_{m}$, 
$m \! \in \! \{1,2,\ldots,N\}$,
\begin{align*}
(\widetilde{\Delta}_{o})_{11} \! =& \, -\dfrac{\overline{\mathfrak{c}_{m}^{
0}}}{\overline{\varsigma_{m}}} \mi \widetilde{\delta}^{-1}(0) \me^{2 \mi 
\sum_{k=1}^{m-1} \phi_{k}} \! + \! \dfrac{\mi \widetilde{\delta}^{-1}(0) 
\me^{2 \mi \sum_{k=1}^{m-1} \phi_{k}}}{\sqrt{\vert t \vert}} \! \left(-
(\widetilde{a}_{1} \! - \! \widetilde{a}_{2}) \dfrac{\overline{\mathfrak{c}_{
m}^{0}}}{\overline{\varsigma_{m}}} \! - \! \left(\dfrac{\mathfrak{b}_{m}^{1}}
{\varsigma_{m}} \! + \! \dfrac{\overline{\mathfrak{c}_{m}^{1}}}{\overline{
\varsigma_{m}}} \right) \right. \\
+& \left. \widetilde{a}_{3} \! \left(1 \! - \! \dfrac{\overline{\mathfrak{
a}_{m}^{0}}}{\overline{\varsigma_{m}}} \right) \! - \! \left(1 \! - \! 
\dfrac{\mathfrak{a}_{m}^{0}}{\varsigma_{m}} \right) \! \dfrac{2 \widetilde{
\delta}(0) \sqrt{\nu (\lambda_{1})} \, \cos (\Theta^{-}(z_{o},t) \! - \! 
\frac{3 \pi}{4})}{\sqrt{(\lambda_{1} \! - \! \lambda_{2})} \, (z_{o}^{2} \! 
+ \! 32)^{1/4}} \right) \! + \! \mathcal{O} \! \left(\dfrac{c^{\mathcal{S}}
(z_{o})}{(z_{o}^{2} \! + \! 32)^{1/2}} \dfrac{\ln \vert t \vert}{t} \right), 
\\
(\widetilde{\Delta}_{o})_{12} \! =& \, -\left(1 \! - \! \dfrac{\mathfrak{
a}_{m}^{0}}{\varsigma_{m}} \right) \! \mi \widetilde{\delta}(0) \me^{-2 \mi 
\sum_{k=1}^{m-1} \phi_{k}} \! + \! \dfrac{\mi \widetilde{\delta}(0) \me^{-2 
\mi \sum_{k=1}^{m-1} \phi_{k}}}{\sqrt{\vert t \vert}} \! \left(-(\widetilde{
a}_{1} \! - \! \widetilde{a}_{2}) \! \left(1 \! - \! \dfrac{\mathfrak{a}_{
m}^{0}}{\varsigma_{m}} \right) \! + \! \left(\dfrac{\mathfrak{a}_{m}^{1}}
{\varsigma_{m}} \! + \! \dfrac{\overline{\mathfrak{d}_{m}^{1}}}{\overline{
\varsigma_{m}}} \right) \right. \\
+& \left. \widetilde{a}_{3} \dfrac{\mathfrak{c}_{m}^{0}}{\varsigma_{m}} \! - 
\! \dfrac{\overline{\mathfrak{c}_{m}^{0}}}{\overline{\varsigma_{m}}} \dfrac{
2 \widetilde{\delta}^{-1}(0) \sqrt{\nu (\lambda_{1})} \, \cos (\Theta^{-}
(z_{o},t) \! - \! \frac{3 \pi}{4})}{\sqrt{(\lambda_{1} \! - \! \lambda_{2})} 
\, (z_{o}^{2} \! + \! 32)^{1/4}} \right) \! + \! \mathcal{O} \! \left(\dfrac{
c^{\mathcal{S}}(z_{o})}{(z_{o}^{2} \! + \! 32)^{1/2}} \dfrac{\ln \vert t 
\vert}{t} \right), \\
(\widetilde{\Delta}_{o})_{21} \! =& \, \left(1 \! - \! \dfrac{\overline{
\mathfrak{a}_{m}^{0}}}{\overline{\varsigma_{m}}} \right) \! \mi \widetilde{
\delta}^{-1}(0) \me^{2 \mi \sum_{k=1}^{m-1} \phi_{k}} \! + \! \dfrac{\mi 
\widetilde{\delta}^{-1}(0) \me^{2 \mi \sum_{k=1}^{m-1} \phi_{k}}}{\sqrt{\vert 
t \vert}} \! \left((\overline{\widetilde{a}_{1}} \! - \! \overline{\widetilde{
a}_{2}}) \! \left(1 \! - \! \dfrac{\overline{\mathfrak{a}_{m}^{0}}}{
\overline{\varsigma_{m}}} \right) \! - \! \left(\dfrac{\overline{\mathfrak{
a}_{m}^{1}}}{\overline{\varsigma_{m}}} \! + \! \dfrac{\mathfrak{d}_{m}^{1}}
{\varsigma_{m}} \right) \right. \\
-& \left. \overline{\widetilde{a}_{3}} \, \dfrac{\overline{\mathfrak{c}_{
m}^{0}}}{\overline{\varsigma_{m}}} \! + \! \dfrac{\mathfrak{c}_{m}^{0}}
{\varsigma_{m}} \dfrac{2 \widetilde{\delta}(0) \sqrt{\nu (\lambda_{1})} \, 
\cos (\Theta^{-}(z_{o},t) \! - \! \frac{3 \pi}{4})}{\sqrt{(\lambda_{1} \! - 
\! \lambda_{2})} \, (z_{o}^{2} \! + \! 32)^{1/4}} \right) \! + \! \mathcal{O} 
\! \left(\dfrac{c^{\mathcal{S}}(z_{o})}{(z_{o}^{2} \! + \! 32)^{1/2}} \dfrac{
\ln \vert t \vert}{t} \right), \\
(\widetilde{\Delta}_{o})_{22} \! =& \, \dfrac{\mathfrak{c}_{m}^{0}}{
\varsigma_{m}} \mi \widetilde{\delta}(0) \me^{-2 \mi \sum_{k=1}^{m-1} \phi_{
k}} \! + \! \dfrac{\mi \widetilde{\delta}(0) \me^{-2 \mi \sum_{k=1}^{m-1} 
\phi_{k}}}{\sqrt{\vert t \vert}} \! \left((\overline{\widetilde{a}_{1}} \! - 
\! \overline{\widetilde{a}_{2}}) \dfrac{\mathfrak{c}_{m}^{0}}{\varsigma_{m}} 
\! + \! \left(\dfrac{\overline{\mathfrak{b}_{m}^{1}}}{\overline{\varsigma_{
m}}} \! + \! \dfrac{\mathfrak{c}_{m}^{1}}{\varsigma_{m}} \right) \right. \\
-& \left. \overline{\widetilde{a}_{3}} \! \left(1 \! - \! \dfrac{\mathfrak{
a}_{m}^{0}}{\varsigma_{m}} \right) \! + \! \left(1 \! - \! \dfrac{\overline{
\mathfrak{a}_{m}^{0}}}{\overline{\varsigma_{m}}} \right) \! \dfrac{2 
\widetilde{\delta}^{-1}(0) \sqrt{\nu (\lambda_{1})} \, \cos (\Theta^{-}(z_{
o},t) \! - \! \frac{3 \pi}{4})}{\sqrt{(\lambda_{1} \! - \! \lambda_{2})} \, 
(z_{o}^{2} \! + \! 32)^{1/4}} \right) \! + \! \mathcal{O} \! \left(\dfrac{
c^{\mathcal{S}}(z_{o})}{(z_{o}^{2} \! + \! 32)^{1/2}} \dfrac{\ln \vert t 
\vert}{t} \right).
\end{align*}
\end{by}

The analogue of Proposition~4.4 is
\setcounter{z0}{1}
\setcounter{z1}{5}
\begin{by}
Let $\phi (x,t)$, $P(\phi_{m},\phi_{k})$, and $Q(\phi_{m})$ be defined by 
Eqs.~{\rm (67)}, {\rm (68)}, and {\rm (69)}, respectively. Then, for 
$\theta_{\gamma_{m}} \! = \! \pm \pi/2$, as $t \! \to \! -\infty$ and $x 
\! \to \! +\infty$ such that $z_{o} \! := \! x/t \! < \! -2$ and $(x,t) \! 
\in \! \daleth_{m}$, $m \! \in \! \{1,2,\ldots,N\}$,
\begin{align*}
\mathfrak{a}_{m}^{0} \! =& -\dfrac{2 \mi \sin (\phi_{m}) \vert \gamma_{m} 
\vert^{2}P^{-2}(\phi_{m},\phi_{k})Q^{-2}(\phi_{m}) \me^{2 \phi (x,t)}}{(1 \! 
- \! \vert \gamma_{m} \vert^{2}P^{-2}(\phi_{m},\phi_{k})Q^{-2}(\phi_{m}) 
\me^{2 \phi (x,t)})}, \\
\mathfrak{c}_{m}^{0} \! =& \mp \dfrac{2 \sin (\phi_{m}) \vert \gamma_{m} 
\vert \widetilde{\delta}^{-1}(0) \me^{\mi (\phi_{m}+s^{-})+\phi (x,t)}
P^{-1}(\phi_{m},\phi_{k})Q^{-1}(\phi_{m})}{(1 \! - \! \vert \gamma_{m} 
\vert^{2}P^{-2}(\phi_{m},\phi_{k})Q^{-2}(\phi_{m}) \me^{2 \phi (x,t)})}, \\
\mathfrak{a}_{m}^{1} \! =& \mp \dfrac{16 \mi \lambda_{1}^{2} \sin^{2}(\phi_{
m}) \vert \gamma_{m} \vert^{3} \sqrt{\nu (\lambda_{1})} \, P^{-3}(\phi_{m},
\phi_{k})Q^{-3}(\phi_{m}) \cos (s^{-}) \me^{3 \phi (x,t)}}{(1 \! - \! \vert 
\gamma_{m} \vert^{2}P^{-2}(\phi_{m},\phi_{k})Q^{-2}(\phi_{m}) \me^{2 \phi (x,
t)})^{2}(\lambda_{1}^{2} \! - \! 2 \lambda_{1} \cos (\phi_{m}) \! + \! 1)^{2} 
\sqrt{(\lambda_{1} \! - \! \lambda_{2})} \, (z_{o}^{2} \! + \! 32)^{1/4}} \\
\times& \left(((\lambda_{1} \! + \! \lambda_{2}) \cos (\phi_{m}) \! - \! 2) 
\cos (\Theta^{-}(z_{o},t) \! - \! \tfrac{3 \pi}{4}) \! - \! (\lambda_{1} \! 
- \! \lambda_{2}) \sin (\phi_{m}) \sin (\Theta^{-}(z_{o},t) \! - \! \tfrac{3 
\pi}{4}) \right) \\
\mp& \, \dfrac{2 \lambda_{1} \sin (\phi_{m}) \vert \gamma_{m} \vert \sqrt{\nu 
(\lambda_{1})} \, P^{-1}(\phi_{m},\phi_{k})Q^{-1}(\phi_{m}) \me^{\phi (x,t)}}
{(1 \! - \! \vert \gamma_{m} \vert^{2}P^{-2}(\phi_{m},\phi_{k})Q^{-2}(\phi_{
m}) \me^{2 \phi (x,t)})(\lambda_{1}^{2} \! - \! 2 \lambda_{1} \cos (\phi_{m}) 
\! + \! 1) \sqrt{(\lambda_{1} \! - \! \lambda_{2})} \, (z_{o}^{2} \! + \! 
32)^{1/4}} \\
\times& \left(2 \cos (s^{-}) \cos (\Theta^{-}(z_{o},t) \! - \! \tfrac{3 \pi}
{4}) \! - \! (\lambda_{1} \! + \! \lambda_{2}) \cos (\phi_{m} \! + \! s^{-}) 
\cos (\Theta^{-}(z_{o},t) \! - \! \tfrac{3 \pi}{4}) \right. \\
+& \left. (\lambda_{1} \! - \! \lambda_{2}) \sin (\phi_{m} \! + \! s^{-}) 
\sin (\Theta^{-}(z_{o},t) \! - \! \tfrac{3 \pi}{4}) \! + \! 2 \mi \sin (s^{
-}) \cos (\Theta^{-}(z_{o},t) \! - \! \tfrac{3 \pi}{4}) \right. \\
-& \left. \mi (\lambda_{1} \! + \! \lambda_{2}) \sin (\phi_{m} \! + \! s^{-}) 
\cos (\Theta^{-}(z_{o},t) \! - \! \tfrac{3 \pi}{4}) \! - \! \mi (\lambda_{1} 
\! - \! \lambda_{2}) \cos (\phi_{m} \! + \! s^{-}) \sin (\Theta^{-}(z_{o},t) 
\! - \! \tfrac{3 \pi}{4}) \right), \\
\mathfrak{b}_{m}^{1} \! =& \, \dfrac{2 \mi \lambda_{1} \sin (\phi_{m}) \vert 
\gamma_{m} \vert^{2} \sqrt{\nu (\lambda_{1})} \, \widetilde{\delta}(0) \me^{
-\mi (\phi_{m}+s^{-})+2\phi (x,t)}P^{-2}(\phi_{m},\phi_{k})Q^{-2}(\phi_{m})}
{(1 \! - \! \vert \gamma_{m} \vert^{2}P^{-2}(\phi_{m},\phi_{k})Q^{-2}(\phi_{
m}) \me^{2 \phi (x,t)})(\lambda_{1}^{2} \! - \! 2 \lambda_{1} \cos (\phi_{m}) 
\! + \! 1) \sqrt{(\lambda_{1} \! - \! \lambda_{2})} \, (z_{o}^{2} \! + \! 
32)^{1/4}} \\
\times& \left(2 \cos (s^{-}) \cos (\Theta^{-}(z_{o},t) \! - \! \tfrac{3 \pi}
{4}) \! - \! (\lambda_{1} \! + \! \lambda_{2}) \cos (\phi_{m} \! + \! s^{-}) 
\cos (\Theta^{-}(z_{o},t) \! - \! \tfrac{3 \pi}{4}) \right. \\
+& \left. (\lambda_{1} \! - \! \lambda_{2}) \sin (\phi_{m} \! + \! s^{-}) \sin 
(\Theta^{-}(z_{o},t) \! - \! \tfrac{3 \pi}{4}) \! + \! 2 \mi \sin (s^{-})\cos 
(\Theta^{-}(z_{o},t) \! - \! \tfrac{3 \pi}{4}) \right. \\
-& \left. \mi (\lambda_{1} \! + \! \lambda_{2}) \sin (\phi_{m} \! + \! s^{-}) 
\cos (\Theta^{-}(z_{o},t) \! - \! \tfrac{3 \pi}{4}) \! - \! \mi (\lambda_{1} 
\! - \! \lambda_{2}) \cos (\phi_{m} \! + \! s^{-}) \sin (\Theta^{-}(z_{o},t) 
\! - \! \tfrac{3 \pi}{4}) \right), \\
\mathfrak{c}_{m}^{1} \! =& -\dfrac{16 \lambda_{1}^{2} \sin^{2}(\phi_{m}) 
\vert \gamma_{m} \vert^{2} \sqrt{\nu (\lambda_{1})} \, \widetilde{\delta}^{-1}
(0) \me^{\mi (\phi_{m}+s^{-})+2\phi (x,t)}P^{-2}(\phi_{m},\phi_{k})Q^{-2}
(\phi_{m}) \cos (s^{-})}{(1 \! - \! \vert \gamma_{m} \vert^{2}P^{-2}(\phi_{
m},\phi_{k})Q^{-2}(\phi_{m}) \me^{2 \phi (x,t)})^{2}(\lambda_{1}^{2} \! - \! 
2 \lambda_{1} \cos (\phi_{m}) \! + \! 1)^{2} \sqrt{(\lambda_{1} \! - \! 
\lambda_{2})} \, (z_{o}^{2} \! + \! 32)^{1/4}} \\
\times& \left(((\lambda_{1} \! + \! \lambda_{2}) \cos (\phi_{m}) \! - \! 2) 
\cos (\Theta^{-}(z_{o},t) \! - \! \tfrac{3 \pi}{4}) \! - \! (\lambda_{1} \! 
- \! \lambda_{2}) \sin (\phi_{m}) \sin (\Theta^{-}(z_{o},t) \! - \! \tfrac{
3 \pi}{4}) \right)
\end{align*}
\begin{align*}
-& \, \dfrac{2 \mi \lambda_{1} \sin (\phi_{m}) \vert \gamma_{m} \vert^{2} 
\sqrt{\nu (\lambda_{1})} \, \widetilde{\delta}^{-1}(0) \me^{\mi (\phi_{m}+
s^{-})+2\phi (x,t)}P^{-2}(\phi_{m},\phi_{k})Q^{-2}(\phi_{m})}{(1 \! - \! 
\vert \gamma_{m} \vert^{2}P^{-2}(\phi_{m},\phi_{k})Q^{-2}(\phi_{m}) \me^{2 
\phi (x,t)})(\lambda_{1}^{2} \! - \! 2 \lambda_{1} \cos (\phi_{m}) \! + \! 
1) \sqrt{(\lambda_{1} \! - \! \lambda_{2})} \, (z_{o}^{2} \! + \! 32)^{1/4}} 
\\
\times& \left(2 \cos (s^{-}) \cos (\Theta^{-}(z_{o},t) \! - \! \tfrac{3 \pi}
{4}) \! - \! (\lambda_{1} \! + \! \lambda_{2}) \cos (\phi_{m} \! + \! s^{-}) 
\cos (\Theta^{-}(z_{o},t) \! - \! \tfrac{3 \pi}{4}) \right. \\
+& \left. (\lambda_{1} \! - \! \lambda_{2}) \sin (\phi_{m} \! + \! s^{+}) 
\sin (\Theta^{-}(z_{o},t) \! - \! \tfrac{3 \pi}{4}) \! - \! 2 \mi \sin (s^{
-}) \cos (\Theta^{-}(z_{o},t) \! - \! \tfrac{3 \pi}{4}) \right. \\
+& \left. \mi (\lambda_{1} \! + \! \lambda_{2}) \sin (\phi_{m} \! + \! s^{-}) 
\cos (\Theta^{-}(z_{o},t) \! - \! \tfrac{3 \pi}{4}) \! + \! \mi (\lambda_{1} 
\! - \! \lambda_{2}) \cos (\phi_{m} \! + \! s^{-}) \sin (\Theta^{-}(z_{o},t) 
\! - \! \tfrac{3 \pi}{4}) \right) \\
+& \, \dfrac{8 \lambda_{1}^{2} \sin^{2}(\phi_{m}) \vert \gamma_{m} \vert^{2} 
\sqrt{\nu (\lambda_{1})} \, \widetilde{\delta}^{-1}(0) \me^{\mi (\phi_{m}
+s^{-})+2\phi (x,t)}P^{-2}(\phi_{m},\phi_{k})Q^{-2}(\phi_{m})}{(1 \! - \! 
\vert \gamma_{m} \vert^{2}P^{-2}(\phi_{m},\phi_{k})Q^{-2}(\phi_{m}) \me^{2 
\phi (x,t)})(\lambda_{1}^{2} \! - \! 2 \lambda_{1} \cos (\phi_{m}) \! + \! 
1)^{2} \sqrt{(\lambda_{1} \! - \! \lambda_{2})} \, (z_{o}^{2} \! + \! 32)^{
1/4}} \\
\times& \left(\left(((\lambda_{1} \! + \! \lambda_{2}) \cos (\phi_{m}) \! - 
\! 2) \cos (\Theta^{-}(z_{o},t) \! - \! \tfrac{3 \pi}{4}) \! - \! (\lambda_{
1} \! - \! \lambda_{2}) \sin (\phi_{m}) \sin (\Theta^{-}(z_{o},t) \! - \! 
\tfrac{3 \pi}{4}) \right) \cos (s^{-}) \right. \\
-& \left.\mi \! \left(((\lambda_{1} \! + \! \lambda_{2}) \cos (\phi_{m}) \! - 
\! 2) \cos (\Theta^{-}(z_{o},t) \! - \! \tfrac{3 \pi}{4}) \! - \! (\lambda_{
1} \! - \! \lambda_{2}) \sin (\phi_{m}) \sin (\Theta^{-}(z_{o},t) \! - \! 
\tfrac{3 \pi}{4}) \right) \sin (s^{-}) \right), \\
\mathfrak{d}_{m}^{1} \! =& \pm \dfrac{2 \lambda_{1} \sin (\phi_{m}) \vert 
\gamma_{m} \vert \sqrt{\nu (\lambda_{1})} \, P^{-1}(\phi_{m},\phi_{k})Q^{-1}
(\phi_{m}) \me^{\phi (x,t)}}{(1 \! - \! \vert \gamma_{m} \vert^{2}P^{-2}
(\phi_{m},\phi_{k})Q^{-2}(\phi_{m}) \me^{2 \phi (x,t)})(\lambda_{1}^{2} \! 
- \! 2 \lambda_{1} \cos (\phi_{m}) \! + \! 1) \sqrt{(\lambda_{1} \! - \! 
\lambda_{2})} \, (z_{o}^{2} \! + \! 32)^{1/4}} \\
\times& \left(2 \cos (s^{-}) \cos (\Theta^{-}(z_{o},t) \! - \! \tfrac{3 \pi}
{4}) \! - \! (\lambda_{1} \! + \! \lambda_{2}) \cos (\phi_{m} \! + \! s^{-}) 
\cos (\Theta^{-}(z_{o},t) \! - \! \tfrac{3 \pi}{4}) \right. \\
+& \left. (\lambda_{1} \! - \! \lambda_{2}) \sin (\phi_{m} \! + \! s^{-}) 
\sin (\Theta^{-}(z_{o},t) \! - \! \tfrac{3 \pi}{4}) \! + \! 2 \mi \sin (s^{
-}) \cos (\Theta^{-}(z_{o},t) \! - \! \tfrac{3 \pi}{4}) \right. \\
-& \left. \mi (\lambda_{1} \! + \! \lambda_{2}) \sin (\phi_{m} \! + \! s^{-}) 
\cos (\Theta^{-}(z_{o},t) \! - \! \tfrac{3 \pi}{4}) \! - \! \mi (\lambda_{1} 
\! - \! \lambda_{2}) \cos (\phi_{m} \! + \! s^{-}) \sin (\Theta^{-}(z_{o},t) 
\! - \! \tfrac{3 \pi}{4}) \right),
\end{align*}
where $s^{-}$ is given in Theorem~{\rm 2.2.1}, Eq.~{\rm (11)}.
\end{by}

The analogue of Proposition~4.5 is
\setcounter{z0}{1}
\setcounter{z1}{6}
\begin{by}
As $t \! \to \! -\infty$ and $x \! \to \! +\infty$ such that $z_{o} \! := \! 
x/t \! < \! -2$ and $(x,t) \! \in \! \daleth_{m}$, $m \! \in \! \{1,2,\ldots,
N\}$,
\begin{align}
u(x,t) \! =& \, \mi \! \left((\widetilde{\Delta}_{o})_{12} \! + \! \widetilde{
a}_{3}^{-} \! + \! \mathfrak{b}_{m} \! + \! \overline{\mathfrak{c}_{m}} \! + 
\! \tfrac{\sqrt{\nu (\lambda_{1})} \, \me^{\frac{\mi \Xi^{-}(0)}{2}} \lambda_{
1}^{-2 \mi \nu (\lambda_{1})}}{\sqrt{\vert t \vert (\lambda_{1}-\lambda_{2})} 
\, (z_{o}^{2}+32)^{1/4}} \! \left(\lambda_{1} \me^{\mi (\Theta^{-}(z_{o},t)-
\frac{3 \pi}{4})} \! + \! \lambda_{2} \me^{-\mi (\Theta^{-}(z_{o},t)-\frac{3 
\pi}{4})} \right) \right) \nonumber \\
+& \, \mathcal{O} \! \left(\tfrac{c^{\mathcal{S}}(z_{o})}{(z_{o}^{2}+32)^{
1/2}} \tfrac{\ln \vert t \vert}{t} \right),
\end{align}
\begin{align}
\int_{+\infty}^{x}(\vert u(x^{\prime},t) \vert^{2} \! - \! 1) \, \md x^{
\prime} \! =& \, -\mi \! \left((\widetilde{\Delta}_{o})_{11} \! + \! 
\widetilde{a}_{1}^{-} \! - \! \widetilde{a}_{2}^{-} \! + \! \mathfrak{a}_{m} 
\! + \! \overline{\mathfrak{d}_{m}} \! + \! 2 \mi \sum_{k=1}^{m-1} \sin 
(\phi_{k}) \right. \nonumber \\
+&\left. \, \mi \! \left(\int\nolimits_{0}^{\lambda_{2}} \! + \! 
\int\nolimits_{\lambda_{1}}^{+\infty} \right) \! \ln (1 \! - \! \vert r(\mu) 
\vert^{2}) \, \tfrac{\md \mu}{2 \pi} \right) \! + \! \mathcal{O} \! \left(
\tfrac{c^{\mathcal{S}}(z_{o})}{(z_{o}^{2}+32)^{1/2}} \tfrac{\ln \vert t 
\vert}{t} \right),
\end{align}
\begin{align}
\int_{-\infty}^{x}(\vert u(x^{\prime},t) \vert^{2} \! - \! 1) \, \md x^{
\prime} \! = \! \int_{+\infty}^{x}(\vert u(x^{\prime},t) \vert^{2} \! - 
\! 1) \, \md x^{\prime} \! - \! 2 \sum_{n=1}^{N} \sin (\phi_{n}) \! - \! 
\int\nolimits_{-\infty}^{+\infty} \ln (1 \! - \! \vert r(\mu) \vert^{2}) \, 
\tfrac{\md \mu}{2 \pi}.
\end{align}
\end{by}

The analogue of Proposition~4.6 is
\setcounter{z0}{1}
\setcounter{z1}{7}
\begin{by}
As $t \! \to \! -\infty$ and $x \! \to \! +\infty$ such that $z_{o} \! := \! 
x/t \! < \! -2$ and $(x,t) \! \in \! \daleth_{m}$, $m \! \in \! \{1,2,\ldots,
N\}$, for $\theta_{\gamma_{m}} \! = \! \pm \pi/2$,
\begin{align*}
(\widetilde{\Delta}_{o})_{11}=& \, \mi \! \left(\pm \tfrac{2 \sin (\phi_{m}) 
\vert \gamma_{m} \vert P^{-1}(\phi_{m},\phi_{k})Q^{-1}(\phi_{m}) \me^{\phi 
(x,t)}}{(1-\vert \gamma_{m} \vert^{2}P^{-2}(\phi_{m},\phi_{k})Q^{-2}(\phi_{
m}) \me^{2 \phi (x,t)})} \! + \! \tfrac{\sqrt{\nu (\lambda_{1})}}{\sqrt{
\vert t \vert (\lambda_{1}-\lambda_{2})} \, (z_{o}^{2}+32)^{1/4}} \! \left(
-2 \cos (\Theta^{-}(z_{o},t) \! - \! \tfrac{3 \pi}{4}) \right. \right. \\
\times& \left. \left. \! \cos (s^{-}) \! + \! \tfrac{4 \sin (\phi_{m}) \vert 
\gamma_{m} \vert^{2}P^{-2}(\phi_{m},\phi_{k})Q^{-2}(\phi_{m}) \sin (s^{-}-
\phi_{m}) \cos (\Theta^{-}(z_{o},t)-\tfrac{3 \pi}{4}) \me^{2 \phi (x,t)}}
{(1-\vert \gamma_{m} \vert^{2}P^{-2}(\phi_{m},\phi_{k})Q^{-2}(\phi_{m}) 
\me^{2 \phi (x,t)})} \right. \right. \\
+& \left. \left. \! \tfrac{8 \lambda_{1}^{2} \sin^{2}(\phi_{m}) \vert \gamma_{
m} \vert^{2}P^{-2}(\phi_{m},\phi_{k})Q^{-2}(\phi_{m})(1+\vert \gamma_{m} 
\vert^{2}P^{-2}(\phi_{m},\phi_{k})Q^{-2}(\phi_{m}) \me^{2 \phi (x,t)}) \cos 
(s^{-}) \me^{2 \phi (x,t)}}{(1-\vert \gamma_{m} \vert^{2}P^{-2}(\phi_{m},
\phi_{k})Q^{-2}(\phi_{m}) \me^{2 \phi (x,t)})^{2}(\lambda_{1}^{2}-2 \lambda_{
1} \cos (\phi_{m})+1)^{2}} \right. \right. \\
\times& \left. \left. \! \left(((\lambda_{1} \! + \! \lambda_{2}) \cos 
(\phi_{m}) \! - \! 2) \cos (\Theta^{-}(z_{o},t) \! - \! \tfrac{3 \pi}{4}) \! 
- \! (\lambda_{1} \! - \! \lambda_{2}) \sin (\phi_{m}) \sin (\Theta^{-}(z_{
o},t) \! - \! \tfrac{3 \pi}{4}) \right) \right. \right. \\
+& \left. \left. \! \tfrac{4 \lambda_{1} \sin (\phi_{m}) \cos (\phi_{m}) 
\vert \gamma_{m} \vert^{2}P^{-2}(\phi_{m},\phi_{k})Q^{-2}(\phi_{m}) \me^{2 
\phi (x,t)}}{(1-\vert \gamma_{m} \vert^{2}P^{-2}(\phi_{m},\phi_{k})Q^{-2}
(\phi_{m}) \me^{2 \phi (x,t)})(\lambda_{1}^{2}-2 \lambda_{1} \cos (\phi_{m})
+1)} \! \left(2 \cos (\Theta^{-}(z_{o},t) \! - \! \tfrac{3 \pi}{4}) \sin 
(s^{-} \! - \! \phi_{m}) \right. \right. \right. \\
-& \left. \left. \left. (\lambda_{1} \! + \! \lambda_{2}) \sin (s^{-}) \cos 
(\Theta^{-}(z_{o},t) \! - \! \tfrac{3 \pi}{4}) \! - \! (\lambda_{1} \! - \! 
\lambda_{2}) \cos (s^{-}) \sin (\Theta^{-}(z_{o},t) \! - \! \tfrac{3 \pi}{4})
\right) \right) \right) \\
+& \, \mathcal{O} \! \left(\tfrac{c^{\mathcal{S}}(z_{o})}{(z_{o}^{2}+32)^{
1/2}} \tfrac{\ln \vert t \vert}{t} \right),
\end{align*}
\begin{align*}
(\widetilde{\Delta}_{o})_{12} =& \, -\mi \me^{-\mi (\theta^{-}(1)+s^{-})} \! 
+ \! \dfrac{2 \sin (\phi_{m}) \vert \gamma_{m} \vert^{2}P^{-2}(\phi_{m},
\phi_{k})Q^{-2}(\phi_{m}) \me^{-\mi (\theta^{-}(1)+\phi_{m}+s^{-})+2 \phi 
(x,t)}}{(1 \! - \! \vert \gamma_{m} \vert^{2}P^{-2}(\phi_{m},\phi_{k})Q^{-2}
(\phi_{m}) \me^{2 \phi (x,t)})} \\
+& \dfrac{1}{\sqrt{t}} \! \left(\dfrac{2 \mi \Im (\widetilde{a}_{1} \! - \! 
\widetilde{a}_{2}) \sin (\phi_{m}) \vert \gamma_{m} \vert^{2}P^{-2}(\phi_{m},
\phi_{k})Q^{-2}(\phi_{m}) \me^{-\mi (\theta^{-}(1)+\phi_{m}+s^{-})+2 \phi 
(x,t)}}{(1 \! - \! \vert \gamma_{m} \vert^{2}P^{-2}(\phi_{m},\phi_{k})Q^{-2}
(\phi_{m}) \me^{2 \phi (x,t)})} \right. \\
\pm& \left. \dfrac{4 \mi \sin (\phi_{m}) \vert \gamma_{m} \vert P^{-1}(\phi_{
m},\phi_{k})Q^{-1}(\phi_{m}) \sqrt{\nu (\lambda_{1})} \, \me^{-\mi (\theta^{-}
(1)+2s^{-})+\phi (x,t)} \cos (\Theta^{-}(z_{o},t) \! - \! \tfrac{3 \pi}{4})}{
(1 \! - \! \vert \gamma_{m} \vert^{2}P^{-2}(\phi_{m},\phi_{k})Q^{-2}(\phi_{
m}) \me^{2 \phi (x,t)}) \sqrt{(\lambda_{1} \! - \! \lambda_{2})} \, (z_{o}^{
2} \! + \! 32)^{1/4}} \right. \\
\mp& \left. \dfrac{16 \mi \lambda_{1}^{2} \sin^{2}(\phi_{m}) \vert \gamma_{m} 
\vert^{3}P^{-3}(\phi_{m},\phi_{k})Q^{-3}(\phi_{m}) \sqrt{\nu (\lambda_{1})} 
\, \me^{-\mi (\theta^{-}(1)+s^{-})+3 \phi (x,t)} \cos (s^{-})}{(1 \! - \! 
\vert \gamma_{m} \vert^{2}P^{-2}(\phi_{m},\phi_{k})Q^{-2}(\phi_{m}) \me^{
2 \phi (x,t)})^{2}(\lambda_{1}^{2} \! - \! 2 \lambda_{1} \cos (\phi_{m}) \! 
+ \! 1)^{2} \sqrt{(\lambda_{1} \! - \! \lambda_{2})} \, (z_{o}^{2} \! + \! 
32)^{1/4}} \right. \\
\times& \left. \left(((\lambda_{1} \! + \! \lambda_{2}) \cos (\phi_{m}) \! - 
\! 2) \sin (\phi_{m}) \cos (\Theta^{-}(z_{o},t) \! - \! \tfrac{3 \pi}{4}) \! 
- \! (\lambda_{1} \! - \! \lambda_{2}) \sin^{2}(\phi_{m}) \sin (\Theta^{-}
(z_{o},t) \! - \! \tfrac{3 \pi}{4}) \right. \right. \\
+& \left. \left. \mi (((\lambda_{1} \! + \! \lambda_{2}) \cos (\phi_{m}) \! 
- \! 2) \cos (\phi_{m}) \cos (\Theta^{-}(z_{o},t) \! - \! \tfrac{3 \pi}{4}) 
\! - \! (\lambda_{1} \! - \! \lambda_{2}) \sin (\phi_{m}) \cos (\phi_{m}) 
\right. \right. \\
\times& \left. \left. \sin (\Theta^{-}(z_{o},t) \! - \! \tfrac{3 \pi}{4}) 
\right) \! + \! \Im (\widetilde{a}_{1} \! - \! \widetilde{a}_{2}) \me^{-\mi 
(\theta^{-}(1)+s^{-})} \right. \\
-& \left. \dfrac{4 \lambda_{1} \sin (\phi_{m}) \vert \gamma_{m} \vert P^{
-1}(\phi_{m},\phi_{k})Q^{-1}(\phi_{m}) \sqrt{\nu (\lambda_{1})} \, \me^{-\mi 
(\theta^{-}(1)+s^{-}) + \phi (x,t)}}{(1 \! - \! \vert \gamma_{m} \vert^{2}
P^{-2}(\phi_{m},\phi_{k})Q^{-2}(\phi_{m}) \me^{2 \phi (x,t)})(\lambda_{1}^{
2} \! - \! 2 \lambda_{1} \cos (\phi_{m}) \! + \! 1) \sqrt{(\lambda_{1} \! 
- \! \lambda_{2})} \, (z_{o}^{2} \! + \! 32)^{1/4}} \right. \\
\times& \left. \left(\mp 2 \sin (s^{-} \! - \! \phi_{m}) \cos (\Theta^{-}(z_{
o},t) \! - \! \tfrac{3 \pi}{4}) \! \pm \! (\lambda_{1} \! + \! \lambda_{2}) 
\sin (s^{-}) \cos (\Theta^{-}(z_{o},t) \! - \! \tfrac{3 \pi}{4}) \right. 
\right. \\
\pm& \left. \left. (\lambda_{1} \! - \! \lambda_{2}) \cos (s^{-}) \sin 
(\Theta^{-}(z_{o},t) \! - \! \tfrac{3 \pi}{4}) \right) \right) \! + \! 
\mathcal{O} \! \left(\tfrac{c^{\mathcal{S}}(z_{o})}{(z_{o}^{2}+32)^{1/2}} 
\tfrac{\ln \vert t \vert}{t} \right), \\
\Im (\widetilde{a}_{1} \! - \! \widetilde{a}_{2}) =& \, \pm \dfrac{\sqrt{\nu 
(\lambda_{1})} \sin (s^{-}) \cos (\Theta^{-}(z_{o},t) \! - \! \tfrac{3 \pi}
{4})(1 \! - \! \vert \gamma_{m} \vert^{2}P^{-2}(\phi_{m},\phi_{k})Q^{-2}
(\phi_{m}) \me^{2 \phi (x,t)})}{\sqrt{(\lambda_{1} \! - \! \lambda_{2})} \, 
(z_{o}^{2} \! + \! 32)^{1/4} \sin (\phi_{m}) \vert \gamma_{m} \vert P^{-1}
(\phi_{m},\phi_{k})Q^{-1}(\phi_{m}) \me^{\phi (x,t)}} \\
\pm& \, \dfrac{4 \lambda_{1}^{2} \sqrt{\nu (\lambda_{1})} \sin (\phi_{m}) 
\vert \gamma_{m} \vert P^{-1}(\phi_{m},\phi_{k})Q^{-1}(\phi_{m}) \sin (s^{-}) 
\me^{\phi (x,t)}}{(\lambda_{1}^{2} \! - \! 2 \lambda_{1} \cos (\phi_{m}) \! 
+ \! 1)^{2} \sqrt{(\lambda_{1} \! - \! \lambda_{2})} \, (z_{o}^{2} \! + \! 
32)^{1/4}} \\
\times& \left(((\lambda_{1} \! + \! \lambda_{2}) \cos (\phi_{m}) \! - \! 2) 
\cos (\Theta^{-}(z_{o},t) \! - \! \tfrac{3 \pi}{4}) \! - \! (\lambda_{1} \! - 
\! \lambda_{2}) \sin (\phi_{m}) \sin (\Theta^{-}(z_{o},t) \! - \! \tfrac{3 
\pi}{4}) \right) \\
\pm& \, \dfrac{2 \lambda_{1} \sqrt{\nu (\lambda_{1})} \cos (\phi_{m}) \vert 
\gamma_{m} \vert P^{-1}(\phi_{m},\phi_{k})Q^{-1}(\phi_{m}) \me^{\phi (x,t)}}
{(\lambda_{1}^{2} \! - \! 2 \lambda_{1} \cos (\phi_{m}) \! + \! 1) \sqrt{
(\lambda_{1} \! - \! \lambda_{2})} \, (z_{o}^{2} \! + \! 32)^{1/4}} \! \left(
2 \cos (s^{-} \! - \! \phi_{m}) \! \cos (\Theta^{-}(z_{o},t) \! - \! \tfrac{
3 \pi}{4}) \right. \\
-& \left. (\lambda_{1} \! + \! \lambda_{2}) \cos (s^{-}) \cos (\Theta^{-}
(z_{o},t) \! - \! \tfrac{3 \pi}{4}) \! + \! (\lambda_{1} \! - \! \lambda_{2}) 
\sin (s^{-}) \sin (\Theta^{-}(z_{o},t) \! - \! \tfrac{3 \pi}{4}) \right) \\
\pm& \, \dfrac{2 \sqrt{\nu (\lambda_{1})} \vert \gamma_{m} \vert P^{-1}(\phi_{
m},\phi_{k})Q^{-1}(\phi_{m}) \cos (s^{-} \! - \! \phi_{m}) \cos (\Theta^{-}
(z_{o},t) \! - \! \tfrac{3 \pi}{4}) \me^{\phi (x,t)}}{\sqrt{(\lambda_{1} \! 
- \! \lambda_{2})} \, (z_{o}^{2} \! + \! 32)^{1/4}} \\
+& \, \mathcal{O} \! \left(\tfrac{c^{\mathcal{S}}(z_{o})}{(z_{o}^{2}+32)^{
1/2}} \tfrac{\ln \vert t \vert}{t} \right),
\end{align*}
\begin{equation*}
\Re (\widetilde{a}_{1} \! - \! \widetilde{a}_{2}) \! = \! \Re (\widetilde{
a}_{3}) \! = \! \Im (\widetilde{a}_{3}) \! = \! 0,
\end{equation*}
where $\theta^{-}(\cdot)$ is given in Theorem~{\rm 2.2.1}, Eq.~{\rm (9)}.
\end{by}

The analogue of Lemma~4.4 is
\setcounter{z2}{1}
\setcounter{z3}{8}
\begin{ay}
As $t \! \to \! -\infty$ and $x \! \to \! +\infty$ such that $z_{o} \! := \! 
x/t \! < \! -2$ and $(x,t) \! \in \! \daleth_{m}$, $m \! \in \! \{1,2,\ldots,
N\}$, $u(x,t)$, the solution of the Cauchy problem for the 
{\rm D${}_{f}$NLSE}, and $\int_{\pm \infty}^{x}(\vert u(x^{\prime},t) \vert^{
2} \! - \! 1) \, \md x^{\prime}$ have the leading-order asymptotic expansions 
(for the lower sign) stated in Theorem~{\rm 2.2.1}, Eqs.~{\rm (7)--(20)}.
\end{ay}
\clearpage
\section*{Appendix~B}
\setcounter{section}{2}
\setcounter{z2}{1}
\setcounter{z3}{1}
In order to obtain the results of Theorems~2.2.3 and~2.2.4, the following 
Lemma, which is the analogue of Lemmae~4.2 and~A.1.6, is requisite.
\begin{ay}
Let $\varepsilon$ be an arbitrarily fixed, sufficiently small positive real 
number, and, for $\lambda \! \in \! \widetilde{\mathfrak{J}} \! := \! \{(s_{
1})^{\pm 1},(s_{2})^{\pm 1}\}$, where
\begin{align*}
s_{1} \! &= \! -\tfrac{1}{2}(a_{1} \! - \! \mi (4 \! - \! a_{1}^{2})^{1/2}) 
\! = \! \me^{\mi \widehat{\varphi}_{1}}, & \widehat{\varphi}_{1} \! &:= \! 
\arctan \! \left( \tfrac{(4-a_{1}^{2})^{1/2}}{\vert a_{1} \vert} \right) \! 
\in \! (0,\tfrac{\pi}{2}), & a_{1} \! &< \! 0, & \vert a_{1} \vert \! &< \! 
2, \\
s_{2} \! &= \! -\tfrac{1}{2}(a_{2} \! - \! \mi (4 \! - \! a_{2}^{2})^{1/2}) 
\! = \! \me^{\mi \widehat{\varphi}_{2}}, & \widehat{\varphi}_{2} \! &:= \! 
-\arctan \! \left( \tfrac{(4-a_{2}^{2})^{1/2}}{\vert a_{2} \vert} \right) 
\! \in \! (\tfrac{\pi}{2},\pi), & a_{2} \! &> \! 0, & \vert a_{2} \vert \! 
&< \! 2,
\end{align*}
with $a_{1}$ and $a_{2}$ given in Theorem~{\rm 2.2.1}, Eq.~{\rm (10)}, set 
$\mathbb{U}(\lambda;\varepsilon) \! := \! \{\mathstrut z; \, \vert z \! - \! 
\lambda \vert \! < \! \varepsilon\}$. Then, for $r(s_{1}) \! = \! \exp (-\mi 
\varepsilon_{1} \pi/2) \vert r(s_{1}) \vert$, $\varepsilon_{1} \! \in \! \{
\pm 1\}$, $r(\overline{s_{2}}) \! = \! \exp (\mi \varepsilon_{2} \pi/2) \vert 
r(\overline{s_{2}}) \vert$, $\varepsilon_{2} \! \in \! \{\pm 1\}$, $0 \! < 
\! r(s_{2}) \overline{r(\overline{s_{2}})} \! < \! 1$, and $\zeta \! \in \! 
\mathbb{C} \setminus \cup_{\lambda \, \in \, \widetilde{\mathfrak{J}}} \, 
\mathbb{U}(\lambda;\varepsilon)$, as $t \! \to \! +\infty$ and $x \! \to \! 
-\infty$ such that $z_{o} \! := \! x/t \! \in \! (-2,0)$, $m^{c}(\zeta)$ has 
the following asymptotics,
\begin{align*}
m^{c}_{11}(\zeta) \! &= \! 1 \! + \! \mathcal{O} \! \left( \! \left(\dfrac{
\underline{c}(z_{o})}{(\zeta \! - \! s_{1})} \! + \! \dfrac{\underline{c}
(z_{o})}{(\zeta \! - \! \overline{s_{2}})} \right) \! \dfrac{\me^{-4 \alpha 
t}}{\beta t} \right), \\
m^{c}_{12}(\zeta) \! &= \! \dfrac{\varepsilon_{1} \me^{-\left(2a_{0}t+\sin 
(\widehat{\varphi}_{1}) \int_{-\infty}^{0} \! \frac{\ln (1-\vert r(\mu) 
\vert^{2})}{(\mu - \cos \widehat{\varphi}_{1})^{2}+\sin^{2} \widehat{
\varphi}_{1}} \frac{\md \mu}{\pi} \right)} \me^{-\mi \left(\widehat{
\varphi}_{1}+\int_{-\infty}^{0} \! \frac{(\mu - \cos \widehat{\varphi}_{1}) 
\ln (1-\vert r(\mu) \vert^{2})}{(\mu -\cos \widehat{\varphi}_{1})^{2}+\sin^{
2} \widehat{\varphi}_{1}} \frac{\md \mu}{\pi} \right)}}{2 (\vert r(s_{1}) 
\vert)^{-1}(b_{0}t)^{1/2}(\zeta \! - \! \overline{s_{1}})} \\
 &+\dfrac{\varepsilon_{2} \me^{-\left(2\widehat{a}_{0}t-\sin (\widehat{
\varphi}_{3}) \int_{-\infty}^{0} \! \frac{\ln (1-\vert r(\mu) \vert^{2})}{
(\mu - \cos \widehat{\varphi}_{3})^{2}+\sin^{2} \widehat{\varphi}_{3}} 
\frac{\md \mu}{\pi} \right)} \me^{\mi \left(\widehat{\varphi}_{3}-\int_{-
\infty}^{0} \! \frac{(\mu -\cos \widehat{\varphi}_{3}) \ln (1-\vert r(\mu) 
\vert^{2})}{(\mu -\cos \widehat{\varphi}_{3})^{2}+\sin^{2} \widehat{
\varphi}_{3}} \frac{\md \mu}{\pi} \right)}}{2 (\vert r(\overline{s_{2}}) 
\vert)^{-1}(1 \! - \! r(s_{2}) \overline{r(\overline{s_{2}})})(\widehat{
b}_{0}t)^{1/2}(\zeta \! - \! s_{2})} \\
 &+\mathcal{O} \! \left( \! \left( \dfrac{\underline{c}(z_{o})}{(\zeta \! - 
\! \overline{s_{1}})} \! + \! \dfrac{\underline{c}(z_{o})}{(\zeta \! - \! 
s_{2})} \right) \! \dfrac{\me^{-4 \alpha t}}{\beta t} \right), \\
m^{c}_{21}(\zeta) \! &= \! \dfrac{\varepsilon_{1} \me^{-\left(2a_{0}t+\sin 
(\widehat{\varphi}_{1}) \int_{-\infty}^{0} \! \frac{\ln (1-\vert r(\mu) 
\vert^{2})}{(\mu - \cos \widehat{\varphi}_{1})^{2}+\sin^{2} \widehat{
\varphi}_{1}} \frac{\md \mu}{\pi} \right)} \me^{\mi \left(\widehat{
\varphi}_{1}+\int_{-\infty}^{0} \! \frac{(\mu - \cos \widehat{\varphi}_{1}) 
\ln (1-\vert r(\mu) \vert^{2})}{(\mu - \cos \widehat{\varphi}_{1})^{2}+
\sin^{2} \widehat{\varphi}_{1}} \frac{\md \mu}{\pi} \right)}}{2 (\vert 
r(s_{1}) \vert)^{-1}(b_{0}t)^{1/2}(\zeta \! - \! s_{1})} \\
 &+\dfrac{\varepsilon_{2} \me^{-\left(2\widehat{a}_{0}t- \sin (\widehat{
\varphi}_{3}) \int_{-\infty}^{0} \! \frac{\ln (1-\vert r(\mu) \vert^{2})}{
(\mu - \cos \widehat{\varphi}_{3})^{2}+\sin^{2} \widehat{\varphi}_{3}} 
\frac{\md \mu}{\pi} \right)} \me^{-\mi \left(\widehat{\varphi}_{3}-\int_{
-\infty}^{0} \! \frac{(\mu -\cos \widehat{\varphi}_{3}) \ln (1-\vert r(\mu) 
\vert^{2})}{(\mu - \cos \widehat{\varphi}_{3})^{2}+\sin^{2} \widehat{
\varphi}_{3}} \frac{\md \mu}{\pi} \right)}}{2(\vert r(\overline{s_{2}}) 
\vert)^{-1}(1 \! - \! r(s_{2}) \overline{r(\overline{s_{2}})})(\widehat{
b}_{0}t)^{1/2}(\zeta \! - \! \overline{s_{2}})} \\
 &+\mathcal{O} \! \left( \! \left( \dfrac{\underline{c}(z_{o})}{(\zeta \! - 
\! s_{1})} \! + \! \dfrac{\underline{c}(z_{o})}{(\zeta \! - \! \overline{s_{
2}})} \right) \! \dfrac{\me^{-4 \alpha t}}{\beta t} \right), \\
m^{c}_{22}(\zeta) \! &= \! 1 \! + \! \mathcal{O} \! \left( \! \left(\dfrac{
\underline{c}(z_{o})}{(\zeta \! - \! \overline{s_{1}})} \! + \! \dfrac{
\underline{c}(z_{o})}{(\zeta \! - \! s_{2})} \right) \! \dfrac{\me^{-4\alpha 
t}}{\beta t} \right),
\end{align*}
where
\begin{gather*}
a_{0} \! = \! \tfrac{1}{2}(z_{o} \! - \! a_{1})(4 \! - \! a_{1}^{2})^{1/2} 
\, \, \, \, \, (>0), \qquad \quad \widehat{a}_{0} \! = \! -\tfrac{1}{2}
(z_{o} \! - \! a_{2})(4 \! - \! a_{2}^{2})^{1/2} \, \, \, \, \, (>0), \\
b_{0} \! = \! \tfrac{1}{2}(z_{o}^{2} \! + \! 32)^{1/2}(4 \! - \! a_{1}^{
2})^{1/2} \, \, \, \, \, (>0), \qquad \quad \widehat{b}_{0} \! = \! \tfrac{
1}{2}(z_{o}^{2} \! + \! 32)^{1/2}(4 \! - \! a_{2}^{2})^{1/2} \, \, \, \, \, 
(>0), \\
\alpha \! := \! \min \{a_{0},\widehat{a}_{0}\}, \qquad  \quad \beta \! := \! 
\min\{b_{0},\widehat{b}_{0}\},
\end{gather*}
and, for $r(\overline{s_{1}}) \! = \! \exp (\mi \varepsilon_{1} \pi/2) 
\vert r(\overline{s_{1}}) \vert$, $\varepsilon_{1} \! \in \! \{\pm 1\}$, 
$r(s_{2}) \! = \! \exp (-\mi \varepsilon_{2} \pi/2) \vert r(s_{2}) \vert$, 
$\varepsilon_{2} \! \in \! \{\pm 1\}$, $0 \! < \! r(s_{1}) \overline{
r(\overline{s_{1}})} \! < \! 1$, and $\zeta \! \in \! \mathbb{C} \setminus 
\cup_{\lambda \, \in \, \widetilde{\mathfrak{J}}} \, \mathbb{U}(\lambda;
\varepsilon)$, as $t \! \to \! -\infty$ and $x \! \to \! +\infty$ such that 
$z_{o} \! \in \! (-2,0)$,
\begin{align*}
m^{c}_{11}(\zeta) \! &= \! 1 \! + \! \mathcal{O} \! \left( \! \left(\dfrac{
\underline{c}(z_{o})}{(\zeta \! - \! \overline{s_{1}})} \! + \! \dfrac{
\underline{c}(z_{o})}{(\zeta \! - \! s_{2})} \right) \! \dfrac{\me^{-4
\alpha \vert t \vert}}{\beta t} \right), \\
m^{c}_{12}(\zeta) \! &= \! -\dfrac{\varepsilon_{1} \me^{-\left(2a_{0} \vert 
t \vert -\sin (\widehat{\varphi}_{1}) \int_{0}^{+\infty} \! \frac{\ln (1-
\vert r(\mu) \vert^{2})}{(\mu - \cos \widehat{\varphi}_{1})^{2}+\sin^{2} 
\widehat{\varphi}_{1}} \frac{\md \mu}{\pi} \right)} \me^{\mi \left(\widehat{
\varphi}_{1}-\int_{0}^{+\infty} \! \frac{(\mu - \cos \widehat{\varphi}_{1}) 
\ln (1-\vert r(\mu) \vert^{2})}{(\mu - \cos \widehat{\varphi}_{1})^{2}+
\sin^{2} \widehat{\varphi}_{1}} \frac{\md \mu}{\pi} \right)}}{2(\vert 
r(\overline{s_{1}}) \vert)^{-1}(1 \! - \! r(s_{1}) \overline{r(\overline{
s_{1}})})(b_{0} \vert t \vert)^{1/2}(\zeta \! - \! s_{1})} \\
 &-\dfrac{\varepsilon_{2} \me^{-\left(2\widehat{a}_{0} \vert t \vert +\sin 
(\widehat{\varphi}_{3}) \int_{0}^{+\infty} \! \frac{\ln (1-\vert r(\mu) 
\vert^{2})}{(\mu - \cos \widehat{\varphi}_{3})^{2}+\sin^{2} \widehat{
\varphi}_{3}} \frac{\md \mu}{\pi} \right)} \me^{-\mi \left(\widehat{
\varphi}_{3}+\int_{0}^{+\infty} \! \frac{(\mu - \cos \widehat{\varphi}_{3}) 
\ln (1-\vert r(\mu) \vert^{2})}{(\mu - \cos \widehat{\varphi}_{3})^{2}+
\sin^{2} \widehat{\varphi}_{3}} \frac{\md \mu}{\pi} \right)}}{2(\vert 
r(s_{2}) \vert)^{-1}(\widehat{b}_{0} \vert t \vert)^{1/2}(\zeta \! - \! 
\overline{s_{2}})} \\
 &+\mathcal{O} \! \left( \! \left( \dfrac{\underline{c}(z_{o})}{(\zeta \! - 
\! s_{1})} \! + \! \dfrac{\underline{c}(z_{o})}{(\zeta \! - \! \overline{
s_{2}})} \right) \! \dfrac{\me^{-4 \alpha \vert t \vert}}{\beta t} \right),
\end{align*}
\begin{align*}
m^{c}_{21}(\zeta) \! &= \! -\dfrac{\varepsilon_{1} \me^{-\left(2a_{0} \vert 
t \vert -\sin (\widehat{\varphi}_{1}) \int_{0}^{+\infty} \! \frac{\ln (1-
\vert r(\mu) \vert^{2})}{(\mu - \cos \widehat{\varphi}_{1})^{2}+\sin^{2} 
\widehat{\varphi}_{1}} \frac{\md \mu}{\pi} \right)} \me^{-\mi \left(\widehat{
\varphi}_{1}-\int_{0}^{+\infty} \! \frac{(\mu - \cos \widehat{\varphi}_{1}) 
\ln (1-\vert r(\mu) \vert^{2})}{(\mu - \cos \widehat{\varphi}_{1})^{2}+
\sin^{2} \widehat{\varphi}_{1}} \frac{\md \mu}{\pi} \right)}}{2(\vert 
r(\overline{s_{1}}) \vert)^{-1}(1 \! - \! r(s_{1}) \overline{r(\overline{
s_{1}})})(b_{0} \vert t \vert)^{1/2}(\zeta \! - \! \overline{s_{1}})} \\
 &-\dfrac{\varepsilon_{2} \me^{-\left(2\widehat{a}_{0} \vert t \vert +\sin 
(\widehat{\varphi}_{3}) \int_{0}^{+\infty} \! \frac{\ln (1-\vert r(\mu) 
\vert^{2})}{(\mu - \cos \widehat{\varphi}_{3})^{2}+\sin^{2} \widehat{
\varphi}_{3}} \frac{\md \mu}{\pi} \right)} \me^{\mi \left(\widehat{
\varphi}_{3}+\int_{0}^{+\infty} \! \frac{(\mu - \cos \widehat{\varphi}_{3}) 
\ln (1-\vert r(\mu) \vert^{2})}{(\mu - \cos \widehat{\varphi}_{3})^{2}+
\sin^{2} \widehat{\varphi}_{3}} \frac{\md \mu}{\pi} \right)}}{2(\vert 
r(s_{2}) \vert)^{-1}(\widehat{b}_{0} \vert t \vert)^{1/2}(\zeta \! - \! 
s_{2})} \\
&+\mathcal{O} \! \left( \! \left( \dfrac{\underline{c}(z_{o})}{(\zeta \! - 
\! \overline{s_{1}})} \! + \! \dfrac{\underline{c}(z_{o})}{(\zeta \! - \! 
s_{2})} \right) \! \dfrac{\me^{-4 \alpha \vert t \vert}}{\beta t} \right), 
\\
m^{c}_{22}(\zeta) \! &= \! 1 \! + \! \mathcal{O} \! \left( \! \left(\dfrac{
\underline{c}(z_{o})}{(\zeta \! - \! s_{1})} \! + \! \dfrac{\underline{c}
(z_{o})}{(\zeta \! - \! \overline{s_{2}})} \right) \! \dfrac{\me^{-4\alpha 
\vert t \vert}}{\beta t} \right),
\end{align*}
where $\sup_{\zeta \in \mathbb{C} \, \setminus \cup_{\lambda \in \widetilde{
\mathfrak{J}}} \mathbb{U}(\lambda;\varepsilon)} \vert (\zeta \! - \! (s_{
n})^{\pm 1})^{-1} \vert \! < \! \infty$, and $m^{c}(\zeta) \! = \! \sigma_{
1} \overline{m^{c}(\overline{\zeta})} \, \sigma_{1}$.
\end{ay}
\clearpage
\section*{Appendix~C. Matrix Riemann-Hilbert Theory in the $L^{2}$ Sobolev 
Space}
\setcounter{section}{3}
\setcounter{z0}{1}
\setcounter{z1}{1}
In this Appendix, the theoretical foundation for this paper is presented. 
Beginning {}from the Lax-pair isospectral deformation formulation for a 
completely integrable NLEE, in the sense of the ISM, a succinct review of 
several basic and key facts {}from the $2 \times 2$ matrix RH factorisation 
theory on unbounded self-intersecting contours is presented: for complete 
details and proofs, see \cite{a40,a44,a45,a46,a47,a50}. For simplicity, one 
begins with the solitonless sector, $\sigma_{d} \! \equiv \! \emptyset$, 
leading to the so-called ``regular'' RHP: inclusion of the (non-empty and 
finitely denumerable) discrete spectrum, $\sigma_{d}$, is known as the 
``singular'' RHP, and is discussed below Theorem~C.1.4.

For a completely integrable system of NLEEs, in the sense of the ISM, 
write the spatial part of the associated Lax pair (see, for example, 
Proposition~2.1.1) as $\partial_{x} \widetilde{\Psi}(x,t;\lambda) \! = \! 
(\widetilde{J}(\lambda) \! + \! \widetilde{R}(x,t;\lambda)) \widetilde{\Psi}
(x,t;\lambda)$, where $(x,t) \! \in \! \mathbb{R} \times [-T,T]$, $\lambda \! 
\in \! \mathbb{C}$, $\widetilde{J}(\lambda) \! := \! \mathrm{diag}(z_{1}
(\lambda),z_{2}(\lambda))$ is rational with distinct entries, and $\widetilde{
R}(x,t;\lambda)$ is off-diagonal. The orders of the poles of $\widetilde{J}
(\lambda)$ and $\widetilde{R}(x,t;\lambda)$ must satisfy the following 
requirements (denote by $\mathrm{P}_{\widetilde{J}}$ the set of poles of 
$\widetilde{J}(\lambda)$, and let $k(\lambda^{\prime})$ denote the order of 
the pole of $\lambda^{\prime} \! \in \! \mathrm{P}_{\widetilde{J}})$: (1) 
every pole of $\widetilde{R}(x,t;\lambda)$ is a pole of $\widetilde{J}
(\lambda)$; (2) if $\infty$ is a pole of $\widetilde{J}(\lambda)$ of order $k
(\infty)$, then it is a pole of $\widetilde{R}(x,t;\lambda)$ of order not 
greater than $k(\infty) \! - \! 1$; and (3) if $\lambda^{\prime}$ is a finite 
pole of $\widetilde{J}(\lambda)$ of order $k(\lambda^{\prime})$, then it is a 
pole of $\widetilde{R}(x,t;\lambda)$ of order not greater than $k(\lambda^{
\prime})$. Hence, one has the following representations for $\widetilde{J}
(\lambda)$ and $\widetilde{R}(x,t;\lambda)$: (1)
\begin{equation*}
\widetilde{J}(\lambda) \! = \! \sum_{\lambda^{\prime} \in \, \mathrm{P}_{
\widetilde{J}} \setminus \{\infty\}} \! \sum_{j=1}^{k(\lambda^{\prime})} 
\widetilde{J}_{\lambda^{\prime},j}(\lambda \! - \! \lambda^{\prime})^{-j} 
\! + \! \sum_{l=0}^{k(\infty)} \widetilde{J}_{\infty,l} \lambda^{l},
\end{equation*}
where $\widetilde{J}_{\lambda^{\prime},j}$ and $\widetilde{J}_{\infty,l}$ 
are $\mathrm{M}_{2}(\mathbb{C})$-valued, diagonal matrices with distinct 
elements; and (2)
\begin{equation*}
\widetilde{R}(x,t;\lambda) \! = \! \sum_{\lambda^{\prime} \in \, \mathrm{P}_{
\widetilde{J}} \setminus \{\infty\}} \! \sum_{j=1}^{k(\lambda^{\prime})}r_{
\lambda^{\prime},j}(x,t)(\lambda \! - \! \lambda^{\prime})^{-j} \! + \! 
\sum_{l=0}^{k(\infty)-1} r_{\infty,l}(x,t) \lambda^{l}.
\end{equation*}
\begin{ey}
Hereafter, for economy of notation, all explicit $x,t$ dependencies are 
suppressed.
\end{ey}
Denote by $\widetilde{\Lambda}$ the closure of $\{\mathstrut \lambda \! \in 
\! \mathbb{C}; \, \Re (z_{1}(\lambda) \! - \! z_{2}(\lambda)) \! = \! 0\}$. 
Decompose $\widetilde{\Lambda}$ into a finite union of piecewise smooth, 
simple, closed curves, $\widetilde{\Lambda} \! := \! \cup_{l \in L} \widetilde{
\Lambda}_{l}$ $(\mathrm{card}(L) \! < \! \infty)$. Denote by $\varpi$ the set 
of all self-intersections of $\widetilde{\Lambda}$, $\varpi \! := \! \{
\mathstrut \lambda; \, \widetilde{\Lambda}_{l} \cap \widetilde{\Lambda}_{m} 
\! \not= \! \emptyset, \, l \! \not= \! m \! \in \! \{1,2,\ldots,\mathrm{card}
(L)\}\}$ (it is assumed throughout that $\mathrm{card}(\varpi) \! < \! 
\infty)$. Divide the complement of $\widetilde{\Lambda}$ into two disjoint 
open subsets of $\mathbb{C}$, $\boldsymbol{\Omega}^{+}$ and $\boldsymbol{
\Omega}^{-}$, each of which have finitely many components, $\boldsymbol{
\Omega}^{\pm} \! := \! \cup_{l^{\pm} \in L^{\pm}} \boldsymbol{\Omega}^{\pm}_{
l^{\pm}}$ $(\mathrm{card}(L^{\pm}) \! < \! \infty)$, such that $\widetilde{
\Lambda}$ admits an orientation so that it can be viewed either as a 
positively (counter-clockwise) oriented boundary, $\widetilde{\Lambda}^{
+}$, for $\boldsymbol{\Omega}^{+}$, or as a negatively (clockwise) oriented 
boundary, $\widetilde{\Lambda}^{-}$, for $\boldsymbol{\Omega}^{-}$; moreover, 
for each component $\boldsymbol{\Omega}^{\pm}_{l^{\pm}}$, $\partial 
\boldsymbol{\Omega}^{\pm}_{l^{\pm}}$ has no self-intersections.
\begin{cy}
For an $\mathrm{M}_{2}(\mathbb{C})$-valued function, $f(\lambda)$, say, 
denote by $f_{\pm}(\lambda)$, respectively, the non-tangential limits, if 
they exist, of $f(\lambda)$ taken {}from $\boldsymbol{\Omega}^{\pm}$. For $f
(\lambda) \colon \widetilde{\Lambda} \! \to \! \mathrm{M}_{2}(\mathbb{C})$, 
define $f^{(0)}(\lambda) \! := \! f(\lambda)$, and, for $k \! \in \! \mathbb{
Z}_{\geqslant 1}$, $f^{(j)}(\lambda) \! := \! \partial_{\lambda}^{j}f
(\lambda)$, $j \! \in \! \{1,2,\ldots,k\}$. For the piecewise smooth simple 
closed curve $\widetilde{\Lambda} \! = \! \cup_{l \in L} \widetilde{\Lambda}_{
l}$, and $k \! \in \! \mathbb{Z}_{\geqslant 1}$, define the $\mathcal{L}^{
2}_{\mathrm{M}_{2}(\mathbb{C})}(\widetilde{\Lambda})$ Sobolev space $H^{k}
(\widetilde{\Lambda},\mathrm{M}_{2}(\mathbb{C}))$ as the set of all $\mathrm{
M}_{2}(\mathbb{C})$-valued functions on $\widetilde{\Lambda}$ satisfying: 
{\rm (1)} for $l \! \in \! \{1,2,\ldots,\mathrm{card}(L)\}$, $f^{(j)} \! \! 
\upharpoonright_{\widetilde{\Lambda}_{l}}$, $j \! \in \! \{0,1,\ldots,k \! - 
\! 1\}$, exist pointwise and $\in \! \mathcal{L}^{2}_{\mathrm{M}_{2}(\mathbb{
C})}(\widetilde{\Lambda}_{l})$; and {\rm (2)} for $l \! \in \! \{1,2,\ldots,
\mathrm{card}(L)\}$, $f^{(k-1)} \! \! \upharpoonright_{\widetilde{\Lambda}_{
l}}$ is locally absolutely continuous and $f^{(k)} \! \! \upharpoonright_{
\widetilde{\Lambda}_{l}} \in \! \mathcal{L}^{2}_{\mathrm{M}_{2}(\mathbb{
C})}(\widetilde{\Lambda}_{l})$. For $k \! = \! 0$, denote $H^{0}(\widetilde{
\Lambda},\mathrm{M}_{2}(\mathbb{C}))$ by $\mathcal{L}^{2}_{\mathrm{M}_{2}
(\mathbb{C})}(\widetilde{\Lambda})$. Define $H^{k}(\widetilde{\Lambda}^{\pm},
\mathrm{M}_{2}(\mathbb{C})) \! := \! \{\mathstrut f \colon \widetilde{\Lambda} 
\! \to \! \mathrm{M}_{2}(\mathbb{C}); \, f \! \! \upharpoonright_{\partial 
\boldsymbol{\Omega}_{l^{\pm}}^{\pm}} \! \in \! H^{k}(\partial \boldsymbol{
\Omega}_{l^{\pm}}^{\pm},\mathrm{M}_{2}(\mathbb{C})), \, l^{\pm} \! \in \! 
\{1,2,\ldots,\mathrm{card}(L^{\pm})\}, k \! \in \! \mathbb{Z}_{\geqslant 1}
\}:$ the norm on $H^{k}(\widetilde{\Lambda}^{\pm},\mathrm{M}_{2}(\mathbb{
C}))$, $k \! \in \! \mathbb{Z}_{\geqslant 1}$, is defined as $\vert \vert f
(\cdot) \vert \vert_{H^{k}(\widetilde{\Lambda},\mathrm{M}_{2}(\mathbb{C}))} 
\! := \! \vert \vert f(\cdot) \vert \vert_{2,k} \! := \! (\sum_{l \in L} \! 
\sum_{j=0}^{k} \vert \vert f^{(j)}(\cdot) \vert \vert^{2}_{\mathcal{L}^{
2}_{\mathrm{M}_{2}(\mathbb{C})}(\widetilde{\Lambda}_{l})})^{1/2}$. With this 
norm, $H^{k}(\widetilde{\Lambda}^{\pm},\mathrm{M}_{2}(\mathbb{C}))$ is a 
Hilbert space: for $k \! = \! 0$, $\vert \vert f(\cdot) \vert \vert_{2,0} 
\! = \! (\sum_{l \in L} \vert \vert f(\cdot) \vert \vert_{\mathcal{L}^{
2}_{\mathrm{M}_{2}(\mathbb{C})}(\widetilde{\Lambda}_{l})}^{2})^{1/2}$.
\end{cy}
The Cauchy integral operators on $\mathcal{L}^{2}_{\mathrm{M}_{2}(\mathbb{C}
)}(\widetilde{\Lambda})$ are defined as $(C_{\pm}f)(\lambda) \! := \! \lim_{
\genfrac{}{}{0pt}{2}{\lambda^{\prime} \to \lambda}{\lambda^{\prime} \in 
\boldsymbol{\Omega}^{\pm}}} \int_{\widetilde{\Lambda}} \! \tfrac{f(z)}{(z-
\lambda^{\prime})} \, \tfrac{\md z}{2 \pi \mi}$: note that $C_{+} \! - \! 
C_{-} \! = \! \id$, where $\id$ is the identity operator on $\mathcal{L}^{
2}_{\mathrm{M}_{2}(\mathbb{C})}(\widetilde{\Lambda})$. Since $\widetilde{
\Lambda} \! = \! \cup_{l \in L} \widetilde{\Lambda}_{l}$, where $\widetilde{
\Lambda}_{l}$, $l \! \in \! \{1,2,\ldots,\mathrm{card}(L)\}$, are piecewise 
smooth and simple, the Cauchy integral operators are bounded {}from $\mathcal{
L}^{2}_{\mathrm{M}_{2}(\mathbb{C})}(\widetilde{\Lambda})$ into $\mathcal{
L}^{2}_{\mathrm{M}_{2}(\mathbb{C})}(\widetilde{\Lambda})$; moreover, the 
aforementioned orientation for $\widetilde{\Lambda}$, that is, $\widetilde{
\Lambda}^{\pm}$, provides the Cauchy integral operators on $\mathcal{L}^{2}_{
\mathrm{M}_{2}(\mathbb{C})}(\widetilde{\Lambda})$ with the crucial property 
that $\pm C_{\pm}$ are complementary projections, that is, $C_{+}^{2} \! = \! 
C_{+}$, $C_{-}^{2} \! = \! - C_{-}$, $C_{+}C_{-} \! = \! C_{-}C_{+} \! = \! 
\underline{\mathbf{0}}$, where $\underline{\mathbf{0}}$ is the null operator 
on $\mathcal{L}^{2}_{\mathrm{M}_{2}(\mathbb{C})}(\widetilde{\Lambda})$. Even 
though $C_{\pm}$ are not bounded in operator norm on $H^{k}(\widetilde{
\Lambda},\mathrm{M}_{2}(\mathbb{C}))$, $C_{\pm}$ are bounded on $\oplus_{
\alpha \in \{\pm\}}H^{k}(\widetilde{\Lambda}^{\alpha},\mathrm{M}_{2}(\mathbb{
C}))$; moreover, injectively, $C_{\pm} \colon H^{k}(\widetilde{\Lambda}^{
\pm},\mathrm{M}_{2}(\mathbb{C})) \! \to \! H^{k}(\widetilde{\Lambda}^{\pm},
\mathrm{M}_{2}(\mathbb{C}))$, and $C_{\pm} \colon H^{k}(\widetilde{\Lambda}^{
\mp},\mathrm{M}_{2}(\mathbb{C})) \! \to \! \widetilde{H}^{k}(\widetilde{
\Lambda},\mathrm{M}_{2}(\mathbb{C})) \! := \! \cap_{\alpha \in \{\pm\}}H^{k}
(\widetilde{\Lambda}^{\alpha},\mathrm{M}_{2}(\mathbb{C}))$. Since, in the 
ISM, $\widetilde{\Lambda}$ is (usually) unbounded, the function $f \! \! 
\upharpoonright_{\widetilde{\Lambda}^{\pm}}= \! \mathrm{I} \! \not\in \! 
H^{k}(\widetilde{\Lambda}^{\pm},\mathrm{M}_{2}(\mathbb{C}))$, $k \! \in \! 
\mathbb{Z}_{\geqslant 0}$; hence, for $D \! \in \! \{\widetilde{\Lambda},
\widetilde{\Lambda}^{\pm}\}$, embed $H^{k}(D,\mathrm{M}_{2}(\mathbb{C}))$, 
$k \! \in \! \mathbb{Z}_{\geqslant 0}$, into a larger Hilbert space $H^{k}_{
I}(D,\mathrm{M}_{2}(\mathbb{C}))$ consisting of $\mathrm{M}_{2}(\mathbb{
C})$-valued functions $f(\lambda)$ on $\widetilde{\Lambda} \cup (\cup_{\alpha 
\in \{\pm\}} \boldsymbol{\Omega}^{\alpha})$ with the limit $f(\infty)$ at 
$\infty$ such that $f(\lambda) \! - \! f(\infty) \! \in \! H^{k}(D,\mathrm{
M}_{2}(\mathbb{C}))$, with the norm defined by $\vert \vert f(\cdot) \vert 
\vert_{H^{k}_{I}(D,\mathrm{M}_{2}(\mathbb{C}))} \! := \! \vert \vert f(\cdot) 
\vert \vert_{I,2,k} \! := \! (\vert f(\infty) \vert^{2} \! + \! \vert \vert 
f(\cdot) \! - \! f (\infty) \vert \vert_{2,k}^{2})^{1/2}$. $H^{k}_{I}(D,
\mathrm{M}_{2}(\mathbb{C}))$, $k \! \in \! \mathbb{Z}_{\geqslant 0}$, is 
isomorphic to the Hilbert space direct sum of $\mathrm{M}_{2}(\mathbb{C})$ 
and $H^{k}(D,\mathrm{M}_{2}(\mathbb{C}))$ $(H^{k}_{I}(D,\mathrm{M}_{2}
(\mathbb{C})) \! \approx \! \mathrm{M}_{2}(\mathbb{C}) \oplus H^{k}(D,
\mathrm{M}_{2}(\mathbb{C})))$.

Define: (1) $GH^{k}_{I}(\widetilde{\Lambda}^{\pm},\mathrm{M}_{2}(\mathbb{C})) 
\! := \! \{\mathstrut f(\lambda) \! \in \! H^{k}_{I}(\widetilde{\Lambda}^{
\pm},\mathrm{M}_{2}(\mathbb{C})); \, \det (f(\lambda)) \! \not\equiv \! 0\}$; 
and (2) $SH^{k}_{I}(\widetilde{\Lambda}^{\pm},\mathrm{M}_{2}(\mathbb{C})) 
\linebreak[4]
:= \! \{\mathstrut f(\lambda) \! \in \! H^{k}_{I}(\widetilde{\Lambda}^{\pm},
\mathrm{M}_{2}(\mathbb{C})); \, \det (f(\lambda)) \! = \! 1\}$. If $\chi^{c}_{
\pm}(\lambda) \! - \! \chi^{c}_{\pm}(\infty) \! \in \! \mathrm{ran} \, C_{
\pm}$ $(\subset \! H^{k}(\widetilde{\Lambda}^{\pm},\mathrm{M}_{2}(\mathbb{
C})))$, where $\chi^{c}_{\pm}(\infty) \! := \! \lim_{\genfrac{}{}{0pt}{2}{
\lambda \to \infty}{\lambda \in \boldsymbol{\Omega}^{\pm}}} \chi^{c}
(\lambda)$, denote by $\chi^{c}(\lambda)$ the sectionally holomorphic 
function on $\cup_{\alpha \in \{\pm\}} \boldsymbol{\Omega}^{\alpha}$ with 
boundary values $\chi^{c}_{\pm}(\lambda)$. Define: (1) $\mathcal{H}^{k}
(\mathbb{C} \setminus \widetilde{\Lambda},\mathrm{M}_{2}(\mathbb{C})) \! := 
\! \{\mathstrut \chi^{c}(\lambda); \, \chi^{c}_{\pm}(\lambda) \! - \! \chi^{
c}_{\pm}(\infty) \! \in \! \mathrm{ran} \, C_{\pm}\}$; (2) $\mathcal{GH}^{k}
(\mathbb{C} \setminus \widetilde{\Lambda},\mathrm{M}_{2}(\mathbb{C})) \! 
:= \! \{\mathstrut \chi^{c}(\lambda) \! \in \! \mathcal{H}^{k}(\mathbb{C} 
\setminus \widetilde{\Lambda},\mathrm{M}_{2}(\mathbb{C})); \, \det (\chi^{
c}(\lambda)) \! \not\equiv \! 0\}$; and (3) $\mathcal{SH}^{k}(\mathbb{C} 
\setminus \widetilde{\Lambda},\mathrm{M}_{2}(\mathbb{C})) \! := \! \{
\mathstrut \chi^{c}(\lambda) \! \in \! \mathcal{H}^{k}(\mathbb{C} \setminus 
\widetilde{\Lambda},\mathrm{M}_{2}(\mathbb{C})); \, \det (\chi^{c}(\lambda)) 
\! = \! 1\}$.
\begin{dy}
Every $v(\lambda) \! \in \! GH^{k}_{I}(\widetilde{\Lambda}^{-},\mathrm{M}_{2}
(\mathbb{C})) \ast GH^{k}_{I}(\widetilde{\Lambda}^{+},\mathrm{M}_{2}(\mathbb{
C}))$ $(A \ast B \! := \! \{\mathstrut xy; \, x\! \in \! A, \, y\! \in \! 
B\})$, $\lambda \! \in \! \widetilde{\Lambda}$, admits an {\rm RH} 
factorisation, $v(\lambda) \! = \! (\chi^{c}_{-}(\lambda))^{-1} \blacklozenge 
(\lambda) \chi^{c}_{+}(\lambda)$, where $\blacklozenge (\lambda) \! := \! 
\mathrm{diag} \! \left((\tfrac{\lambda -\lambda_{+}}{\lambda -\lambda_{-}})^{
k_{1}},(\tfrac{\lambda -\lambda_{+}}{\lambda -\lambda_{-}})^{k_{2}} \right)$, 
$\lambda_{\pm} \! \in \! \boldsymbol{\Omega}^{\pm}$, and $\chi^{c}(\lambda) 
\! \in \! \mathcal{GH}^{k}(\mathbb{C} \setminus \widetilde{\Lambda},\mathrm{
M}_{2}(\mathbb{C}))$ $(k_{i}$, $i \! \in \! \{1,2\}$, are called the partial 
indices (uniquely determined by $v(\cdot)$ up to a permutation) of 
$v(\lambda))$; moreover, if $\det(v(\lambda)) \! = \! 1$, $\chi^{c}(\lambda)$ 
can be chosen to be in $\mathcal{SH}^{k}(\mathbb{C} \setminus \widetilde{
\Lambda},\mathrm{M}_{2}(\mathbb{C}))$, and $\sum_{j=1}^{2}k_{j} \! = \! 0$. 
The matrix $\chi^{c}(\lambda) \! \in \! \widetilde{H}^{j}_{I}(\widetilde{
\Lambda},\mathrm{M}_{2}(\mathbb{C}))$, for some $j \! \in \! \{0,1,\ldots,
k\}$, $k \! \in \! \mathbb{Z}_{\geqslant 1}$, is said to be a solution of 
the {\rm RH} factorisation problem of $v(\lambda)$ if $\chi^{c}_{\pm} 
(\lambda) \! - \! \chi^{c}_{\pm}(\infty) \! \in \! \mathrm{ran} \, C_{\pm} 
\! \subset \! H^{k}(\widetilde{\Lambda}^{\pm},\mathrm{M}_{2}(\mathbb{C}))$. 
When $v(\infty) \! = \! \mathrm{I}$ and $\blacklozenge (\lambda) \! = \! 
\mathrm{I}$, $\chi^{c}_{\pm}(\lambda)$ can be uniquely determined by letting 
$\chi^{c}_{\pm}(\infty) \! = \! \mathrm{I}$ (canonical normalisation), in 
which case, $\chi^{c}_{\pm}(\lambda)$, or $\chi^{c}(\lambda)$ $(\chi^{c}
(\infty) \! = \! \mathrm{I})$, is called the fundamental solution of the 
{\rm RHP} of $v(\lambda)$. For the {\rm ISM}, $v(\infty) \! = \! \mathrm{
I}$. Conversely, if $v(\lambda)$ admits a factorisation $v(\lambda) \! = \! 
(\chi^{c}_{-}(\lambda))^{-1} \blacklozenge (\lambda) \chi^{c}_{+}(\lambda)$, 
then $v(\lambda) \! \in \! GH^{k}_{I}(\widetilde{\Lambda}^{-},\mathrm{M}_{
2}(\mathbb{C})) \ast GH^{k}_{I}(\widetilde{\Lambda}^{+},\mathrm{M}_{2}
(\mathbb{C}))$.
\end{dy}
\begin{by}
$\mathrm{tr}(\widetilde{R}(\lambda)) \! = \! 0 \! \Rightarrow \! \det(\chi^{
c}(\lambda)) \! = \! \mathrm{const.}$.
\end{by}
\setcounter{z0}{1}
\setcounter{z1}{2}
\begin{cy}
A linear operator $\mathscr{L}$ on $H^{k}_{I}(D,\mathrm{M}_{2}(\mathbb{C}))$ 
is Fredholm if: {\rm (1)} the complement of $\mathrm{ran} \, \mathscr{L}$ is 
open in $H^{k}_{I}(D;\mathrm{M}_{2}(\mathbb{C}));$ and {\rm (2)} $\dim \ker 
({\mathscr L})$ and $\dim \mathrm{coker}({\mathscr L})$ are finite. For ${
\mathscr L}$ linear and Fredholm, $i({\mathscr L}) \! := \! \dim \ker ({
\mathscr L}) \! - \! \dim \mathrm{coker}({\mathscr L})$ is called the 
(Fredholm) index of ${\mathscr L}$.
\end{cy}
\setcounter{z0}{1}
\setcounter{z1}{2}
\begin{dy}
Let $k \! \in \! \mathbb{Z}_{\geqslant 1}$. If $v(\lambda)$ in 
Theorem~{\rm C.1.1} can be represented as the following (algebraic) block 
triangular factorisation, $v(\lambda) \! := \! (\mathrm{I} \! - \! w^{-}
(\lambda))^{-1}(\mathrm{I} \! + \! w^{+}(\lambda))$, $\lambda \! \in \! 
\widetilde{\Lambda}$, where $w^{\pm}(\lambda) \! \in \! H^{k}(\widetilde{
\Lambda}^{\pm},\mathrm{M}_{2}(\mathbb{C}))$, $\mathrm{I} \! \pm \! w^{\pm}
(\lambda) \! \in \! GH^{k}_{I}(\widetilde{\Lambda}^{\pm},\mathrm{M}_{2}
(\mathbb{C}))$, and $w^{\pm}(\lambda)$ are nilpotent, with degree of 
nilpotency 2, and if, as a linear operator on $\widetilde{H}^{k}_{I}
(\widetilde{\Lambda},\mathrm{M}_{2}(\mathbb{C})) \! := \! \cap_{\alpha \in 
\{\pm\}}H^{k}_{I}(\widetilde{\Lambda}^{\alpha},\mathrm{M}_{2}(\mathbb{C}))$, 
$C_{w} \colon \widetilde{H}^{k}_{I}(\widetilde{\Lambda},\mathrm{M}_{2}
(\mathbb{C})) \! \to \! \widetilde{H}^{k}_{I}(\widetilde{\Lambda},\mathrm{
M}_{2}(\mathbb{C}))$ is defined as $(f \! \in \! \widetilde{H}^{k}_{I}
(\widetilde{\Lambda},\mathrm{M}_{2}(\mathbb{C})))$ $f \! \mapsto \! C_{+}
(fw^{-}) \! + \! C_{-}(fw^{+})$, then $\id \! - \! C_{w}$, where $\id$ is the 
identity operator on $\widetilde{H}^{k}_{I}(\widetilde{\Lambda},\mathrm{M}_{
2}(\mathbb{C}))$, is Fredholm, that is, $i(\id \! - \! C_{w}) \! = \! \dim 
\ker (\id \! - \! C_{w}) \! - \! \dim \mathrm{coker}(\id \! - \! C_{w}) 
\! = \! 0$, $\dim \ker (\id \! - \! C_{w}) \! = \! 2 \sum_{k_{j}>0}k_{j}$, 
and $\dim \mathrm{coker}(\id \! - \! C_{w}) \! = \! - 2 \sum_{k_{j}<0}
k_{j}$, where $k_{i}$, $i \! \in \! \{1,2\}$, are the partial indices of 
$v(\lambda)$; moreover, $i(\id \! - \! C_{w}) \! = \! 2 \, \mathrm{ind} \det(
v(\lambda)) \! = \! \tfrac{1}{\pi} \! \int_{\widetilde{\Lambda}} \! d(\arg 
\det (v(\cdot))) \! = \! 0$, where $\mathrm{ind} \det (v(\lambda))$, the 
index of $\det(v(\lambda))$, equals $\sum_{j=1}^{2} \! k_{j}$. Define $\chi^{
c}_{o}(\lambda) \! := \! ((\id \! - \! C_{w})^{-1} \mathrm{I})(\lambda)$: 
then the boundary values $\chi^{c}_{\pm}(\lambda) \! := \! \chi^{c}_{o}
(\lambda)(\mathrm{I} \! \pm \! w^{\pm}(\lambda)) \! \in \! (\mathrm{I} \! 
+ \! \mathrm{ran} \, C_{\pm}) \! \cap \! GH^{k}_{I}(\widetilde{\Lambda}^{
\pm},\mathrm{M}_{2}(\mathbb{C})) \! \subset \! (\mathrm{I} \! + \! H^{k}
(\widetilde{\Lambda}^{\pm},\mathrm{M}_{2}(\mathbb{C}))) \! \cap \! GH^{k}_{I}
(\widetilde{\Lambda}^{\pm},\mathrm{M}_{2}(\mathbb{C}))$ give the fundamental 
solution of the {\rm RH} factorisation problem for $v(\lambda)$.
\end{dy}
\setcounter{z0}{1}
\setcounter{z1}{3}
\begin{dy}
If all the partial indices of $v(\lambda)$ are zero $(k_{i} \! = \! 0$, $i 
\! \in \! \{1,2\})$, then the Fredholm operator $\id \! - \! C_{w}$ is 
invertible on $\widetilde{H}^{k}_{I}(\widetilde{\Lambda},\mathrm{M}_{2}
(\mathbb{C}))$, namely, $\ker(\id \! - \! C_{w}) \! = \! \emptyset$ $(\dim 
\ker (\id \! - \! C_{w}) \! = \! 0)$.
\end{dy}
\setcounter{z0}{1}
\setcounter{z1}{1}
\begin{ay}
The {\rm RHP} of $v(\lambda) \! := \! (\mathrm{I} \! - \! w^{-}(\lambda))^{-
1}(\mathrm{I} \! + \! w^{+}(\lambda)) \! = \! (\chi^{c}_{-}(\lambda))^{-1} 
\chi^{c}_{+}(\lambda)$, $\lambda \! \in \! \widetilde{\Lambda}$, where $w^{
\pm}(\lambda) \! \in \! H^{k}(\widetilde{\Lambda}^{\pm},\mathrm{M}_{2}
(\mathbb{C}))$, has a fundamental solution $(\chi^{c}(\infty) \! = \! 
\mathrm{I}$, $\chi^{c}(\lambda) \! \not\equiv \! 0)$ only if $\tfrac{1}{2 
\pi} \! \int_{\widetilde{\Lambda}} \! d(\arg \det (v(\cdot))) \! = \! 0$. 
Conversely, if $\chi^{c}(\lambda) \! \in \! \widetilde{H}^{k}_{I}(\widetilde{
\Lambda},\mathrm{M}_{2}(\mathbb{C}))$, $k \! \in \! \mathbb{Z}_{\geqslant 
1}$, $\chi^{c}(\infty) \! = \! \mathrm{I}$ is a solution of the {\rm RHP} of 
$v(\lambda)$ on $\widetilde{\Lambda}$, and $\tfrac{1}{2 \pi} \! \int_{
\widetilde{\Lambda}} \! d(\arg \det(v(\cdot))) \! = \! 0$, then $\chi^{c}
(\lambda)$ is a fundamental solution; furthermore, $\det(v(\lambda)) \! = \! 
1 \! \Rightarrow \! \det (\chi^{c}(\lambda)) \! = \! 1$.
\end{ay}
\setcounter{z0}{1}
\setcounter{z1}{2}
\begin{by}
If the {\rm RHP} of $v(\lambda) \! := \! (\mathrm{I} \! - \! w^{-}(\lambda))^{
-1}(\mathrm{I} \! + \! w^{+}(\lambda)) \! = \! (\chi^{c}_{-}(\lambda))^{-1} 
\chi^{c}_{+}(\lambda)$, $\lambda \! \in \! \widetilde{\Lambda}$, where $w^{
\pm}(\lambda) \! \in \! H^{k}(\widetilde{\Lambda}^{\pm},\mathrm{M}_{2}
(\mathbb{C}))$, admits a fundamental solution $\chi^{c}(\lambda) \! \in \! 
\widetilde{H}^{j}_{I}(\widetilde{\Lambda},\mathrm{M}_{2}(\mathbb{C}))$ for 
some $j \! \in \! \mathbb{Z}_{\geqslant 1}$, then it is unique in $\mathcal{
L}^{2}_{I}(\widetilde{\Lambda},\mathrm{M}_{2}(\mathbb{C})) \! := \! 
\widetilde{H}^{0}_{I}(\widetilde{\Lambda},\mathrm{M}_{2}(\mathbb{C}))$.
\end{by}
\setcounter{z0}{1}
\setcounter{z1}{3}
\begin{by}
If the {\rm RHP} of $v(\lambda) \! := \! (\mathrm{I} \! - \! w^{-}(\lambda))^{
-1}(\mathrm{I} \! + \! w^{+}(\lambda)) \! = \! (\chi^{c}_{-}(\lambda))^{-1} 
\chi^{c}_{+}(\lambda)$, $\lambda \! \in \! \widetilde{\Lambda}$, where $w^{
\pm}(\lambda) \! \in \! H^{k}(\widetilde{\Lambda}^{\pm},\mathrm{M}_{2}
(\mathbb{C}))$, admits a fundamental solution $\chi^{c}(\lambda) \! \in \! 
\widetilde{H}^{j}_{I}(\widetilde{\Lambda},\mathrm{M}_{2}(\mathbb{C}))$ for 
some $j \! \in \! \mathbb{Z}_{\geqslant 0}$, then $\id \! - \! C_{w}$ is 
invertible on $\widetilde{H}^{j^{\prime}}_{I}(\widetilde{\Lambda},\mathrm{
M}_{2}(\mathbb{C}))$ $\forall \, j^{\prime} \! \in \! \{0,1,\ldots,k\}$, $k 
\! \in \! \mathbb{Z}_{\geqslant 1}$.
\end{by}
\setcounter{z0}{1}
\setcounter{z1}{4}
\begin{by}
Suppose that $w^{\pm}(\lambda) \! \in \! H^{k}(\widetilde{\Lambda}^{\pm},
\mathrm{M}_{2}(\mathbb{C}))$. If $\id \! - \! C_{w}$ is invertible on 
$\widetilde{H}^{j}_{I}(\widetilde{\Lambda},\mathrm{M}_{2}(\mathbb{C}))$ for 
any $j \! \leqslant \! k$, $k \! \in \! \mathbb{Z}_{\geqslant 1}$, then it 
is invertible $\forall \, j \! \leqslant \! k$.
\end{by}
Denote the Schwarz reflection of an $\mathrm{M}_{2}(\mathbb{C})$-valued 
function by $f^{\mathcal{S}}(\lambda) \! := \! (f(\overline{\lambda}))^{
\dagger}$, where ${}^{\dagger}$ denotes Hermitian conjugation, and, for a 
subset of $\mathbb{C}$, as the reflection about $\mathbb{R}$.
\setcounter{z0}{1}
\setcounter{z1}{4}
\begin{dy}
If $\widetilde{\Lambda}$ is a Schwarz reflection invariant contour about 
$\mathbb{R}$, $v(\lambda) \! \in \! SH^{k}_{I}(\widetilde{\Lambda}^{-},
\mathrm{M}_{2}(\mathbb{C})) \ast SH^{k}_{I}(\widetilde{\Lambda}^{+},\mathrm{
M}_{2}(\mathbb{C}))$, $v(\infty) \! = \! \mathrm{I}$, $v(\cdot)$ is positive 
definite on $\mathbb{R}$, $\Re(v(\lambda)) \! \! \upharpoonright_{\mathbb{R}}
>0$, and $v(\lambda) \! \! \upharpoonright_{\widetilde{\Lambda} \setminus 
\mathbb{R}}=\! \sigma^{-1}v^{\mathcal{S}}(\lambda) \! \! 
\upharpoonright_{\widetilde{\Lambda} \setminus \mathbb{R}} \cdot \sigma$, 
where $\sigma$ is a constant, invertible, finite-order matrix involution 
which changes the sign(s) of some (or all) of the elements of the matrix on 
which it (and its inverse) is multiplied, then all the partial indices of 
$v(\lambda)$ are zero, $k_{i} \! = \! 0$, $i \! \in \! \{1,2\}$. In this 
case, the {\rm RHP} for $v(\lambda)$ is solvable.
\end{dy}
The singular RHP, that is, the RH factorisation problem with isolated 
singularities (in this work, first-order poles), is now introduced. Let 
$\zeta \! \in \! \mathbb{C}$. For the remainder of this Appendix, the same 
symbol is used to denote an $\mathrm{M}_{2}(\mathbb{C})$-valued function 
analytic in a punctured neighbourhood of $\zeta$ and the germ (the set of 
equivalence classes of analytic continuations) at $\zeta$ it represents, 
with the algebra of all such germs denoted by $\mathscr{A}_{\zeta}$, and 
$S\mathscr{A}_{\zeta} \! := \! \{\mathstrut \varphi_{\zeta}(\lambda) \! \in 
\! \mathscr{A}_{\zeta}; \, \det(\varphi_{\zeta}(\lambda)) \! = \! 1\}$. Let 
$\mathscr{D} \! \subset \! \mathbb{C}$, with $\mathrm{card}(\mathscr{D}) \! 
< \! \infty$. Set $\mathscr{D}^{\pm} \! := \! \mathscr{D} \cap \boldsymbol{
\Omega}^{\mp}$. Define $H^{k}(\widetilde{\Lambda}^{\pm} \cup \mathscr{D},
\mathrm{M}_{2}(\mathbb{C})) \! := \! H^{k}(\widetilde{\Lambda}^{\pm},\mathrm{
M}_{2}(\mathbb{C})) \oplus (\oplus_{\zeta \in \mathscr{D}} \mathscr{A}_{
\zeta})$. An element in $\oplus_{\alpha \in \{\pm\}}H^{k}(\widetilde{
\Lambda}^{\alpha} \cup \mathscr{D},\mathrm{M}_{2}(\mathbb{C}))$ is 
represented either as $\varphi(\lambda) \! := \! (\varphi_{c}(\lambda),
\varphi_{\zeta}(\lambda))_{\zeta \in \mathscr{D}}$, or
\begin{equation*}
\varphi (\lambda) \! := \! 
\begin{cases}
\varphi_{c}(\lambda), &\text{$\lambda \! \in \! \widetilde{\Lambda}$,} \\
\varphi_{\zeta}(\lambda), &\text{$\lambda \! \approx \! \zeta, \quad \zeta 
\! \in \! \mathscr{D}$,}
\end{cases}
\end{equation*}
where $\varphi_{c}(\lambda) \! \in \! \oplus_{\alpha \in \{\pm\}}H^{k}
(\widetilde{\Lambda}^{\alpha},\mathrm{M}_{2}(\mathbb{C}))$, and $\varphi_{
\zeta}(\lambda) \! \in \! \mathscr{A}_{\zeta}$ (in the above, the subscript 
$c$ is used to connote ``continuous'', while the subscript $\zeta$ (for 
$\zeta \! \in \! \mathscr{D})$ is used to connote ``discrete''). The Cauchy 
integral operators, $C_{\pm}$, are defined on $\oplus_{\alpha \in \{\pm\}}
H^{k}(\widetilde{\Lambda}^{\alpha} \cup \mathscr{D},\mathrm{M}_{2}(\mathbb{
C}))$ in the following sense: construct the \emph{augmented} contour 
$\widetilde{\Lambda}_{\mathrm{aug}} \! := \! \widetilde{\Lambda} \cup (\cup_{
\zeta \in \mathscr{D}} \widetilde{\Lambda}_{\zeta})$, where $\widetilde{
\Lambda}_{\zeta}$ are sufficiently small, mutually disjoint, and disjoint 
with respect to $\widetilde{\Lambda}$, disks oriented counter-clockwise 
(respectively, clockwise) $\forall \, \, \zeta \! \in \! \mathscr{D}^{+}$ 
(respectively, $\forall \, \, \zeta \! \in \! \mathscr{D}^{-})$. Since, with 
the above-given conditions on $\widetilde{\Lambda}_{\zeta}$, $\zeta \! \in \! 
\mathscr{D}$, and, for each $\varphi(\lambda) \! \in \! \oplus_{\alpha \in 
\{\pm\}}H^{k}(\widetilde{\Lambda}^{\alpha} \cup \mathscr{D},\mathrm{M}_{
2}(\mathbb{C}))$, $\varphi(\lambda) \! \! \upharpoonright_{\lambda \in 
\widetilde{\Lambda}_{\mathrm{aug}}} \, \exists$, it represents an element 
in $\oplus_{\alpha \in \{\pm\}}H^{k}(\widetilde{\Lambda}^{\alpha}_{\mathrm{
aug}},\mathrm{M}_{2}(\mathbb{C}))$; hence, $(C_{\pm} \varphi)(\lambda)$ are 
defined, and $(C_{\pm} \varphi)(\lambda) \! \in \! H^{k}(\widetilde{\Lambda}^{
\pm}_{\mathrm{aug}},\mathrm{M}_{2}(\mathbb{C}))$. Hereafter, $(C_{\pm} 
\varphi)(\lambda)$ are to be understood as elements in $H^{k}(\widetilde{
\Lambda}^{\pm} \cup \mathscr{D},\mathrm{M}_{2}(\mathbb{C}))$. For $\zeta \! 
\in \! \mathscr{D}^{+}$, $(C_{+} \varphi)(\lambda)$ extends analytically 
into the disk bounded by $\widetilde{\Lambda}_{\zeta}$, and $(C_{-} \varphi)
(\lambda) \! := \! (C_{+} \varphi)(\lambda) \! - \! \varphi(\lambda)$ extends 
analytically into the punctured disk; therefore, they represent germs in 
$\mathscr{A}_{\zeta}$, denoted by $f^{\pm}_{\zeta}$, respectively. Similarly, 
for $\zeta \! \in \! \mathscr{D}^{-}$, $(C_{-} \varphi)(\lambda)$ extends 
analytically into the disk bounded by $\widetilde{\Lambda}_{\zeta}$, and 
$(C_{+} \varphi)(\lambda) \! := \! (C_{-} \varphi)(\lambda) \! + \! \varphi
(\lambda)$ extends analytically into the punctured disk; therefore, they 
represent germs in $\mathscr{A}_{\zeta}$, denoted by $f^{\mp}_{\zeta}$, 
respectively. Write $f_{c}^{\pm} \! := \! (C_{\pm} \varphi)(\lambda) \! \! 
\upharpoonright_{\lambda \in \widetilde{\Lambda}}$, and define $f^{\pm} \! 
:= \! (f_{c}^{\pm},f^{\pm}_{\zeta})_{\zeta \in \mathscr{D}} \! \in \! H^{k}
(\widetilde{\Lambda}^{\pm} \cup \mathscr{D},\mathrm{M}_{2}(\mathbb{C}))$. 
{}From the construction above, $C_{\pm} \colon \oplus_{\alpha \in \{\pm\}}
H^{k}(\widetilde{\Lambda}^{\alpha} \cup \mathscr{D},\mathrm{M}_{2}(\mathbb{
C})) \! \to \! H^{k}(\widetilde{\Lambda}^{\pm} \cup \mathscr{D},\mathrm{
M}_{2}(\mathbb{C}))$, and $(C_{\pm} \varphi)(\lambda) \! = \! f^{\pm}$. In 
this sense, $C_{\pm}$ are called the Cauchy integral operators with singular 
support $\widetilde{\Lambda} \cup \mathscr{D}$. The following notion of 
piecewise-holomorphic matrix-valued function has been used throughout this 
paper. For an $\mathrm{M}_{2}(\mathbb{C})$-valued function, $\Psi(\lambda)$, 
say, the ``symbol'' $\Psi(\lambda) \! := \! (\Psi_{c}(\lambda),\Psi_{\zeta}
(\lambda))_{\zeta \in \mathscr{D}}$ is said to be a piecewise-holomorphic 
matrix-valued function with respect to the contour $\widetilde{\Lambda} 
\cup \mathscr{D}$ if $\Psi_{c}(\lambda)$ is a piecewise-holomorphic 
matrix-valued function on $\boldsymbol{\Omega} \setminus \mathscr{D}$ and 
$\Psi_{\zeta}(\lambda) \! \in \! \mathscr{A}_{\zeta}$ is analytic at each 
$\zeta \! \in \! \mathscr{D}$. The boundary values $\Psi_{\pm}(\lambda)$, 
if they exist, of the (generalised) holomorphic matrix-valued function 
$\Psi(\lambda) \! := \! (\Psi_{c}(\lambda),\Psi_{\zeta}(\lambda))_{\zeta 
\in \mathscr{D}}$ are defined by
\begin{equation}
\Psi_{+}(\lambda) \! := \! 
\begin{cases}
(\Psi_{c}(\lambda))_{+}, &\text{$\lambda \! \in \! \widetilde{\Lambda}$,} \\
\Psi_{c}(\lambda), &\text{$\lambda \! \approx \! \zeta, \quad \zeta \! \in \! 
\mathscr{D}^{-}$,} \\
\Psi_{\zeta}(\lambda), &\text{$\lambda \! \approx \! \zeta, \quad \zeta \! 
\in \! \mathscr{D}^{+}$,}
\end{cases} \qquad \, \, \Psi_{-}(\lambda) \! := \! 
\begin{cases}
(\Psi_{c}(\lambda))_{-}, &\text{$\lambda \! \in \! \widetilde{\Lambda}$,} \\
\Psi_{c}(\lambda), &\text{$\lambda \! \approx \! \zeta, \quad \zeta \! \in 
\! \mathscr{D}^{+}$,} \\
\Psi_{\zeta}(\lambda), &\text{$\lambda \! \approx \! \zeta, \quad \zeta \! 
\in \! \mathscr{D}^{-}$,} \tag{C.1}
\end{cases}
\end{equation}
where $(\Psi_{c}(\lambda))_{\pm} \! := \! \lim_{\genfrac{}{}{0pt}{2}{\lambda^{
\prime} \to \lambda}{\lambda^{\prime} \in \boldsymbol{\Omega}^{\pm}}} \! \Psi_{
c}(\lambda^{\prime})$. Define $\mathcal{H}^{k}(\mathbb{C} \setminus \widetilde{
\Lambda} \cup \mathscr{D},\mathrm{M}_{2}(\mathbb{C})) \! := \! \{\mathstrut 
\Psi(\lambda); \, \Psi_{\pm}(\lambda) \! - \! \Psi_{\pm}(\infty) \! \in \! 
\mathrm{ran}C_{\pm}\}$, and $\mathcal{SH}^{k}(\mathbb{C} \setminus \widetilde{
\Lambda} \cup \mathscr{D},\mathrm{M}_{2}(\mathbb{C})) \! := \! \{\mathstrut 
\Psi(\lambda) \! \in \! \mathcal{H}^{k}(\mathbb{C} \setminus \widetilde{
\Lambda} \cup \mathscr{D},\mathrm{M}_{2}(\mathbb{C})); \, \det(\Psi 
(\lambda)) \! = \! 1\}$.
\setcounter{z0}{1}
\setcounter{z1}{5}
\begin{dy}
Every $v(\lambda) \! \in \! SH^{k}_{I}(\widetilde{\Lambda}^{-} \cup \mathscr{
D},\mathrm{M}_{2}(\mathbb{C})) \ast SH^{k}_{I}(\widetilde{\Lambda}^{+} \cup 
\mathscr{D},\mathrm{M}_{2}(\mathbb{C}))$ admits an {\rm RH} factorisation 
$v(\lambda) \! := \! (\chi_{-}(\lambda))^{-1} \blacklozenge (\lambda) \chi_{
+}(\lambda)$, where $\chi(\lambda) \! \in \! \mathcal{SH}^{k}(\mathbb{C} 
\setminus \widetilde{\Lambda} \cup \mathscr{D},\mathrm{M}_{2}(\mathbb{C}))$, 
$\blacklozenge (\lambda)$ is defined in Theorem~{\rm C.1.1}, and $\lambda_{
\pm} \! \in \! \mathscr{D}^{\pm} \cup (\boldsymbol{\Omega}^{\pm} \setminus 
\mathscr{D}^{\mp})$.
\end{dy}
\setcounter{z0}{1}
\setcounter{z1}{6}
\begin{dy}
If $\widetilde{\Lambda} \! \cup \! \mathscr{D}$ is Schwarz reflection 
invariant with respect to $\mathbb{R}$, $v(\lambda) \! \in \! SH^{k}_{I}
(\widetilde{\Lambda}^{-} \! \cup \! \mathscr{D},\mathrm{M}_{2}(\mathbb{C})) 
\linebreak[4]
\ast SH^{k}_{I}(\widetilde{\Lambda}^{+} \cup \mathscr{D},\mathrm{M}_{2}
(\mathbb{C}))$, $v(\infty) \! = \! \mathrm{I}$, $\Re(v(\lambda)) \! \! 
\upharpoonright_{\lambda \in \mathbb{R}}> \! 0$, and $v(\lambda) \! \! 
\upharpoonright_{\lambda \in (\widetilde{\Lambda} \cup \mathscr{D}) \setminus 
\mathbb{R}} = \! \sigma^{-1}v^{\mathcal{S}}(\lambda) \! \! \upharpoonright_{
\lambda \in (\widetilde{\Lambda} \cup \mathscr{D}) \setminus \mathbb{R}} \! 
\sigma$, where $\sigma$ is a constant, invertible, finite-order matrix 
involution which changes the sign(s) of some (or all) of the elements of the 
matrix on which it (and its inverse) is multiplied, then all the partial 
indices of $v(\lambda)$ are zero, $k_{i} \! = \! 0$, $i \! \in \! \{1,2\}$. 
In this case, the {\rm RHP} for $v(\lambda)$ is solvable.
\end{dy}
Note that, for $\mathscr{D} \! \equiv \! \emptyset$, Theorem~C.1.6 reduces to 
Theorem~C.1.4. The asymptotic analysis of the latter part of the above-given 
paradigm, related to the singular RHP (when $\mathscr{D} \! \not\equiv \! 
\emptyset$ and $\mathrm{card}(\mathscr{D}) \! < \! \infty)$, is the subject 
of the present asymptotic study. Using the results of this subsection, the 
very important Lemma~2.4 of \cite{a50}, and the Deift-Zhou non-linear 
steepest descent method \cite{a55}, the (rigorous) asymptotic analysis, as 
$\vert t \vert \! \to \! \infty$ and $\vert x \vert \! \to \! \infty$ such 
that $z_{o} \! := \! x/t \! \sim \! \mathcal{O}(1)$ and $\in \! \mathbb{R} 
\setminus \{-2,0,2\}$, of the RHP for $m(\zeta)$ formulated in Lemma~2.1.2, 
for $\sigma_{d} \! \equiv \! \emptyset$, was completed in \cite{a38}.

\vspace{3.0cm}

\textbf{\Large Acknowledgements}

The author is very grateful to X.~Zhou for the invitation to Duke University 
and for the opportunity to complete this work. The author is also grateful 
to the referees for helpful suggestions.


\clearpage
\begin{thebibliography}{111}
\bibitem{a1} Agrawal, G. P.: \emph{Nonlinear Fiber Optics} (2nd edn), Academic 
Press, San Diego, 1995.
\bibitem{a2} Kodama, Y.: The Whitham Equations for Optical Communications: 
Mathematical Theory of NRZ, \emph{SIAM J. Appl. Math.} 59 (1999), 2162--2192.
\bibitem{a3} Weiner, A. M.: Dark optical solitons, in: Taylor, J. R. (ed), 
\emph{Optical Solitons - Theory and Experiment}, Cambridge Studies in Modern 
Optics, Vol 10, Cambridge University Press, Cambridge, 1992, pp. 378--408.
\bibitem{a4} Lundquist, P. B., Andersen, D. R. and Swartzlander, G. A., Jr.: 
Asymptotic behavior of the self-defocusing nonlinear Schr\"{o}dinger equation 
for piecewise constant initial conditions, \emph{J. Opt. Soc. Am. B} 12 
(1995), 698--703.
\bibitem{a5} Lundina, D. Sh. and Marchenko, V. A.: Compactness of the Set 
of Multisoliton Solutions of the Nonlinear Schr\"{o}dinger Equation, 
\emph{Russian Acad. Sci. Sb. Math.} 75 (1993), 429--443.
\bibitem{a6} Boutet de Monvel, A. and Marchenko, V.: The Cauchy problem for 
nonlinear Schr\"{o}dinger equation with bounded initial data, \emph{Mat. Fiz. 
Anal. Geom.} 4 (2000), 3--45.
\bibitem{a7} Novikov, S. P., Manakov, S. V., Pitaevskii, L. P. and Zakharov, 
V. E.: \emph{Theory of Solitons: The Inverse Scattering Method}, Plenum, New 
York, 1984.
\bibitem{a8} Ablowitz, M. J. and Clarkson P. A.: \emph{Solitons, Nonlinear 
Evolution Equations and Inverse Scattering}, LMS 149, Cambridge University 
Press, Cambridge, 1991.
\bibitem{a9} Cherednik, I.: \emph{Basic Methods of Soliton Theory}, Advanced 
Series in Mathematical Physics, Vol 25, World Scientific, Singapore, 1996.
\bibitem{a10} Schuur, P. C.: \emph{Asymptotic Analysis of Soliton Problems}, 
LNM 1232, Springer-Verlag, Berlin, 1986.
\bibitem{a11} Zakharov, V. E. and Manakov, S. V.: On the Complete 
Integrability of a Nonlinear Schr\"{o}dinger Equation, \emph{Theor. Math. 
Phys.} 19 (1974), 551--559.
\bibitem{a12} Faddeev, L. D. and Takhtajan, L. A.: \emph{Hamiltonian Methods 
in the Theory of Solitons}, Springer-Verlag, Berlin, 1987.
\bibitem{a13} Zakharov, V. E. and Shabat, A. B.: Interaction between solitons 
in a stable medium, \emph{Sov. Phys. JETP} 37 (1973), 823--828.
\bibitem{a14} Asano, N. and Kato, Y.: Non-self-adjoint Zakharov-Shabat 
operator with a potential of the finite asymptotic values. I. Direct spectral 
and scattering problems, \emph{J. Math. Phys.} 22 (1981), 2780--2793. Asano, 
N. and Kato, Y.: Non-self-adjoint Zakharov-Shabat operator with a potential 
of the finite asymptotic values. II. Inverse problem, \emph{J. Math. Phys.} 
25 (1984), 570--588.
\bibitem{a15} Kawata, T. and Inoue, H.: Inverse Scattering Method for the 
Nonlinear Evolution Equations under Nonvanishing Conditions, \emph{J. Phys. 
Soc. Japan} 44 (1978), 1722--1729.
\bibitem{a16} Frolov, I. S.: Inverse Scattering Problem for a Dirac System 
on the Whole Axis, \emph{Soviet Math. Dokl.} 13 (1972), 1468--1472.
\bibitem{a17} Anders, I. A. and Kotlyarov, V. P.: Characterization of the 
Scattering Data of the Schr\"{o}dinger and Dirac Operators, \emph{Theor. 
Math. Phys.} 88 (1991), 725--734.
\bibitem{a18} Boiti, M. and Pempinelli, F.: The Spectral Transform for the 
NLS Equation with Left-Right Asymmetric Boundary Conditions, \emph{Il Nuovo 
Cimento} 69B (1982), 213--227.
\bibitem{a19} Kawata, T. and Inoue, H.: Exact Solutions of the Derivative 
Nonlinear Schr\"{o}dinger Equation under the Nonvanishing Conditions, 
\emph{J. Phys. Soc. Japan} 44 (1978), 1968--1976. Kawata, T., Sakai, J. and 
Kobayashi, N.: Inverse Method for the Mixed Nonlinear Schr\"{o}dinger 
Equation and Soliton Solutions, \emph{J. Phys. Soc. Japan} 48 (1980), 
1371--1379.
\bibitem{a20} Marchenko, V. A.: The Cauchy Problem for the KdV Equation with 
Non-Decreasing Initial Data, in:  Zakharov, V. E. (ed), \emph{What is 
Integrability?}, Springer Series in Nonlinear Dynamics, Springer-Verlag, 
Berlin, 1991, pp. 273--318.
\bibitem{a21} Boutet de Monvel, A., Khruslov, E. Ya. and Kotlyarov, V. P.: 
The Cauchy problem for the sinh-Gordon equation and regular solitons, 
\emph{Inverse Problems} 14 (1998), 1403--1427.
\bibitem{a22} Boutet de Monvel, A., Egorova, I. and Khruslov, E.: Soliton 
asymptotics of the Cauchy problem solution for the Toda lattice, 
\emph{Inverse Problems} 13 (1997), 223--237. Boutet de Monvel, A. and 
Egorova, I.: The Toda lattice with step-like initial data. Soliton 
asymptotics, \emph{Inverse Problems} 16 (2000), 955--977.
\bibitem{a23} Kotlyarov, V. P. and Khruslov, E. Ya.: Solitons of the 
Nonlinear Schr\"{o}dinger Equation Generated by the Continuum, \emph{Theor. 
Math. Phys.} 68 (1986), 751--761. Kotlyarov, V. P. and Khruslov, E. Ya.: 
Asymptotic solitons of the modified Korteweg-de Vries equation, \emph{Inverse 
Problems} 5 (1989), 1075--1088. Kotlyarov, V. P.: Asymptotic Solitons of 
the Sine-Gordon Equation, \emph{Theor. Math. Phys.} 80 (1989), 679--689.
\bibitem{a24} Khruslov, E. Ya.: Asymptotic Solution of the Cauchy Problem for 
the Korteweg-de Vries Equation with Step-Type Initial Data, \emph{Math. USSR 
Sbornik} 99 (1976), 261--281 (in Russian).
\bibitem{a25} Kirsch, W. and Kotlyarov, V.: Soliton Asymptotics of Solutions 
of the Sine-Gordon Equation, \emph{Math. Phys. Anal. Geom.} 2 (1999), 25--51.
\bibitem{a26} Khruslov, E. Ya. and Stephan, H.: Splitting of some 
non-localized solutions of the Korteweg-de Vries equation into solitons, 
\emph{Mat. Fiz. Anal. Geom.} 5 (1998), 49--67.
\bibitem{a27} Borisov, A. B. and Kiseliev, V. V.: Inverse problem for an 
elliptic sine-Gordon equation with an asymptotic behaviour of the 
cnoidal-wave type, \emph{Inverse Problems} 5 (1989), 959--982.
\bibitem{a28} Vekslerchik, V. E. and Konotop, V. V.: Discrete nonlinear 
Schr\"{o}dinger equation under non-vanishing boundary conditions, \emph{
Inverse Problems} 8 (1992), 889--909.
\bibitem{a29} Vekslerchik, V. E.: Inverse scattering transform for the $O(3,
1)$ nonlinear $\sigma$-model, \emph{Inverse Problems} 12 (1996), 517--534.
\bibitem{a30} Cohen, A. and Kappeler, T.: Scattering and Inverse Scattering 
for Steplike Potentials in the Schr\"{o}dinger Equation, \emph{Indiana Univ. 
Math. J.} 34 (1985), 127--180.
\bibitem{a31} Grebert, B.: Inverse scattering for the Dirac operator on the 
real line, \emph{Inverse Problems} 8 (1992), 787--807.
\bibitem{a32} Bikbaev, R. F. and Sharipov, R. A.: Asymptotics as $t \! \to 
\! \infty$ of the Solution to the Cauchy Problem for the Korteweg-de Vries 
Equation in the Class of Potentials with Finite-Gap Behavior as $x \! \to \! 
\pm \infty$, \emph{Theor. Math. Phys.} 78 (1989), 244--252.
\bibitem{a33} Zakharov, V. E. and Shabat, A. B.: Integration of the 
non-linear equations of mathematical physics by the method of the inverse 
scattering transform. II, \emph{Funct. Anal. Appl.} 13 (1980), 166--173.
\bibitem{a34} Deift, P. A., Kamvissis, S., Kriecherbauer, T. and Zhou, X.: 
The Toda Rarefaction Problem, \emph{Comm. Pure Appl. Math.} 49 (1996), 
35--83.
\bibitem{a35} Gesztesy, F. and Svirsky, R.: \emph{(m)KdV Solitons on the 
Background of Quasi-Periodic Finite-Gap Solutions}, Mem. Amer. Math. Soc., 
Vol 118, AMS, Providence, 1995.
\bibitem{a36} Renger, W.: Toda Soliton Limits on General Backgrounds, 
\emph{J. Differential Equations} 151 (1999), 191--230.
\bibitem{a37} Kitaev, A. V. and Vartanian, A. H.: Asymptotics of Solutions 
to the Modified Nonlinear Schr\"{o}dinger Equation: Solitons on a 
Nonvanishing Continuous Background, \emph{SIAM J. Math. Anal.} 30 (1999), 
787--832.
\bibitem{a38} Vartanian, A. H.: Long-Time Asymptotics of Solutions to the 
Cauchy Problem for the Defocusing Non-Linear Schr\"{o}dinger Equation 
with Finite-Density Initial Data. I. Solitonless Sector, (2001) 
\texttt{arXiv:nlin.SI/0110024}.
\bibitem{a39} Its, A. R.: Asymptotics of Solutions of the Nonlinear 
Schr\"{o}dinger Equation and Isomonodromic Deformations of Systems of 
Linear Differential Equations, \emph{Soviet Math. Dokl.} 24 (1981), 452--456.
\bibitem{a40} Clancey, K. and Gohberg, I.: \emph{Factorization of Matrix 
Functions and Singular Integral Operators}, Operator Theory: Advances and 
Applications, Vol 3, Birkh\"{a}user, Basel, 1981.
\bibitem{a41} Beals, R. and Coifman, R. R.: Scattering and Inverse Scattering 
for First Order Systems, \emph{Comm. Pure Appl. Math.} 37 (1984), 39--90.
\bibitem{a42} Deift, P.: \emph{Orthogonal Polynomials and Random Matrices: 
A Riemann-Hilbert Approach}, Courant Lecture Notes in Mathematics, Vol 3, 
CIMS, New York, 1999.
\bibitem{a43} Fokas, A. S.: On the integrability of linear and nonlinear 
partial differential equations, \emph{J. Math. Phys.} 41 (2000), 4188--4237.
\bibitem{a44} Zhou, X.: The Riemann-Hilbert Problem and Inverse Scattering, 
\emph{SIAM J. Math. Anal.} 20 (1989), 966--986.
\bibitem{a45} Zhou, X.: Direct and Inverse Scattering Transforms with 
Arbitrary Spectral Singularities, \emph{Comm. Pure Appl. Math.} 42 (1989), 
895--938.
\bibitem{a46} Zhou, X.: Inverse Scattering Transform for Systems with 
Rational Spectral Dependence, \emph{J. Differential Equations} 115 (1995), 
277--303.
\bibitem{a47} Zhou, X.: Strong Regularizing Effect of Integrable Systems, 
\emph{Comm. Partial Differential Equations} 22 (1997), 503--526.
\bibitem{a48} Its, A. R. and Ustinov, A. F.: The time asymptotics of the 
solution of the Cauchy problem for the nonlinear Schr\"{o}dinger equation 
with finite density boundary conditions, \emph{Dokl. Akad. Nauk SSSR} 291 
(1986), 91--95 (in Russian).
\bibitem{a49} Its, A. R. and Ustinov, A. F.: Formulation of Scattering Theory 
for the Nonlinear Schr\"{o}dinger Equation with Boundary Conditions of the 
Finite Density Type in a Soliton-Free Sector, \emph{J. Sov. Math.} 54 (1991), 
900--905.
\bibitem{a50} Zhou, X.: $L^{2}$-Sobolev Space Bijectivity of the Scattering 
and Inverse Scattering Transforms, \emph{Comm. Pure Appl. Math.} 51 (1998), 
697--731.
\bibitem{a51} Gradshteyn, I. S. and Ryzhik, I. M.: \emph{Tables of Integrals, 
Series, and Products} (5th edn), Jeffrey, A. (ed), Academic Press, San Diego, 
1994.
\bibitem{a52} Zakharov, V. E. and Shabat, A. B.: Exact theory of 
two-dimensional self-focusing and one-dimensional self-modulation of waves 
in nonlinear media, \emph{Sov. Phys. JETP} 34 (1972), 62--69.
\bibitem{a53} Deift, P. and Zhou, X.: Direct and Inverse Scattering on the 
Line with Arbitrary Singularities, \emph{Comm. Pure Appl. Math.} 44 (1991), 
485--533.
\bibitem{a54} Deift, P. and Zhou, X.: Long-time Asymptotics for Integrable 
Systems. Higher Order Theory, \emph{Comm. Math. Phys.} 165 (1994), 175--191.
\bibitem{a55} Deift, P. and Zhou, X.: A Steepest descent method for 
oscillatory Riemann-Hilbert problems. Asymptotics for the MKdV equation, 
\emph{Ann. of Math.} 137 (1993), 295--368.
\bibitem{a56} Vartanian, A. H.: Higher Order Asymptotics of the Modified 
Non-Linear Schr\"{o}dinger Equation, \emph{Comm. Partial Differential 
Equations} 25 (2000), 1043--1098.
\end{thebibliography}
\end{document}